\documentclass[review]{elsarticle}

\usepackage[margin=3cm]{geometry}
\usepackage{epstopdf}
\usepackage[tight,normalsize,sf,SF]{subfigure}
\usepackage{algorithm}
\usepackage[noend]{algorithmic}
\usepackage{url}
\usepackage{framed}
\usepackage[table]{xcolor}
\usepackage{amsthm}
\usepackage{amsfonts,amssymb}
\usepackage{mathtools}
\usepackage{amssymb}
\usepackage{booktabs}
\usepackage{color}

\graphicspath{{./eps/capExp/}{./pdf/capRInClose/}}

\newtheorem{mydef}{Definition}[section]

\newtheorem{property}{Property}[section]

\journal{Information Sciences}

\begin{document}

\begin{frontmatter}

\title{Enumerating all maximal biclusters in numerical datasets}

\author[myaddress1]{Rosana~Veroneze\corref{mycorrespondingauthor}}
\cortext[mycorrespondingauthor]{Corresponding author}
\ead{veroneze@dca.fee.unicamp.br}

\author[myaddress2]{Arindam~Banerjee}

\author[myaddress1]{Fernando~J.~Von~Zuben}

\address[myaddress1]{University of Campinas (DCA/FEEC), 400 Albert Einstein Street, Campinas, SP, Brazil}
\address[myaddress2]{University of Minnesota, 200 Union Street, Minneapolis, MN, USA}

\begin{abstract}
Biclustering has proved to be a powerful data analysis technique due to its wide success in various application domains. However, the existing literature presents efficient solutions only for enumerating maximal biclusters with constant values, or heuristic-based approaches which can not find all biclusters or even support the maximality of the obtained biclusters. Here, we present a general family of biclustering algorithms for enumerating all maximal biclusters with ($i$) constant values on rows, ($ii$) constant values on columns, or ($iii$) coherent values. Versions for perfect and for perturbed biclusters are provided. Our algorithms have four key properties (just the algorithm for perturbed biclusters with coherent values fails to exhibit the first property): they are (1) efficient (take polynomial time per pattern), (2) complete (find all maximal biclusters), (3) correct (all biclusters attend the user-defined measure of similarity), and (4) non-redundant (all the obtained biclusters are maximal and the same bicluster is not enumerated twice). They are based on a generalization of an efficient formal concept analysis algorithm called In-Close2. Experimental results point to the necessity of having efficient enumerative biclustering algorithms and provide a valuable insight into the scalability of our family of algorithms and its sensitivity to user-defined parameters.
\end{abstract}


\end{frontmatter}


\section{Introduction}
\label{sec:intro}

Biclustering is a local approach for clustering that operates simultaneously over the set of objects and attributes of a data matrix. It looks for submatrices constituted of subsets of objects that have a highly consistent pattern across a subset of attributes. Biclustering methods are thus able to consider coherence measures which are more general than distance functions, such as Euclidean and Manhattan distances, and hence are going to find biclusters supporting more general affinities than conventional numerical proximity of elements \cite{WangEtAl2002}.

In the literature, biclustering has great value in finding interesting patterns in microarray expression data \cite{ZhaoZaki2005}. Indeed, the application of biclustering is fully disseminated and not limited to biological data. For instance, we can mention: dimensionality reduction \cite{AgrawalEtAl1998}, text mining \cite{BanerjeeEtAl2007, Dhillon2001}, collaborative filtering \cite{AlqadahEtAl2014, deCastroEtAl2007a, SymeonidisEtAL2007, SymeonidisEtAL2008}, and treatment of missing data \cite{BanerjeeEtAl2007, ColantonioEtAl2010, deFrancaEtAl2013, Veroneze2011}. Moreover, the importance of biclustering continues to increase, as researchers are (\textit{i}) finding new applications in scientific and commercial domains, including bioinformatics, social network analysis, and text mining; and (\textit{ii}) unveiling the connection between biclustering and several other important problems, including subspace clustering \cite{KriegelEtAl2009}, frequent pattern mining (FPM) \cite{Goethals2003}, and formal concept analysis (FCA) \cite{Ganter1997}.

Biclustering may be interpreted as a hard combinatorial optimization problem. The more flexible the bicluster structure, the more complex the problem, and we are considering the most flexible structure in this work: arbitrarily positioned overlapping biclusters \cite{MadeiraOliveira2004}. Thereby, an object/attribute can belong to none, one, or more than one bicluster. In this scenario, finding all maximal biclusters in a data matrix is an NP-hard problem \cite{MadeiraOliveira2009}. Due to this, most of the proposed algorithms are heuristic-based \cite{MadeiraOliveira2004}, and many of them consider an inflexible bicluster structure and mine a number of biclusters defined a priori. The heuristics-based algorithms potentially produce sub-optimal solutions, missing important biclusters and not guaranteeing the maximality of the identified ones. Some examples of well-known heuristic-based biclustering algorithms are: CC \cite{ChengChurch2000}, FLOC \cite{Yang2003}, ROCC \cite{DeodharEtAL2009}, ISA \cite{IhmelsEtAl2002}, Plaid \cite{Lazzeroni2002}, and OPSM \cite{BenDorEtal2003}. For surveys, refer to Madeira and Oliveira \cite{MadeiraOliveira2004} and Busygin et al. \cite{BusyginEtAL2008}.

In FCA and related areas, such as FPM and graph theory, we have plenty of algorithms for enumerating all maximal biclusters with constant values (CTV) in a binary dataset. These maximal CTV biclusters are called formal concepts in FCA, closed frequent itemsets (or patterns) in FPM \footnote{Being more specific, a closed frequent itemset corresponds to the column-set of a bicluster.} , and maximal bicliques in graph theory (for more details about the connection of these areas, see Section~\ref{fca_related}). Some examples of algorithms are: Makino and Uno \cite{MakinoEtAl2004}, Eppstein \textit{et al.} \cite{EppsteinEtAL2010}, Close-by-One (CbO) \cite{Kuznetsov1999}, In-Close \cite{Andrews2009}, In-Close2 \cite{Andrews2011}, FCbO \cite{Krajca2010}, CHARM \cite{ZakiEtAL2002}, and LCM \cite{UnoEtAL2004}. Their enumeration process is characterized by being:

\begin{enumerate}
	\item Efficient: it takes polynomial time per pattern, i.e., it takes polynomial time to enumerate the first bicluster and takes polynomial time between enumerating two consecutive biclusters. It is the best one can computationally do in such scenario. If done properly, such algorithm will have time complexity linear in the number of biclusters and polynomial in the input size. Moreover, if the number of maximal biclusters is polynomial in the input size, the overall algorithm will be a polynomial time algorithm.
	\item Complete: it finds all maximal biclusters. A complete enumeration guarantees to include the results produced by any other biclustering solution (given the same restrictions of similarity and size). So, such biclustering solution is at least of equal quality, but probably of better quality, when compared with the solution provided by any other contender. 
	\item Correct: all found biclusters attend the user-defined measure of similarity. For instance, in the case of the aforementioned algorithms, all biclusters are submatrices of ones. Complete and correct enumerators are crucial for some applications such as the identification of biological indicators \cite{LiuEtAl2014} and classification based on associations \cite{LiuEtAl1998}.
	\item Non-redundant: all biclusters are maximal and it does not enumerate the same maximal bicluster twice. Property number 4 is very important because the number of biclusters produced from a dataset can be very large. So, it is useful to identify the smallest representative set of biclusters from which all other biclusters can be derived \cite{TanEtAl2005}. The set of all maximal biclusters is necessary and sufficient to capture all the information about the biclusters, and has a much smaller cardinality than the set of all attainable biclusters \cite{Zaki2000}. It is important to note that the algorithm must have a smart solution to avoid redundancy, otherwise it will not be efficient. For instance, a procedure to be avoid is to check if a new bicluster is not redundant by comparing with all previously mined biclusters.
\end{enumerate}

Once the researchers found the link between these areas and biclustering, many algorithms have been proposed to deal with numerical (not only binary, but also integer or real-valued) datasets and other types of biclusters. In fact, nowadays the state-of-the-art biclustering algorithms are based on FPM \cite{HenriquesMadeira2014}. Many proposals, such as \cite{HenriquesMadeira2014, MadeiraOliveira2009, MartinezEtAl2007, OkadaEtAl2007b, OkadaEtAl2007a, SerinVingron2011}, binarize the data and apply the aforementioned algorithms. However, binarizing the dataset leads to loss of information, and guides to the necessity of tedious Boolean property encoding phases \cite {Besson2007}. Therefore, there are also proposals to deal directly with numerical datasets, such as \cite{Besson2007, kaytoueEtAl2014, PandeyEtAl2009, PeiEtAL2003, ZhaoZaki2005}. Without binarizing the numerical dataset, we are going to show in Section~\ref{sec:relWork} that there is no proposal in the literature able to enumerate biclusters with ($i$) constant value on columns (CVC), ($ii$) constant values on rows (CVR), or ($iii$) coherent values (CHV) (see definitions in Section~\ref{sec:bic}), so that the aforementioned four properties are preserved in this extended scenario (not only binary values in the dataset). Although some authors claim that their proposals do preserve these four properties, a more careful analysis to be presented in Section~\ref{sec:relWork} shows that they all fail to exhibit one or more of these four properties. So, the aim of this paper is to cover some of these gaps. In fact, we are proposing algorithms capable of preserving these four properties when enumerating perfect CVC biclusters, perturbed CVC biclusters, and perfect CHV biclusters. The problem of enumerating CVR biclusters is equivalent to enumerating CVC biclusters. We are also proposing an algorithm with the last three of these properties to enumerate perturbed CHV biclusters. Note that CVC, CVR and CHV biclusters are a generalization of CTV biclusters \cite{MadeiraOliveira2004} (for more details see Section~\ref{sec:bicTypes}). We call our family of algorithms \textit{RIn-Close} because they are generalizations of the FCA algorithm In-Close2 \cite{Andrews2011}. Tables~\ref{tab:comparison_cvc} and \ref{tab:comparison_chv} (see Section~\ref{sec:relWork}) show a technical comparison between our proposals and the competitors, attesting that we are proposing a number of improvements when enumerating biclusters from numerical datasets.

The remainder of the paper is organized as follows. Section \ref{sec:bic} introduces definitions and mathematical formulation for biclustering. Section \ref{sec:fca} reviews FCA and the algorithm In-Close2. Section \ref{sec:relWork} is devoted to related works.  Section \ref{sec:rinclose} presents the main contributions of this paper, more specifically the RIn-Close family of efficient enumerative algorithms for maximal CVC, CVR, or CHV biclusters. Experimental results are discussed in Section \ref{sec:exp}, and we conclude in Section \ref{sec:conclusion}.

\section{Biclustering}
\label{sec:bic}

The term \textit{biclustering} was introduced by Mirkin \cite{Mirkin1996} to describe the simultaneous clustering of the sets of rows and columns of a data matrix. More recently, the term was used in the analysis of gene expression data \cite{ChengChurch2000}. Cheng and Church \cite{ChengChurch2000} were responsible for the popularization of biclustering techniques with their algorithm called CC. However, Hartigan \cite{Hartigan1972} was the first one to propose an algorithm for biclustering, using the term \textit{direct clustering}. Other terms that are found in literature are: co-clustering, two-way clustering and bidimensional clustering, among others \cite{MadeiraOliveira2004}.

\subsection{Definitions and Taxonomy of Biclusters}

Let $\mathbf{A}_{n \times m}$ be a data matrix with the row index set $X = \left \{ 1, 2,..., n \right \}$ and the column index set $Y = \left \{ 1, 2, ...,m \right \}$. Each element $a_{ij} \in \mathbf{A}$ represents the relationship between row $i$ and column $j$. We use $(X,Y)$ to denote the entire matrix $\mathbf{A}$. Considering that $I \subseteq X$ and $J \subseteq Y$, $\mathbf{A}_{IJ} = (I, J)$ denotes the submatrix of $\mathbf{A}$ with the row index subset $I$ and column index subset $J$.

\begin{mydef}
A bicluster is a submatrix $(I,J)$ of the data matrix $\mathbf{A}_{n \times m}$ such that the rows in the index subset $I = \left \{ i_1,..., i_k \right \}$ ($I \subseteq X$ and $k \leq n$) exhibits similar behavior across the columns in the index subset $J = \left \{ j_1,..., j_s \right \}$ ($J \subseteq Y$ and $s \leq m$), and vice-versa.
\label{def:bic}
\end{mydef}

Thus, a bicluster $(I,J)$ is a $k \times s$ submatrix of the matrix $\mathbf{A}$, with not necessarily contiguous rows and columns, such that it meets a certain homogeneity criterion. A biclustering algorithm looks for a set of biclusters $\mathfrak{B} = (I_l, J_l)$, $l = 1, ..., q$, such that each bicluster $(I_l, J_l)$ satisfies some specific characteristics of homogeneity \cite{MadeiraOliveira2004}. Considering these characteristics, there are four major types of biclusters \cite{MadeiraOliveira2004}: ($i$) biclusters with constant values (CTV), ($ii$) biclusters with constant values on columns (CVC) or rows (CVR), ($iii$) biclusters with coherent values (CHV), and ($iv$) biclusters with coherent evolutions (CHE). The total number of biclusters, $q$, will depend on the features of the selected biclustering algorithm, on the constraints imposed, and on the behavior of the dataset being analyzed. 

In this paper, we will provide the definition of CTV, CVC, CVR, and CHV biclusters. Please, refer to Madeira and Oliveira \cite{MadeiraOliveira2004} for a complete survey with definitions and examples of all bicluster types.

\subsection{Types of Biclusters}
\label{sec:bicTypes}

Although perfect biclusters can be found in some data matrices, in real data, they are usually masked by noise. Therefore, we will define the perfect and the perturbed case for all types of biclusters. The perturbed case is always a generalization of the perfect case. An user-defined parameter $\epsilon \geq 0$ determines the maximum residue (perturbation) allowed in a bicluster.

\begin{mydef}[CTV biclusters]
A perfect CTV bicluster is a submatrix $(I, J)$ of a data matrix $\mathbf{A}_{n \times m}$ such that $a_{ij} = a_{kl}$, $\forall i, k \in I$ and $\forall j, l \in J$. A perturbed CTV bicluster is a submatrix $(I, J)$ of a data matrix $\mathbf{A}_{n \times m}$ such that $|a_{ij} - a_{kl}| \leq \epsilon$, $\forall i, k \in I$ and $\forall j, l \in J$, i.e.,

\begin{equation}
  \max_{i \in I, j \in J} (a_{ij}) - \min_{i \in I, j \in J} (a_{ij}) \leq \epsilon.
	\label{eq:ctvbic}
\end{equation}

\label{def:ctvbic}
\end{mydef}

\begin{mydef}[CVC biclusters]
A perfect CVC bicluster is a submatrix $(I, J)$ of a data matrix $\mathbf{A}_{n \times m}$ such that $a_{ij} = a_{kj}$, $\forall i, k \in I$ and $\forall j \in J$. A perturbed CVC bicluster is a submatrix $(I, J)$ such that $|a_{ij} - a_{kj}| \leq \epsilon$, $\forall i, k \in I$ and $\forall j \in J$, i.e.,

\begin{equation}
  \max_{i \in I} (a_{ij}) - \min_{i \in I} (a_{ij}) \leq \epsilon, \forall j \in J.
	\label{eq:cvcbic}
\end{equation}

\label{def:cvcbic}
\end{mydef}

The definition of a \textit{CVR bicluster} is the equivalent transpose of the definition of a CVC bicluster.

There are two perspectives for CHV biclusters: (\textit{i}) additive model, and (\textit{ii}) multiplicative model. Biclusters based on the additive model are called \textit{shifting biclusters}. Biclusters based on the multiplicative model are called \textit{scaling biclusters}. Any row (column) of a perfect shifting bicluster can be obtained by adding a constant value to any other row (column) of the bicluster. Similarly, any row (column) of a perfect scaling bicluster can be obtained by multiplying a constant value to any other row (column) of the bicluster. The problems of mining shifting and scaling biclusters are equivalent. Using an algorithm to mine shifting (scaling) biclusters, we can mine scaling (shifting) biclusters by previously taking the logarithm (exponential) of all entries of the data matrix (see a proof in Zhao and Zaki \cite{ZhaoZaki2005}). Therefore, we are going to focus only in the additive model in this paper. Another interesting property of mining CHV biclusters is that it is a symmetric problem. The CHV biclusters are fully preserved when rows become columns and columns become rows of the matrix (see a proof in Zhao and Zaki \cite{ZhaoZaki2005}).

\begin{mydef}[CHV biclusters - additive model]
Let $Z^{jl} = \{a_{ij} - a_{il}\}_{i \in I}$, $j,l \in J$, i.e., the set of values of the difference between two attributes for the subset of rows $I$. A perfect shifting bicluster is a submatrix $(I,J)$ of a data matrix $\mathbf{A}_{n \times m}$ such that all elements of the set $Z^{jl}$, $\forall j,l \in J$, are equal, i.e., $z = w$, $\forall z, w \in Z^{jl}$, $\forall j,l \in J$. A perturbed shifting bicluster is a submatrix $(I,J)$ such that $|z - w| \leq \epsilon$, $\forall z, w \in Z^{jl}$, $\forall j,l \in J$, i.e.,

\begin{equation}
  \max(Z^{jl}) - \min(Z^{jl}) \leq \epsilon, \forall j,l \in J.
	\label{eq:chvabic}
\end{equation}

\label{def:chvabic}
\end{mydef}

\subsection{Metrics and indices}

Here, we will outline some bicluster metrics and indices to ease the reading of this work.

The \textit{volume} of a bicluster $(I,J)$ is given by $|I| \times |J|$. The \textit{overlap} between two biclusters $(I,J)$ and $(I',J')$ is given by
\begin{equation}
ove((I,J),(I',J')) = \frac{|I \cap I'| \times |J \cap J'|}{\min(|I \times J|, |I' \times J'|)}. 
\end{equation}

\noindent
Let $\mathfrak{B} = (I_l, J_l)$, $l = 1, ..., k$, be a biclustering solution. The \textit{span} of the solution $\mathfrak{B}$ is given by
\begin{equation}
span(\mathfrak{B}) = \bigcup_{(I_l,J_l)} I_l \times J_l.
\end{equation}

\noindent
The \textit{coverage} of the solution $\mathfrak{B}$ is given by $cov(\mathfrak{B}) = | span(\mathfrak{B}) |$, i.e., the number of cells of the data matrix covered by at least one bicluster. It is more usual to present the coverage in terms of percentage, i.e, $cov(\mathfrak{B}) / (n \times m)$. The \textit{global overlap} of the solution $\mathfrak{B}$ is given by
\begin{equation}
oveg(\mathfrak{B}) = \frac{\sum_{(I_l \times J_l)} |I_l \times J_l| - cov(\mathfrak{B})}{cov(\mathfrak{B})}.
\end{equation}

\noindent
If we have a reference bicluster solution $\mathfrak{\dot{B}}$, we can measure how good is a found bicluster solution $\mathfrak{B}$ by means of an external evaluation \cite{HortaCampello2014}. We will use \textit{precision} and \textit{recall} to this end, respectively given by $prec(\mathfrak{B},\mathfrak{\dot{B}}) = |span(\mathfrak{B}) \cap span(\mathfrak{\dot{B}})| / cov(\mathfrak{B})$ and $rec(\mathfrak{B},\mathfrak{\dot{B}}) = prec(\mathfrak{\dot{B}},\mathfrak{B})$.

\subsection{Maximality and Monotonicity}
\label{sec:bicMaxProp}

\begin{mydef}[Maximal bicluster]
Given the desired characteristics of homogeneity, a bicluster $(I,J)$ is called a maximal bicluster iff:
\begin{itemize}
	\item $\forall x \in X \setminus I$, $(I \cup \{x\}, J)$ is not a (valid) bicluster, and
	\item $\forall y \in Y \setminus J$, $(I, J \cup \{y\})$ is not a (valid) bicluster.
\end{itemize}
\end{mydef}

\noindent It means that a bicluster is maximal if we can not add any object/attribute to it without violating the desired characteristics of homogeneity.

For all bicluster definitions given in Subsection~\ref{sec:bicTypes}, we have the following properties.

\begin{property}[Anti-Monotonicity]
Let $(I,J)$ be a bicluster. Any submatrix $(I', J')$, where $I' \subseteq I$ and $J' \subseteq J$, is also a bicluster.
\end{property}

\begin{property}[Monotonicity]
Let $(I,J)$ be a maximal bicluster. Any submatrix $(I', J')$, where $I' \supseteq I$ and $J' \supseteq J$, is not a bicluster.
\end{property}

Usually, the efficient algorithms for enumerating CTV biclusters of ones from binary data are based on the monotonicity and anti-monotonicity properties \cite{Besson2007}. In fact, we do not know any efficient algorithm for this task that is not based on these properties. RIn-Close (see Section~\ref{sec:rinclose}), in turn, also considers these properties, but now in the context of numerical datasets. Any definition of a bicluster type that meets these properties can be used in our framework. For instance, our definition of a CVC bicluster considers the same maximum residue $\epsilon$ for all attributes, however it is possible to use a different maximum residue for each attribute. If we are analyzing a dataset where the attributes have different range values, we could, for example, use the percentage of the variation of each attribute as a maximum residue for each attribute. But, we also could normalize / scale the domain of the attributes and use the same maximum residue for all of them.

\begin{property}
Let $\mathfrak{B}_{\epsilon}$ be an enumerative bicluster solution with maximum perturbation $\epsilon$, and $\mathfrak{B}_{\epsilon'}$ be an enumerative bicluster solution with maximum perturbation $\epsilon'$, where $\epsilon > \epsilon'$. Both, $\mathfrak{B}_{\epsilon}$ and $\mathfrak{B}_{\epsilon'}$, with the same restrictions in the minimum number of rows and columns of the biclusters. In this case, $span(\mathfrak{B}_{\epsilon}) \supseteq span(\mathfrak{B}_{\epsilon}')$, and therefore $cov(\mathfrak{B}_{\epsilon}) \geq cov(\mathfrak{B}_{\epsilon}')$.
\label{prop:cov2}
\end{property}

Property~\ref{prop:cov2} states that the coverage is a monotonic function with respect to $\epsilon$. However, it does not indicate that the number of biclusters will always increase with $\epsilon$. With an increasing in $\epsilon$, new biclusters tend to be found, but it is not a rule. For example, two biclusters found with $\epsilon = x$ can be merged into one with some $\epsilon > x$. In fact, if $\epsilon$ is too high, the entire dataset will be considered a single valid bicluster.

\section{Formal Concept Analysis}
\label{sec:fca}

Formal Concept Analysis (FCA) is a field of applied mathematics based on mathematical order theory, in particular on the theory of complete lattices \cite{Ganter1997}. Here, we will explain the basic principles of FCA. For more details refer to Ganter et al. \cite{Ganter1997}.

\begin{mydef}[Formal Context]
A formal context is a triple $(G, M, I)$ of two sets $G$ and $M$, and a relation $I \subseteq G \times M$. Each $g \in G$ is interpreted as an object, and each $m \in M$ is interpreted as an attribute. In order to express that an object $g$ is in a relation $I$ with an attribute $m$, we write $(g, m) \in I$ or $gIm$. We read it as ``the object $g$ has the attribute $m$''. 
\end{mydef}

Notice that a formal context can be easily represented by a binary matrix $\mathbf{D}$, where rows represent objects, and columns represent attributes. We will have $d_{gm} = 1$ if the object $g$ has the attribute $m$, and we will have $d_{gm} = 0$ otherwise.

For a subset $A \subseteq G$ of objects, we define:
\begin{equation}
  A' = \{m \in M| \forall g \in A : gIm \}
\end{equation}
\noindent (the set of attributes common to the objects in $A$). Similarly, for a subset $B \subseteq M$, we define:
\begin{equation}
  B' = \{g \in G| \forall m \in B : gIm \}
	\label{eq:blinha}
\end{equation}

\noindent (the set of objects common to the attributes in $B$).

\begin{mydef}[Formal Concept]
A formal concept of the formal context $(G, M, I)$ is a pair $(A,B)$ with $A \subseteq G$, $B \subseteq M$, $A'= B$, and $B'=A$. The subset $A$ of a formal concept $(A, B)$ is called \textit{extent}, and the subset $B$ is called \textit{intent}.
\end{mydef}

By the definition, we see that though many subsets $A$ can generate the same subset $B$, only the largest (closed) subset $A$ is part of a formal concept (and vice versa). Formal concepts are partially ordered by $(A_1, B_1) \leq (A_2, B_2) \Leftrightarrow A_1 \subseteq A_2 (\Leftrightarrow B_2 \subseteq B_1)$. With respect to this partial order, the set of all formal concepts forms a complete lattice called the \textit{concept lattice} of the formal context $(G,M,I)$, denoted by $\mathfrak{B}(G, M, I)$. There are several algorithms in the literature that are able to extract the concept lattice of a formal context. Some examples are: Close-by-One (CbO) \cite{Kuznetsov1999}, In-Close \cite{Andrews2009}, In-Close2 \cite{Andrews2011}, and FCbO \cite{Krajca2010}. As the algorithms that we will propose are based on In-Close2, Subsection~\ref{subsec:inclose2} is devoted to its formalization.

\subsection{In-Close2}
\label{subsec:inclose2}

In-Close2 \cite{Andrews2011} and its precursor In-Close \cite{Andrews2009} are based conceptually on Close-By-One \cite{Kuznetsov1999}. These algorithms use the lexicographic approach for mining formal concepts, thus avoiding the discovery of the same formal concept multiple times. Ganter \cite{Ganter1984} showed how the lexicographical order of concepts can be used to avoid the search of repeated results. In the mathematical order theory, combinations have a lexicographical order, for instance, \{1, 2, 3\} comes before \{1, 2, 4\}, and also before \{1, 3\} \cite{Andrews2009}. In-Close and In-Close2 maintain a \textit{current attribute}. The concept next generated is new (\textit{canonical}) if its intent contains no attribute preceding the current attribute. Therefore, to verify canonicity, In-Close/In-Close2 does the following: supposing that $B$ is the current intent, $j$ is the current attribute, and $RW$ is the resulting extent, $RW$ is not canonical if

\begin{equation}
  \exists k \in M \setminus B | [k < j] \: \wedge \: [\forall g \in RW: gIk].
	\label{eq:inc2_iscan}
\end{equation}

\noindent i.e., if there is an attribute $k < j$ where $k \notin B$ and $RW \subseteq \{k\}'$. See Eq.~(\ref{eq:blinha}) for a definition of $\{k\}'$. The concept of canonicity was introduced in Kuznetsov \cite{Kuznetsov1996}.

Algorithm \ref{alg:inclose2} shows In-Close2 pseudocode. When we use $A_r$ and $B_r$, it means the extent and the intent of the $r$-th formal concept, respectively. When we write $J_k$ it means the element of the set $J$ at position $k$, for instance, if $J = \{2, 5, 7, 8, 13\}$ and $k = 2$, $J_2 = 5$. The same for $R_k$. In the main function of In-Close2, we set $(A_1, B_1) \leftarrow (\{1, ..., n\}, \{\})$ (which is called \textit{supremum}), and $r_{new} \leftarrow 1$. Then, we call the function In-Close2 to incrementally close the formal concept $(A_1, B_1)$, beginning at attribute index 1. Thereafter, all formal concepts will be found recursively. During the closure of a formal concept, In-Close2 iterates across the attributes. If the current attribute $j$ is not an inherited attribute, In-Close2 computes the candidate to a new extent $RW$. If the extent $RW$ is the same as the current extent $A_r$, then attribute $j$ is added to the current intent $B_r$. If the extent $RW$ is not the same as the current extent $A_r$, In-Close2 tests if $RW$ is canonical. If yes, the current formal concept $(A_r, B_r)$ will give rise to a child formal concept. After the closure of the current formal concept $(A_r, B_r)$, In-Close2 starts to close its children.

\linespread{1}

\begin{algorithm}
\caption{In-Close2}
\label{alg:inclose2}
\begin{algorithmic}[1]
  \small
	\REQUIRE Binary data matrix $\mathbf{D}_{n \times m}$, minimum number of rows $minRow$, index of the formal concept to be closed $r$, index of the initial attribute $y$
	\STATE $J \leftarrow \{\}$
	\STATE $R \leftarrow \{\}$
  \FOR{$j \leftarrow y$ to $m$}
	  \IF{$j \notin B_r$}
      \STATE $RW \leftarrow A_r \cap \{j\}'$
      \IF{$\left |RW \right| \geq  minRow$}
        \IF{$\left |RW \right | =  \left |A_r \right |$}
          \STATE $B_r \leftarrow B_r \cup \{j\}$
        \ELSIF{$RW$ is canonical}
				  \STATE $r_{new} \leftarrow r_{new} + 1$ \COMMENT{global variable}
	        \STATE $J \leftarrow J \cup \{ j \}$
	        \STATE $R \leftarrow R \cup \{ r_{new} \}$
  				\STATE $A_{r_{new}} \leftarrow RW$
        \ENDIF
      \ENDIF
	  \ENDIF
  \ENDFOR
	\FOR{$k \leftarrow 1$ to $\left | J \right |$}
	 \STATE $B_{R_k} \leftarrow B_r \cup \{J_k\}$
	 \STATE In-Close2$(\mathbf{D}, minRow, R_k, J_k + 1)$
	\ENDFOR
\end{algorithmic}
\end{algorithm}

\linespread{1.5}

The worst-case time of In-Close2 is $O(knm^2)$, where $k$ is the number of biclusters. If $minRow = 1$, In-Close2 mines the concept lattice of the formal context represented by the binary matrix $\mathbf{D}$. Otherwise, if $minRow > 1$, In-Close2 mines the set of all \textit{frequent concepts} for the threshold $minRow$, called the \textit{iceberg concept lattice} \cite{Lazzeroni2002}. In addition to the minimum number of rows $minRow$, we can easily add a minimum number of columns $minCol$ to In-Close2. While In-Close2 loops through the attributes, a formal concept $(A_r,B_r)$ can be discarded if, even adding all remaining attributes to its intent, it will not meet the minimum number of columns $minCol$ (therefore, its next descendants will not meet the minimum number of columns $minCol$ as well). Although this restriction can be checked only during the closure of a formal concept, it will save computational resources because it avoids generating descendants that do not meet the restriction $minCol$.

\subsection{FCA and related areas of research in the literature}
\label{fca_related}

The problem of extracting the concept lattice from a formal context is the same as extracting all maximal CTV biclusters of ones from a binary data matrix. A formal concept is a maximal CTV bicluster of ones. Extent $A$ and intent $B$ are the set of rows (objects) and columns (attributes) that compose a bicluster, respectively.

The association mining problem is also closely related to FCA. This problem is divided in two sub-problems: (\textit{i}) the frequent itemset (pattern) mining problem, and (\textit{ii}) the problem of mining the association rules from these itemsets. As the first sub-problem is the most computationally expensive, almost all researches have been focused on the frequent itemset generation phase. In terms of FCA, the problem of mining all \textit{frequent itemsets} (patterns) can be described as follows. Given a formal context $(G, M, I)$, determine all subsets $B \subseteq M$ such that the support of $B$ ($supp(B) = |B'|$) is above a user-defined parameter \cite{LakhalStumme2005}. Examples of algorithms that perform this task are Apriori \cite{AgrawalEtAl1993} and Eclat \cite{ZakiEtAl1997}. To reduce the computational cost of the frequent pattern mining problem, some algorithms, such as GenMax \cite{GoudaZaki2005}, mine only the \textit{maximal frequent itemsets}, i.e., those frequent itemsets from which all supersets are infrequent and all subsets are frequent. The problem of this approach is that it leads to a loss of information since the supports of the subsets are not available. An option to reduce the computational cost without loss of information is to mine only the \textit{closed frequent itemsets}. A frequent itemset $B$ is called closed if there exists no superset $D \supset B$ with $B' = D'$. The closed frequent itemsets are also called \textit{frequent concept intents}. For any itemset $B$, its concept intent is given by $B''$. Note that this approach is the most closely related to FCA. Remarkably, a concept lattice contains all necessary information to derive the support of all (frequent) itemsets \cite{LakhalStumme2005}. Indeed, the set of closed frequent itemsets uniquely determines the exact frequency of all itemsets, and it can be orders of magnitude smaller than the set of all frequent itemsets \cite{ZakiEtAL2002}. Moreover, this approach drastically reduces the number of rules that have to be presented to the user, without any information loss \cite{LakhalStumme2005}. CHARM \cite{ZakiEtAL2002} is a well-known algorithm to mine all closed frequent itemsets. It exploits the fact that the extents of the formal concepts are irrelevant in the frequent pattern mining problem (just the intents and the cardinality of the extents are relevant). Thus, it drastically cuts down the size of memory required \cite{ZakiEtAL2002}.

The problem of enumerating all maximal bicliques from a bipartite graph is also closely related to FCA. Madeira and Oliveira \cite{MadeiraOliveira2004} stated that in the simplest case of biclustering, where we are looking for CTV biclusters of ones in a binary data matrix $\mathbf{D}$, a bicluster corresponds to a biclique in the corresponding bipartite graph. Rows and columns of the matrix $\mathbf{D}$ correspond, respectively, to the first and second sets of vertices of a bipartite graph. Each element $d_{ij} $ is equal to 1 if vertice $i$ is connected to vertice $j$, and 0 otherwise. Thus, the binary matrix $\mathbf{D}$ is the adjacency matrix of a bipartite graph. In this scenario, a formal concept from the binary matrix $\mathbf{D}$ is equivalent to a maximal biclique (bicluster). So, finding a concept lattice is also equivalent to finding all maximal bicliques of a bipartite graph. The connection between FCA and the problem of enumerating all maximal bicliques from a bipartite graph is explored in several papers \cite{AbelloEtAl2004, GaumeEtAl2010, GelyEtAL2009}. Moreover, G$\acute{\mathrm{e}}$ly \textit{et al.} \cite{GelyEtAL2009} pointed out several algorithms to find all maximal bicliques from a bipartite graph, most of them are from the area of FCA.

\section{Related Works}
\label{sec:relWork}

Due to the inherent computational complexity of the problem of finding all maximal biclusters, most of the proposed algorithms are heuristic-based \cite{MadeiraOliveira2004}. Some relevant heuristics in the literature are: CC \cite{ChengChurch2000}, FLOC \cite{Yang2003}, ROCC \cite{DeodharEtAL2009}, ISA \cite{IhmelsEtAl2002}, Plaid \cite{Lazzeroni2002}, and OPSM \cite{BenDorEtal2003}. CC looks for biclusters with mean squared residue (MSR) below a user-defined threshold $\delta$. It mines one bicluster at each iteration, and performs random perturbations to the data to mask the already discovered bicluster. FLOC is also based on the MSR, but performs simultaneous bicluster identification. Briefly, the goal of CC and FLOC is to mine a set of biclusters with high average volume given the residue limit $\delta$. ROCC is scalable and very versatile because it can be parametrized to mine several types of biclusters. It works in two steps: (\textit{i}) find $k \times l$ biclusters arranged in a grid structure, keeping only the $sr$ rows and $sc$ columns with the lowest error associated with them, and (\textit{ii}) filter out the biclusters with the largest error values, and merge similar biclusters. ISA looks for biclusters where their rows have an average value above a threshold $t_g$, and their columns have an average value above a threshold $t_c$. ISA starts with $nseed$ biclusters, and iteratively updates the columns and rows of the biclusters until convergence. Plaid fits parameters of a generative model of the data iteratively by minimizing the mean squared error between the modeled data and the true data. OPSM is a deterministic greedy algorithm dedicated to find large order-preserving submatrices.


Clearly, even the best heuristics potentially lead to sub-optimal solutions, so there are many proposals of exhaustive bicluster enumeration. Most of the work in this area is designed to mine all maximal CTV biclusters of ones from a binary dataset. In FCA, FPM and graph theory, there are several efficient algorithms able to perform this task. Proposals, such as \cite{HenriquesMadeira2014, MadeiraOliveira2009, MartinezEtAl2007, OkadaEtAl2007b, OkadaEtAl2007a, SerinVingron2011}, binarize the dataset and use FCA, FPM or graph theory algorithms to enumerate the biclusters. To avoid the loss of information of tedious Boolean property encoding phases \cite{Besson2007}, many proposals deal directly with numerical datasets, as the following.

The proposals of \cite{AtluriEtAl2000, Besson2007, kaytoue2013, Kaytoue2011} are dedicated to enumerate CTV biclusters from numerical datasets.

The RCB algorithm \cite{AtluriEtAl2000} is based on an FPM algorithm called Apriori \cite{AgrawalSrikant1994}, which has a worst-case time exponential on the number of attributes. Thus, Apriori and the algorithms based on it are not efficient. It is also noteworthy that Apriori mines frequent itemsets, not closed frequent itemsets. Thus, it produces many redundant biclusters. RCB adopts a two step process. First, all the square submatrices that qualify to be a CTV bicluster are enumerated. Second, these square CTV biclusters are merged to form rectangular CTV biclusters of arbitrary sizes.

The NBS-Miner algorithm \cite{Besson2007} mines all maximal CTV biclusters of a numerical dataset. The algorithm starts with the lattice $((\{\},\{\}),(G,M))$ (whose bottom is $(\{\},\{\})$ and top is $(G,M)$), i.e, the lattice containing all possible biclusters. Then, NBS-Miner explores its sublattices using three functions: enumeration, pruning, and propagation. The enumeration function splits recursively the current sublattice into two new sublattices. The prune function is responsible for pruning the sublattices that do not attend to the restriction of similarity (imposed by $\epsilon$, see Eq. \ref{eq:ctvbic}) or maximality. The propagation function implements the reduction of the size of the search space of a sublattice, not considering the entire current sublattice. The algorithm finds a bicluster when it finds a sublattice whose top is equal to the bottom.

Kaytoue \textit{et al.} \cite{Kaytoue2011} proposed two FCA-based methods to enumerate CTV biclusters. The former is based on the discretization of the numerical data matrix using  \textit{conceptual scaling} \cite{Ganter1997}. Let $W$ be the set of values that an object $g \in G$ can take for an attribute $m \in M$. First of all, they compute all \textit{tolerance classes} \cite{kaytoueEtAl2010} from $W$. Then, they create one formal context for each class of tolerance and use FCA standard algorithms to enumerate the formal concepts from them. Each formal concept corresponds to a maximal CTV bicluster. The formal contexts are created in a way that avoids finding redundant CTV biclusters, but at a price of not finding some biclusters. Since the resulting binary tables may be numerous depending on the number of elements of $W$ and the parameter $\epsilon$, this method is not feasible in many real-world scenarios. The second method is divided in two phases. In the first one, it enumerates all the CVC biclusters using \textit{interval pattern structures} (IPS) \cite{Ganter2001}. It is noteworthy that this method returns redundant CVC biclusters. In the second phase, CTV biclusters are extracted from the CVC biclusters, but this process is not so clear, because a CVC bicluster can give rise to many CTV biclusters.

In \cite{kaytoue2013}, the authors also use tolerance classes over the set of numbers $W$, and create one formal context for each class of tolerance. But they proposed a new algorithm called TriMax to mine the CTV biclusters from these formal contexts. TriMax is able to perform a complete, correct and non-redundat enumeration of all maximal CTV biclusters in a numerical dataset. But due to the scaling process, TriMax may be not feasible in many real-world scenarios too.

In the next two subsections, we highlight the competitors of the algorithms that we are proposing.

\subsection{Enumerating CVC (or CVR) biclusters}


RAP \cite{PandeyEtAl2009} is also based on Apriori \cite{AgrawalSrikant1994}. The authors did not describe their strategy to avoid redundancy, but we conjecture that the best that can be done is a pairwise comparison of biclusters with $k$ and $k+1$ columns.

In \cite{CodocedoNapoli2014a, CodocedoNapoli2014b}, the authors proposed a method based on \textit{partition pattern structure} (PPS) \cite{BaixeriesEtAl2014}. Their proposal is not able to perform a complete enumeration because the components of the partition of the set $G$ must be disjunct. They proposed a strategy based on a lattice to remove the redundancy, which is much faster than to compare one bicluster against all the others. But it is necessary to mine the redundant biclusters to make this verification, so it is not an efficient method.

In \cite{kaytoueEtAl2014}, the authors revisited the proposals of mining CVC biclusters using IPS \cite{Kaytoue2011} and PPS \cite{CodocedoNapoli2014a, CodocedoNapoli2014b}. They also proposed an approach based on Triadic Concept Analysis (TCA) \cite{Lehmann1995}. From a numerical dataset, they derivate a triadic context given by $(M,G,G,Y)$ such that $(m,g_1,g_2) \in Y$ iif $|d_{g_1m} - d_{g_2m}| \leq \epsilon$. Then, they use standard TCA algorithms to enumerate the triadic concepts, but not all triadic concepts are maximal CVC biclusters.

Table~\ref{tab:comparison_cvc} shows a comparison between these proposals and the algorithms that we are proposing to enumerate CVC biclusters in Section~\ref{sec:rinclose}. As we see, our proposals are the only algorithms that are able to perform an efficient, complete, correct and non-redundant enumeration of all maximal CVC biclusters.

\linespread{1}

\begin{table}
  \centering
  \caption{Comparison between RIn-Close\_CVC\_P, RIn-Close\_CVC and their competitors.}
    \begin{tabular}{ccccc}
		\toprule
		& Complete	& Correct & Non-Redundant & Efficient \\
    \midrule
		RIn-Close\_CVC\_P & $\checkmark$	& $\checkmark$ & $\checkmark$ & $\checkmark$ \\
		RIn-Close\_CVC & $\checkmark$	& $\checkmark$ & $\checkmark$ & $\checkmark$ \\
		RAP & $\checkmark$	& $\checkmark$ &  &  \\
		PPS & $\circ$ & $\checkmark$ & $\checkmark$ & \\
		IPS & $\checkmark$ & $\checkmark$  & & \\
		TCA & $\checkmark$ & $\checkmark$  & & \\
    \bottomrule
    \end{tabular}
		\\ \footnotemark[1]{The symbol $\checkmark$ indicates that the algorithm has the property. The symbol $\circ$ indicates that the authors claim their algorithm has the property, but it fails to exhibit the property.}
  \label{tab:comparison_cvc}
\end{table}

\linespread{1.5}

\subsection{Enumerating CHV biclusters}


pCluster \cite{WangEtAl2002} was the first deterministic algorithm with an enumerative approach to mine CHV biclusters. pCluster computes all row-maximal biclusters with two columns and all column-maximal biclusters with two rows, prunes the unpromising biclusters, and stores the remaining column-maximal biclusters in a prefix tree. Then, pCluster makes a depth-first search in this prefix tree in order to mine larger biclusters. However, pCluster has several shortcomings. It does not find all biclusters, can find biclusters that do not attend the user-defined measure of similarity, and returns redundant biclusters. Furthermore, pCluster's computational complexity is exponential w.r.t. the number of attributes.

Maple \cite{PeiEtAL2003} is an improved version of pCluster and it is closer to an actual enumerative algorithm. It returns only non-redundant biclusters, but it does not have an efficient approach to do this: for each possible new bicluster, Maple must look at all previously generated biclusters to avoid redundancy. Besides, there are two scenarios where Maple fails in performing a complete and correct enumeration of all maximal biclusters. If two biclusters have the same set of objects and share some attributes, Maple would return only one bicluster containing both of them (thus violating the user-defined measure of similarity). Maple also may miss biclusters due to its routine of pruning unpromising biclusters: Maple keeps an attribute-list ordered by some criterion. If a bicluster has a subset of objects and a superset of attributes of another bicluster, and its extra attributes are subsequently considering Maple's attribute-list, Maple would prune it incorrectly. The worst-case time of Maple's search is also exponential w.r.t. the number of attributes.

MicroCluster \cite{ZhaoZaki2005} constructs a multigraph that represents all row-maximal biclusters with two columns, where nodes represent attributes and edges represent sets of objects. It uses a depth-first search on the multigraph to mine the biclusters. We tested the authors' MicroCluster implementation (\url{http://www.cs.rpi.edu/\~ zaki/www-new/pmwiki.php/Software/Software}), and we observed that MicroCluster can fail in enumerating all the maximal biclusters and can return biclusters that violate the user-defined measure of similarity. Its worst-case time is also exponential w.r.t. the number of attributes. As Maple, MicroCluster must look at all previously generated biclusters to avoid redundancy, which is always a strategy to be avoided. After mining a CHV bicluster, MicroCluster verifies if the variation in its rows and columns attends user-defined measures of similarity too. If a bicluster does not attend these restrictions, it is discarded. Note that this verification can be done in any biclustering solution.

To reduce computational cost, algorithms such as SeqClus \cite{WangEtAl2004} and CPT \cite{GuanEtAL2009} relaxed the definition of CHV biclusters. Instead of looking for every pair of attributes (or objects), they use a pivot attribute to compute the CHV biclusters. Thus, given a user-defined parameter $\epsilon$, their biclusters will have residue less or equal to $2\epsilon$ \cite{GuanEtAL2009}. The biclustering solution of these algorithms depends on the choices of the pivot attributes. So, we can say that this strategy yields an approximate result for the actual enumeration. A genuine complete and correct enumeration of CHV biclusters would provide the same solution when dealing with the original matrix or with its transpose version. The CHV biclusters are fully preserved when rows become columns and columns become rows of the matrix. But unfortunately, when resorting to the pivot attribute, those techniques are prone to lose this fundamental property. Accordingly, these algorithms are not able to perform a complete and correct enumeration. As SeqClus and CPT are based on algorithms to mine frequent itemset, not closed frequent itemsets, they return redundant biclusters and are not efficient to mine maximal biclusters. So, this idea of mining CHV biclusters using a pivot attribute can be better explored. In fact, we could use this idea in RIn-Close\_CHV (as we used in RIn-Close\_CHV\_P since, in this case, if a new candidate attribute is coherent with any attribute of the current bicluster, it will be coherent with all other attributes), which would lead to an efficient algorithm. But our main goal is to provide an algorithm able to perform a complete and correct enumeration of all maximal CHV biclusters. Moreover, despite not having polynomial delay, RIn-Close\_CHV has good computational performance as we can see with the experimental results in Section~\ref{sec:exp}.

Table~\ref{tab:comparison_chv} shows a comparison between these proposals and the algorithms that we are proposing to enumerate CHV biclusters in Section~\ref{sec:rinclose}. As we see, we are proposing algorithms with a number of additional features that are able to support a wider range of application scenarios, including the identification of biological indicators \cite{LiuEtAl2014} and classification based on associations \cite{LiuEtAl1998}. Moreover, a bicluster solution provided by these competitors  is contained in the RIn-Close's solution, given the adequate parameter $\epsilon$. For instance, if a user parameterizes CPT with $\epsilon = 1$, it would obtain a biclustering solution with biclusters with residue up to 2. Parameterizing RIn-Close with $\epsilon = 2$, a user would obtain a biclustering solution containing all the maximal versions of the biclusters of the CPT's solution, and possibly containing additional biclusters.

\linespread{1}

\begin{table}
  \centering
  \caption{Comparison between RIn-Close\_CHV\_P, RIn-Close\_CHV and their competitors.}
    \begin{tabular}{ccccc}
		\toprule
		& Complete	& Correct & Non-Redundant & Efficient \\
    \midrule
		RIn-Close\_CHV\_P & $\checkmark$	& $\checkmark$ & $\checkmark$ & $\checkmark$ \\
		RIn-Close\_CHV & $\checkmark$	& $\checkmark$ & $\checkmark$ &  \\
		pCluster & $\circ$ & $\circ$ &  &  \\
		Maple & $\circ$ & $\circ$ & $\checkmark$ & \\
		MicroCluster & $\circ$ & $\circ$ & $\checkmark$ & \\
		SeqClus & $\circ$ & $\circ$ & &  \\
		CPT & $\circ$ & $\circ$ & & \\
    \bottomrule
    \end{tabular}
		\\ \footnotemark[1]{The symbol $\checkmark$ indicates that the algorithm has the property. The symbol $\circ$ indicates that the authors claim their algorithm has the property, but it fails to exhibit the property.}
  \label{tab:comparison_chv}
\end{table}

\linespread{1.5}

\section{RIn-Close}
\label{sec:rinclose}

In-Close2 has been specifically designed to extract all maximal CTV biclusters of ones from a binary data matrix. Now, we will propose generalizations of In-Close2 to enumerate other types of biclusters from numerical (not only binary, but also integer or real-valued) matrices. We call our family of algorithms RIn-Close.

We chose to adapt In-Close2 because: (\textit{i}) it is easy to understand; (\textit{ii}) it is one of the fastest algorithms of FCA; (\textit{iii}) it has support to the desired minimum number of rows in a bicluster (the parameter $minRow$); (\textit{iv}) it is easy to incorporate a support to the desired minimum number of columns in a bicluster (the parameter $minCol$); and (\textit{v}) In-Close2 starts with all objects in the extent of a formal concept. This latter aspect of In-Close2 is very important when working with integer or real-valued data matrices. For instance, when finding CVC biclusters, given the current attribute, we can look for the subsets of rows of the extent that accomplish the similarity restriction $\epsilon$.

\subsection{Finding biclusters with constant values on columns (CVC)}
 
The algorithms of this section compute an efficient,  complete, correct and non-redundant enumeration of all maximal CVC biclusters. First, we will show how to extract perfect biclusters, because it is the easiest case. After that, given a user-defined parameter $\epsilon$ ($\epsilon > 0$), we will show how to extract non-perfect CVC biclusters with maximum residue $\epsilon$, as presented in Eq.~(\ref{eq:cvcbic}).

\subsubsection{Perfect Biclusters}

\begin{figure*}
  \centering
	
	\subfigure[Example of the generation of a single descendant by In-Close2.]{
		\includegraphics[trim=3.5cm 13cm 12cm 2.5cm, clip, scale=0.6]{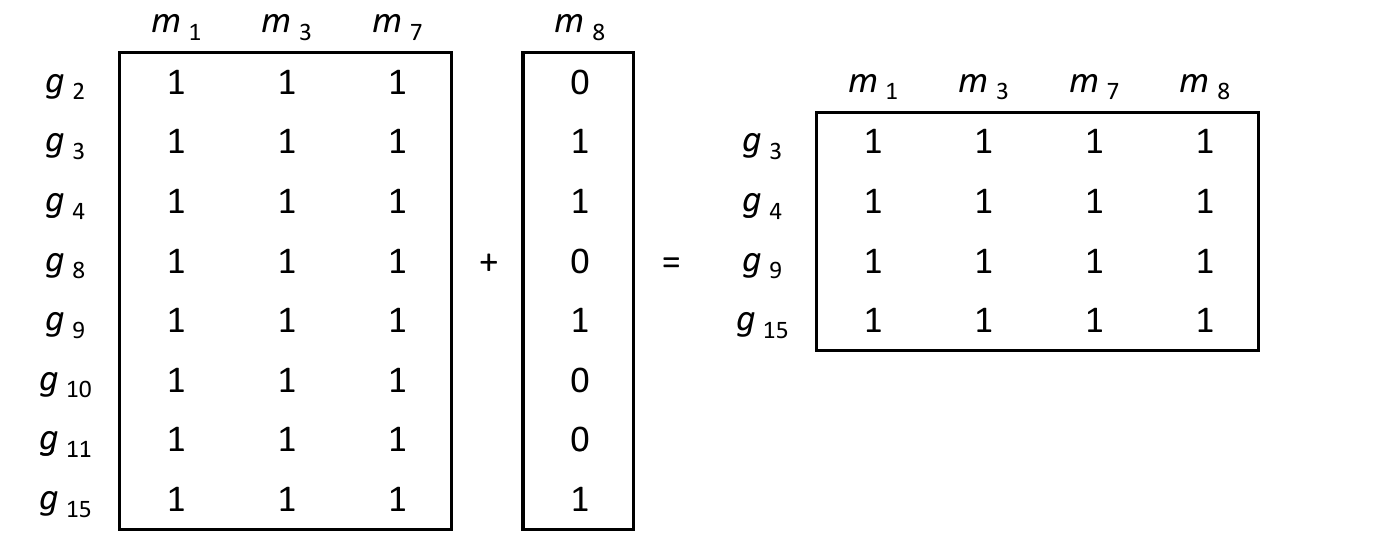}
		\label{fig:variosFilhos0}
	}
	
	\subfigure[Example of the generation of multiple descendants by RIn-Close\_CVC\_P.]{
		\includegraphics[trim=3.5cm 13cm 12cm 2.5cm, clip, scale=0.6]{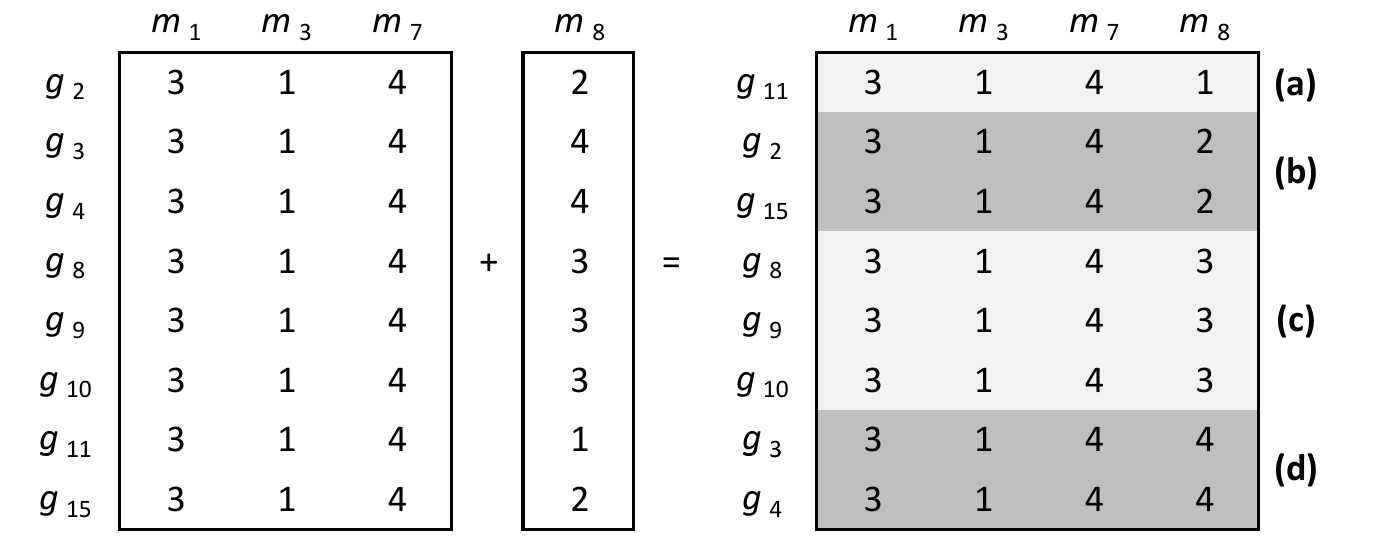}
		\label{fig:variosFilhos1}
	}
	
	\subfigure[Example of the generation of multiple descendants by RIn-Close\_CVC (considering $\epsilon = 1$).]{
		\includegraphics[trim=3.5cm 10cm 12cm 2.5cm, clip, scale=0.6]{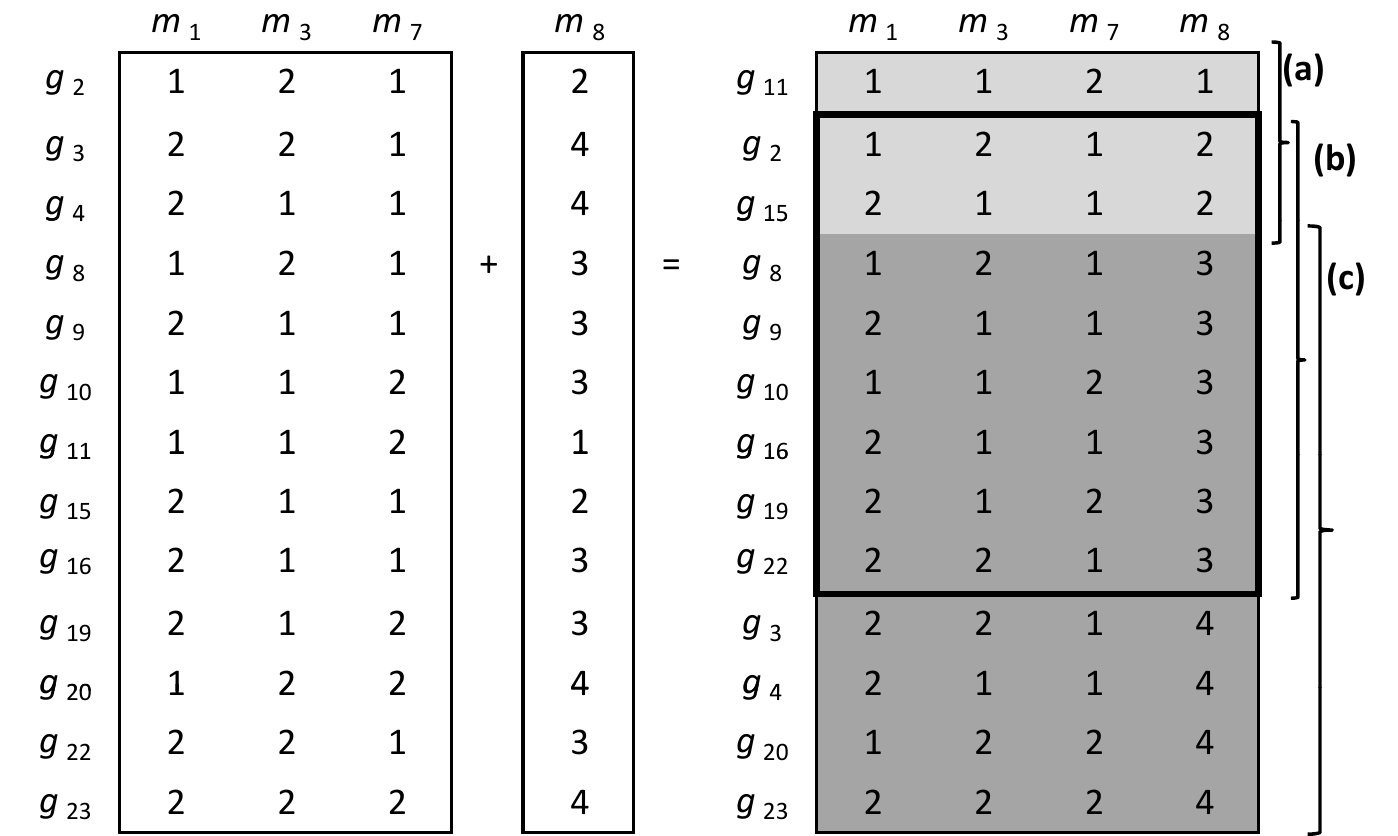}
		\label{fig:variosFilhos2}
	}
	
	\caption{Generation of descendants by In-Close2, and RIn-Close\_CVC\_P and and RIn-Close\_CVC.}
	\label{fig:variosFilhos}
\end{figure*}

The adaptation of In-Close2 to enumerate all maximal perfect CVC biclusters, called RIn-Close\_CVC\_P, is straightforward. We have only one major difference. In In-Close2, each bicluster $(A_r, B_r)$ can generate just one descendant per attribute, whereas in RIn-Close\_CVC\_P, each bicluster $(A_r, B_r)$ can generate multiple descendants per attribute. It happens because In-Close2 just looks for blocks of 1s, whereas RIn-Close\_CVC\_P looks for any blocks of constant values on columns. Fig.~\ref{fig:variosFilhos} illustrates this difference. In Fig.~\ref{fig:variosFilhos0}, In-Close2 is closing the bicluster $(A_r = \{g_2, g_3, g_4, g_8, g_9, g_{10}, g_{11}, g_{15}\}, B_r = \{ m_1, m_3, m_7\})$. When it tries to add the attribute $m_8$, bicluster $(A_r, B_r)$ gives rise to a new bicluster $(A = \{g_3, g_4, g_9, g_{15}\}, B = \{ m_1, m_3, m_7, m_8\})$. In Fig.~\ref{fig:variosFilhos1}, RIn-Close\_CVC\_P is closing the same bicluster $(A_r, B_r)$, but when it tries to add the attribute $m_8$, bicluster $(A_r, B_r)$ gives rise to four new perfect CVC biclusters \emph{without overlap between their extents}: (a) $(A = \{g_{11}\}, B = \{ m_1, m_3, m_7, m_8\})$, (b) $(A = \{g_2, g_{15}\}, B = \{ m_1, m_3, m_7, m_8\})$, (c) $(A = \{g_8, g_9, g_{10}\}, B = \{ m_1, m_3, m_7, m_8\})$, and (d) $(A = \{g_3, g_4\}, B = \{ m_1, m_3, m_7, m_8\})$. Note that the elements of the current attribute, $m_8$, were sorted in order to easily identify all subsets of objects with constant values.

Algorithm~\ref{alg:rinc_cvc_p} shows the pseudocode of RIn-Close\_CVC\_P. Notice that it is almost the same as In-Close2. There are basically two differences. The first one is that the current attribute $j$ is added to the current intent $B_r$ if all values of attribute $j$ and objects $A_r$ are equal. And the second one occurs by the fact that the bicluster $(A_r, B_r)$ can generate multiple descendants. So, RIn-Close\_CVC\_P computes all the possible extents and loops across them. The test of canonicity is also essentially the same as in In-Close2. Let $B$ be the current intent, $j$ be the current attribute, and $RW$ be the extent of the new bicluster, it is not canonical if

\begin{equation}
	\exists k \in M \setminus B | [k < j] \: \wedge \: [\max_{i \in RW}(d_{ik}) - \min_{i \in RW}(d_{ik}) = 0],
	\label{eq:rinc_cvc_p_iscan}
\end{equation}

\noindent i.e., if there is an attribute $k < j$ that we can add to the bicluster $(RW, B)$ and it remains a valid perfect CVC bicluster. The worst-case time of RIn-Close\_CVC\_P is almost the same as In-Close2: $O(knm(\log n + m))$. The difference is due to the use of a sorting algorithm to compute the possible extents.

\linespread{1}

\begin{algorithm}
\caption{RIn-Close\_CVC\_P}
\label{alg:rinc_cvc_p}
\begin{algorithmic}[1]
  \small
	\REQUIRE Data matrix $\mathbf{D}_{n \times m}$, minimum number of rows $minRow$, index of the bicluster to be closed $r$, index of the initial attribute $y$
	\STATE $J \leftarrow \{\}$
	\STATE $R \leftarrow \{\}$
  \FOR{$j \leftarrow y$ to $m$}
	  \IF{$j \notin B_r$}
		  \IF{$\max_{i \in A_r}(d_{ij}) - \min_{i \in A_r}(d_{ij}) = 0$}
			  \STATE $B_r \leftarrow B_r \cup \{j\}$
			\ELSE
			  \STATE Compute the possible extents
			  \FOR{each possible extent $RW$}
					\IF{$|RW| \geq minRow$ \AND $RW$ is canonical}
						\STATE $r_{new} \leftarrow r_{new} + 1$
						\STATE $R \leftarrow R \cup \{r_{new}\}$
						\STATE $J \leftarrow J \cup \{j\}$
						\STATE $A_{r_{new}} \leftarrow RW$
					\ENDIF
				\ENDFOR
			\ENDIF
		\ENDIF
	\ENDFOR
	\FOR{$k \leftarrow 1$ to $|J|$}
	  \STATE $B_{R_k} \leftarrow B_r \cup \{J_k\}$
		\STATE RIn-Close\_CVC\_P$(\mathbf{D}, minRow, R_k, J_k+1)$
	\ENDFOR
\end{algorithmic}
\end{algorithm}

\linespread{1.5}

\subsubsection{Non-Perfect Biclusters}
\label{sec:rinclosecvcnp}

This adaptation of In-Close2, called RIn-Close\_CVC, is significantly more elaborate than RIn-Close\_CVC\_P because, besides a bicluster $(A_r, B_r)$ being able to generate multiple descendants per attribute, there may be overlaps between their extents. For instance, in Fig.~\ref{fig:variosFilhos2}, RIn-Close\_CVC is closing the bicluster $(A_r = \{g_2, g_3, g_4, g_8, g_9, g_{10}, g_{11}, g_{15}, g_{16}, g_{19}, g_{20}, g_{22}, g_{23}\}, B_r = \{ m_1, m_3, m_7\})$, when it tries to add the current attribute $m_8$, bicluster $(A_r, B_r)$ gives rise to three new biclusters \emph{with overlap between their extents} (considering $\epsilon = 1$).
Notice again that the elements of the current attribute were sorted to facilitate the identification of all possible extents.

Due to a bicluster $(A_r, B_r)$ being able to generate multiple descendants per attribute, with overlap between them, it is necessary to take some actions to avoid the generation of duplicate and non-maximal biclusters. In fact, these challenging issues can occur if two descendant biclusters share $minRow$ rows or more in their extents.

Assuming $minRow = 3$, in our example in Fig.~\ref{fig:variosFilhos2}, biclusters (a) and (b) can not generate duplicate biclusters because they share only 2 rows in their extents. But biclusters (b) and (c) can generate duplicate biclusters with extent $A \subseteq \{g_8, g_9, g_{10}, g_{16}, g_{19}, g_{22}\}$ and $|A| \geq minRow$, when adding a new attribute. To solve this problem, we added one more verification on the test of canonicity. This new verification is based on the fact that two distinct CVC biclusters must have two distinct extents. So, we track the extents that have already been generated using efficient symbol table implementations, such as hash tables (HTs) or balanced search trees (BSTs). So, symbol table's keys are given by the extents, in such way that the rows in an extent are in their ascending (or descending) order. The worst-case time to insert and search in a BST is $O(\log k)$, where $k$ is its number of elements. The worst-case time to insert and search in a HT is $O(1)$ and $O(k)$, respectively. However, under reasonable assumptions, the average time to search in a HT is $O(1)$. The remainder of the test of canonicity is again essentially the same as in In-Close2. Supposing that $B$ is the current intent, $j$ is the current attribute, and $RW$ is the extent of the new bicluster, it is not canonical if

\begin{equation}
  \exists k \in M \setminus B | [k < j] \: \wedge \: [\max_{i \in RW}(d_{ik}) - \min_{i \in RW}(d_{ik}) \leq \epsilon],
	\label{eq_rinc_cvc_iscan}
\end{equation} 

\noindent i.e., if there is an attribute $k < j$ that we can add to the bicluster $(RW, B)$ and it remains a valid CVC bicluster.

But even with this new verification on the test of canonicity, we still have the undesirable possibility of generating non-maximal biclusters. For instance, in Fig.~\ref{fig:variosFilhos2}, bicluster (c) can give rise to the bicluster $(A = \{g_4, g_8, g_9, g_{10}\}, B = \{ m_1, m_3, m_7, m_8, m_{11}, m_{16}\})$, and bicluster (b) can give rise to the bicluster $(A = \{g_8, g_9, g_{10}\}, B = \{ m_1, m_3, m_7, m_8, m_{11}, m_{16}\})$, which is clearly non-maximal. So, when two biclusters share $minRow$ rows or more in their extents, we need to verify if their descendants are maximal in their extents (row-maximal). Therefore, the descendants of biclusters (b) and (c), whose extents are contained in the shared rows, need to check if they are row-maximal. Suppose that $RM$ is the set of rows that the bicluster $(A,B)$ must check to find out if it is row-maximal. The bicluster $(A, B)$ is not row-maximal if there is an object $g \in RM$ that we can add to the bicluster and it remains a valid CVC bicluster, i.e.,

\begin{equation}
  \exists g \in RM | [\max_{i \in \{A \cup \{g\}\}}(d_{ij}) - \min_{i \in \{A \cup \{g\}\}}(d_{ij}) \leq \epsilon], \forall \; j \in B.
	\label{eq:cvc_ismaximal}
\end{equation}

\begin{figure}
  \centering 
  \includegraphics[trim=2cm 14.5cm 13cm 2.5cm, clip, scale=0.75]{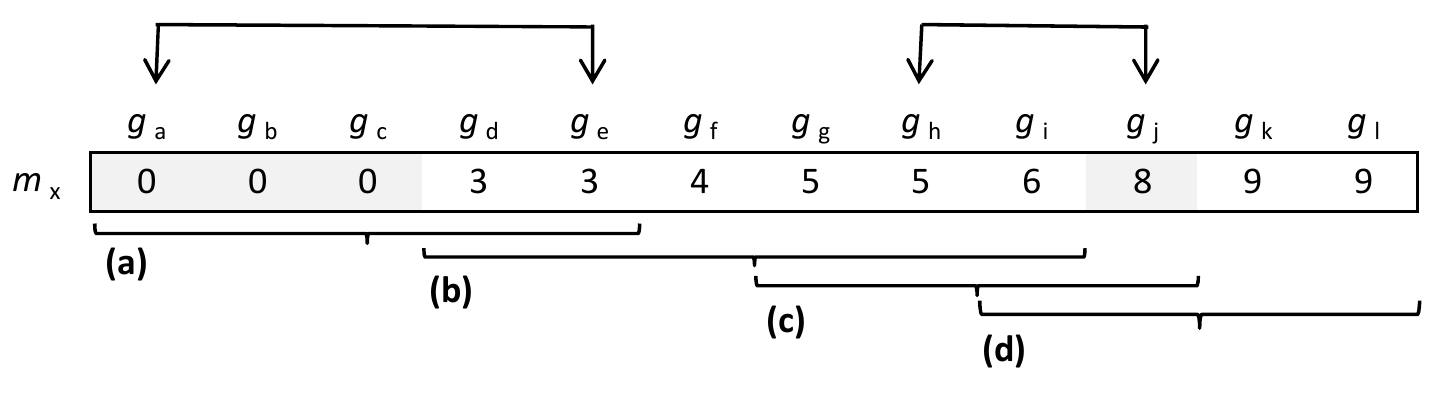}
  \caption{Example of how to find $RM$ (considering $\epsilon = 3$ and $minRow = 2$).}
  \label{fig:RM}
\end{figure}

To explain how to compute $RM$, let us see the example in Fig.~\ref{fig:RM}, which considers $\epsilon = 3$ and $minRow = 2$. Suppose that when adding attribute $m_x$, a bicluster generated four biclusters: (a), (b), (c) and (d), whose extents are highlighted in Fig.~\ref{fig:RM}. Let us compute the set of rows $RM$ that the descendants of the bicluster (b) must check to verify their maximality, i.e., $RM_{(b)}$. As the problem occurs when biclusters share $minRow$ rows or more in their extents, the pivot elements to compute $RM_{(b)}$ are $g_e$ and $g_h$ because they are the $minRow$-$th$ first and last element of (b), respectively. Their values are $g_e = 3$ and $g_h = 5$. Rows with values greater than or equal to 0 ($g_e - \epsilon$) or less than or equal to 8 ($g_h + \epsilon$) must comprise $RM_{(b)}$, so $RM_{(b)} = \{g_a, g_b, g_c, g_j\}$.

Back to our example in Fig.~\ref{fig:variosFilhos2}, $RM_{(b)} = \{g_3, g_4, g_{20}, g_{23}\}$. So, all descendants of the bicluster (b) with extent  $A \subseteq \{g_8, g_9, g_{10}, g_{16}, g_{19}, g_{22}\}$ must test the rows in $RM_{(b)}$ to verify if they are row-maximal. For simplicity, the result will be correct if we implement this verification for all descendants of a bicluster.

It is very important to note that biclusters also need to inherit the set $RM$ of their parents. For instance, suppose that the bicluster (b) of Fig.~\ref{fig:variosFilhos2} gives rise to a bicluster $(A_x = \{g_8, g_9, g_{19}, g_{22}\}, B_x)$. So, we must have $RM_x \supseteq RM_{(b)}$ because the descendants of $(A_x, B_x)$ must test the rows in $RM_{(b)}$ to verify if they are maximal.

\linespread{1}

\begin{algorithm}
\caption{RIn-Close\_CVC}
\label{alg:rinc_cvc}
\begin{algorithmic}[1]
  \small
	\REQUIRE Data matrix $\mathbf{D}_{n \times m}$, minimum number of rows $minRow$, index of the bicluster to be closed $r$, index of the initial attribute $y$, similarity constraint $\epsilon$
	\STATE $J \leftarrow \{\}$
	\STATE $R \leftarrow \{\}$
  \FOR{$j \leftarrow y$ to $m$}
	  \IF{$j \notin B_r$}
		  \IF{$\max_{i \in A_r}(d_{ij}) - \min_{i \in A_r}(d_{ij}) \leq \epsilon$}
			  \STATE $B_r \leftarrow B_r \cup \{j\}$
			\ELSE
			  \STATE Compute the possible extents
			  \FOR{each possible extent $RW$}
					\IF{$|RW| \geq minRow$ \AND $RW$ is canonical \AND $RW$ is row-maximal}
					  \STATE Sort the elements of $RW$
						\STATE $r_{new} \leftarrow r_{new} + 1$
						\STATE $R \leftarrow R \cup \{r_{new}\}$
						\STATE $J \leftarrow J \cup \{j\}$
						\STATE $A_{r_{new}} \leftarrow RW$
						\STATE Set $RM_{r_{new}}$
						\STATE Update the symbol table
					\ENDIF					
				\ENDFOR
			\ENDIF
		\ENDIF
	\ENDFOR
	\FOR{$k \leftarrow 1$ to $|J|$}
	  \STATE $B_{R_k} \leftarrow B_r \cup \{J_k\}$
		\STATE RIn-Close\_CVC$(\mathbf{D}, minRow, R_k, J_k+1, \epsilon)$
	\ENDFOR
\end{algorithmic}
\end{algorithm}

\linespread{1.5}

Algorithm~\ref{alg:rinc_cvc} shows the pseudocode of RIn-Close\_CVC. The worst-case time of RIn-Close\_CVC is $O(kmn(mn + x))$, where $x$ is the worst-case time of searching in the symbol table, so $x = O(\log k)$ for BSTs and $x = O(k)$ for HTs. But recall that HTs have a much better computational cost on average: $O(1)$. 

\subsection{Finding biclusters with coherent values (CHV)}

Once again, we will first show how to enumerate perfect CHV biclusters. We named this algorithm RIn-Close\_CHV\_P. It it is very similar to RIn-Close\_CVC\_P. Secondly, we will show how to enumerate non-perfect CHV biclusters. We named this algorithm RIn-Close\_CHV.

\subsubsection{Perfect Biclusters}
\label{secsec:chvp}

When we are looking for CVC or CVR biclusters, we look directly to the values of the data matrix. But when we are looking for CHV biclusters, we need to check if there is coherence (additive or multiplicative) between each pair of columns (or rows) of the data matrix. For this, in RIn-Close\_CHV\_P, a bicluster starts with one column in its intent, which we call \textit{pivot column}. Then, RIn-Close\_CHV\_P mines all biclusters that have this pivot column in their intents. Algorithm~\ref{alg:rinc_chv_p_main} shows this procedure. At the first iteration of the loop, RIn-Close\_CHV\_P will find all biclusters that have column 1 in their intents; at the second iteration, RIn-Close\_CHV\_P will find all biclusters that have column 2 and do not have column 1 in their intents; at the third iteration, RIn-Close\_CHV\_P will find all biclusters that have column 3 and do not have column 1 and 2 in their intents; and so on.

RIn-Close\_CHV\_P exploits the fact that for mining perfect CHV biclusters it is not necessary to check if there is coherence between all pair of columns in an intent. Note that in RIn-Close\_CHV\_P pseudocode, Algorithm~\ref{alg:rinc_chv_p}, we just compute the difference between the current attribute $j$ and the pivot column of the current intent $B_r$, i.e., $B_{r_1}$. If the current attribute $j$ matches perfectly the pivot column, it will match perfectly the other columns of the intent $B_r$ as well.

\linespread{1}

\begin{algorithm}
\caption{Main\_RIn-Close\_CHV\_P}
\label{alg:rinc_chv_p_main}
\begin{algorithmic}[1]
	\small
	\REQUIRE Data matrix $\mathbf{D}_{n \times m}$, minimum number of rows $minRow$
  \STATE $r_{new} \leftarrow 0$ \COMMENT{global variable}
	\FOR{$atr \leftarrow 1$ to $m-1$}
	  \STATE $r_{new} \leftarrow r_{new} + 1$
    \STATE $A_{r_{new}} \leftarrow \{1, ..., n\}$
	  \STATE $B_{r_{new}} \leftarrow \{atr\}$
    \STATE RIn-Close\_CHV\_P$(\mathbf{D}, minRow, r_{new}, atr+1)$
	\ENDFOR
\end{algorithmic}
\end{algorithm}

\linespread{1.5}

\linespread{1}

\begin{algorithm}
\caption{RIn-Close\_CHV\_P}
\label{alg:rinc_chv_p}
\begin{algorithmic}[1]
  \small
	\REQUIRE Data matrix $\mathbf{D}_{n \times m}$, minimum number of rows $minRow$, index of the bicluster to be closed $r$, index of the initial attribute $y$
	\STATE $J \leftarrow \{\}$
	\STATE $R \leftarrow \{\}$
  \FOR{$j \leftarrow y$ to $m$}
	  \IF{$j \notin B_r$}
		  \STATE $Z \leftarrow \{d_{iB_{r_1}} - d_{ij}\}_{i \in A_r}$
			\IF{$\max(Z) - \min(Z) = 0$}
			  \STATE $B_r \leftarrow B_r \cup \{j\}$
			\ELSE
			  \STATE Compute the possible extents
			  \FOR{each possible extent $RW$}
					\IF{$|RW| \geq minRow$ \AND $RW$ is canonical}
						\STATE $r_{new} \leftarrow r_{new} + 1$
						\STATE $R \leftarrow R \cup \{r_{new}\}$
						\STATE $J \leftarrow J \cup \{j\}$
						\STATE $A_{r_{new}} \leftarrow RW$
					\ENDIF
				\ENDFOR
			\ENDIF
		\ENDIF
	\ENDFOR
	\FOR{$k \leftarrow 1$ to $|J|$}
	  \STATE $B_{R_k} \leftarrow B_r \cup \{J_k\}$
		\STATE RIn-Close\_CHV\_P$(\mathbf{D}, minRow, R_k, J_k+1)$
	\ENDFOR
\end{algorithmic}
\end{algorithm}

\linespread{1.5}

The test of canonicity is also essentially the same as in In-Close2. Let $B$ be the current intent, $j$ be the current attribute, and $RW$ be the extent of the new bicluster. It is not canonical if

\begin{equation}
	\exists k \in M \setminus B | [k < j] \: \wedge \: [\max(Z^{B_{r_1}k}) - \min(Z^{B_{r_1}k}) = 0],
	\label{eq:rinc_chv_p_iscan}
\end{equation}

\noindent where $Z^{B_{r_1}k} \leftarrow \{d_{iB_{r_1}} - d_{ik}\}_{i \in RW}$, i.e., if there is an attribute $k < j$ that we can add to the bicluster and it remains a valid perfect CHV bicluster.

The worst-case time of RIn-Close\_CHV\_P is the same as that of RIn-Close\_CVC\_P: $O(knm(\log n + m))$.

\subsubsection{Non-Perfect Biclusters}
\label{secsec:chv}

Now, we will explain how to perform a complete, correct and non-redundant enumeration of all maximal perturbed CHV biclusters, given a similarity constraint determined by the user-defined parameter $\epsilon$ ($\epsilon > 0$), as presented in Eq.~(\ref{eq:chvabic}). 

To achieve this goal, we can not simply apply to RIn-Close\_CHV\_P the same adaptations that we have made in RIn-Close\_CVC\_P to achieve RIn-Close\_CVC. First of all, if RIn-Close\_CHV computed the set $Z$ considering only the pivot column $B_{r_1}$ and the current attribute $j$, it could occur a difference up to $2\epsilon$ between any other two columns of $B_r$. Besides, the order of choice of the pivot columns interfere in the outcome in this scenario. In this way RIn-Close\_CHV would yield just an approximate result of an actual enumeration (as SeqClus \cite{WangEtAl2004} and CPT \cite{GuanEtAL2009} do), and this is not the case here. Also, RIn-Close\_CHV could not simply verify if the current attribute $j$ fits all the attributes in the current intent $B_r$. An example of a problem that could happen is that: if the data matrix has two biclusters with the same extent $A$, but different intents $B_x = \{m_1, m_3, m_5\}$ and $B_y = \{m_1, m_3, m_6, m_8\}$, a naive procedure would find just the first one because it loops through the attributes in its sequential order and the attribute $m_5$ is not coherent with attributes $m_6$ and $m_8$ (considering the rows in extent $A$). Again, this is not the case here. In addition, it would be quite difficult to define the possible new extents. Another challenging issue of this approach is to determine when a bicluster is canonical or not. For instance, if the data matrix has two biclusters with the same extent $A$, but different intents $B_x = \{m_1, m_2, m_3, m_4\}$ and $B_y = \{m_2, m_3, m_4, m_5\}$, a naive procedure would discard the second because the attributes $m_2$, $m_3$ and $m_4$ are coherent with attribute $m_1$. Not to mention that a non-canonical bicluster could give rise to a canonical bicluster. 

\begin{figure*}
  \centering 
	
	\subfigure[All CVC biclusters found in the data matrix of Table~\ref{tab:augmented} (using $minRow = 2$ and $\epsilon = 1$).]{
		\includegraphics[trim=2cm 12.2cm 12cm 2.2cm, clip, scale=0.75]{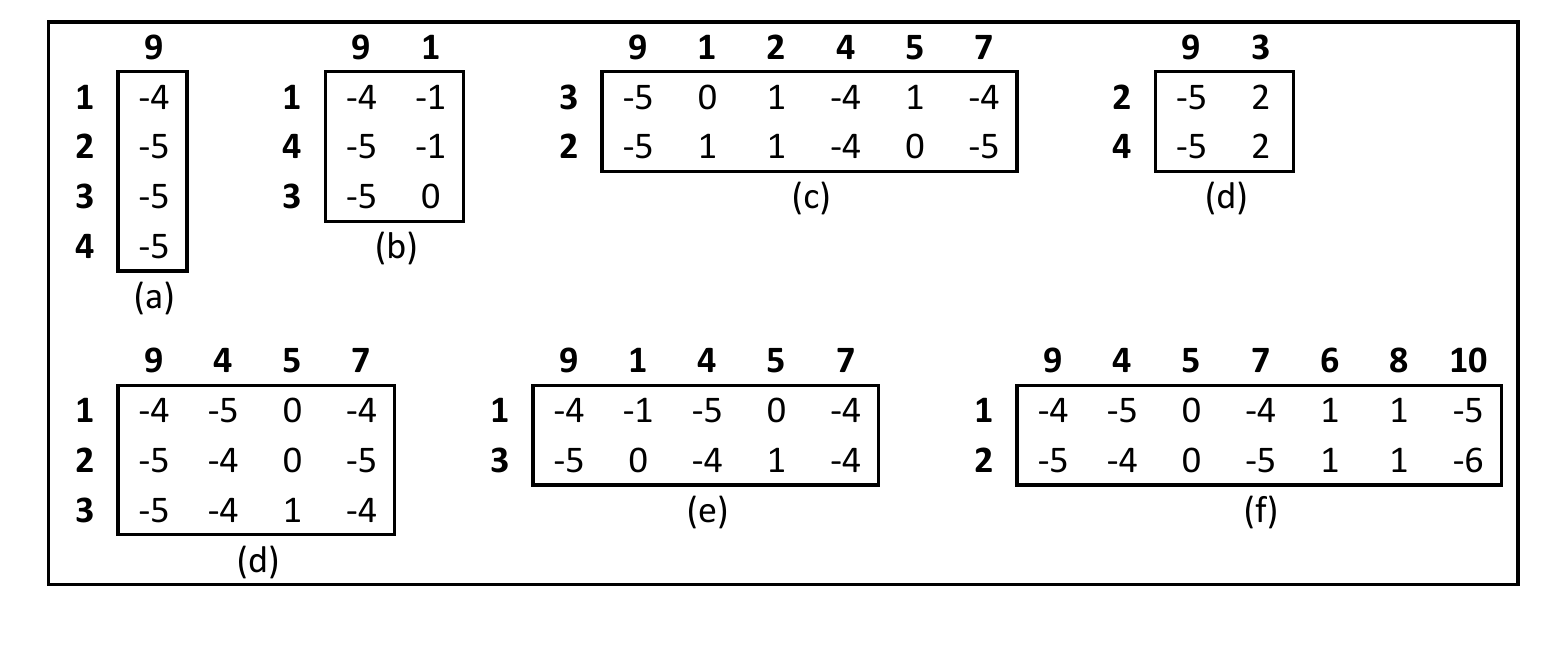}
		\label{fig:Bics_matAug}
	}
	\subfigure[Illustrative scheme to show how a CVC bicluster is processed by RIn-Close\_CHV.]{
		\includegraphics[trim=2.3cm 10cm 9.5cm 2cm, clip, scale=0.75]{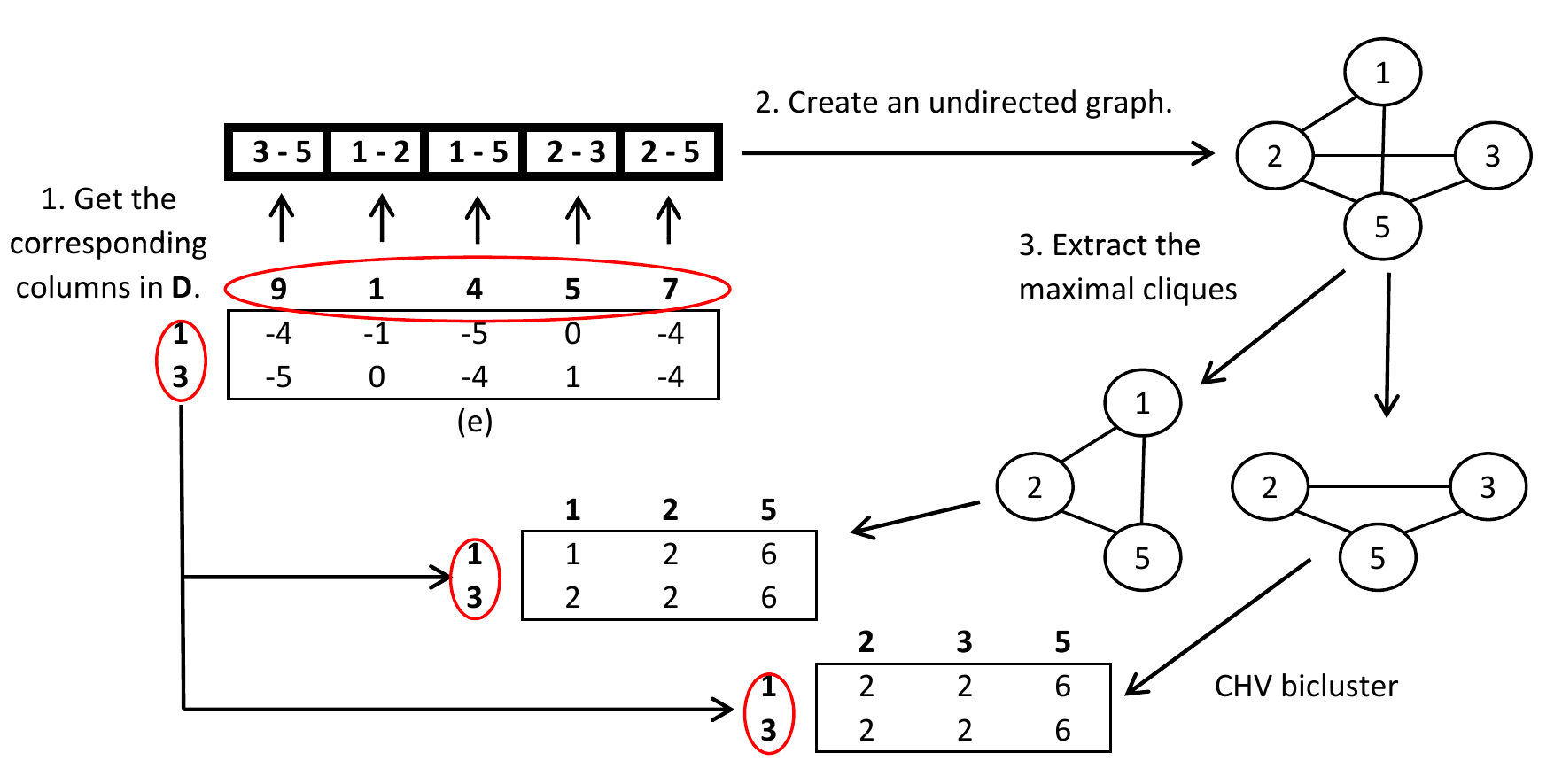}
		\label{fig:process}
	}
	\caption{RIn-Close\_CHV's framework.}
	\label{fig:RInCloseCHVfw}
\end{figure*}

To avoid all these undesired scenarios, RIn-Close\_CHV uses the following procedure with three steps:

\begin{enumerate}
	\item Compute the augmented matrix of the data matrix $\mathbf{D}$, denoted $\mathbf{D}_a$. $\mathbf{D}_a$ is a matrix with the difference between all pairs of columns of $\mathbf{D}$. For instance, the augmented matrix $\mathbf{D}_a$ of the data matrix $\mathbf{D}$ in Table~\ref{tab:exampleDataset} is illustrated in Table~\ref{tab:augmented}. The first column of $\mathbf{D}_a$ is the difference between columns 1 and 2 of $\mathbf{D}$, the second column of $\mathbf{D}_a$ is the difference between columns 1 and 3 of $\mathbf{D}$, the third column of $\mathbf{D}_a$ is the difference between columns 1 and 4 of $\mathbf{D}$, and so on for all combinations of pairs of columns. 
	\item Apply RIn-Close\_CVC to the augmented matrix $\mathbf{D}_a$. To illustrate, Fig.~\ref{fig:Bics_matAug} shows all maximal CVC biclusters found by RIn-Close\_CVC when applied to the data matrix of Table~\ref{tab:augmented} (using $minRow = 2$ and $\epsilon = 1$).
	\item Extract all maximal CHV biclusters from the maximal CVC biclusters found by RIn-Close\_CVC (see the pseudocode in Algorithm~\ref{alg:rinc_chv_process}).
\end{enumerate}

\linespread{1}

\begin{table}[htbp]
  \begin{minipage}[t]{.5\textwidth}
  \centering
  \caption{Example of a numerical dataset \cite{Besson2007})}
    \begin{tabular}{c|ccccc}
    \hline
          & $m_1$    & $m_2$    & $m_3$    & $m_4$    & $m_5$ \\
    \hline
    $g_1$    & 1     & 2     & 2     & 1     & 6 \\
    $g_2$    & 2     & 1     & 1     & 0     & 6 \\
    $g_3$    & 2     & 2     & 1     & 7     & 6 \\
    $g_4$    & 8     & 9     & 2     & 6     & 7 \\
    \hline
    \end{tabular}
  \label{tab:exampleDataset}
	\end{minipage}
	\begin{minipage}[t]{.5\textwidth}
  \centering
  \caption{Augmented matrix of the data matrix in Table~\ref{tab:exampleDataset}.}
    \begin{tabular}{r|rrrrrrrrrr}
    \hline
          & \textbf{1} & \textbf{2} & \textbf{3} & \textbf{4} & \textbf{5} & \textbf{6} & \textbf{7} & \textbf{8} & \textbf{9} & \textbf{10} \\
    \hline
    \textbf{1}     & -1    & -1    & 0     & -5    & 0     & 1     & -4    & 1     & -4    & -5 \\
    \textbf{2}     & 1     & 1     & 2     & -4    & 0     & 1     & -5    & 1     & -5    & -6 \\
    \textbf{3}     & 0     & 1     & -5    & -4    & 1     & -5    & -4    & -6    & -5    & 1 \\
    \textbf{4}     & -1    & 6     & 2     & 1     & 7     & 3     & 2     & -4    & -5    & -1 \\
    \hline
    \end{tabular}%
  \label{tab:augmented}%
	\end{minipage}
\end{table}

\linespread{1.5}

Algorithm~\ref{alg:rinc_chv_process} shows the pseudocode of the procedure to process each CVC bicluster. Steps in lines 2 to 5 are illustrated in Fig.~\ref{fig:process}. In this illustrative scheme, we are extracting CHV biclusters from the CVC bicluster (e) of Fig.~\ref{fig:Bics_matAug}. The first step is to get the corresponding columns in $\mathbf{D}$ of the intent of bicluster (e). For instance, column 9 of $\mathbf{D}_a$ corresponds to the difference between columns 3 and 5 of $\mathbf{D}$. Therefore, columns 3 and 5 are coherent with each order considering the extent $\{1, 3\}$ and $\epsilon=1$. Let us name this corresponding set of columns as $B2$. The second step is to create an undirected graph, in which the nodes represent $B2$, and the edges represent the columns that are coherent with each other. The third step is to find all maximal cliques from this undirected graph. Each one of these cliques indicates the subsets of $B2$ in which all columns are coherent with each other considering the user-defined parameter $\epsilon$. Thus, the CHV biclusters generated by the CVC bicluster (e) are: $(\{1,3\}, \{1, 2, 5\})$ and $(\{1, 3\}, \{2, 3, 5\})$. Lines 7 and 8 of the pseudocode in Algorithm~\ref{alg:rinc_chv_process} verify if the CHV bicluster $(C,D)$ is new and, if so, store it in the list of CHV biclusters. A CHV bicluster $(C,D)$ is new if (\textit{i}) its intent $D$ is equal to $B2$, or (\textit{ii}) $(C,D)$ is row-maximal (there is no object $g$ that we can add to its extent $C$), i.e.,

\begin{equation}
  \nexists g \in G \setminus C | [\max(Z^{jl}) - \min(Z^{jl}) \leq \epsilon], \forall j, l \in D,
	\label{eq:chv_ismaximal}
\end{equation}

\noindent where $Z^{jl} = \{d_{ij}-d_{il}\}_{i \in C \cup \{g\}}$. The worst-case time of verifying if a CHV bicluster is row-maximal is $O(mn)$. Makino and Uno \cite{MakinoEtAl2004} proved that all maximal cliques of a $v$ vertices graph can be enumerated with time delay $O(M(v))$, where $M(v)$ is the cost of multiplying two $v \times v$ matrices. As stated in Subsection~\ref{sec:rinclosecvcnp}, each CVC bicluster can be enumerated with time delay $O(m_an(m_an + x))$, where $m_a = m(m-1)/2$ is the number of columns of the augmented matrix $\mathbf{D}_a$.

\linespread{1}

\begin{algorithm}
\caption{Mining CHV biclusters from CVC biclusters}
\label{alg:rinc_chv_process}
\begin{algorithmic}[1]
  \small
	\REQUIRE List of CVC biclusters $lbCVC$
	\ENSURE List of CHV biclusters
	\FOR{\textbf{each} $(A,B)$ in $lbCVC$}
		\STATE Compute $B2$
		\STATE Create an undirected graph
		\STATE Find all maximal cliques
		\STATE Compute the CHV biclusters
		\FOR{\textbf{each} CHV bicluster $(C,D)$}
	    \IF{$D = B2$ \OR $(C,D)$ is row-maximal}
			  \STATE Store $(C,D)$ in the list of CHV biclusters
			\ENDIF
		\ENDFOR
	\ENDFOR
\end{algorithmic}
\end{algorithm}

\linespread{1.5}

\section{Experimental Results}
\label{sec:exp}

We evaluated the RIn-Close family of algorithms on both synthetic and real datasets. Our goals are to highlight various practicalities in the usage of RIn-Close, and to outline the advantages and distinct aspects of an enumerative algorithm when compared to heuristics.

We implemented RIn-Close using C++. For the third step of RIn-Close\_CHV, we implemented the BK algorithm \cite{BronKerbosch1973} with the I. Koch’s pivot selection strategy \cite{Koch2001}, because it is the best one in practice \cite{CazalsKarande2008}. For the heuristics, we used the MTBA toolbox \cite{VermaEtAl2013}. The only exception was ROCC's implementation that was obtained from the authors. The experiments were carried out on a PC Intel(R) Core(TM) i7-4770K CPU @ 3.5 GHz, 32 GB of RAM, and running under Ubuntu 14.04. The codes of RIn-Close algorithms are available at \url{https://sourceforge.net/projects/rinclose/}.

\subsection{RIn-Close's scalability}
\label{sec:exp_scalability}

This experiment aims to test RIn-Close's performance when varying ($i$) the number $n$ of rows of the dataset, ($ii$) the number $m$ of columns of the dataset, ($iii$) the number of biclusters in the dataset, ($iv$) the bicluster row size, ($v$) the bicluster column size, ($vi$) the overlap, and ($vii$) the noise in the dataset. For this purpose, we created synthetic datasets that let us embed biclusters and then test how RIn-Close performs when varying each one of these parameters in isolation. The default parameters used in the synthetic data generator were: $n = 5000$, $m = 60$, number of biclusters $= 10$, bicluster row size $= 200$, bicluster column size $= 8$, overlap $= 0.2$, and Gaussian noise with $\mu = 0$ and $\sigma = 0.01$. The synthetic data generator creates the biclusters and assigns random values to the other regions of the dataset. Then, it adds the Gaussian noise and shuffles the rows and columns of the dataset. Therefore, the generator creates arbitrarily positioned overlapping biclusters, so that the resulting biclusters are usually non-contiguous. For each configuration, we created 50 synthetic datasets to compute the average runtimes. RIn-Close\_CVC\_P and RIn-Close\_CHV\_P, that look for perfect biclusters, were applied to datasets without noise.

\begin{figure*}
  \centering
	\subfigure[]{
		\includegraphics[trim=0.2cm 0.1cm 0.6cm 0.4cm, clip, scale=0.3]{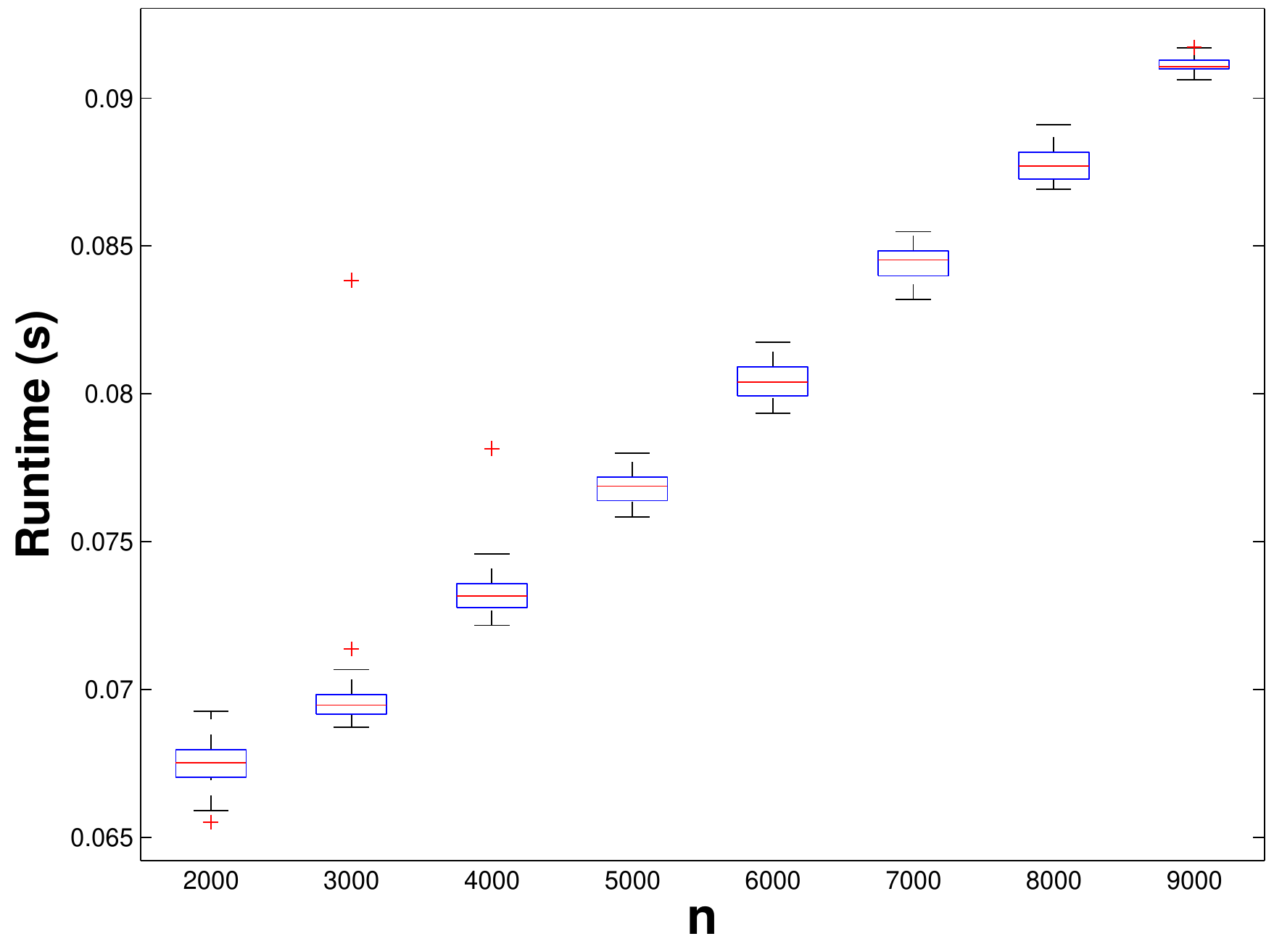}
	}
	\subfigure[]{
		\includegraphics[trim=0.2cm 0.1cm 0.6cm 0.4cm, clip, scale=0.3]{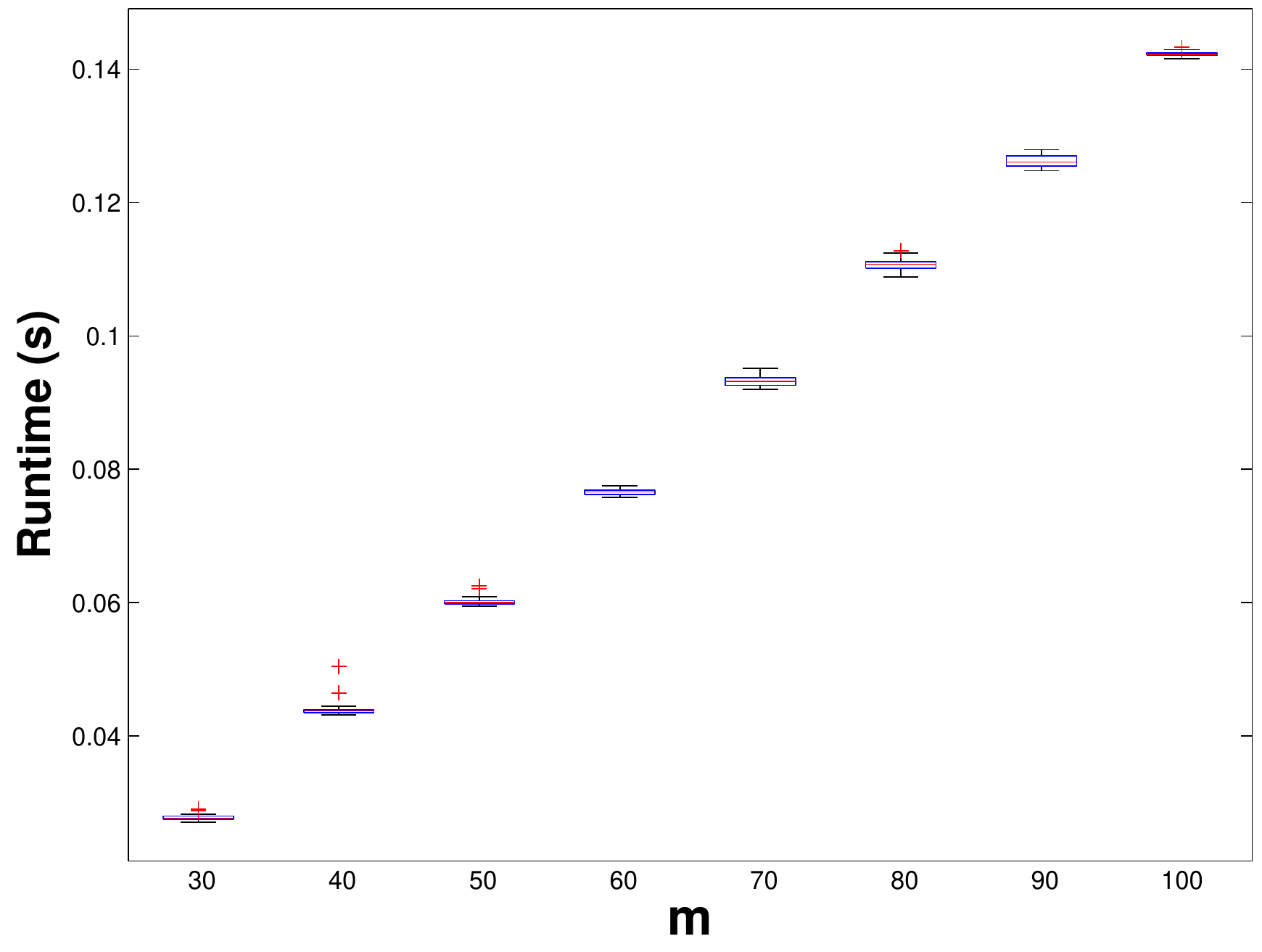}
	}
	\subfigure[]{
		\includegraphics[trim=0.2cm 0.1cm 0.6cm 0.4cm, clip, scale=0.3]{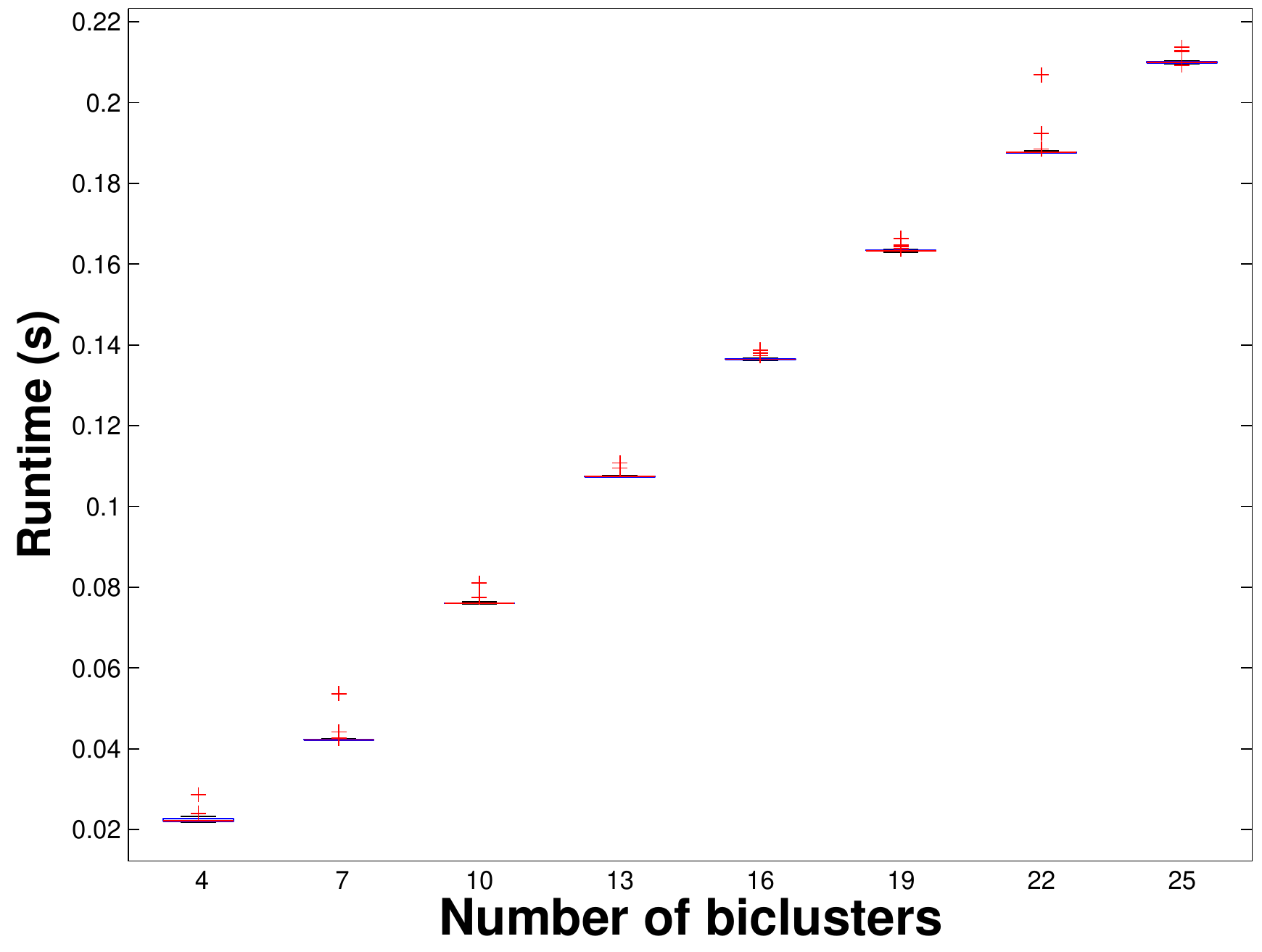}
	}
	\subfigure[]{
		\includegraphics[trim=0.2cm 0.1cm 0.6cm 0.4cm, clip, scale=0.3]{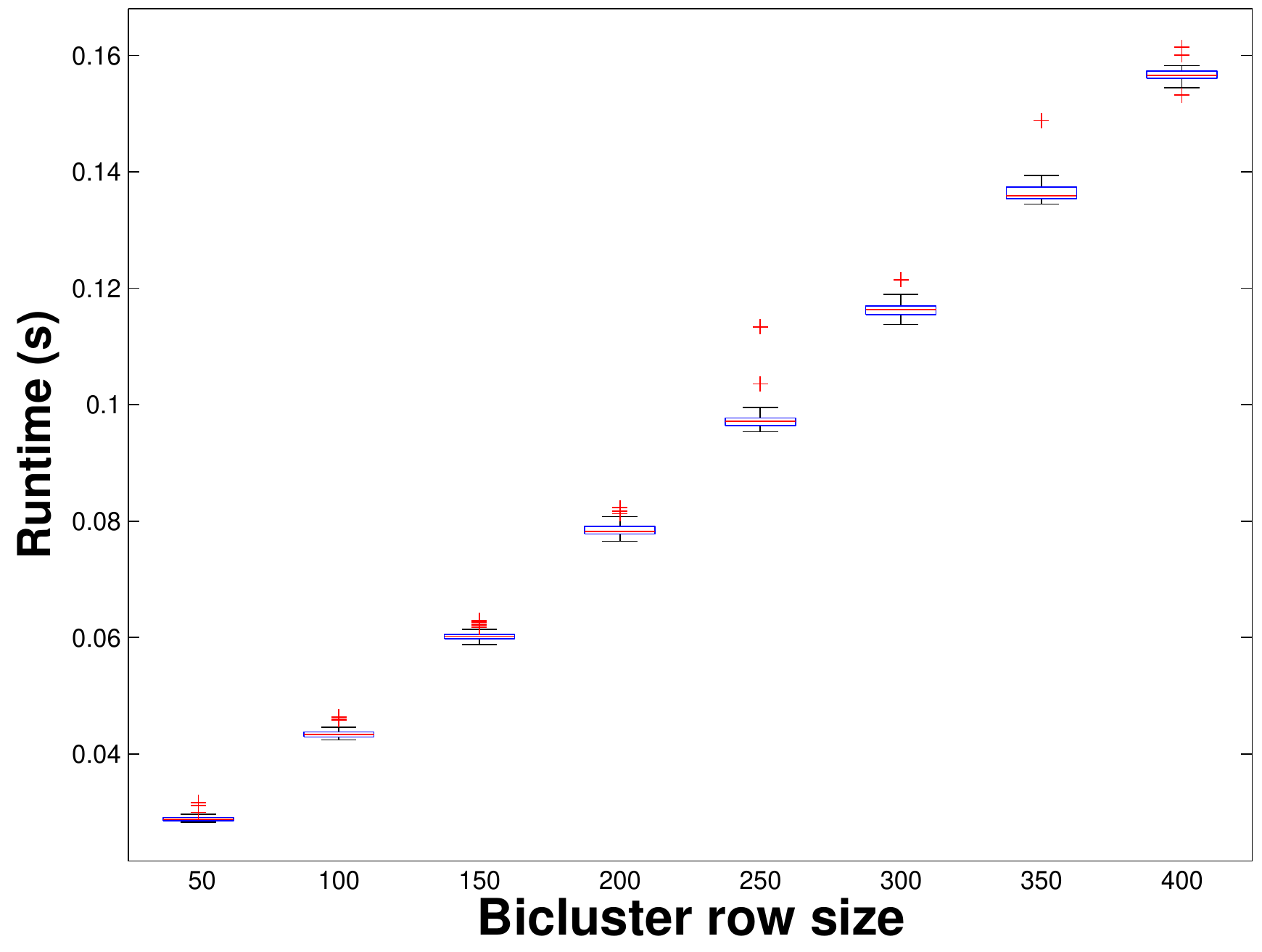}
	}
	\subfigure[]{
		\includegraphics[trim=0.2cm 0.1cm 0.6cm 0.4cm, clip, scale=0.3]{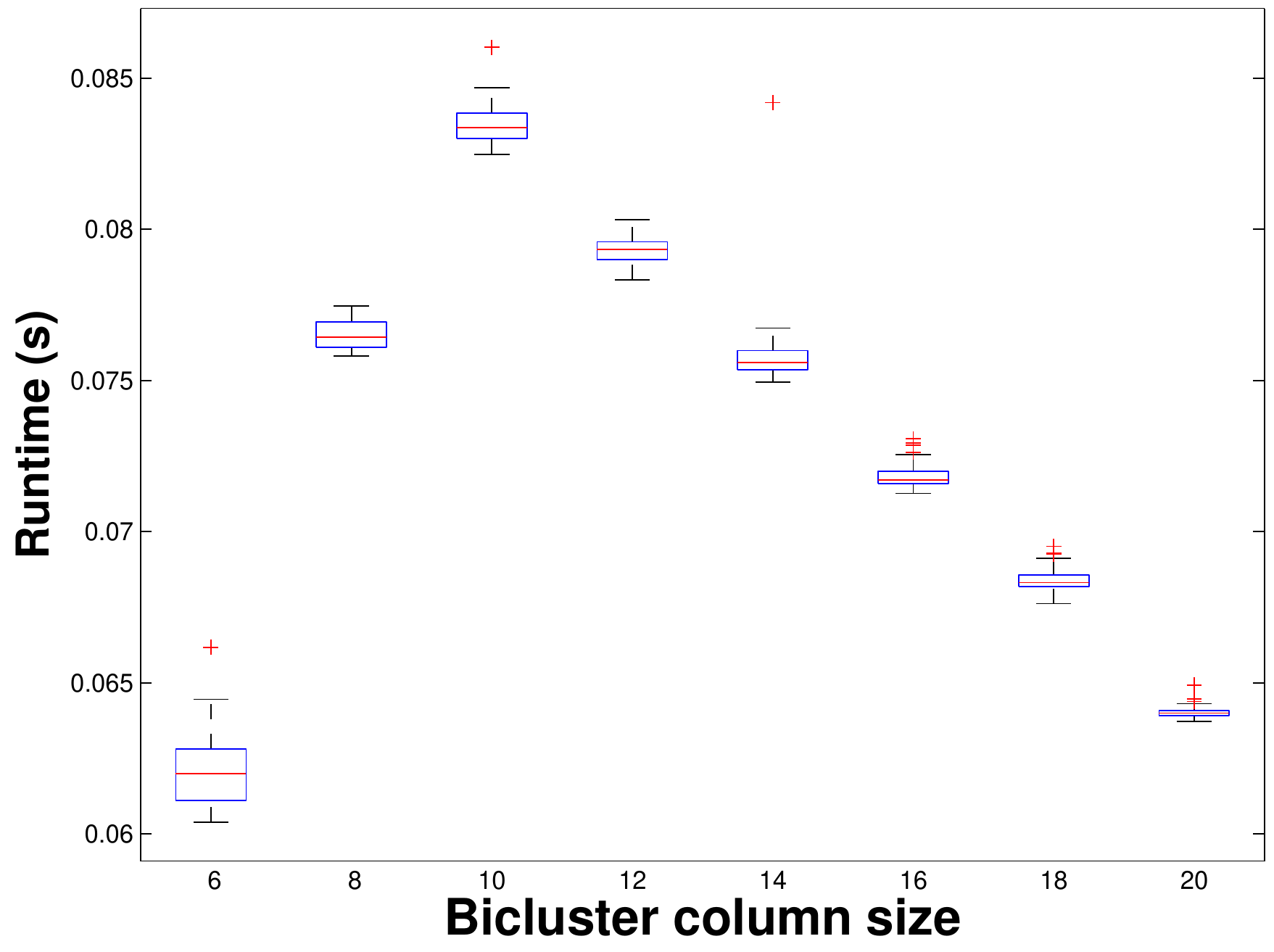}
	}
	\subfigure[]{
		\includegraphics[trim=0.2cm 0.1cm 0.6cm 0.4cm, clip, scale=0.3]{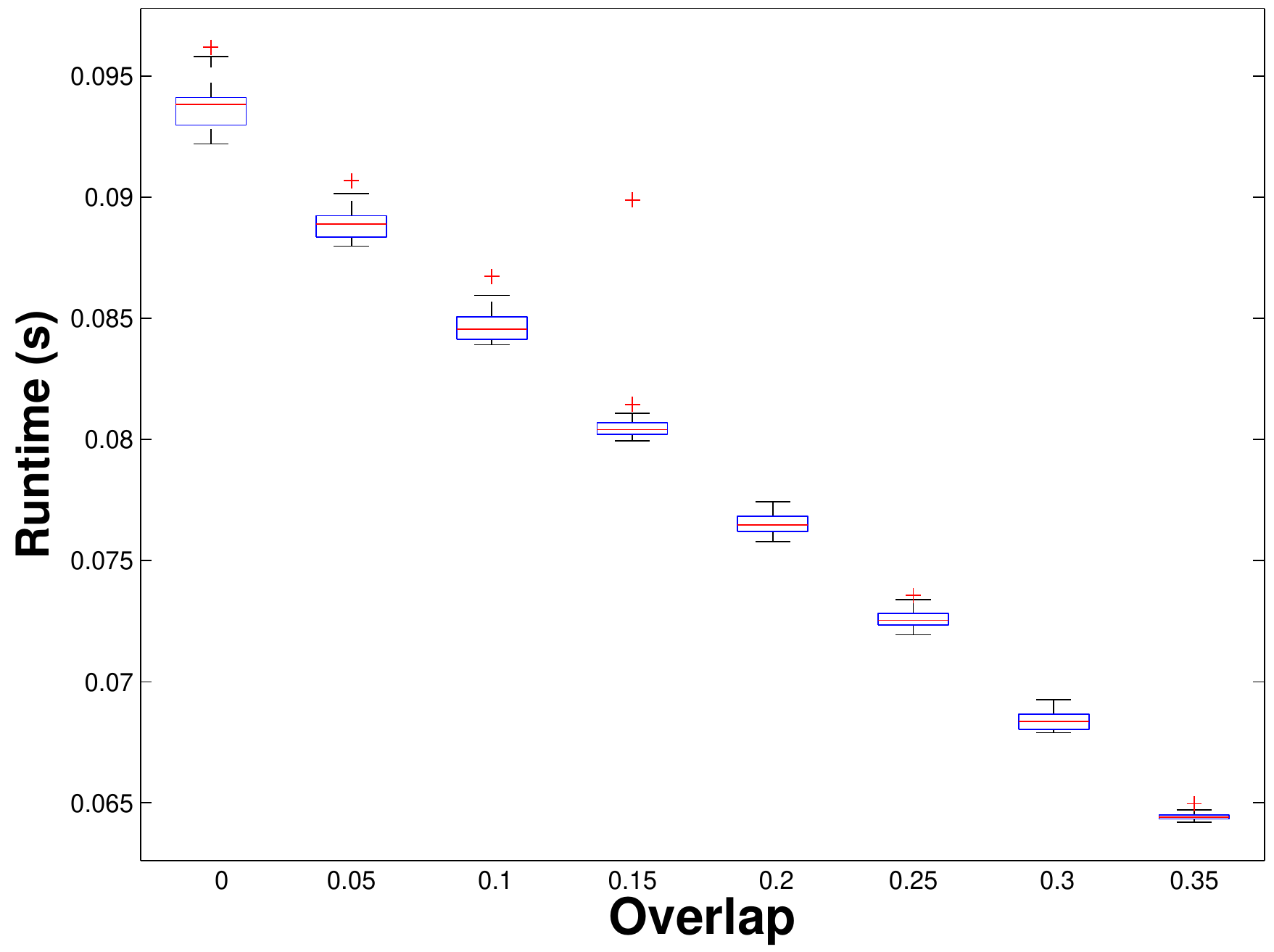}
	}
  \caption{Results of the performance of RIn-Close\_CVC\_P when varying (a) the number $n$ of rows of the dataset, (b) the number $m$ of columns of the dataset, (c) the number of biclusters in the dataset, (d) the bicluster row size, (e) the bicluster column size, and (f) the overlap.}
  \label{fig:expSynDataRTcvcp}
\end{figure*}

\begin{figure*}
  \centering
	\subfigure[]{
		\includegraphics[trim=0.2cm 0.1cm 0.6cm 0.4cm, clip, scale=0.3]{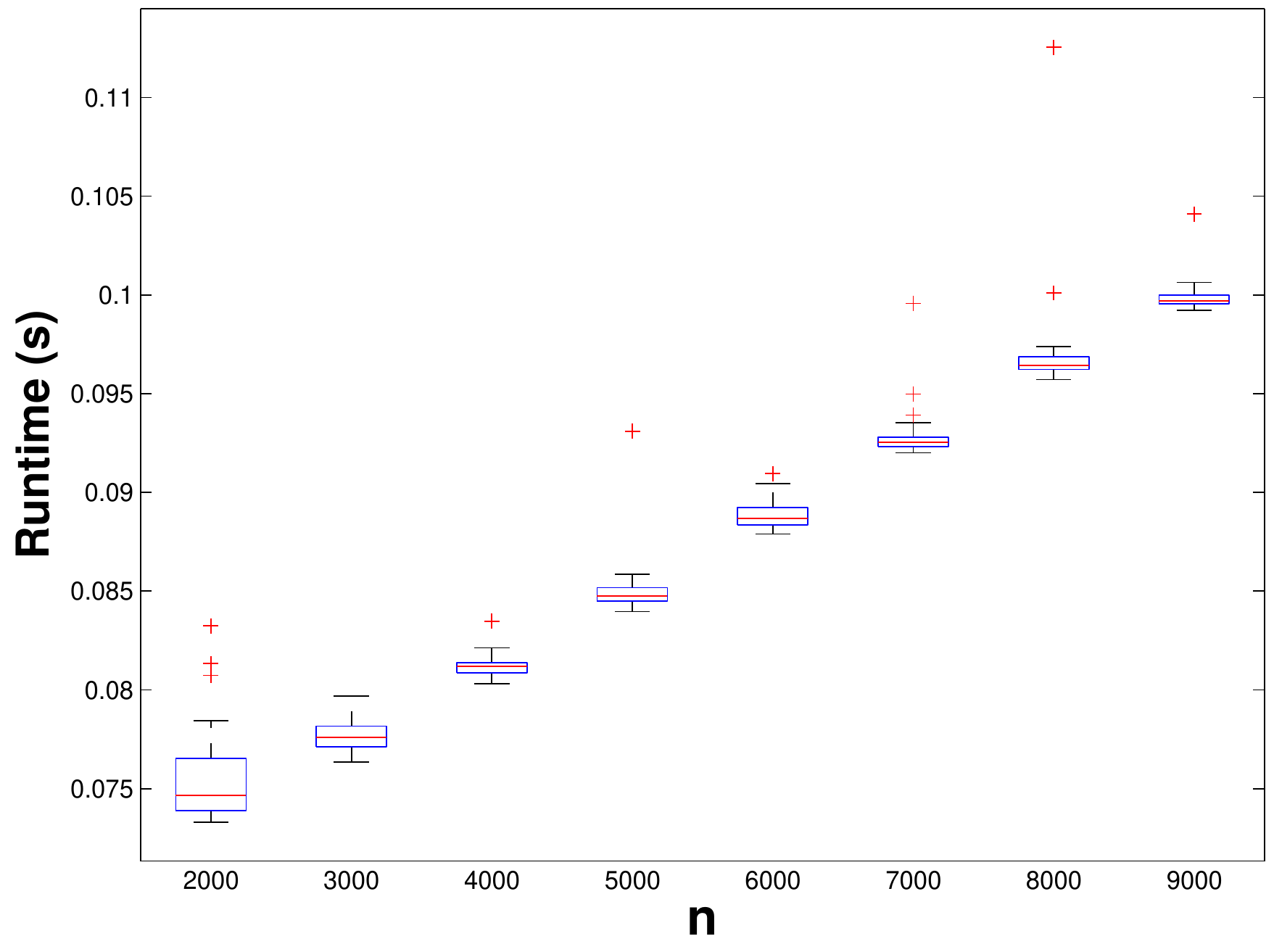}
	}
	\subfigure[]{
		\includegraphics[trim=0.2cm 0.1cm 0.6cm 0.4cm, clip, scale=0.3]{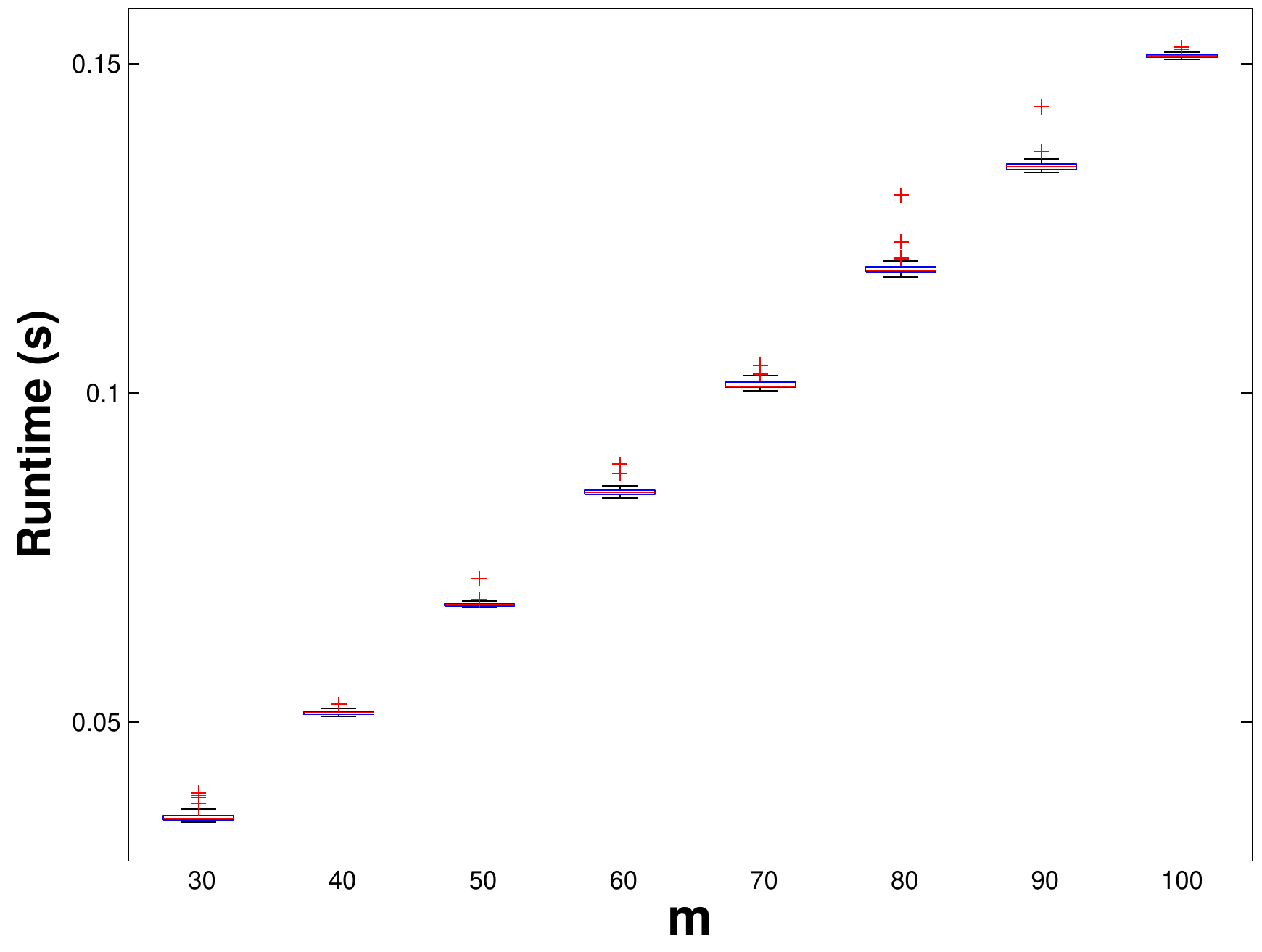}
	}
	\subfigure[]{
		\includegraphics[trim=0.2cm 0.1cm 0.6cm 0.4cm, clip, scale=0.3]{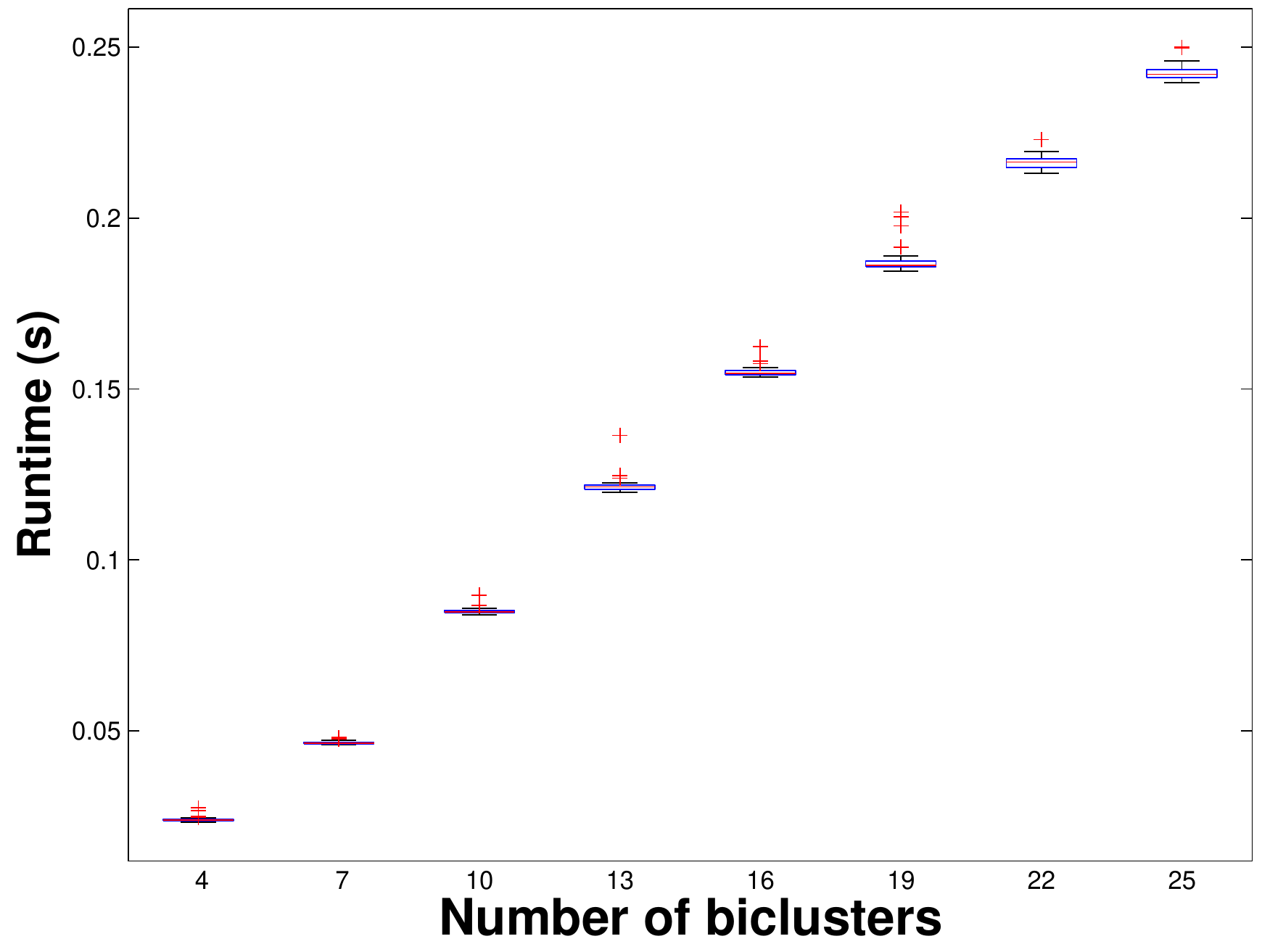}
	}
	\subfigure[]{
		\includegraphics[trim=0.2cm 0.1cm 0.6cm 0.4cm, clip, scale=0.3]{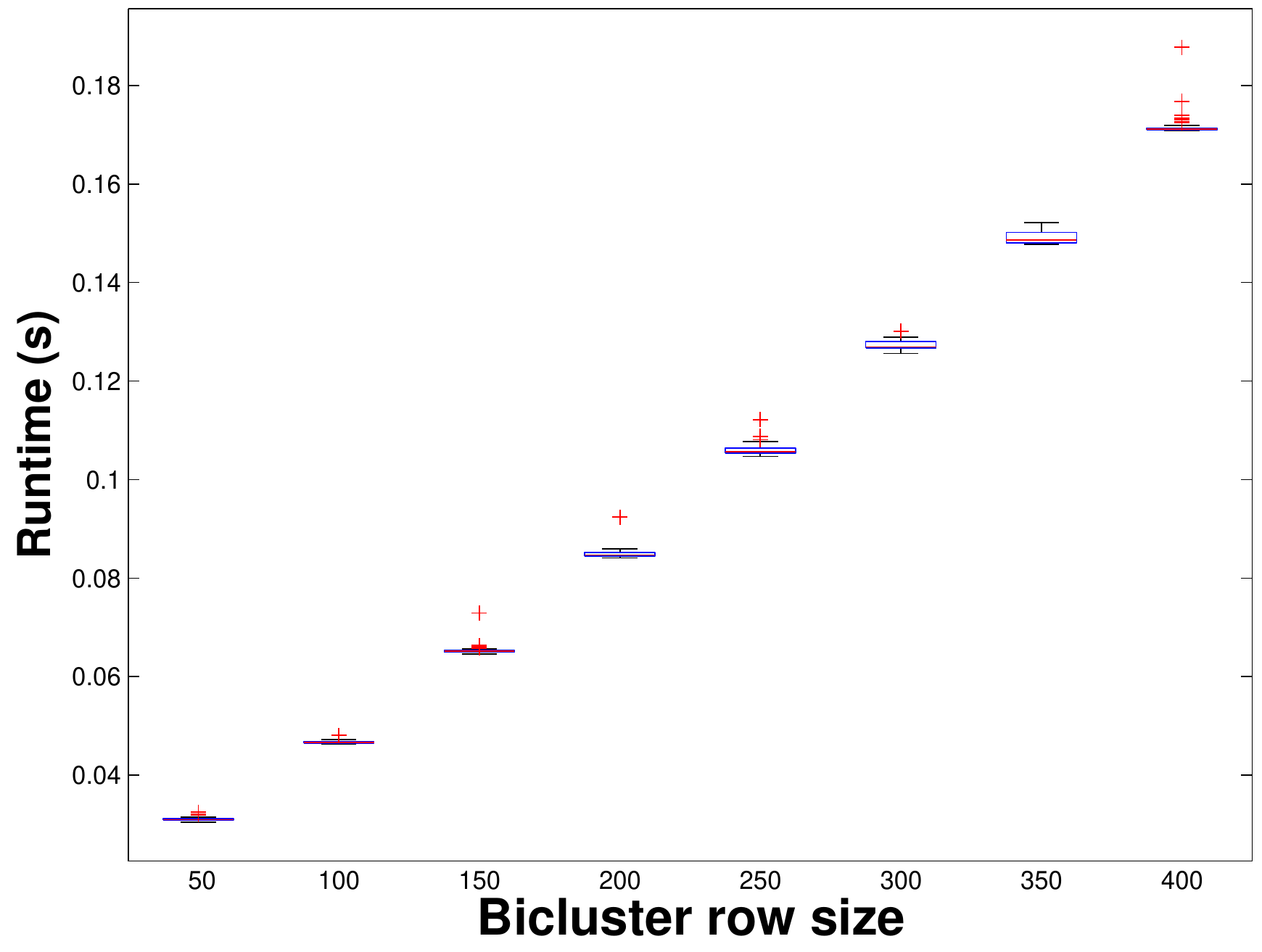}
	}
	\subfigure[]{
		\includegraphics[trim=0.2cm 0.1cm 0.6cm 0.4cm, clip, scale=0.3]{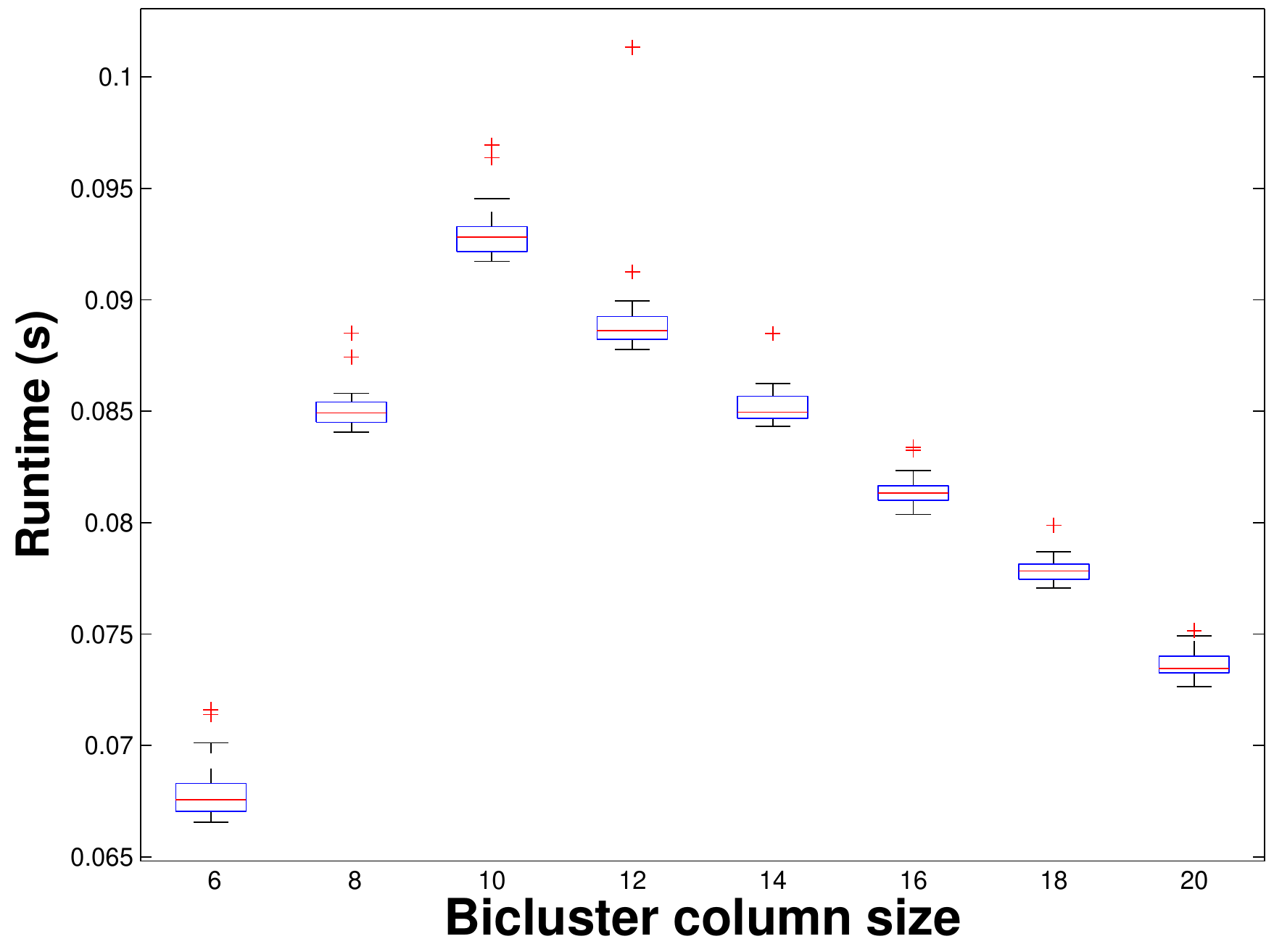}
	}
	\subfigure[]{
		\includegraphics[trim=0.2cm 0.1cm 0.6cm 0.4cm, clip, scale=0.3]{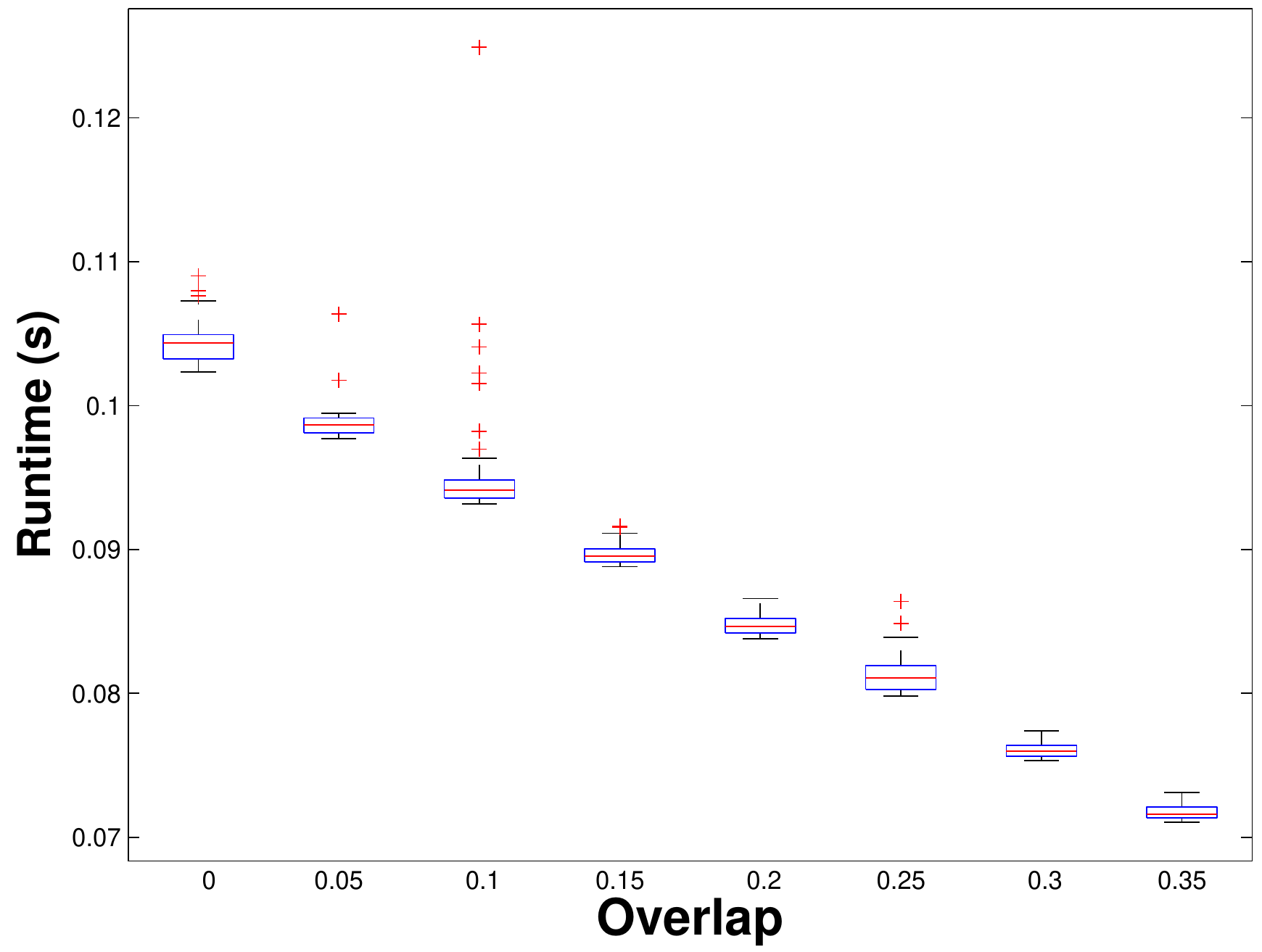}
	}
	\subfigure[]{
		\includegraphics[trim=0.2cm 0.1cm 0.6cm 0.4cm, clip, scale=0.3]{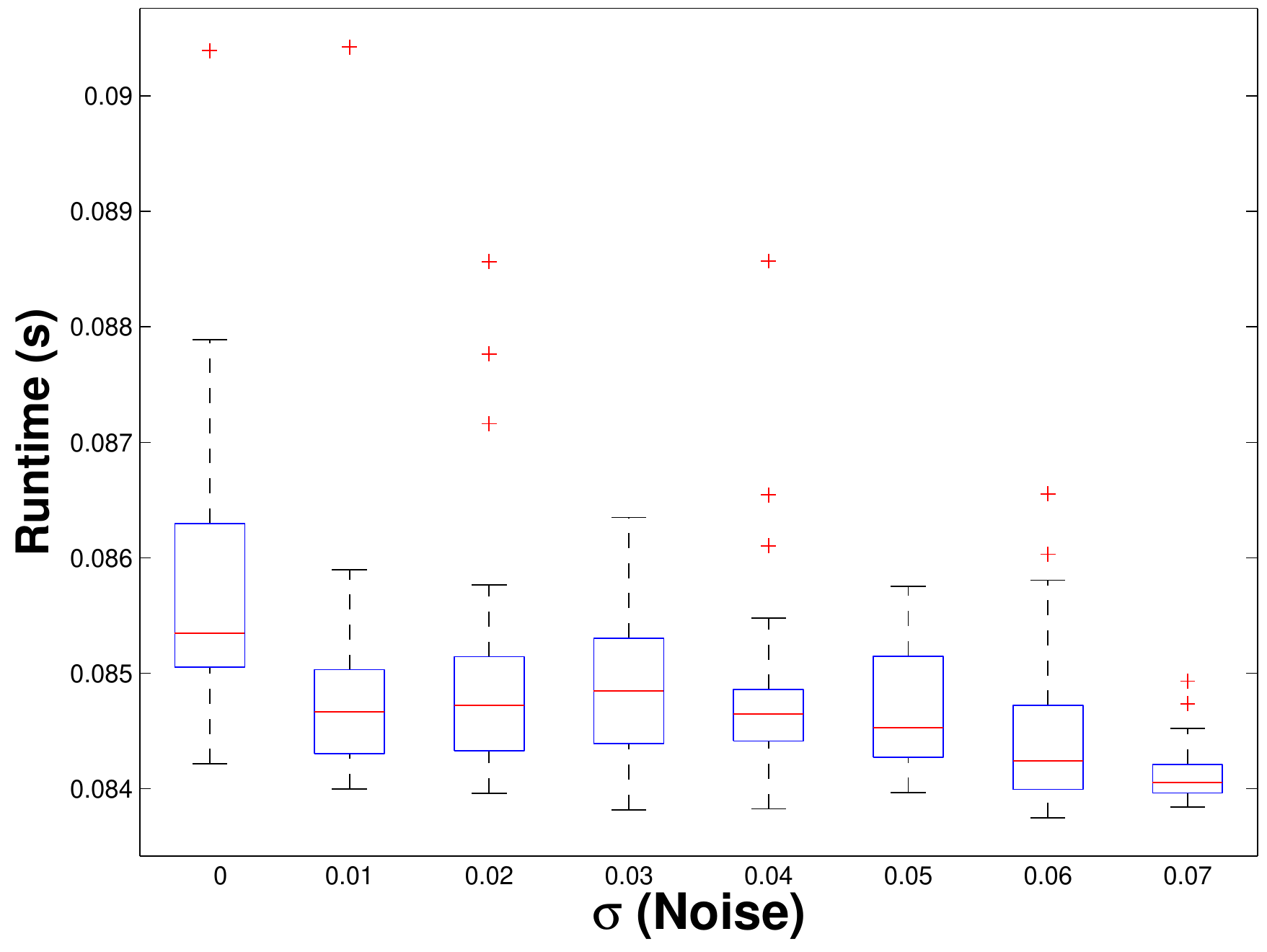}
	}
  \caption{Results of the performance of RIn-Close\_CVC when varying (a) the number $n$ of rows of the dataset, (b) the number $m$ of columns of the dataset, (c) the number of biclusters in the dataset, (d) the bicluster row size, (e) the bicluster column size, (f) the overlap, and (g) the noise.}
  \label{fig:expSynDataRTcvc}
\end{figure*}

\begin{figure*}
  \centering
	\subfigure[]{
		\includegraphics[trim=0.2cm 0.1cm 0.6cm 0.4cm, clip, scale=0.3]{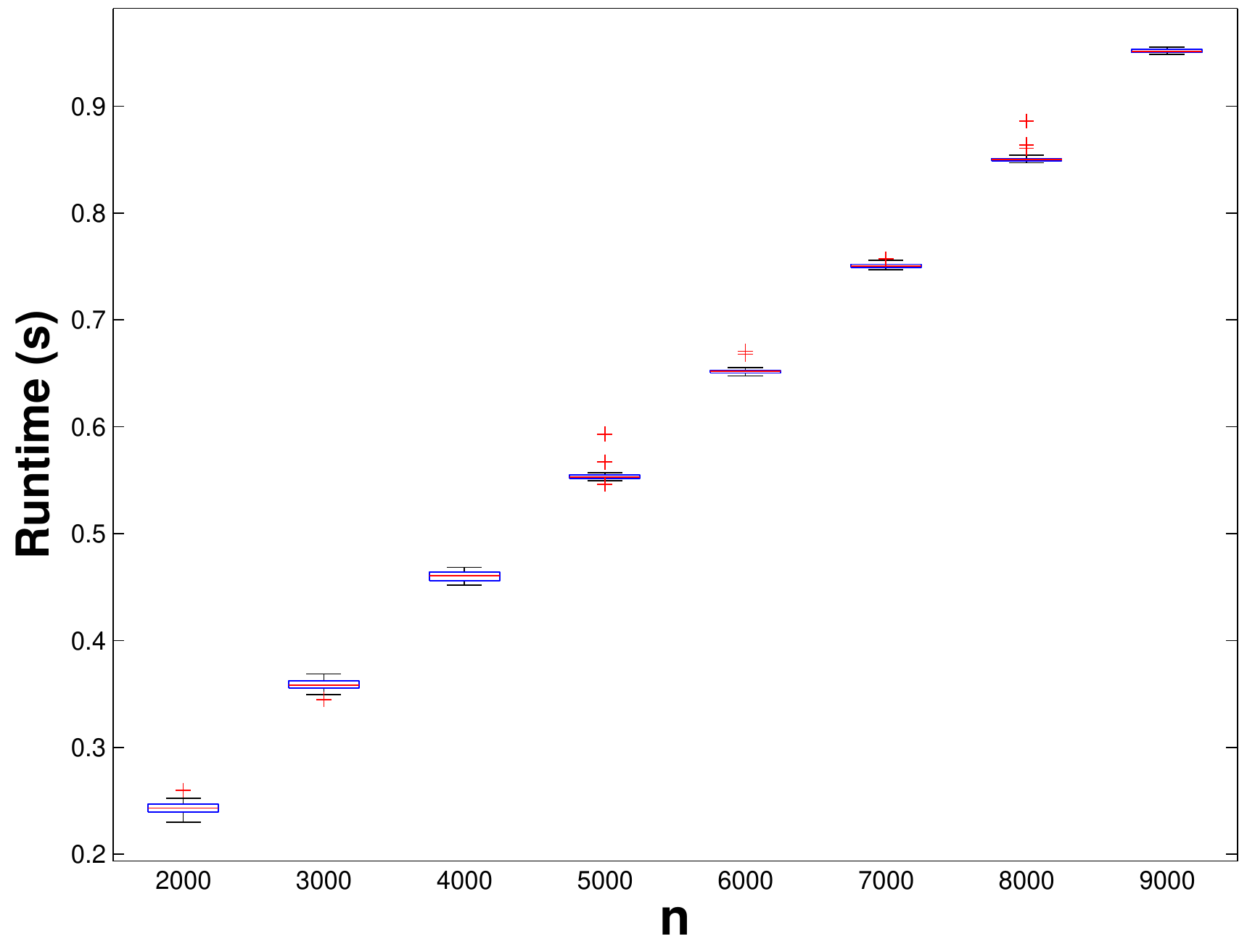}
	}
	\subfigure[]{
		\includegraphics[trim=0.2cm 0.1cm 0.6cm 0.4cm, clip, scale=0.3]{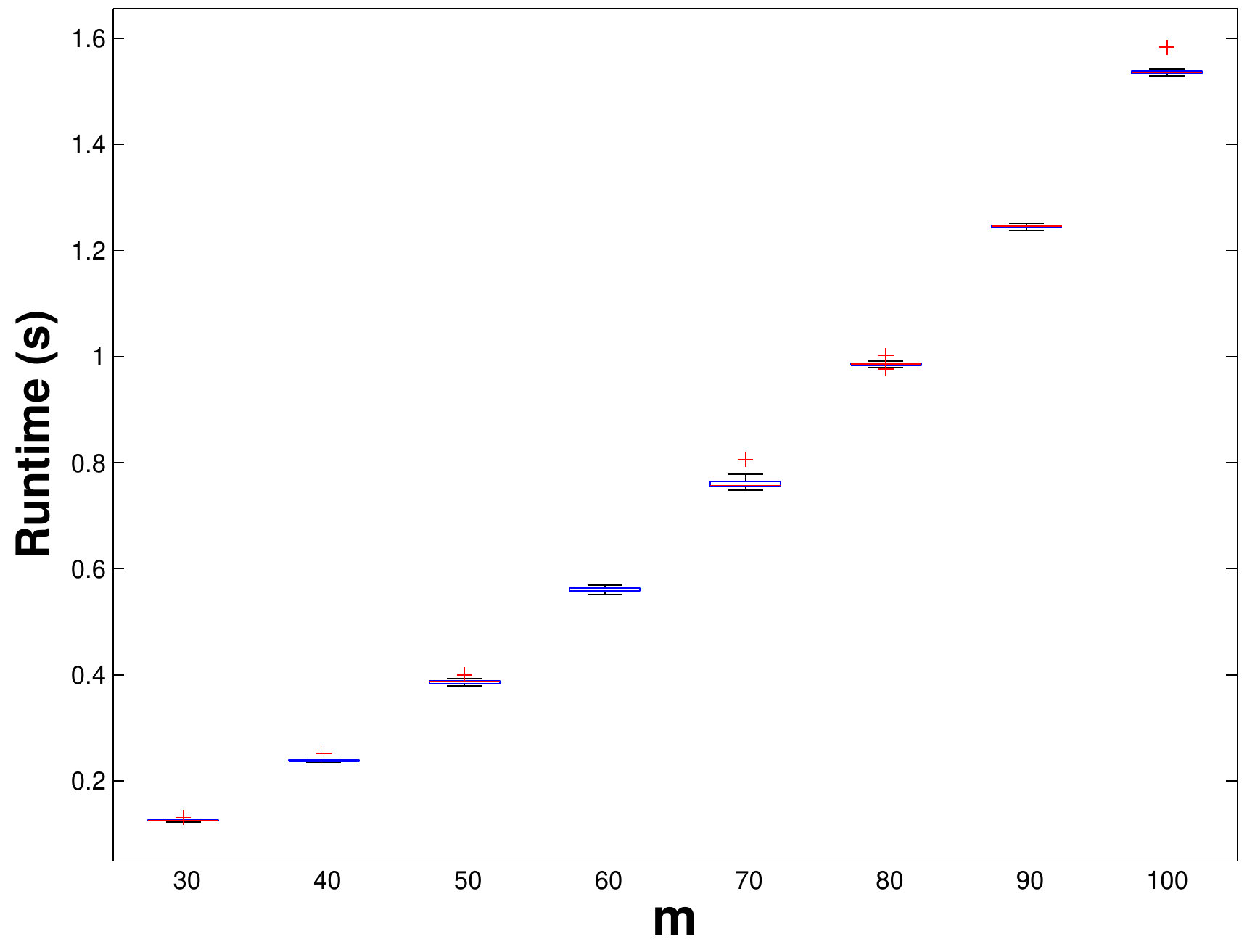}
	}
	\subfigure[]{
		\includegraphics[trim=0.2cm 0.1cm 0.6cm 0.4cm, clip, scale=0.3]{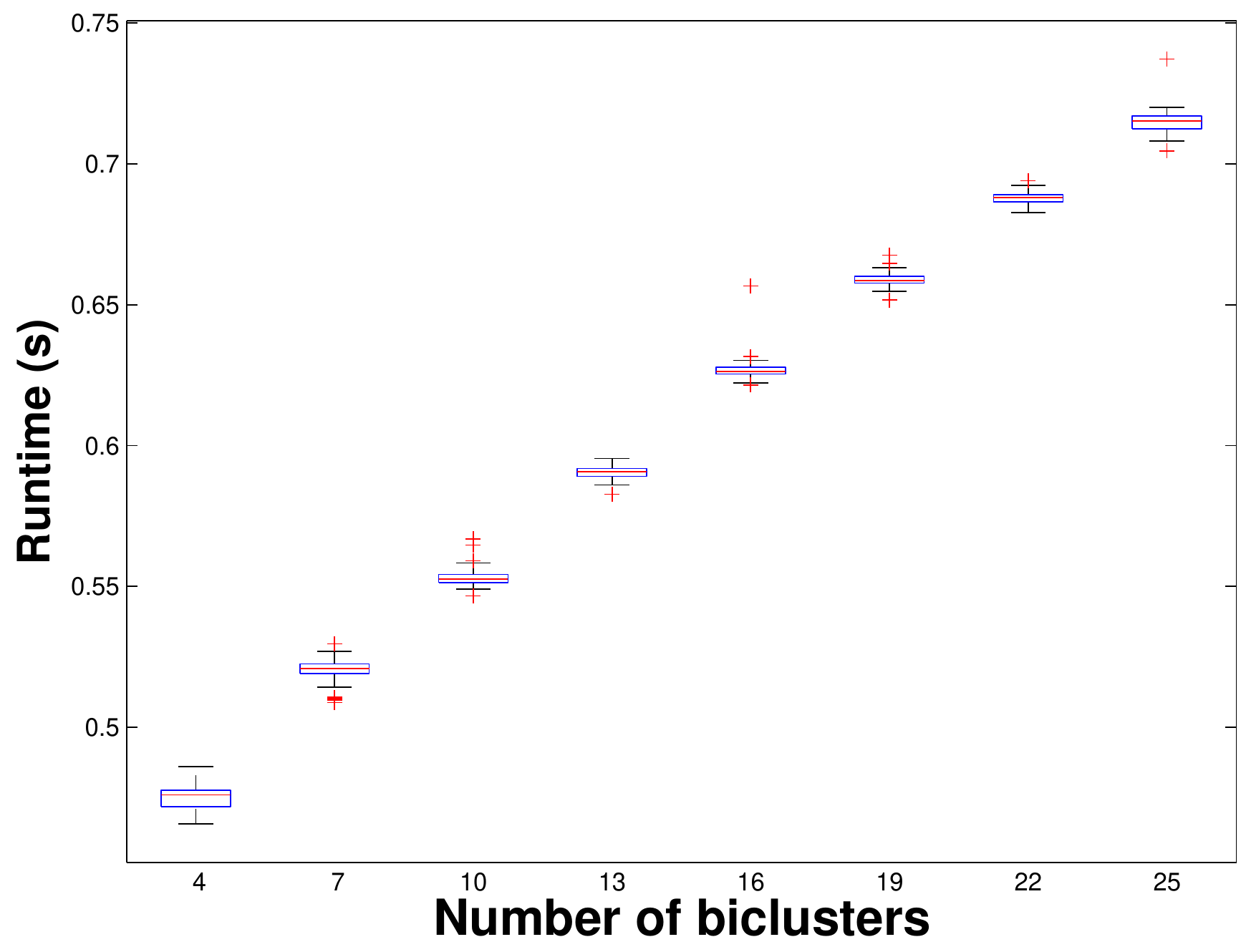}
	}
	\subfigure[]{
		\includegraphics[trim=0.2cm 0.1cm 0.6cm 0.4cm, clip, scale=0.3]{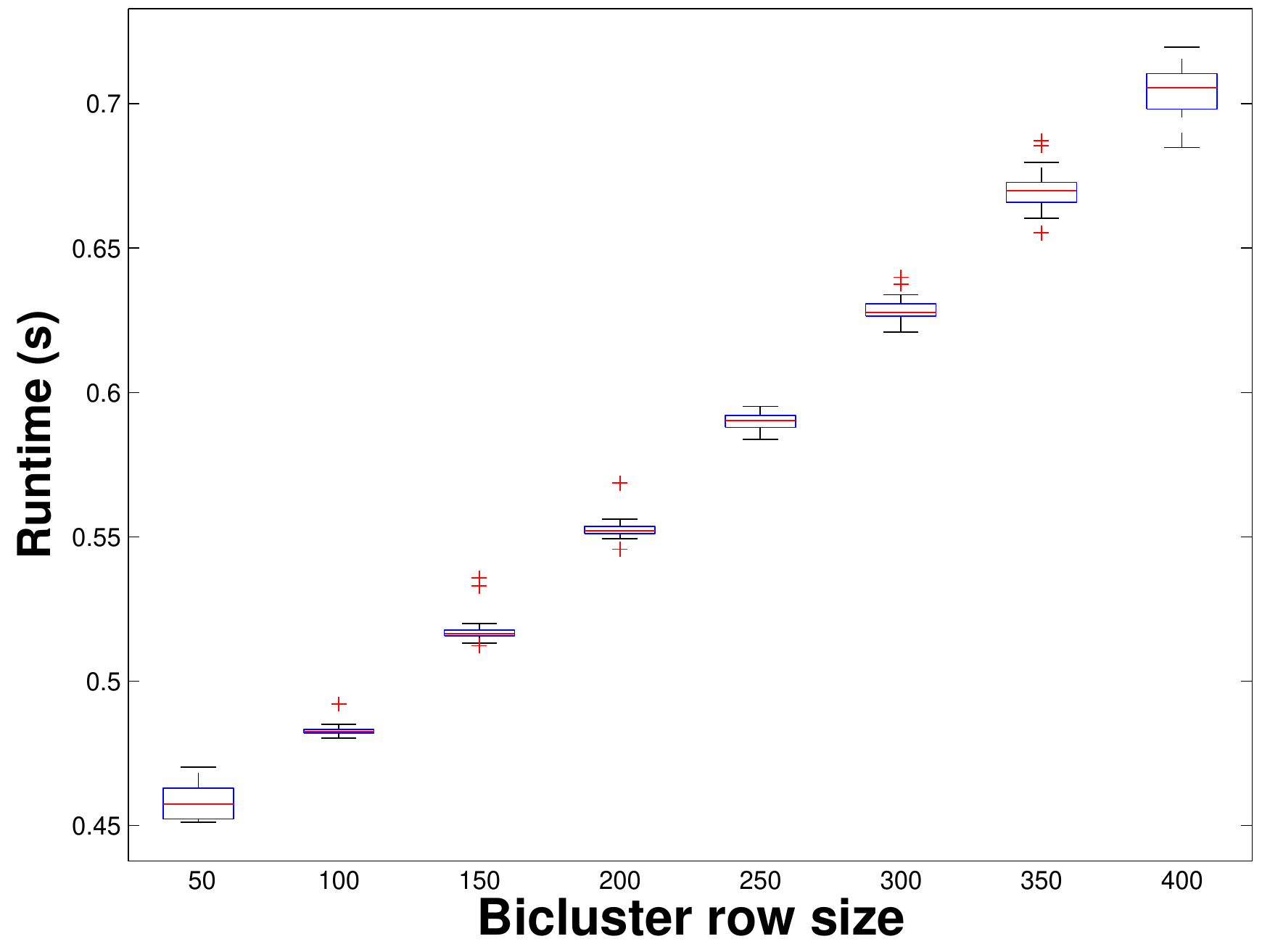}
	}
	\subfigure[]{
		\includegraphics[trim=0.2cm 0.1cm 0.6cm 0.4cm, clip, scale=0.3]{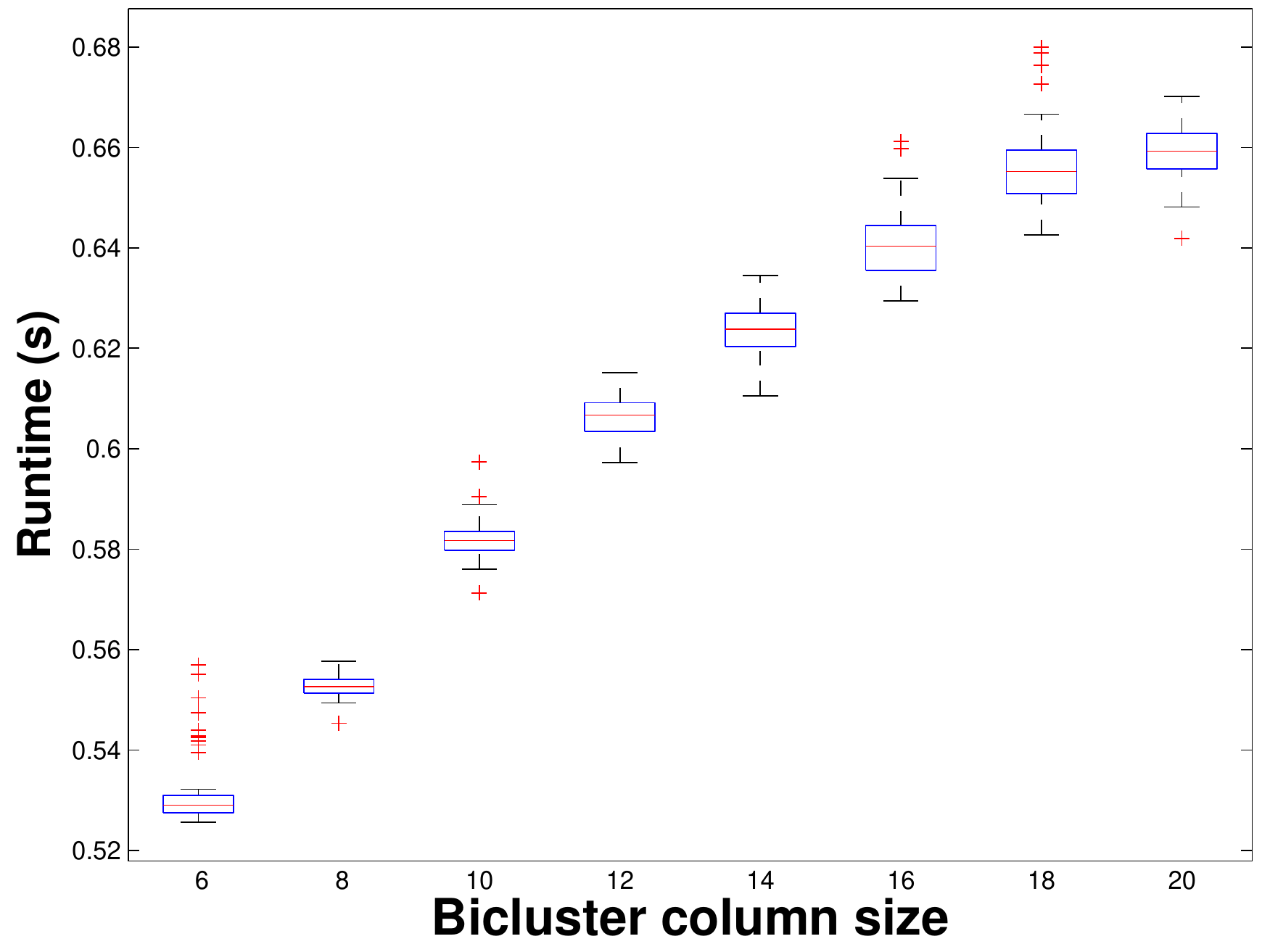}
	}
	\subfigure[]{
		\includegraphics[trim=0.2cm 0.1cm 0.6cm 0.4cm, clip, scale=0.3]{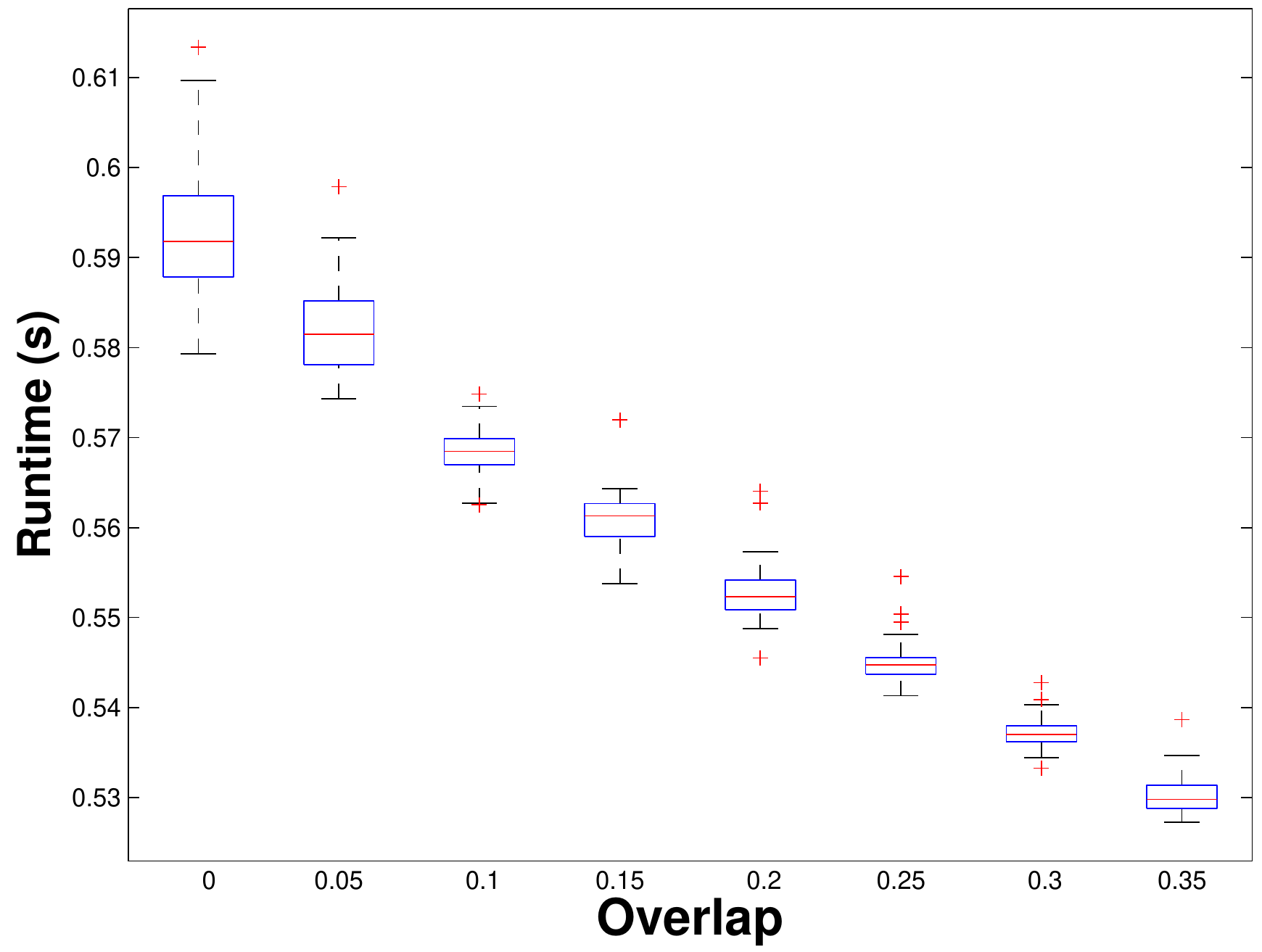}
	}
  \caption{Results of the performance of RIn-Close\_CHV\_P when varying (a) the number $n$ of rows of the dataset, (b) the number $m$ of columns of the dataset, (c) the number of biclusters in the dataset, (d) the bicluster row size, (e) the bicluster column size, and (f) the overlap.}
  \label{fig:expSynDataRTchvp}
\end{figure*}

\begin{figure*}
  \centering
	\subfigure[]{
		\includegraphics[trim=0.2cm 0.1cm 0.6cm 0.4cm, clip, scale=0.3]{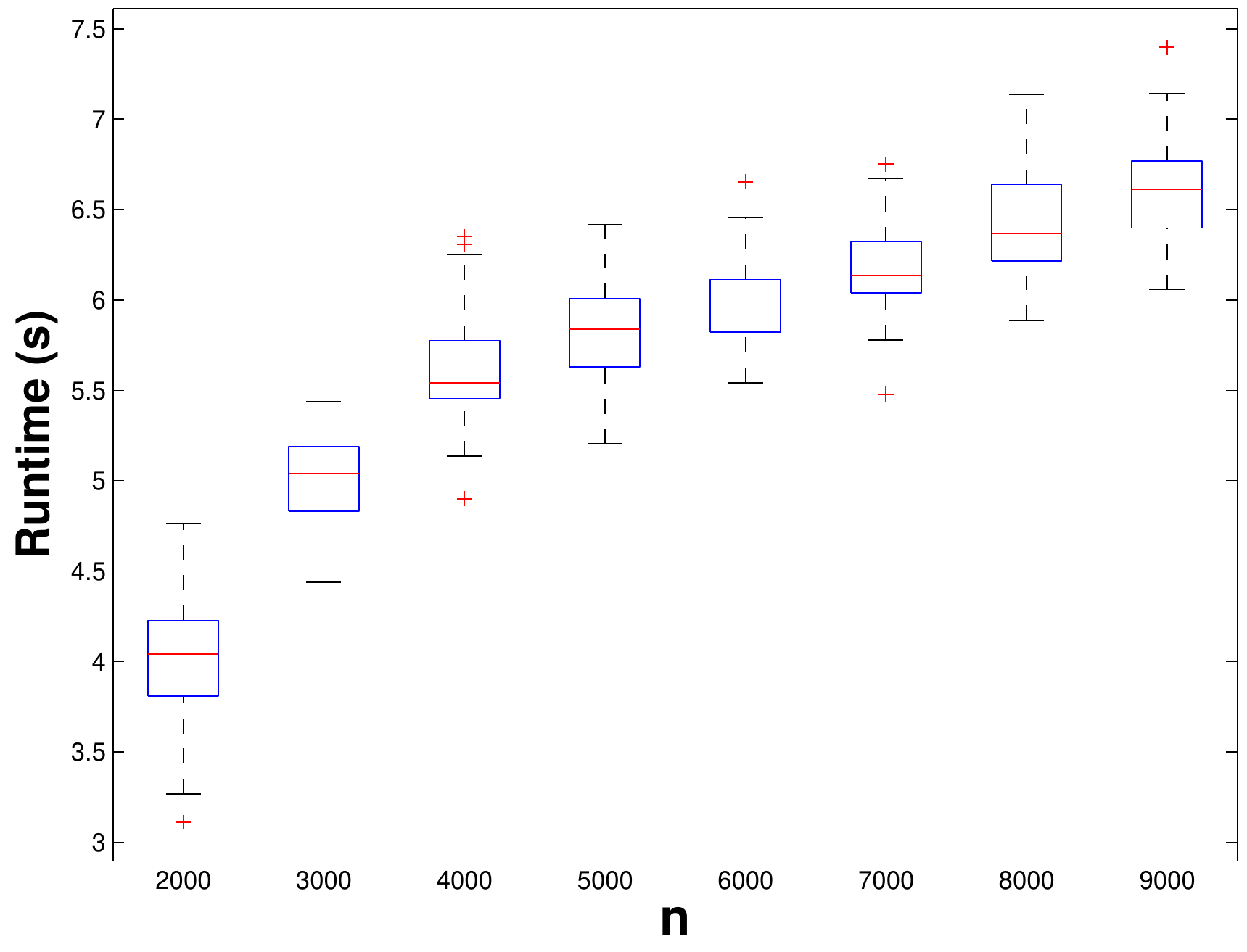}
	}
	\subfigure[]{
		\includegraphics[trim=0.2cm 0.1cm 0.6cm 0.4cm, clip, scale=0.3]{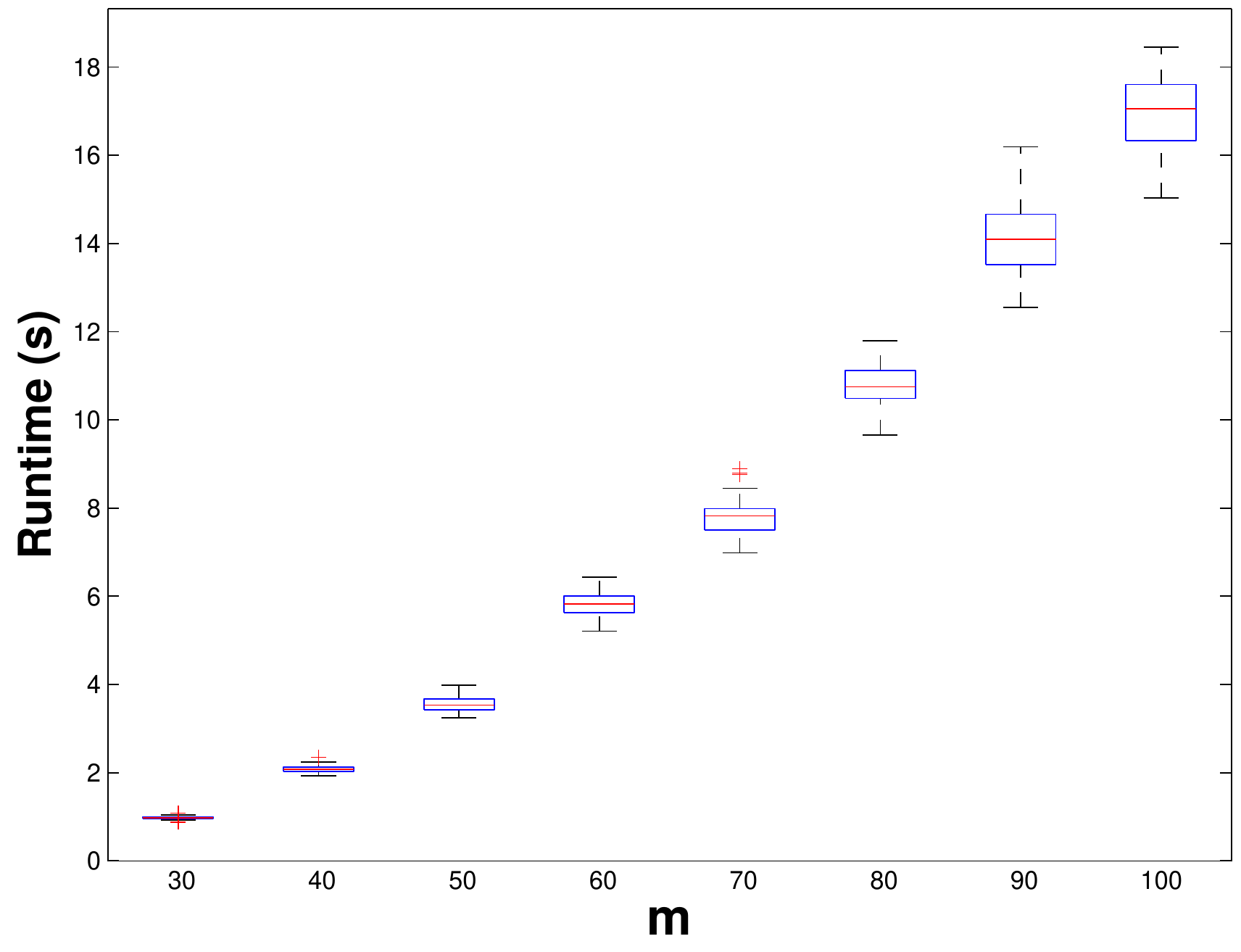}
	}
	\subfigure[]{
		\includegraphics[trim=0.2cm 0.1cm 0.6cm 0.4cm, clip, scale=0.3]{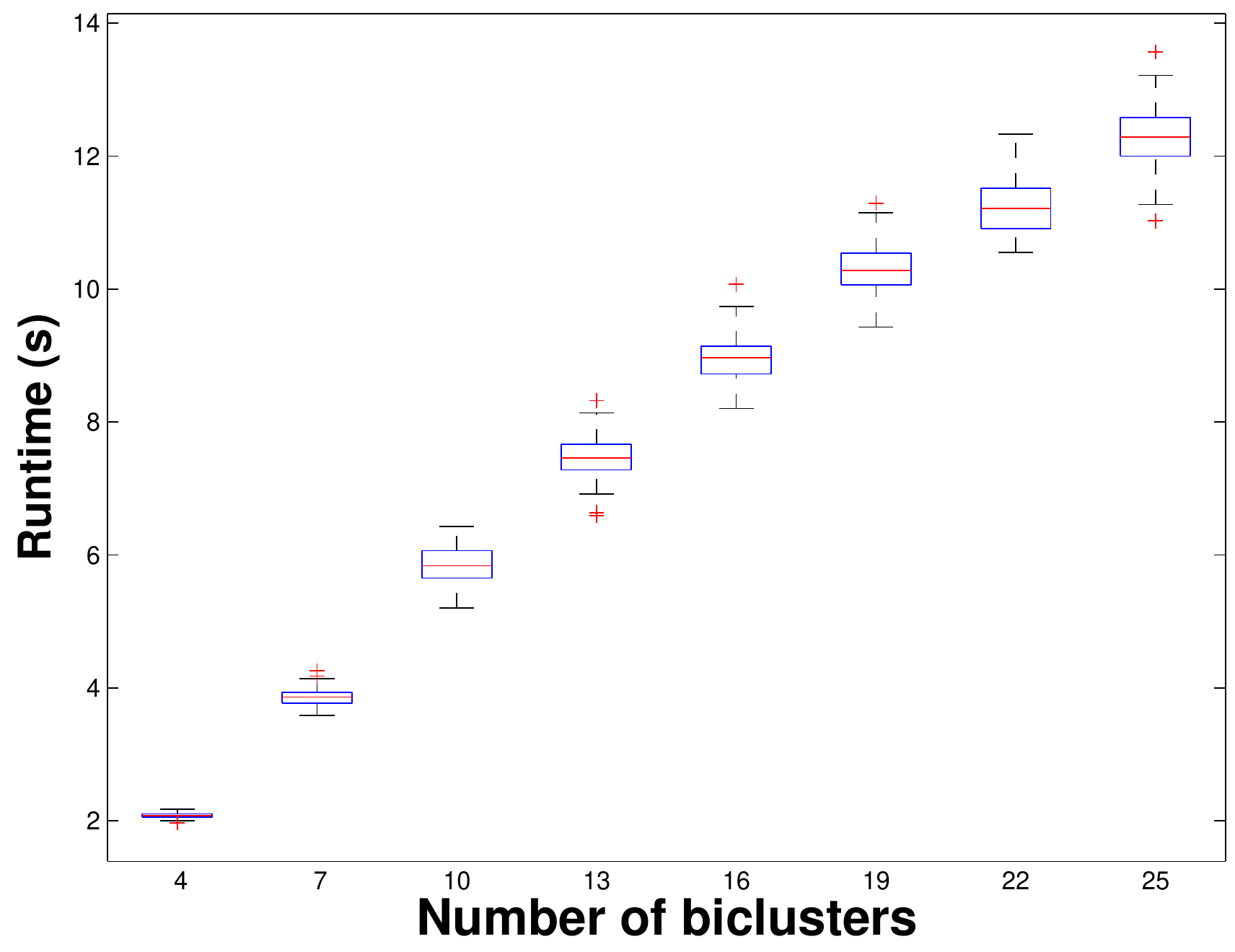}
	}
	\subfigure[]{
		\includegraphics[trim=0.2cm 0.1cm 0.6cm 0.4cm, clip, scale=0.3]{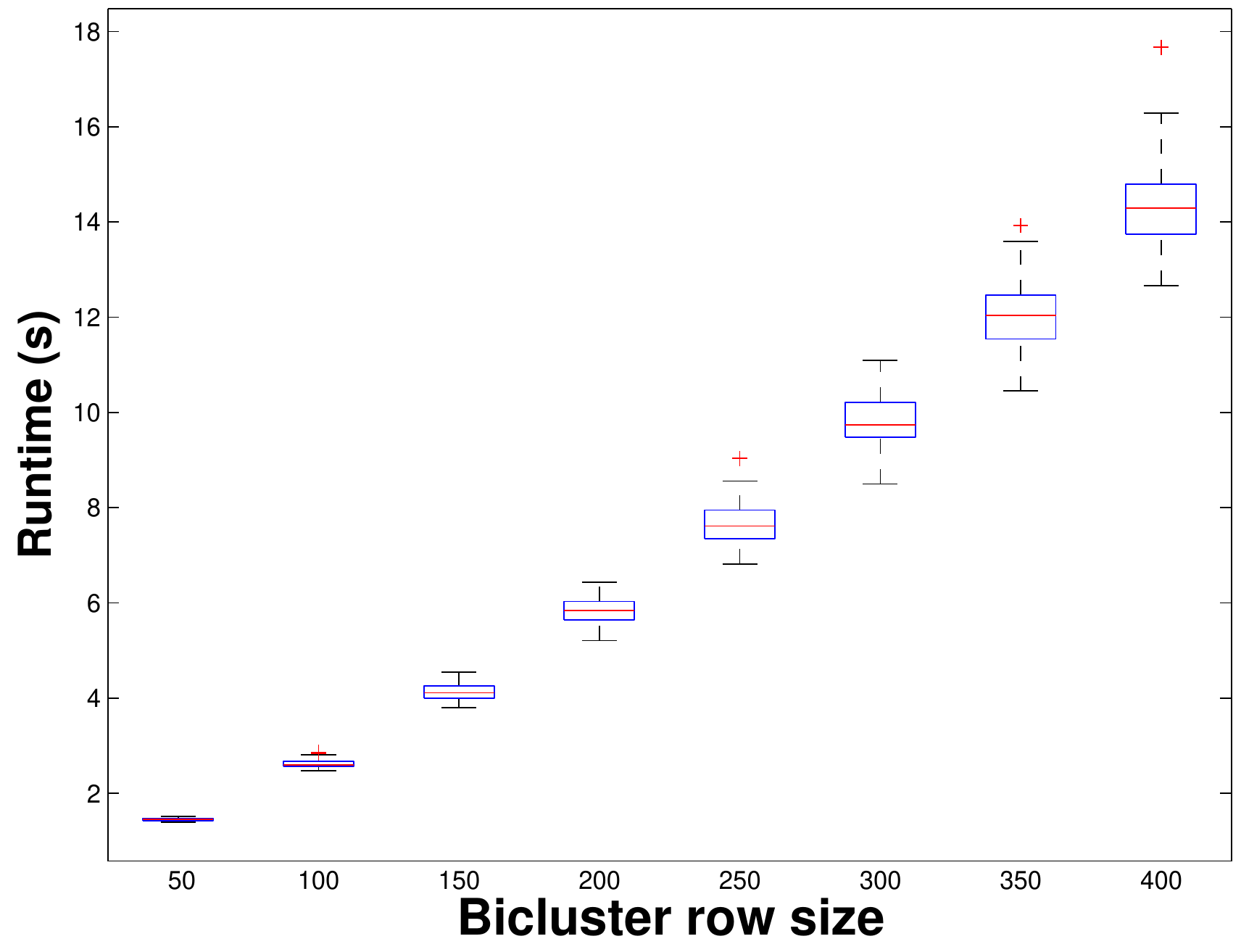}
	}
	\subfigure[]{
		\includegraphics[trim=0.2cm 0.1cm 0.6cm 0.4cm, clip, scale=0.3]{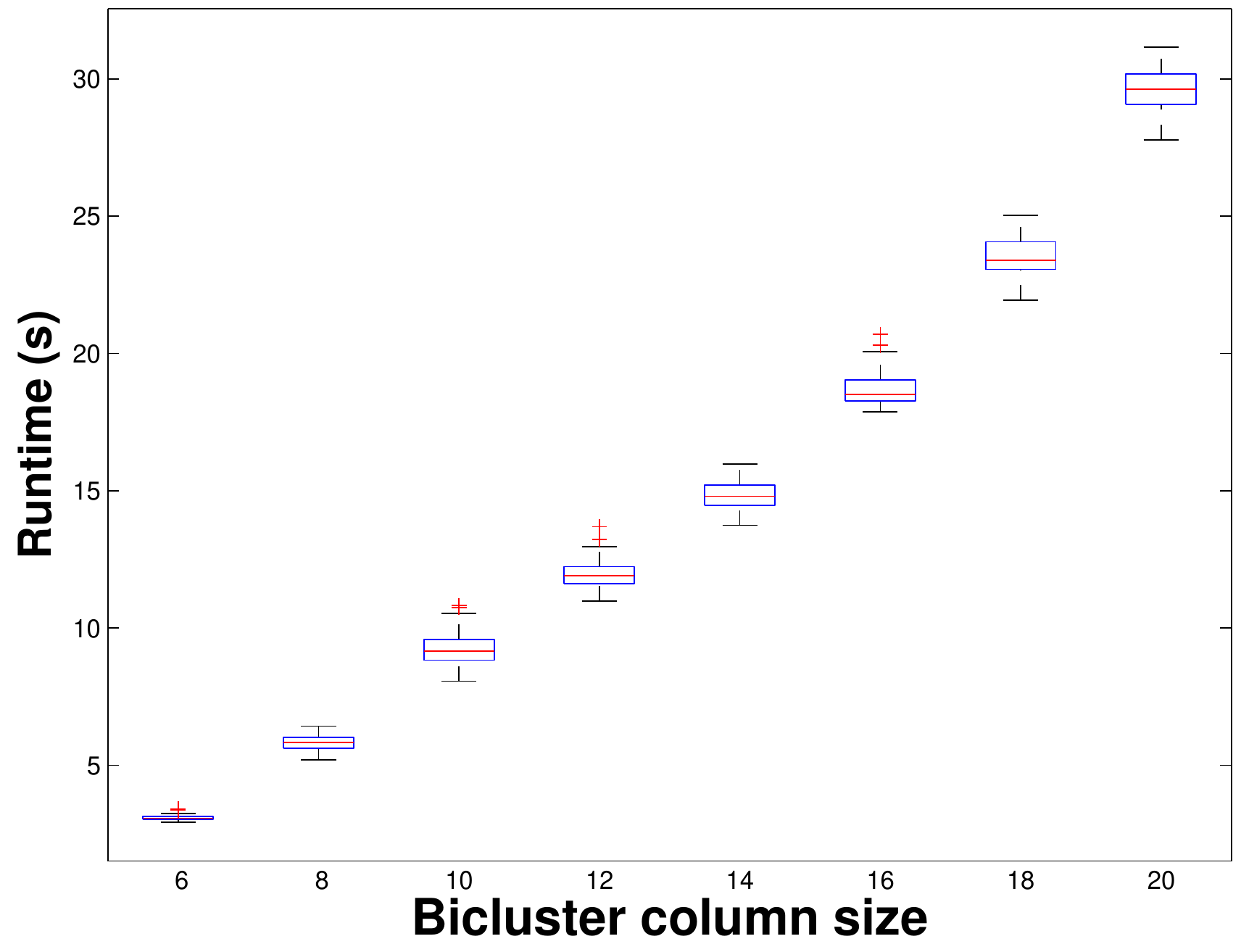}
	}
	\subfigure[]{
		\includegraphics[trim=0.2cm 0.1cm 0.6cm 0.4cm, clip, scale=0.3]{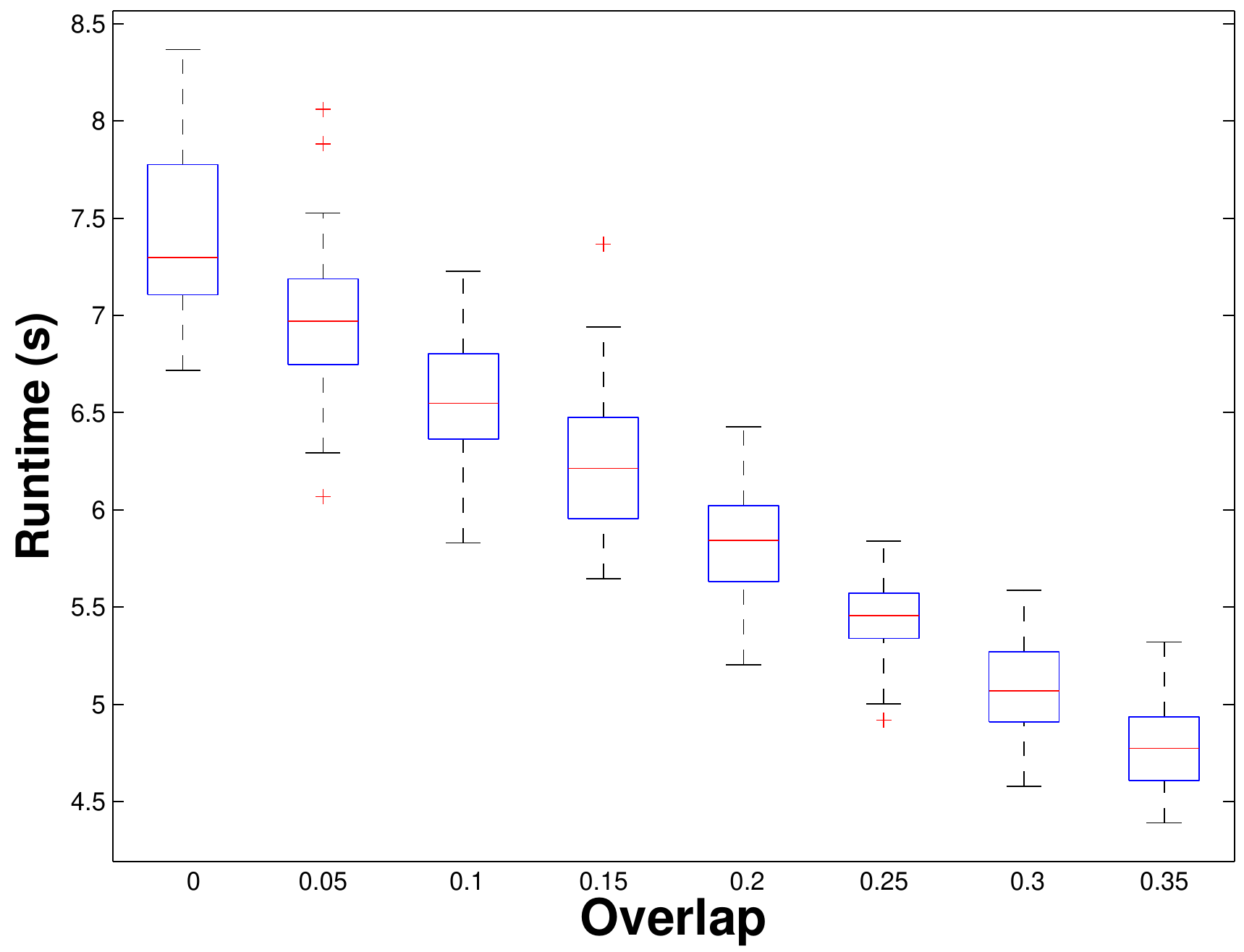}
	}
	\subfigure[]{
		\includegraphics[trim=0.2cm 0.1cm 0.6cm 0.4cm, clip, scale=0.3]{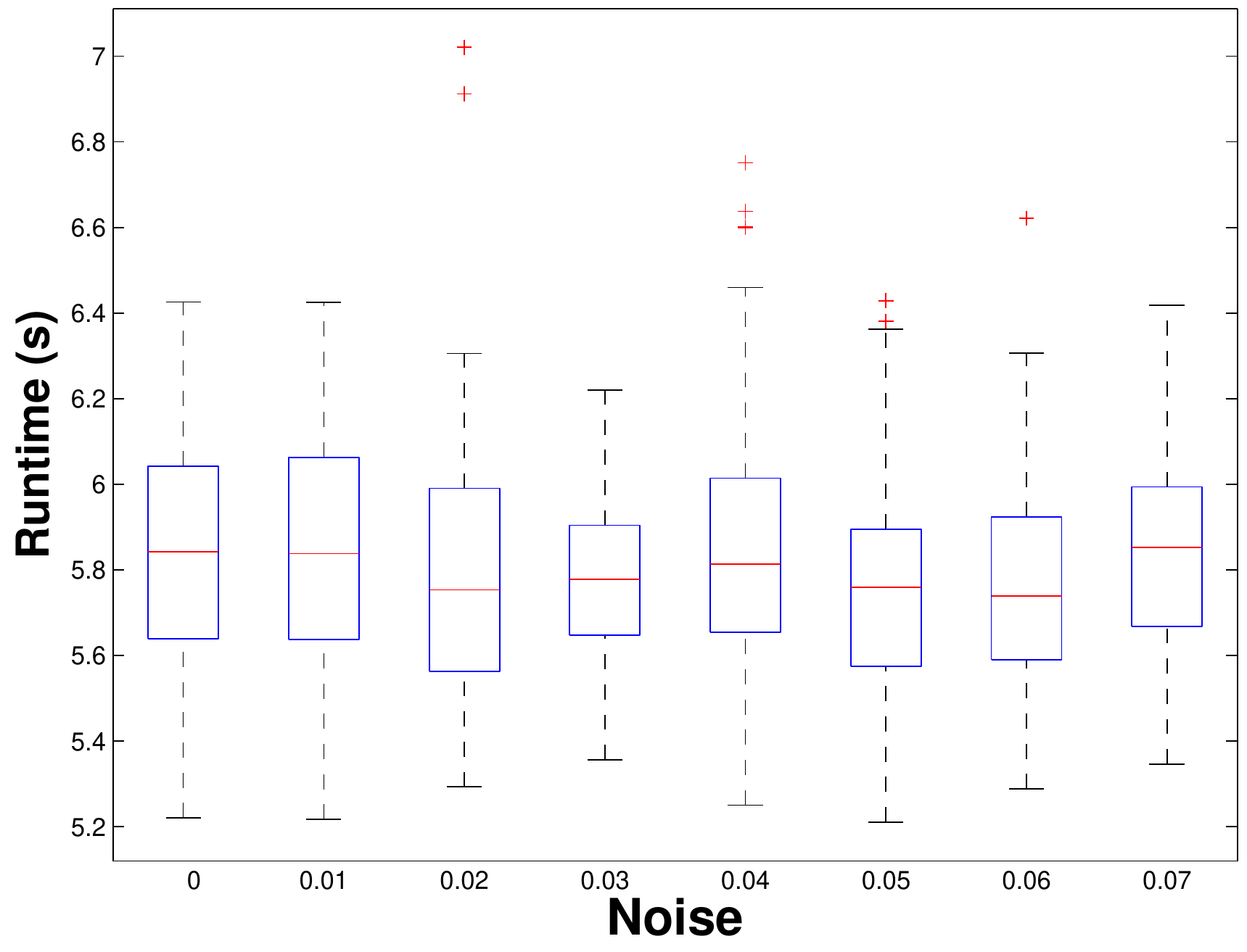}
	}
  \caption{Results of the performance of RIn-Close\_CHV when varying (a) the number $n$ of rows of the dataset, (b) the number $m$ of columns of the dataset, (c) the number of biclusters in the dataset, (d) the bicluster row size, (e) the bicluster column size, (f) the overlap, and (g) the noise.}
  \label{fig:expSynDataRTchv}
\end{figure*}

Figs.~\ref{fig:expSynDataRTcvcp} to \ref{fig:expSynDataRTchv} show, respectively, the sensitivity to different datasets' configurations of RIn-Close\_CVC\_P,  RIn-Close\_CVC,  RIn-Close\_CHV\_P,  and RIn-Close\_CHV. The first note on these results is that the runtime of Step 1 and 3 of RIn-Close\_CHV was negligible. Therefore, the results presented in Fig.~\ref{fig:expSynDataRTchv} are basically the runtime of the Step 2 of the algorithm, which is to mine the CVC biclusters from the augmented matrix $\mathbf{D}_a$.

The runtime increased linearly with $n$, except for RIn-Close\_CHV, for which it increased logarithmically. But for all algorithms without exception, the runtime growth rate was better than their worst-case time complexities (see Section~\ref{sec:rinclose}). For the variable $m$, the runtime increased linearly for RIn-Close\_CVC\_P and RIn-Close\_CVC, which is better than their worst-case time complexities. For RIn-Close\_CHV\_P and RIn-Close\_CHV, the runtime increased polynomially with $m$, which coincides with their worst-case time complexities. The runtime increased linearly with the number of biclusters, except for RIn-Close\_CHV. For RIn-Close\_CHV, we can notice a tendency of a logarithmically growth, which is good news because, due to its worst-case time complexity, we expected a linear growth too. For the bicluster row size, we had a linear growth of the runtime, except for RIn-Close\_CHV, for which we had a smooth polynomial growth. RIn-Close\_CVC\_P and RIn-Close\_CVC had the same behavior for the bicluster column size: the runtime increased logarithmically, but saturates and started to decrease linearly. With the increase in the bicluster column size, the coverage of the dataset increases, and it seemed to help these algorithms to find the biclusters more quickly. With more columns in the biclusters, more the inheritance of the columns tends to have a positive effect, and also less columns tend to be tested in the canonicity function. For the RIn-Close\_CHV\_P and RIn-Close\_CHV, the runtime increased logarithmically and polynomially with the bicluster column size, respectively. For all algorithms, the runtime decreased linearly with the overlap. The noise did not greatly affect the runtime. 

The variations presented in the boxplots are due to two main reasons: ($i$) an intrinsic variation due to the machine; and ($ii$) some characteristics of the synthetic datasets, more specifically, due to the arrangement of the biclusters and the noise in the datasets. The arrangement of the biclusters in the datasets impacts in the heritage of the columns, in the number of times that the canonicity function is called, and in the number of columns that are tested in the canonicity function. The noise in the datasets impacts in the computation of the possible extents, and also in the number of times that the canonicity function is called.

\subsection{Sensitivity Analysis}

This experiment aims to test RIn-Close's sensitivity to the parameters $\epsilon$ and $minRow$. We ran RIn-Close for four microarray gene expression datasets: Yeast\footnote{http://arep.med.harvard.edu/biclustering, last accessed 04/09/2015} \cite{ChoEtAL1998}, GDS232\footnote{\label{footnote-GDS}http://www.ncbi.nlm.nih.gov, last accessed 04/09/2015} \cite{MacdonaldEtAl2001}, GDS750\textsuperscript{\ref{footnote-GDS}} \cite{LeberEtAl2004} and GDS4085\textsuperscript{\ref{footnote-GDS}} \cite{JulienEtAl2011}. The former dataset, Yeast, were preprocessed by Cheng and Church \cite{ChengChurch2000}, we just threw away the two genes with missing values. The last three datasets were preprocessed by us. For each one of them, we remove the empty spots; we threw away the data for any genes where one or more expression levels were not measured; we filtered out genes with small variance over time; and we scale the data of each column to integers between 0 and 1000. Table~\ref{tab:realdatasets} shows more information about these datasets. We ran RIn-Close 50 times to compute the average runtime, and we looked for biclusters with at least 3 columns.

\linespread{1}

\begin{table}
  \centering
  \caption{Datasets description.}
    \begin{tabular}{lrlrrrr}
		\toprule
		Name	& Dimension	& Organism\\
    \midrule
			Yeast	& $2882\times17$ & Saccharomyces cerevisiae\\
			GDS232 & $589\times23$ & Homo sapiens\\
			GDS750 & $3456\times13$ & Saccharomyces cerevisiae\\
			GDS4085 & $1133\times19$ &	Homo sapiens\\
    \bottomrule
    \end{tabular}
  \label{tab:realdatasets}
\end{table}

\linespread{1.5}

\begin{figure*}
  \centering
	
	\subfigure[]{
		\includegraphics[trim=0.3cm 0.1cm 0.9cm 0.1cm, clip, scale=0.26]{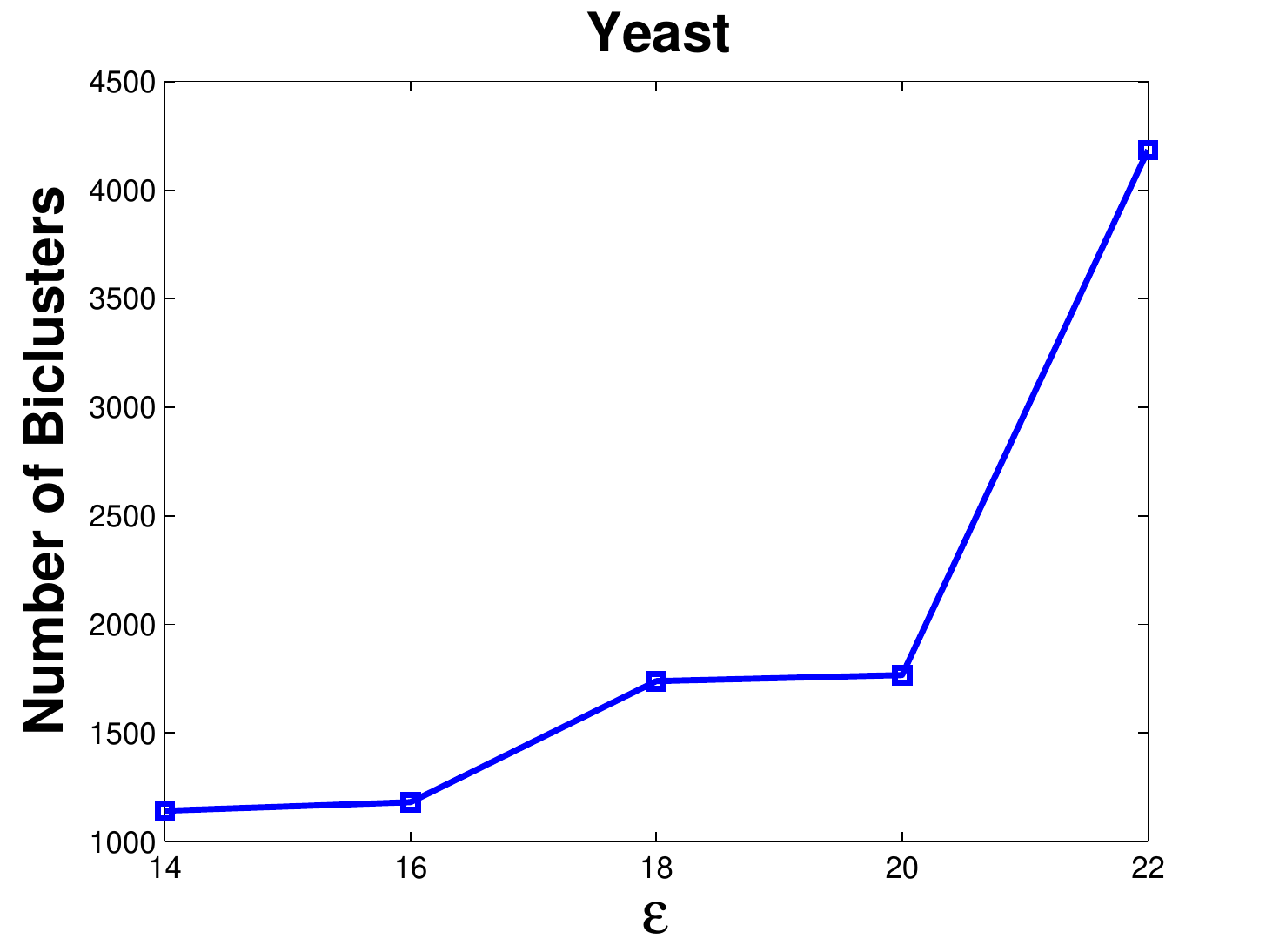}
	}
	\subfigure[]{
		\includegraphics[trim=0.3cm 0.1cm 0.9cm 0.1cm, clip, scale=0.26]{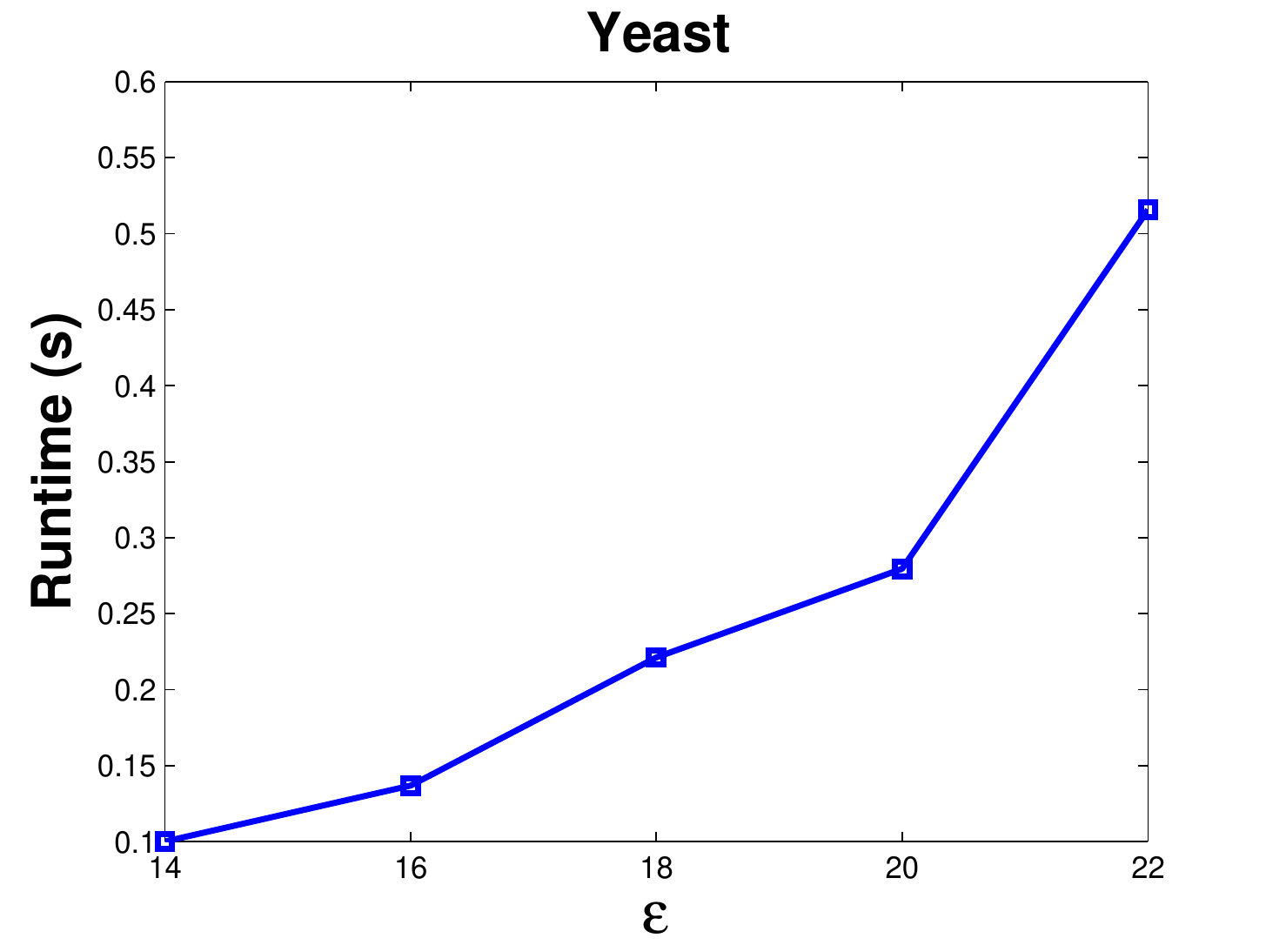}
	}
	\subfigure[]{
		\includegraphics[trim=0.3cm 0.1cm 0.9cm 0.1cm, clip, scale=0.26]{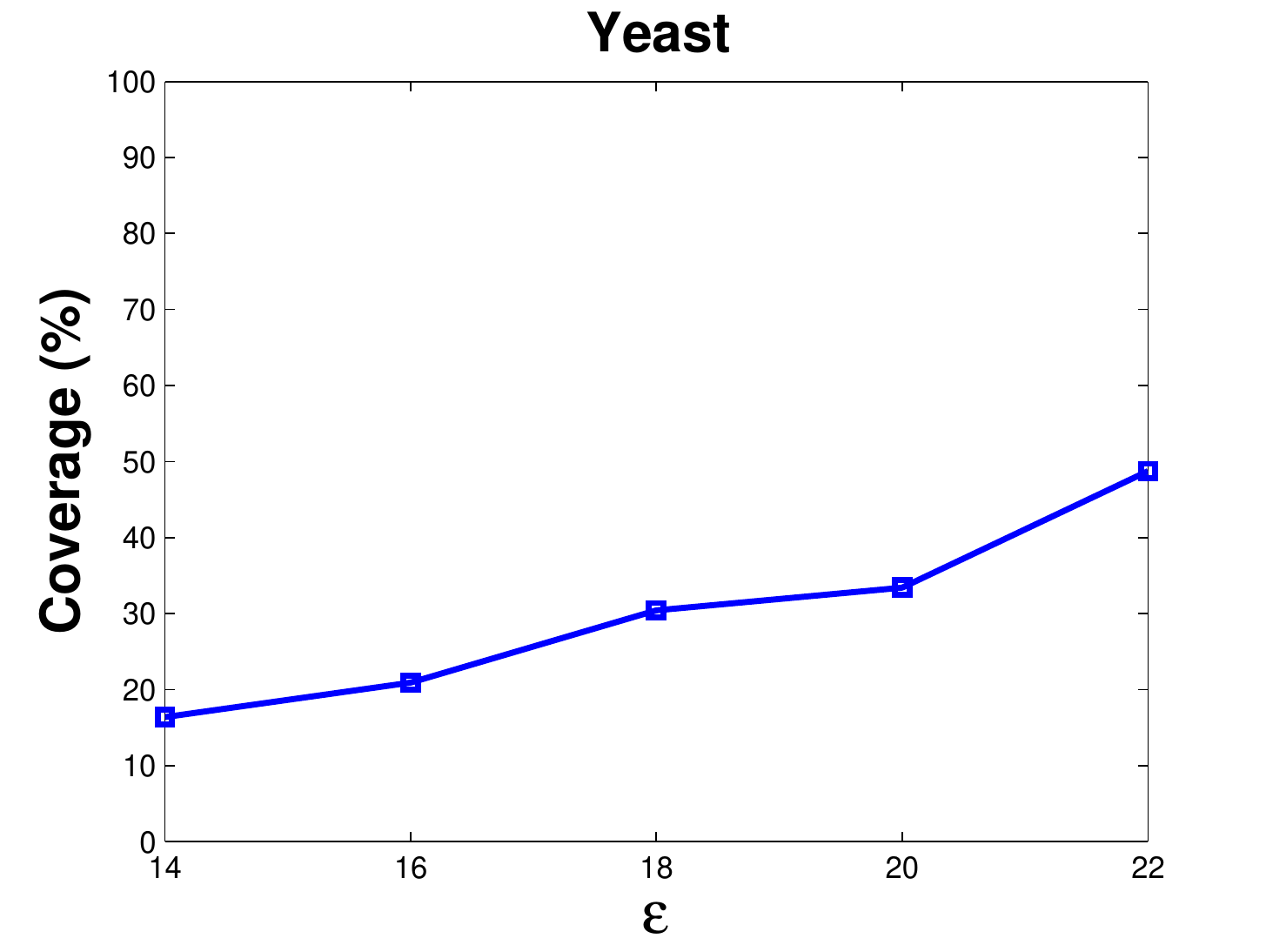}
	}
	\subfigure[]{
		\includegraphics[trim=0.3cm 0.1cm 0.9cm 0.1cm, clip, scale=0.26]{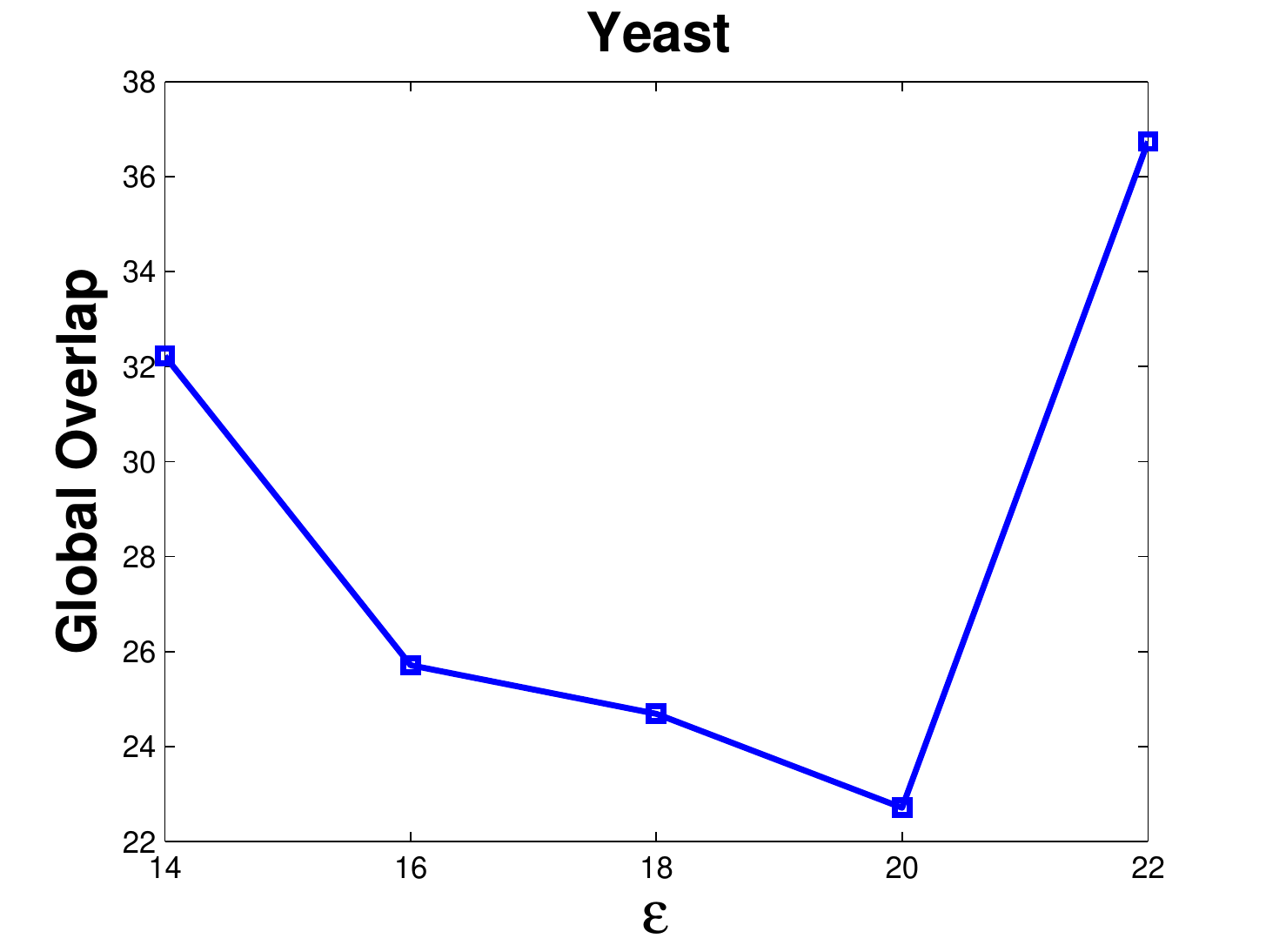}
	}
	
	\subfigure[]{
		\includegraphics[trim=0.3cm 0.1cm 0.9cm 0.1cm, clip, scale=0.26]{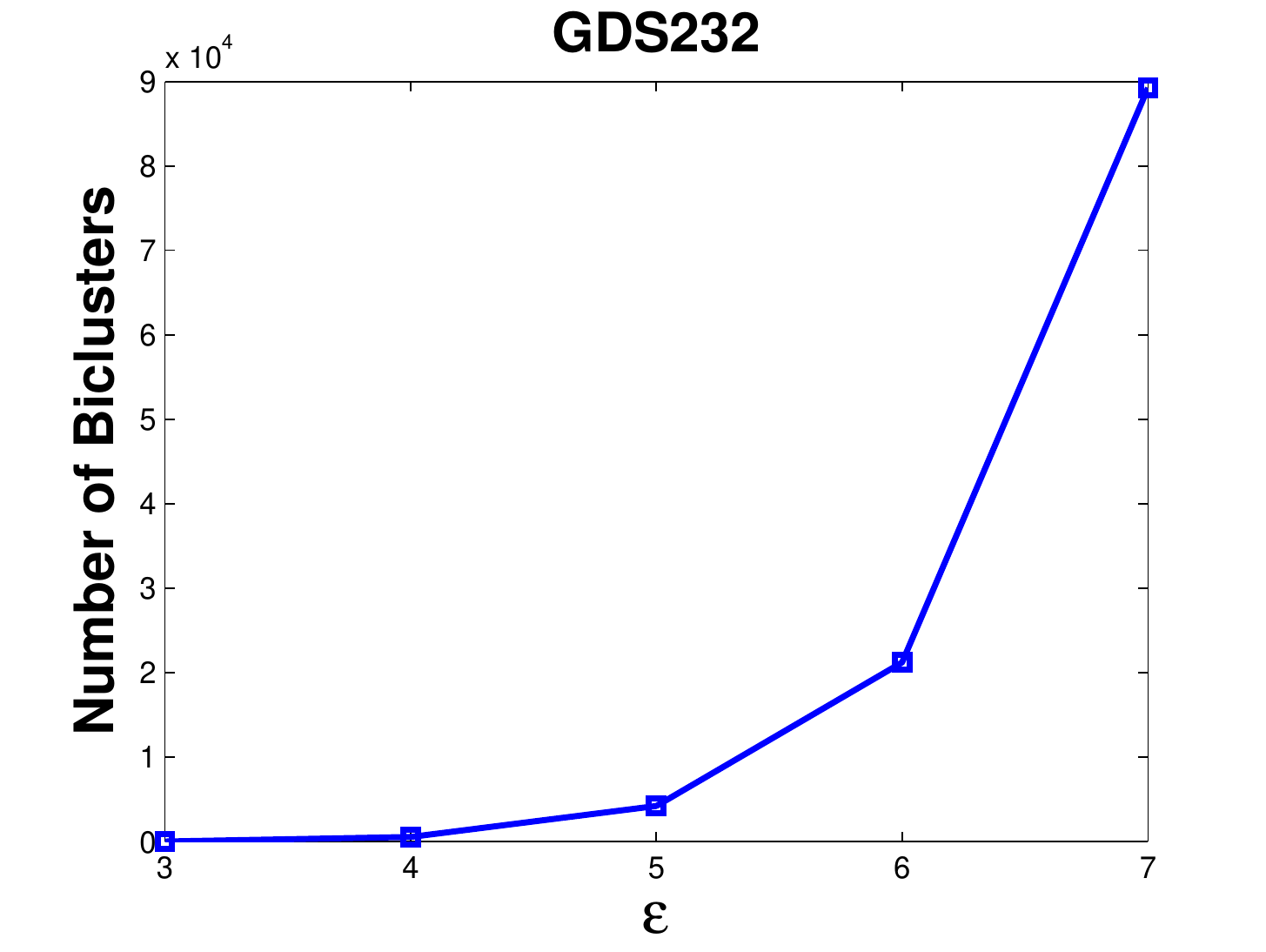}
	}
	\subfigure[]{
		\includegraphics[trim=0.3cm 0.1cm 0.9cm 0.1cm, clip, scale=0.26]{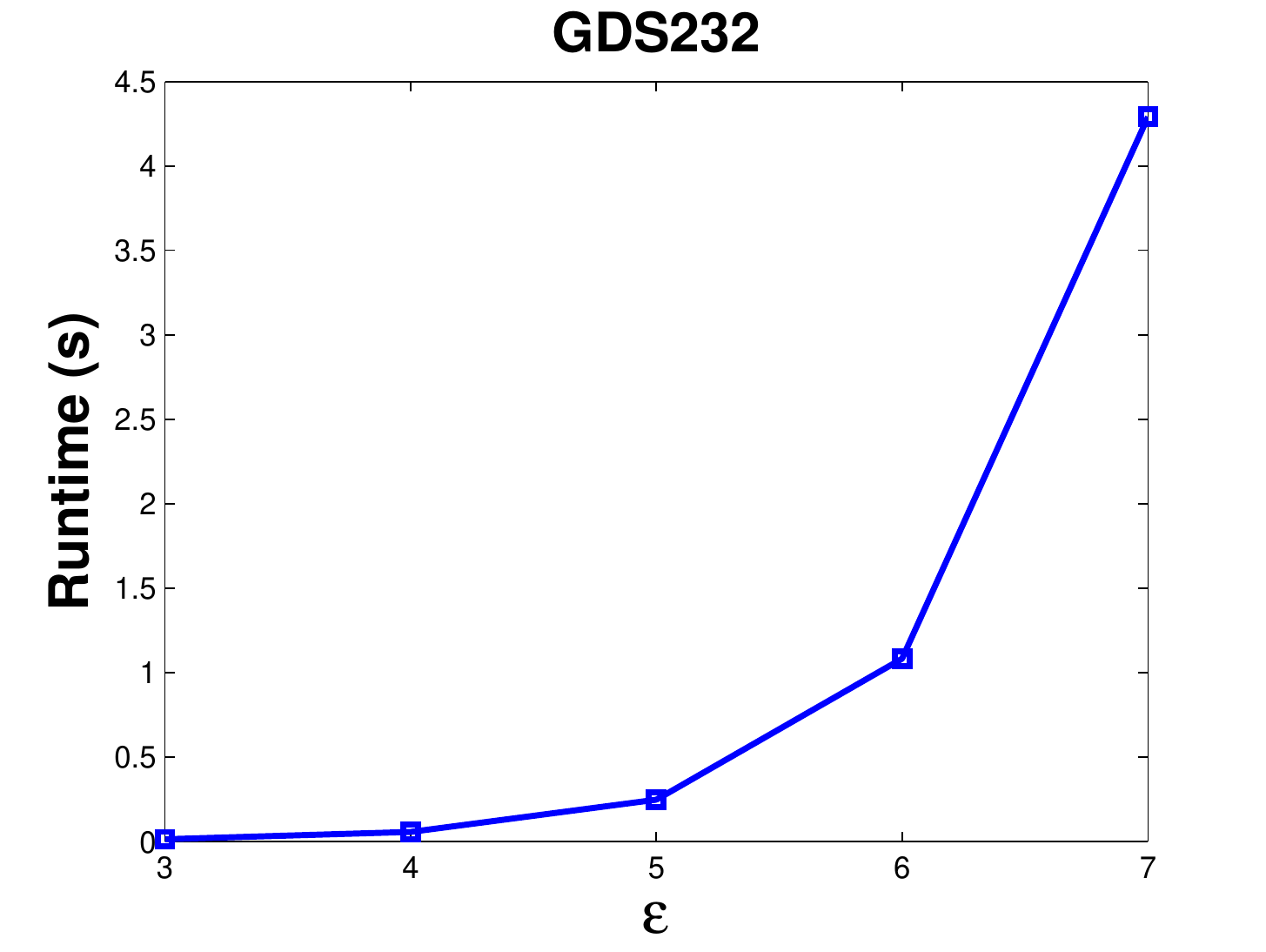}
	}
	\subfigure[]{
		\includegraphics[trim=0.3cm 0.1cm 0.9cm 0.1cm, clip, scale=0.26]{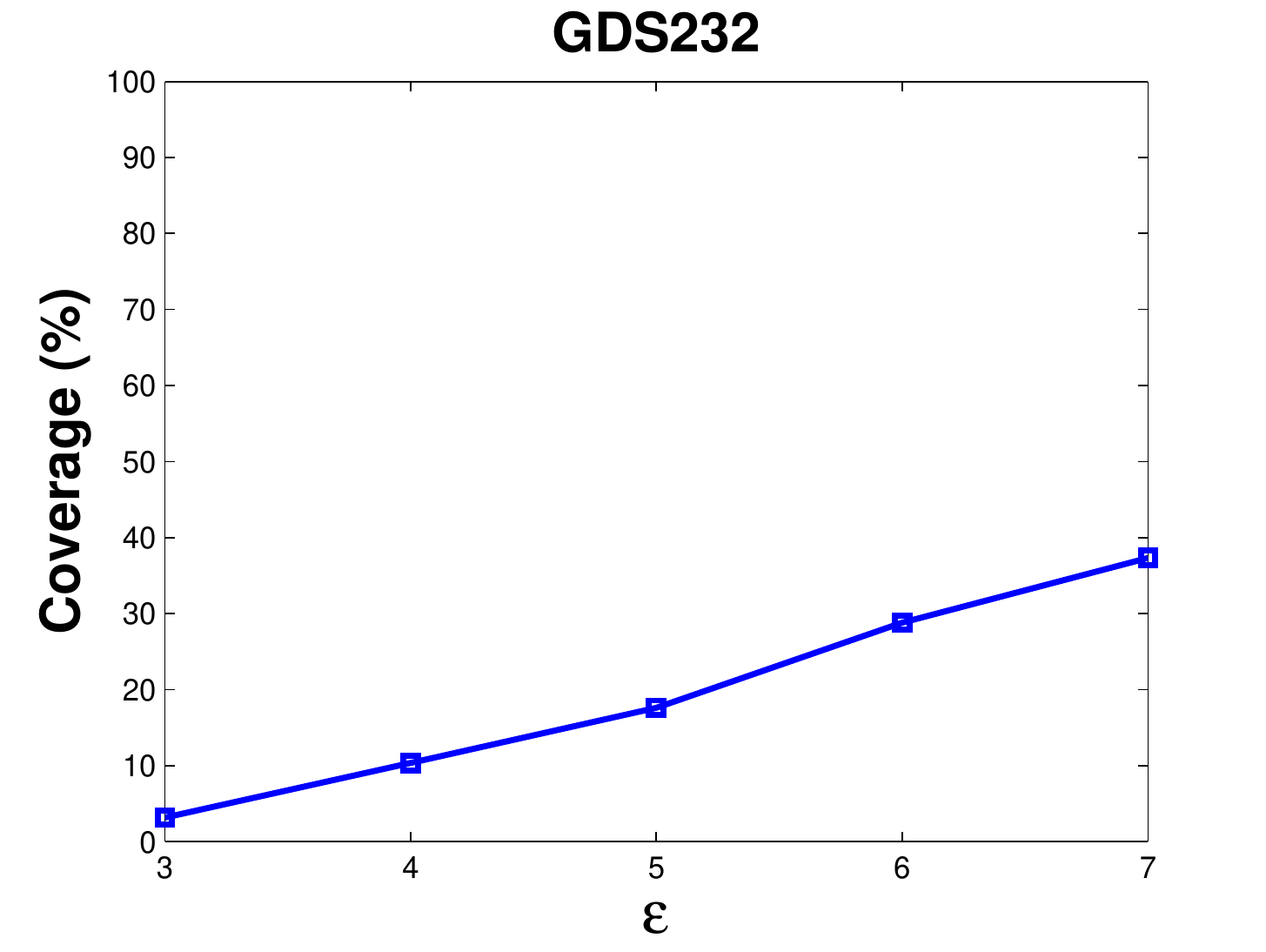}
	}
	\subfigure[]{
		\includegraphics[trim=0.3cm 0.1cm 0.9cm 0.1cm, clip, scale=0.26]{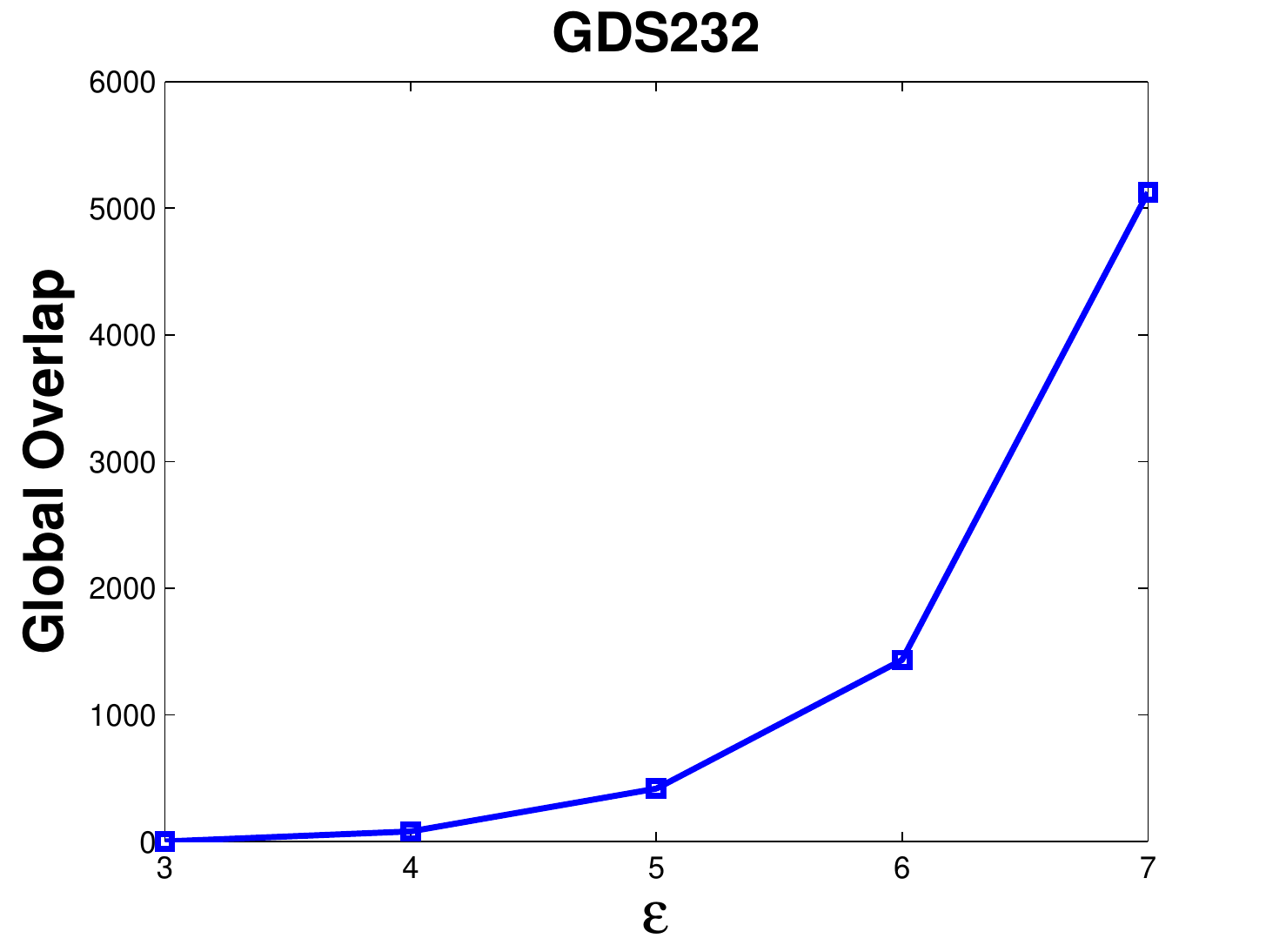}
	}
	
	\subfigure[]{
		\includegraphics[trim=0.3cm 0.1cm 0.9cm 0.1cm, clip, scale=0.26]{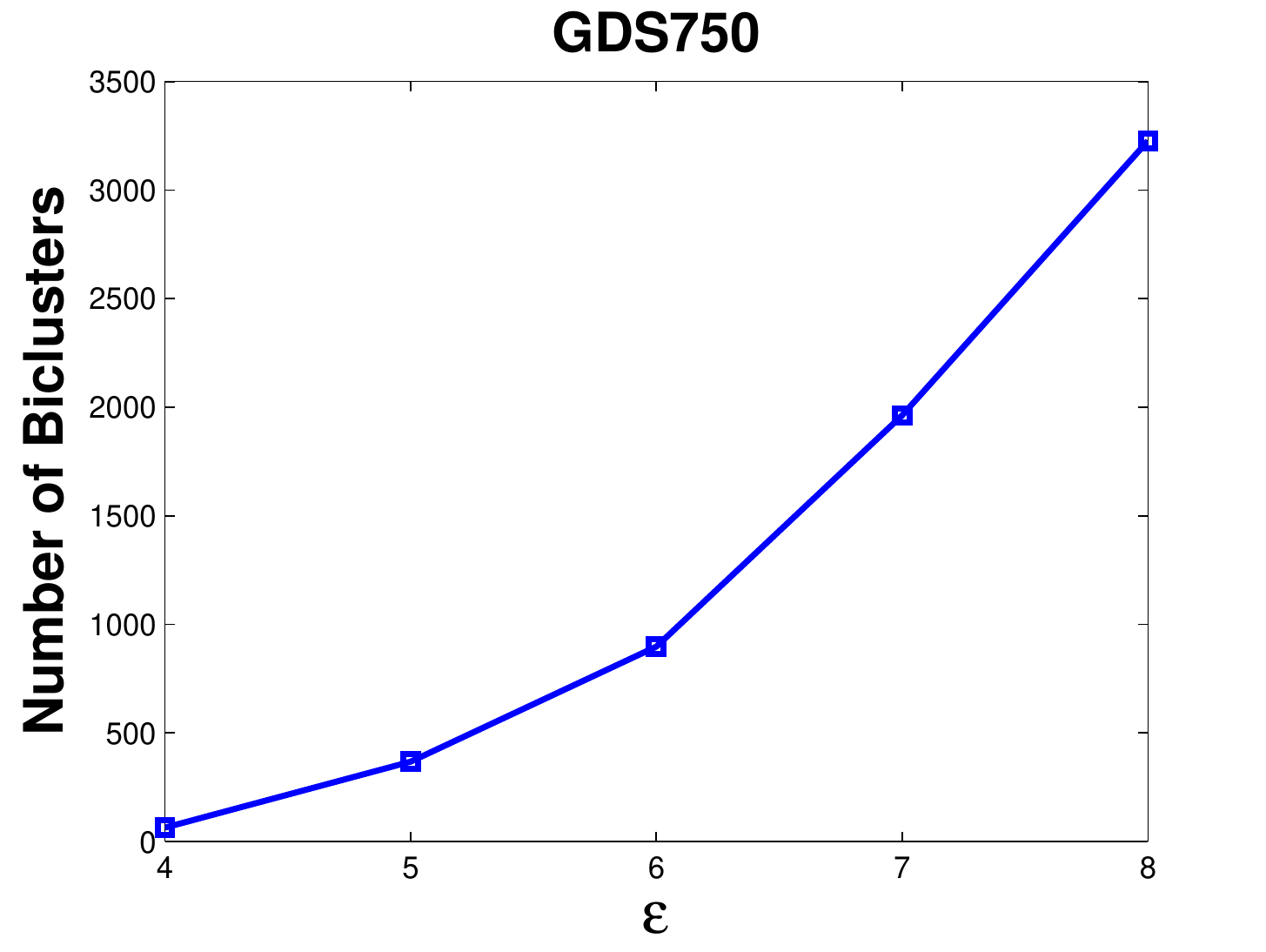}
	}
	\subfigure[]{
		\includegraphics[trim=0.3cm 0.1cm 0.9cm 0.1cm, clip, scale=0.26]{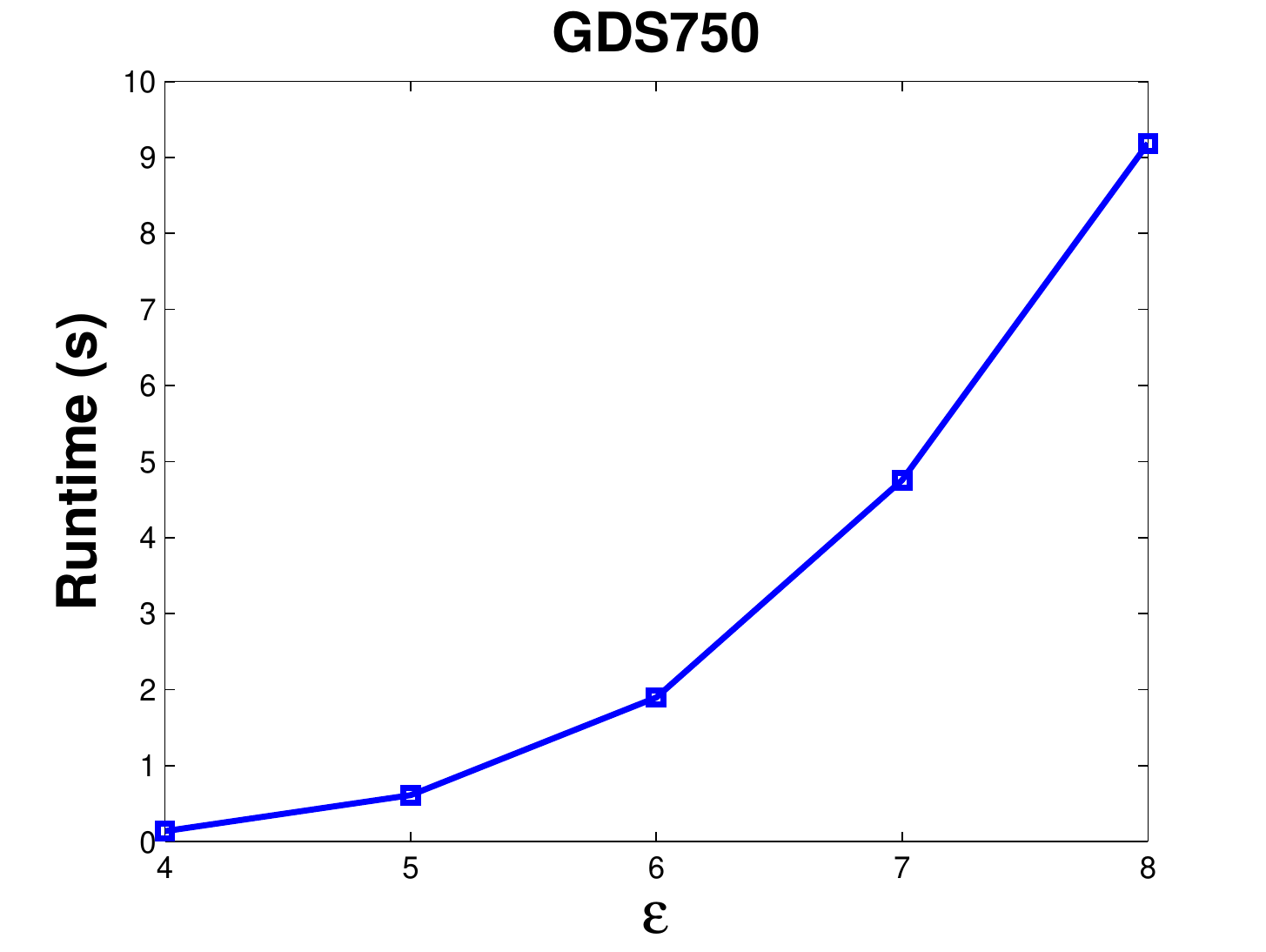}
	}
	\subfigure[]{
		\includegraphics[trim=0.3cm 0.1cm 0.9cm 0.1cm, clip, scale=0.26]{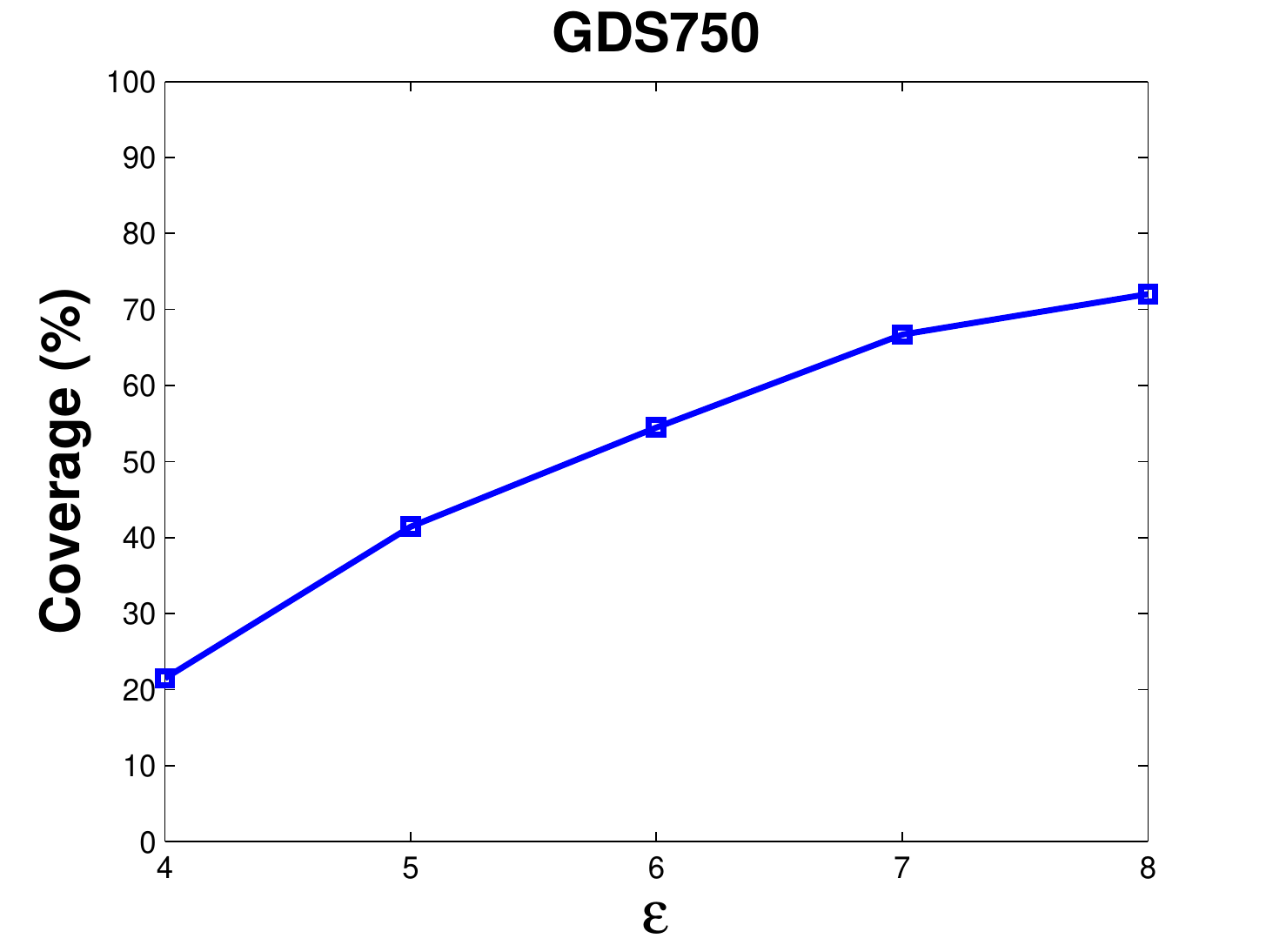}
	}
	\subfigure[]{
		\includegraphics[trim=0.3cm 0.1cm 0.9cm 0.1cm, clip, scale=0.26]{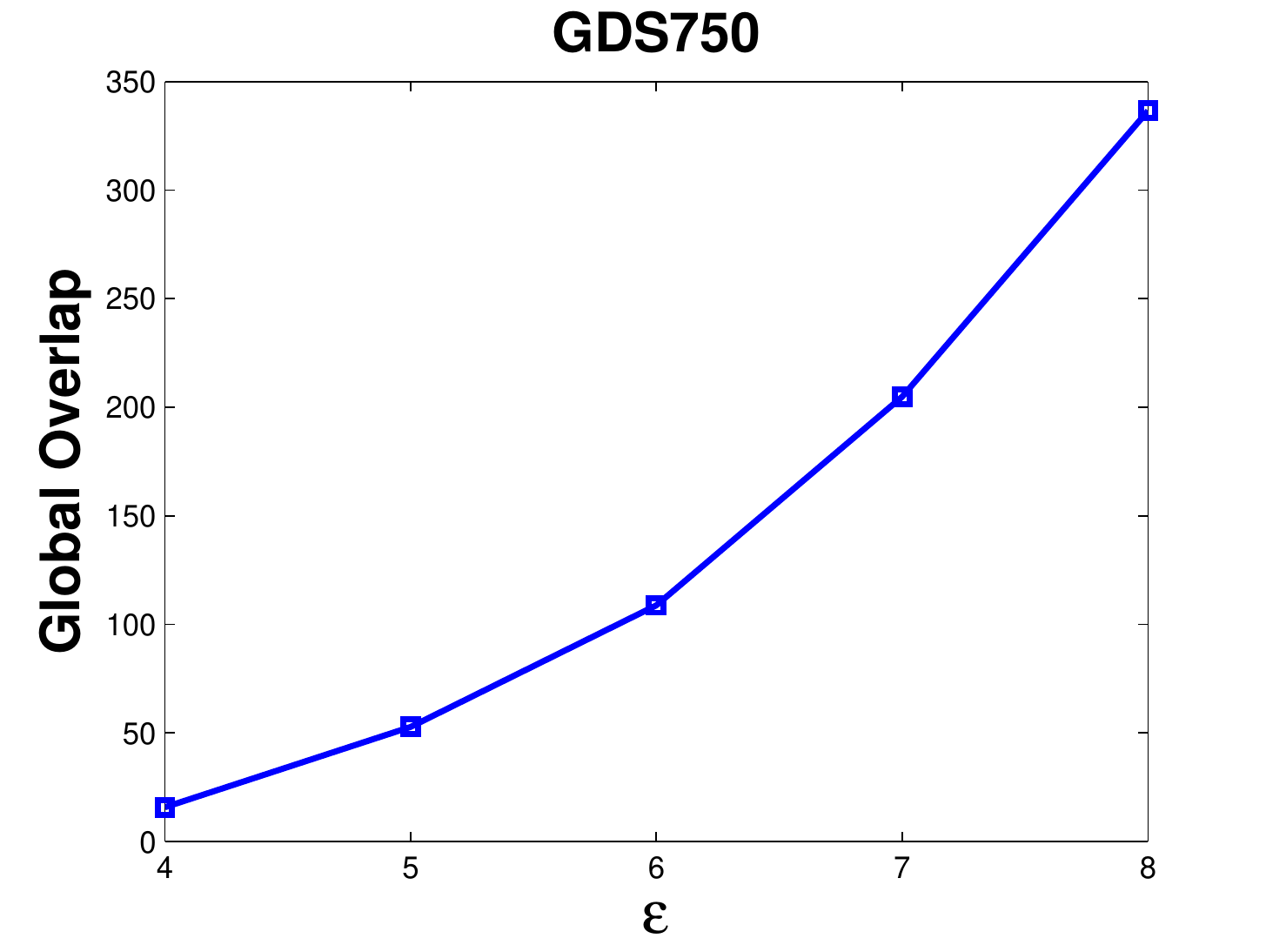}
	}

	\subfigure[]{
		\includegraphics[trim=0.3cm 0.1cm 0.9cm 0.1cm, clip, scale=0.26]{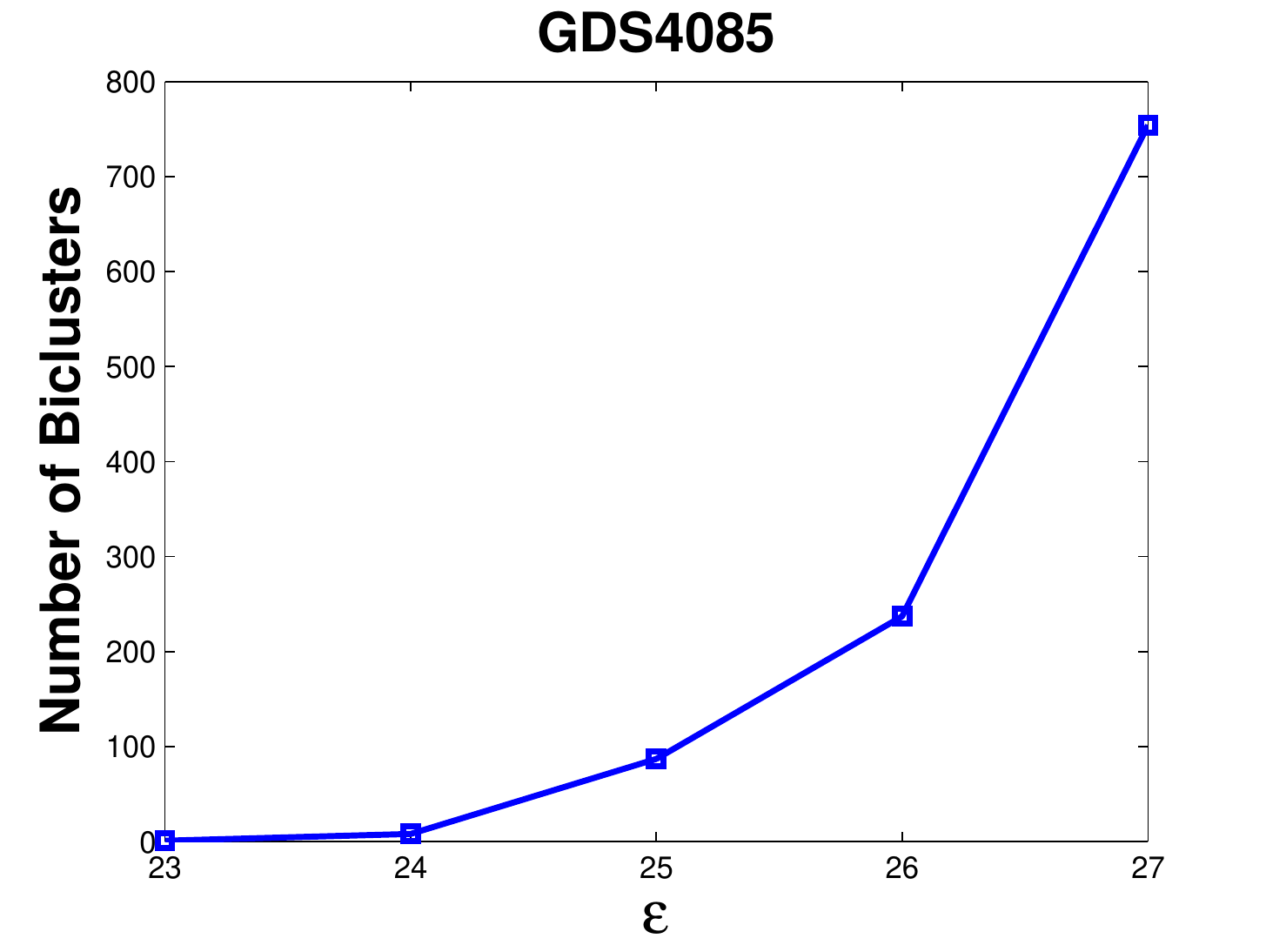}
	}
	\subfigure[]{
		\includegraphics[trim=0.3cm 0.1cm 0.9cm 0.1cm, clip, scale=0.26]{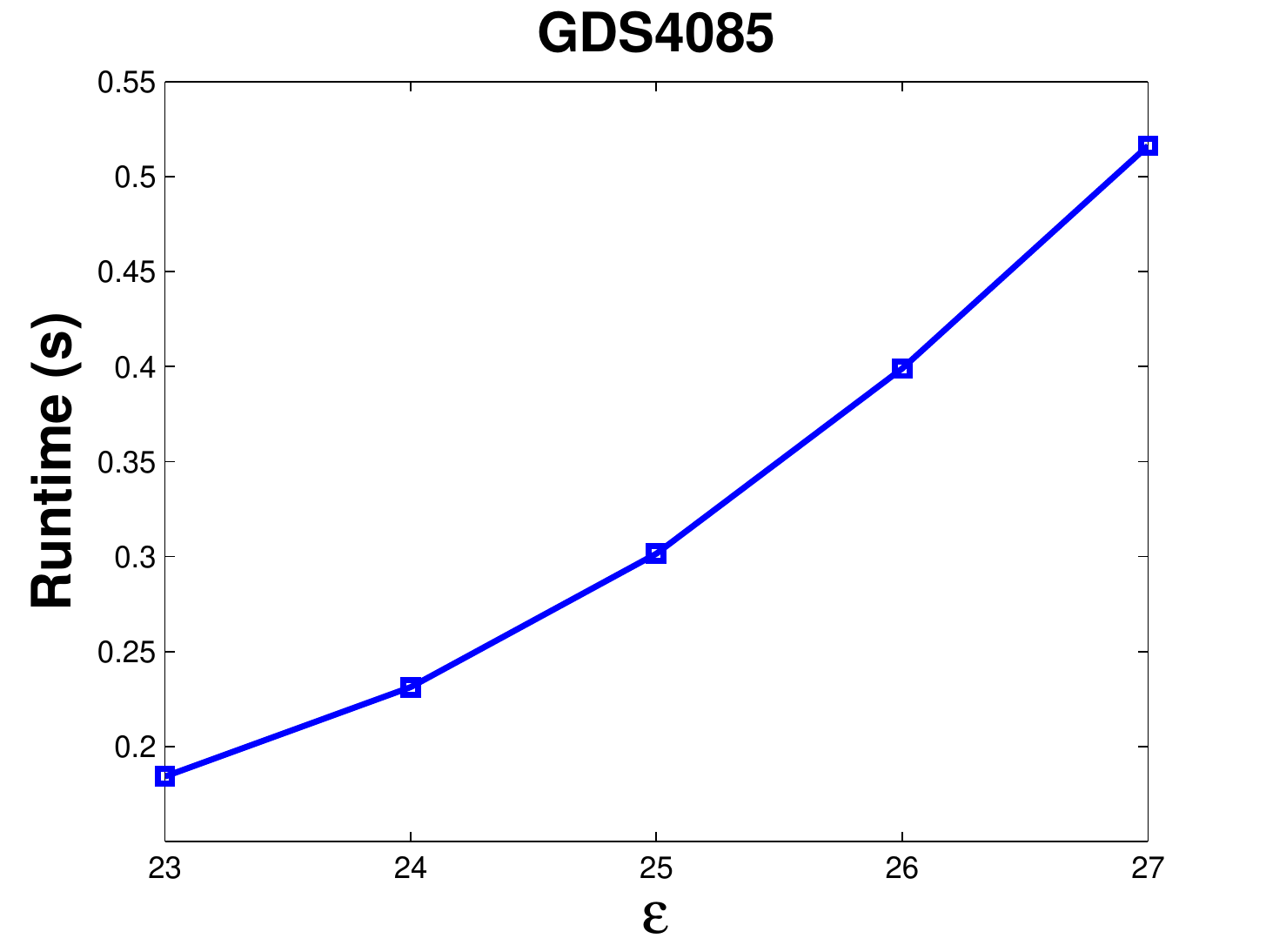}
	}
	\subfigure[]{
		\includegraphics[trim=0.3cm 0.1cm 0.9cm 0.1cm, clip, scale=0.26]{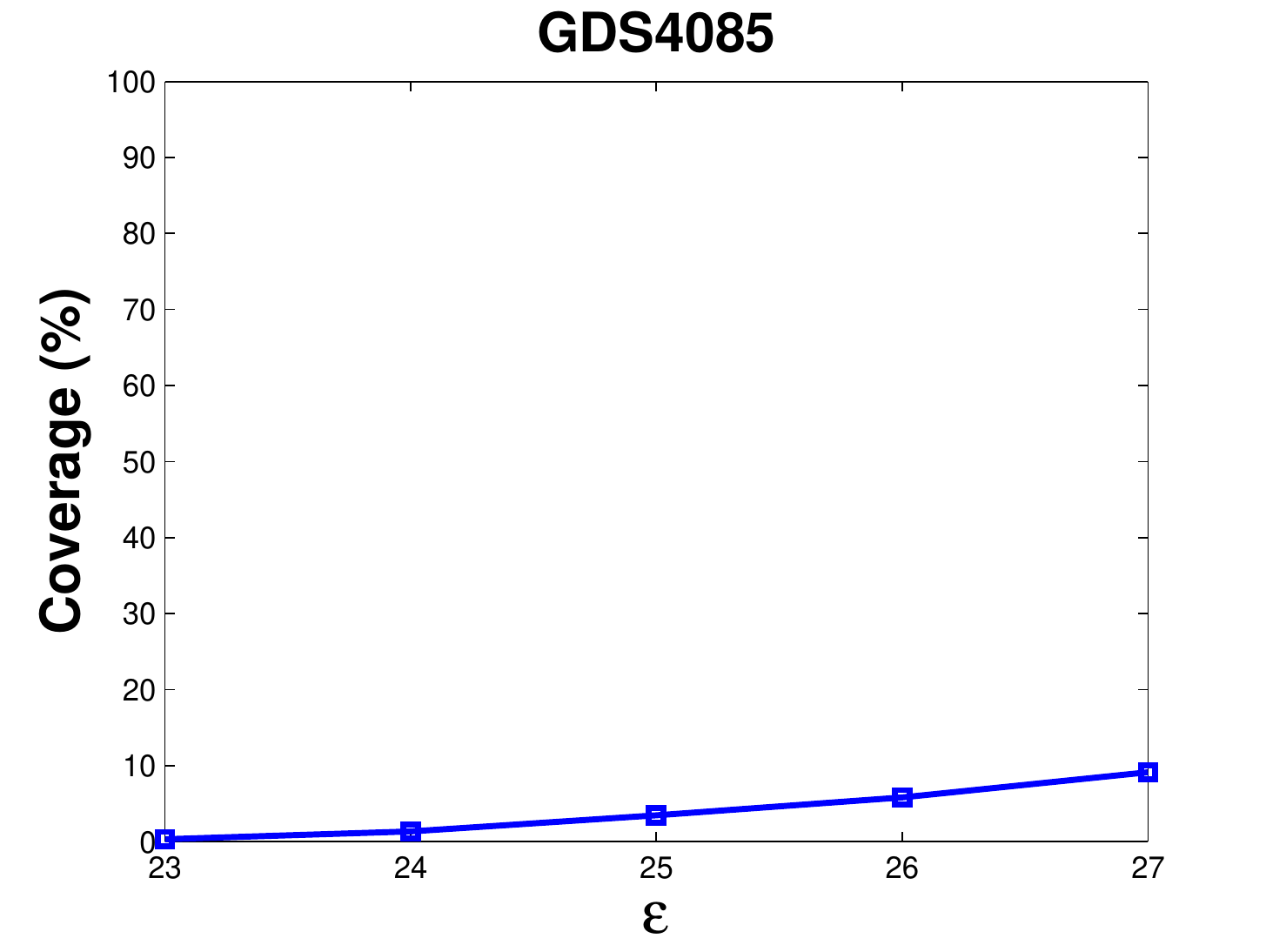}
	}
	\subfigure[]{
		\includegraphics[trim=0.3cm 0.1cm 0.9cm 0.1cm, clip, scale=0.26]{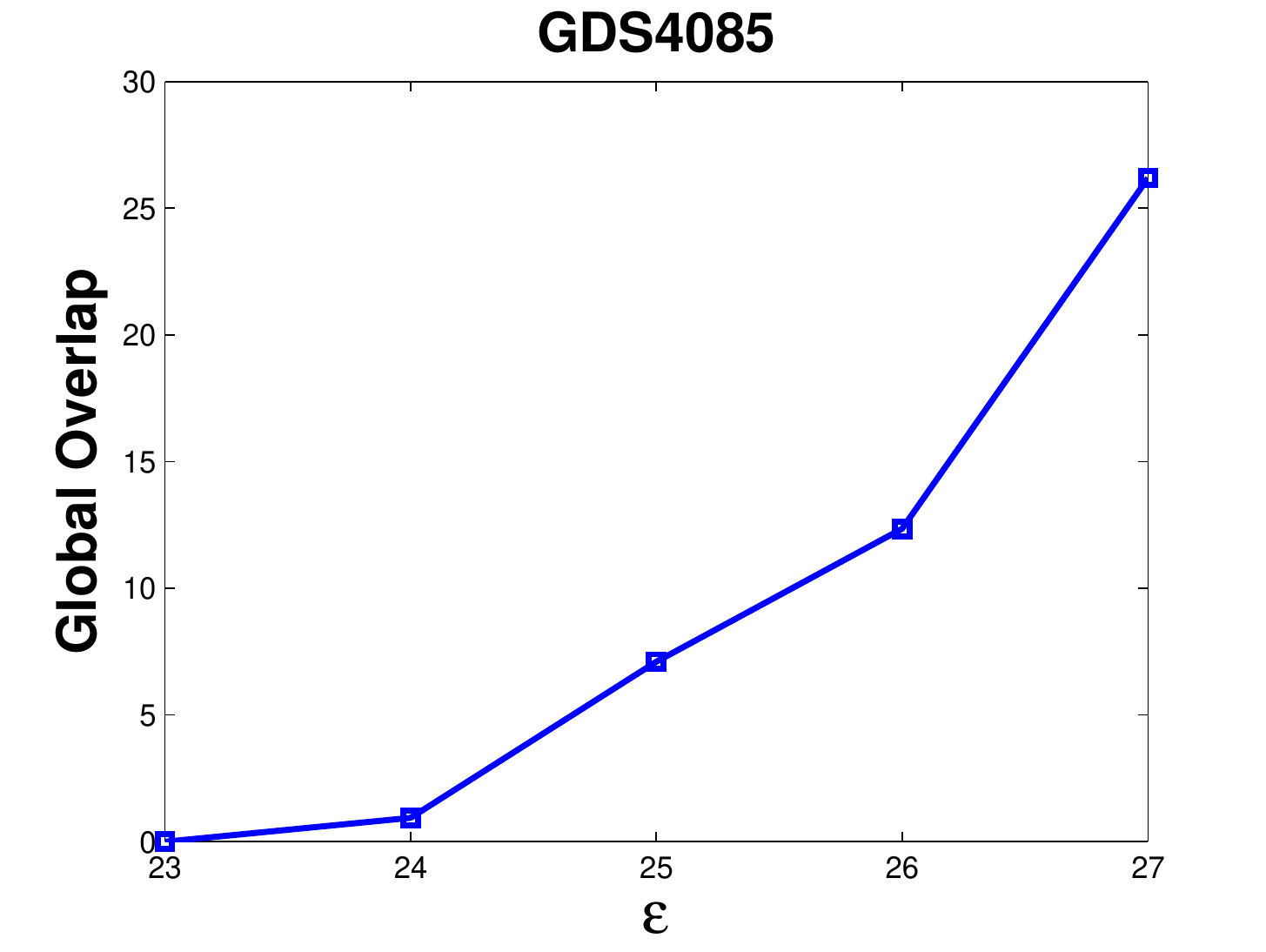}
	}
		
  \caption{Results of RIn-Close\_CVC's sensitivity to the parameter $\epsilon$. The parameter $minRow$ was set to: 144 for Yeast; 59 for GDS232; 795 for GDS750; and 23 for GDS4085.}
  \label{fig:expSensAnalCVC_e}
\end{figure*}

\begin{figure*}
  \centering
	
	\subfigure[]{
		\includegraphics[trim=0.3cm 0.1cm 0.9cm 0.1cm, clip, scale=0.26]{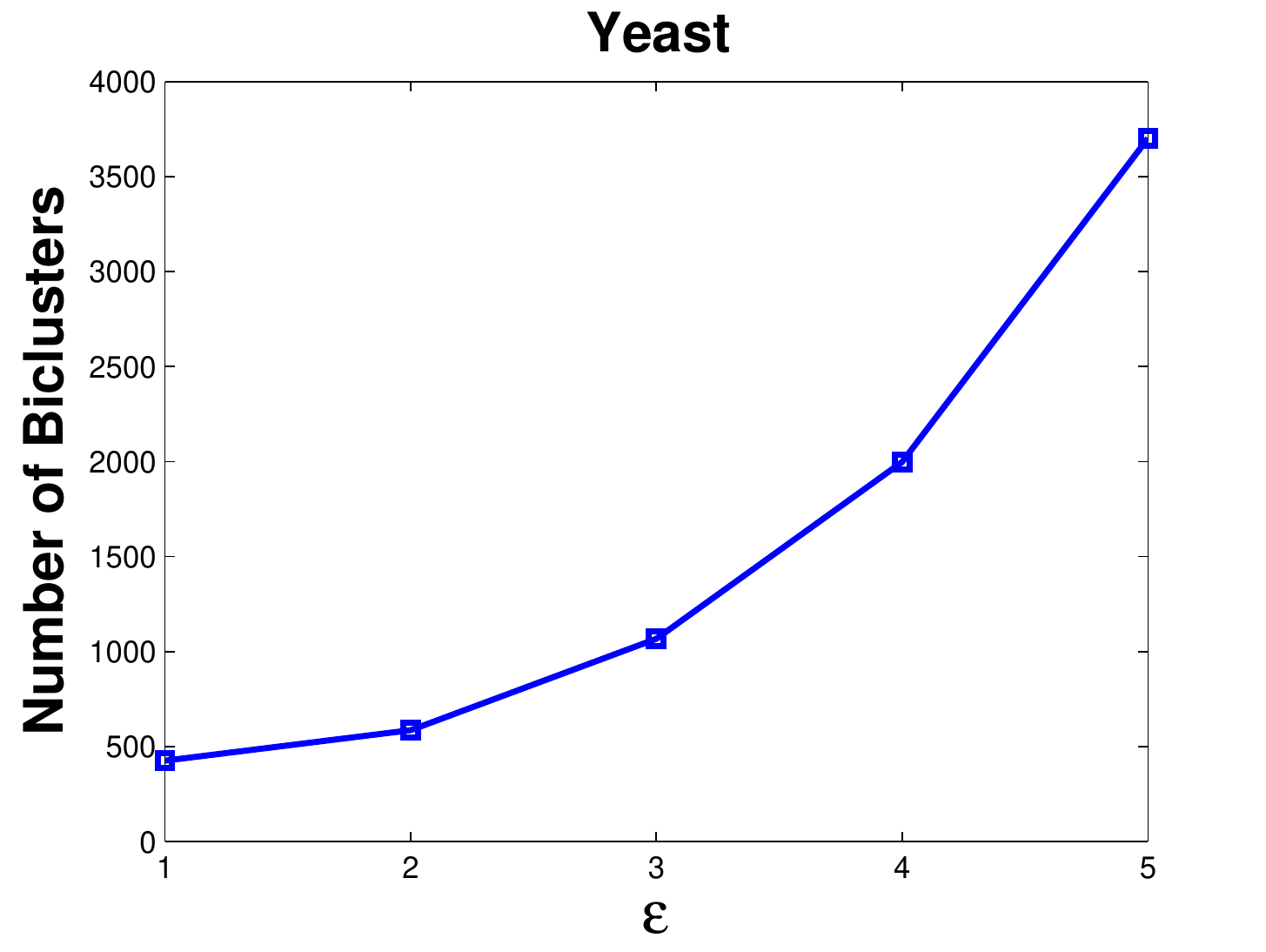}
	}
	\subfigure[]{
		\includegraphics[trim=0.3cm 0.1cm 0.9cm 0.1cm, clip, scale=0.26]{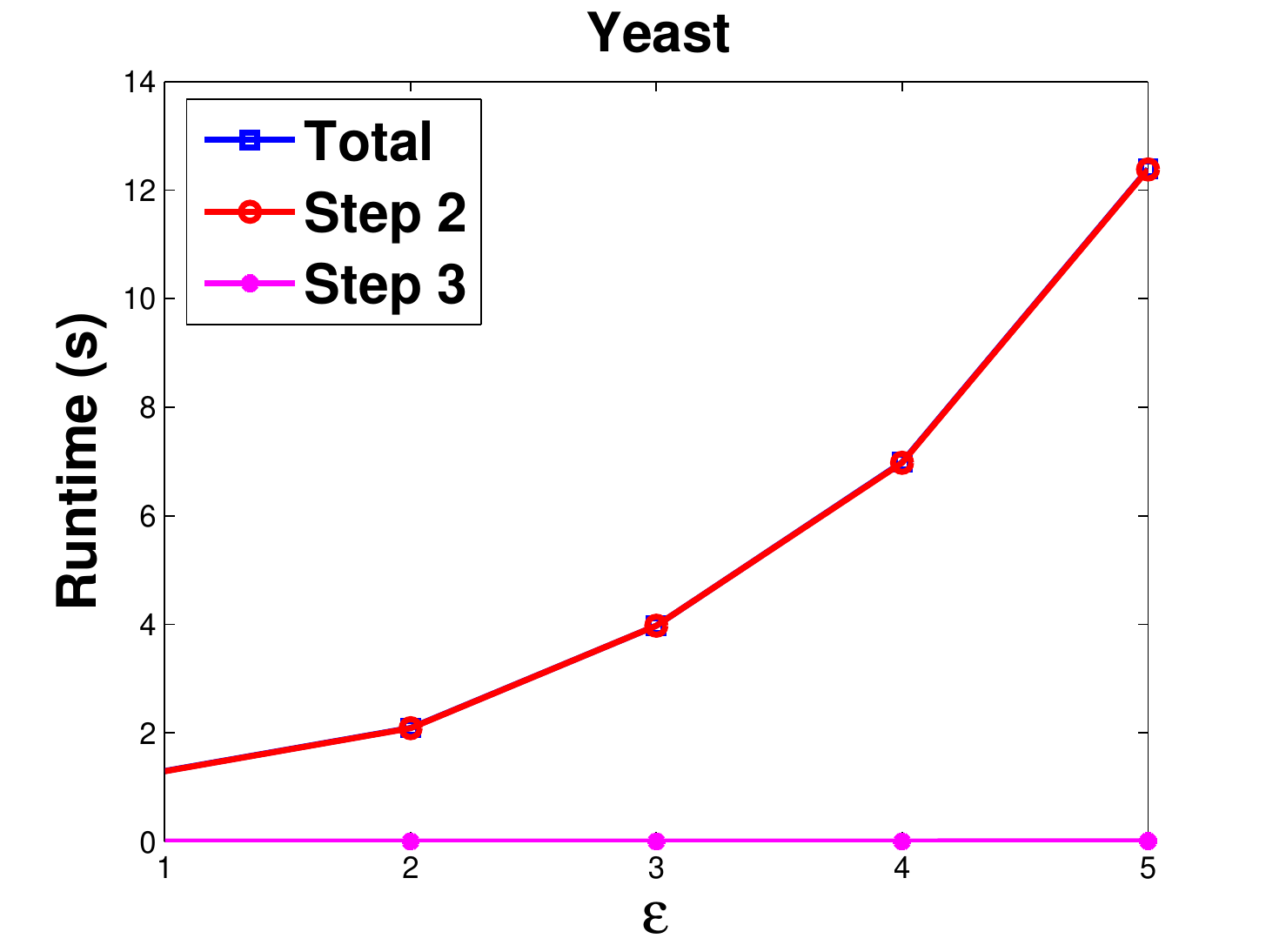}
	}
	\subfigure[]{
		\includegraphics[trim=0.3cm 0.1cm 0.9cm 0.1cm, clip, scale=0.26]{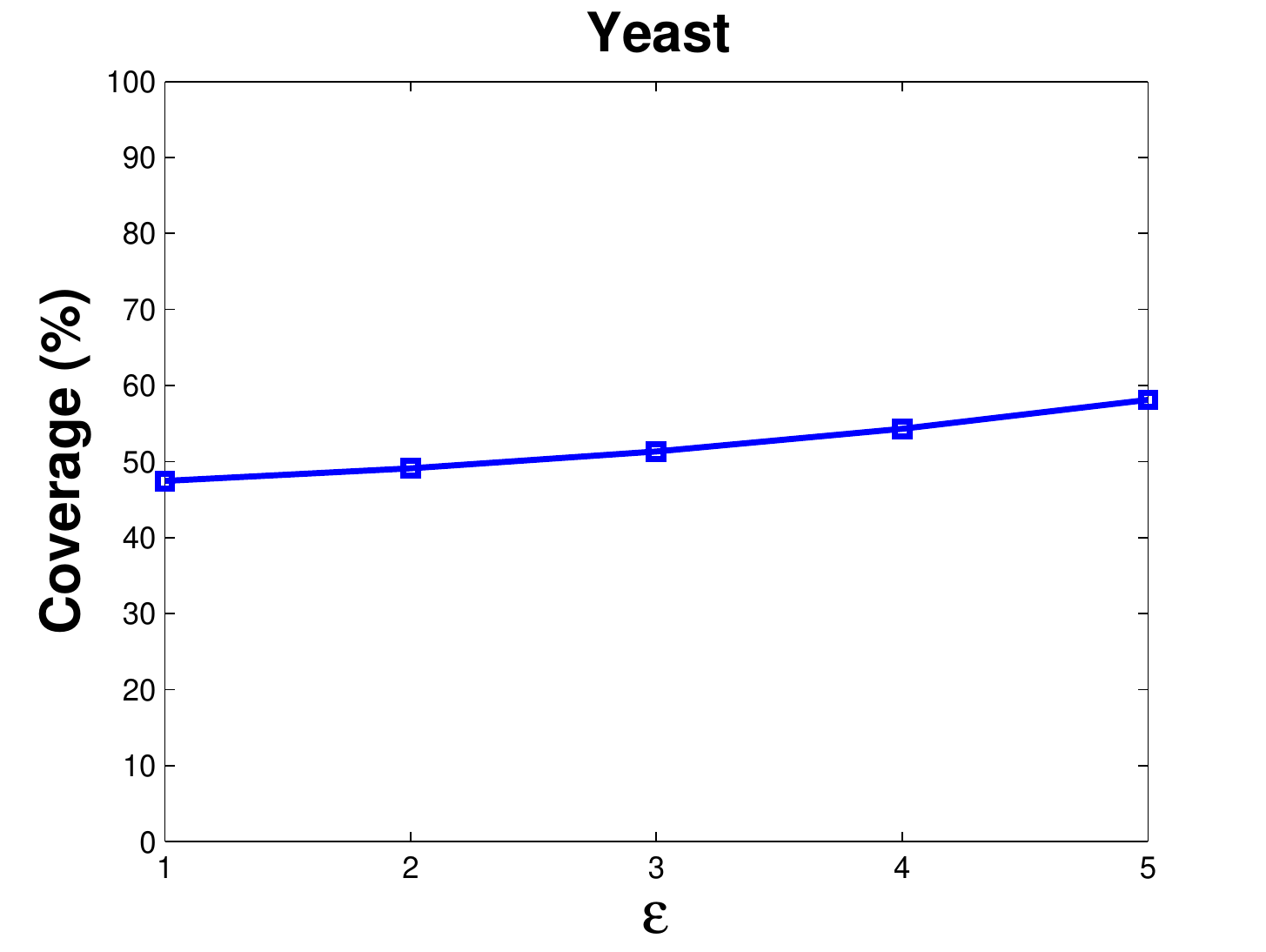}
	}
	\subfigure[]{
		\includegraphics[trim=0.3cm 0.1cm 0.9cm 0.1cm, clip, scale=0.26]{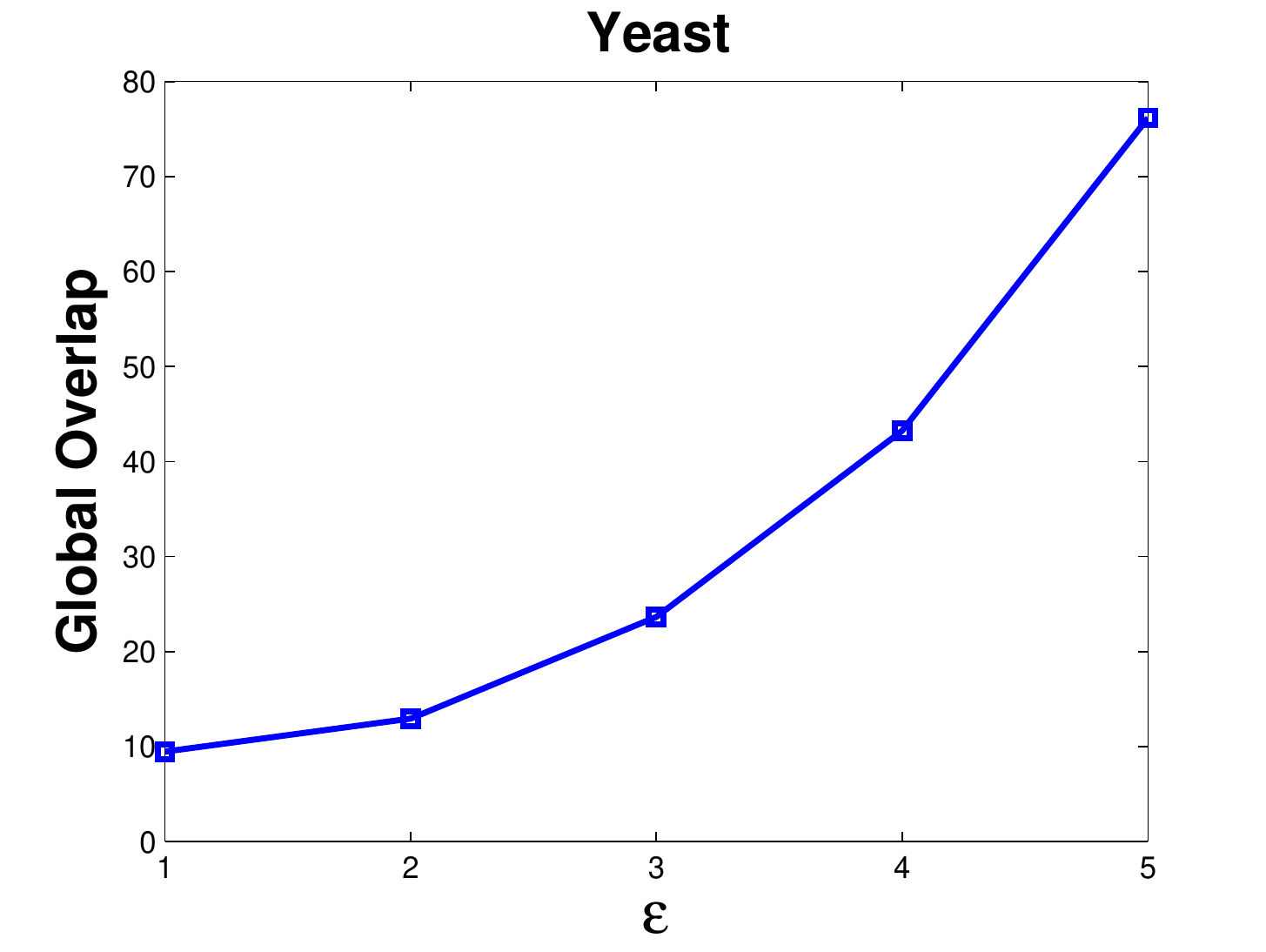}
	}
	
	\subfigure[]{
		\includegraphics[trim=0.3cm 0.1cm 0.9cm 0.1cm, clip, scale=0.26]{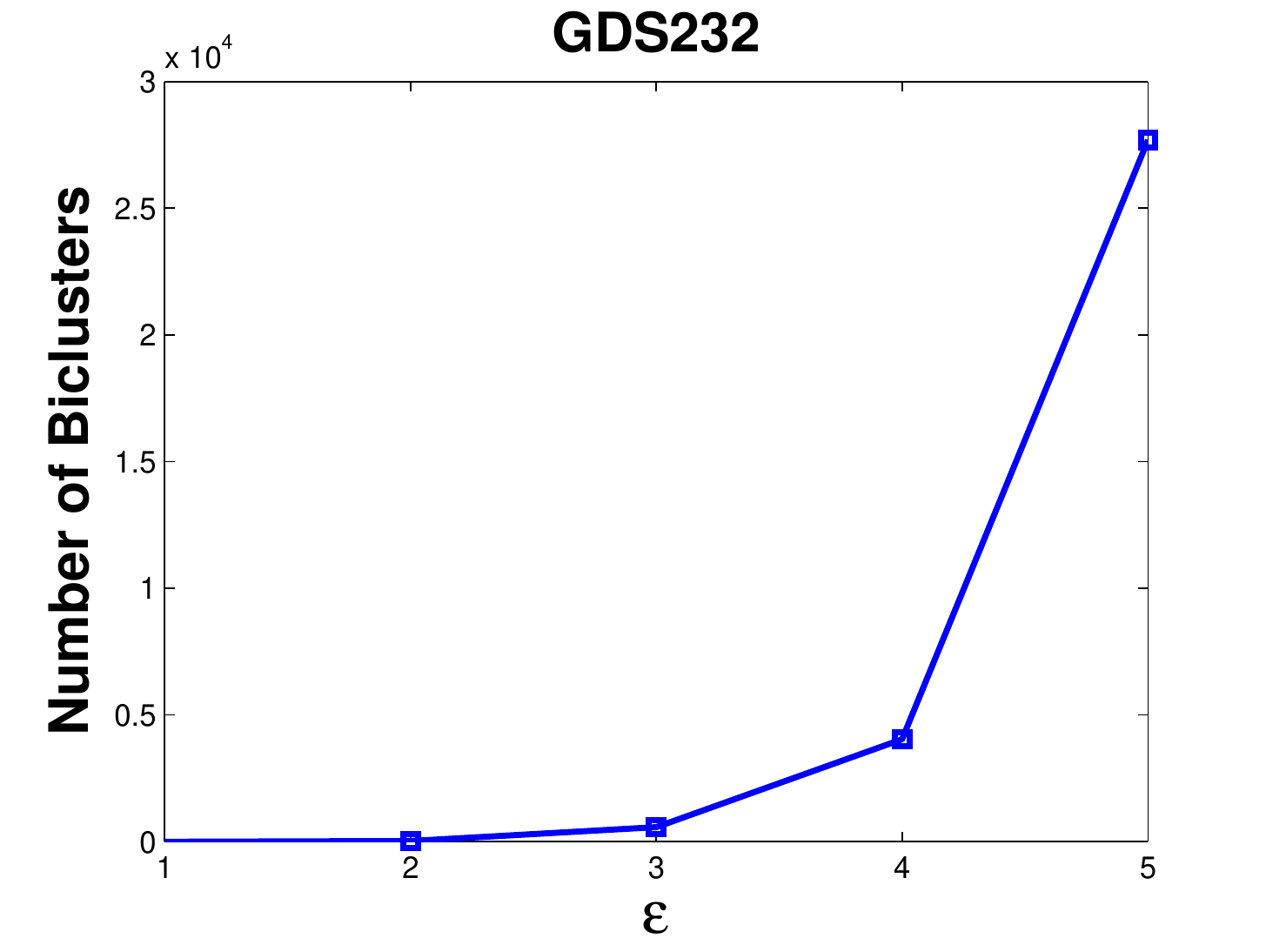}
	}
	\subfigure[]{
		\includegraphics[trim=0.3cm 0.1cm 0.9cm 0.1cm, clip, scale=0.26]{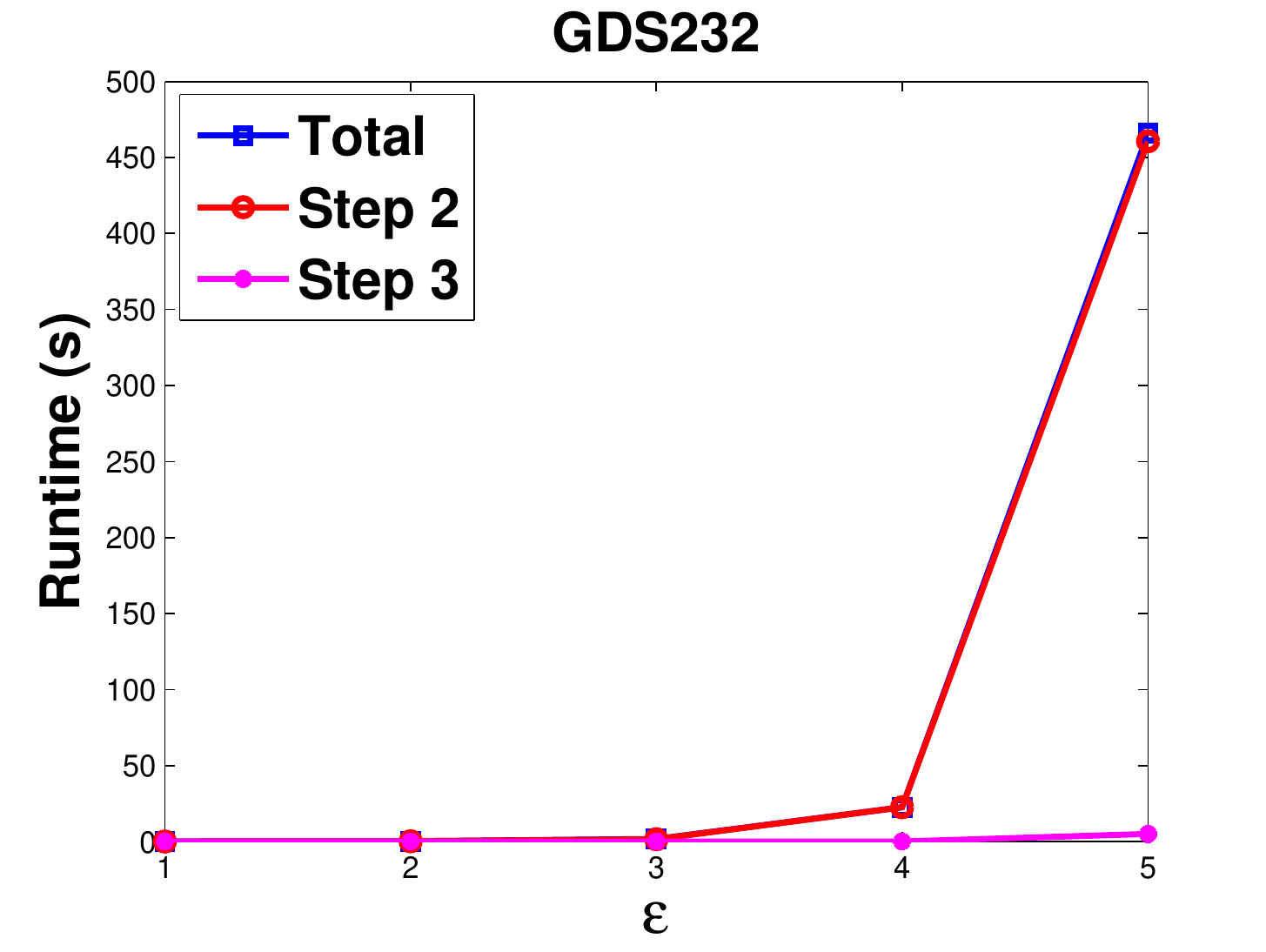}
	}
	\subfigure[]{
		\includegraphics[trim=0.3cm 0.1cm 0.9cm 0.1cm, clip, scale=0.26]{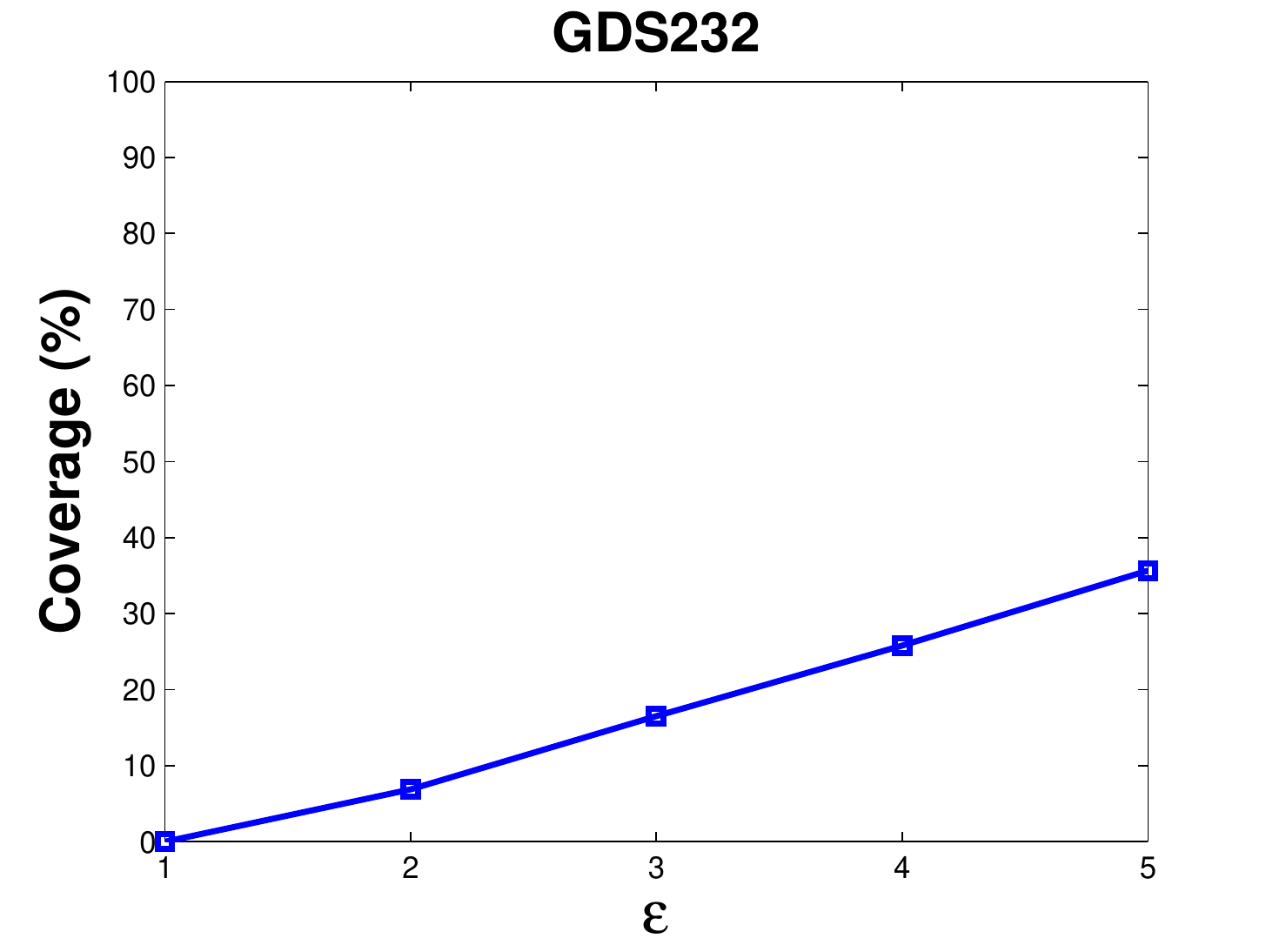}
	}
	\subfigure[]{
		\includegraphics[trim=0.3cm 0.1cm 0.9cm 0.1cm, clip, scale=0.26]{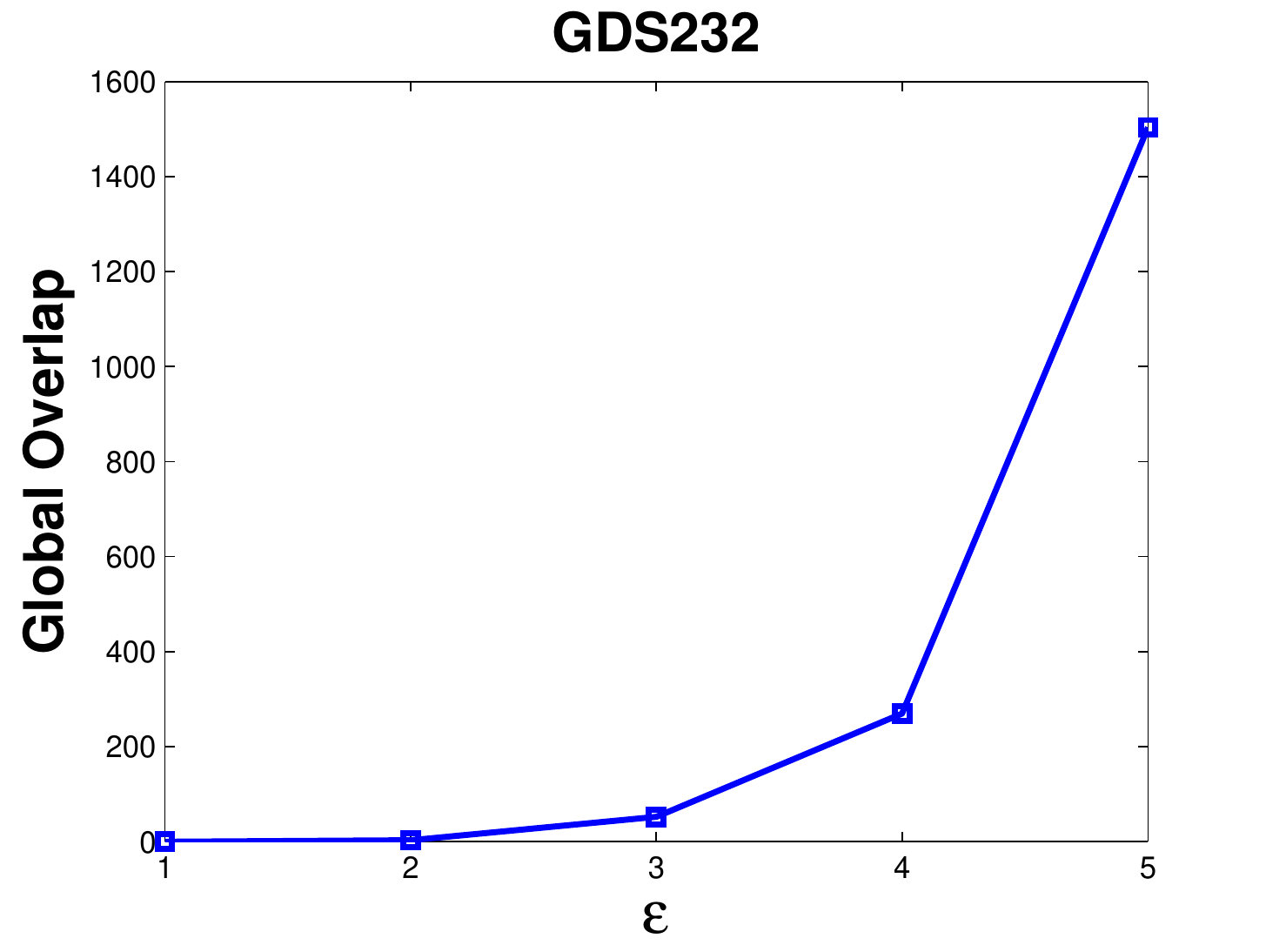}
	}
	
	\subfigure[]{
		\includegraphics[trim=0.3cm 0.1cm 0.9cm 0.1cm, clip, scale=0.26]{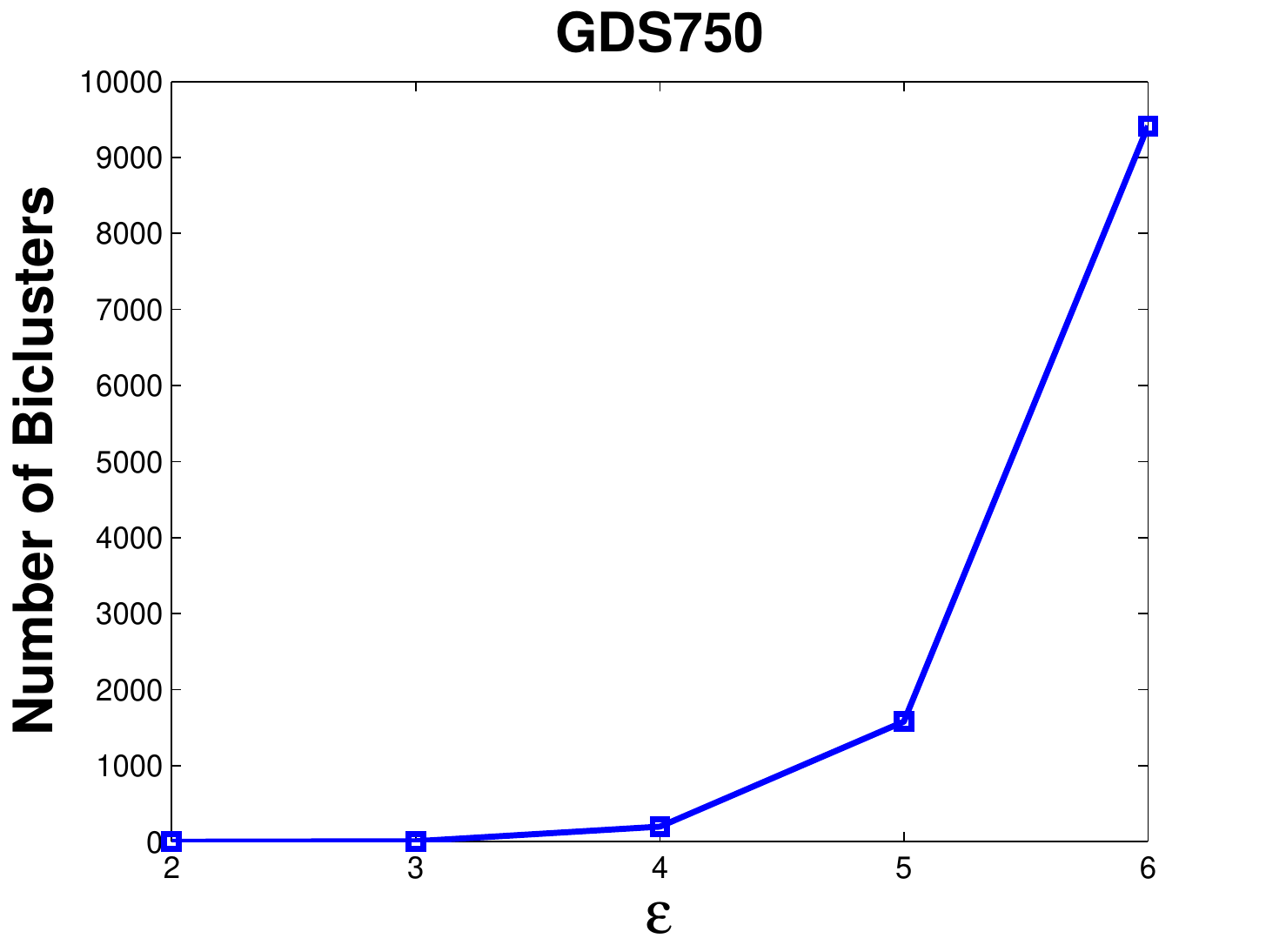}
	}
	\subfigure[]{
		\includegraphics[trim=0.3cm 0.1cm 0.9cm 0.1cm, clip, scale=0.26]{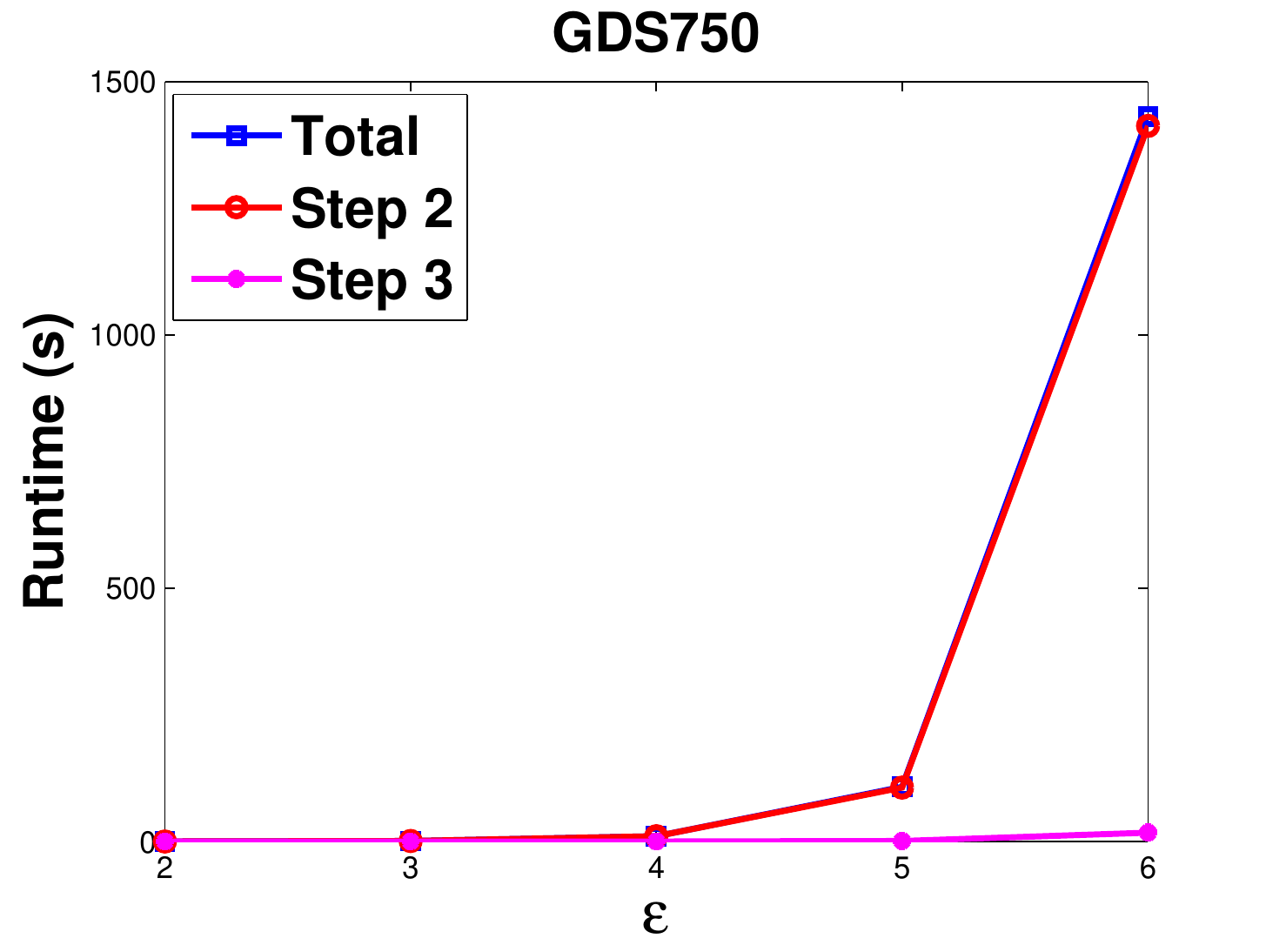}
	}
	\subfigure[]{
		\includegraphics[trim=0.3cm 0.1cm 0.9cm 0.1cm, clip, scale=0.26]{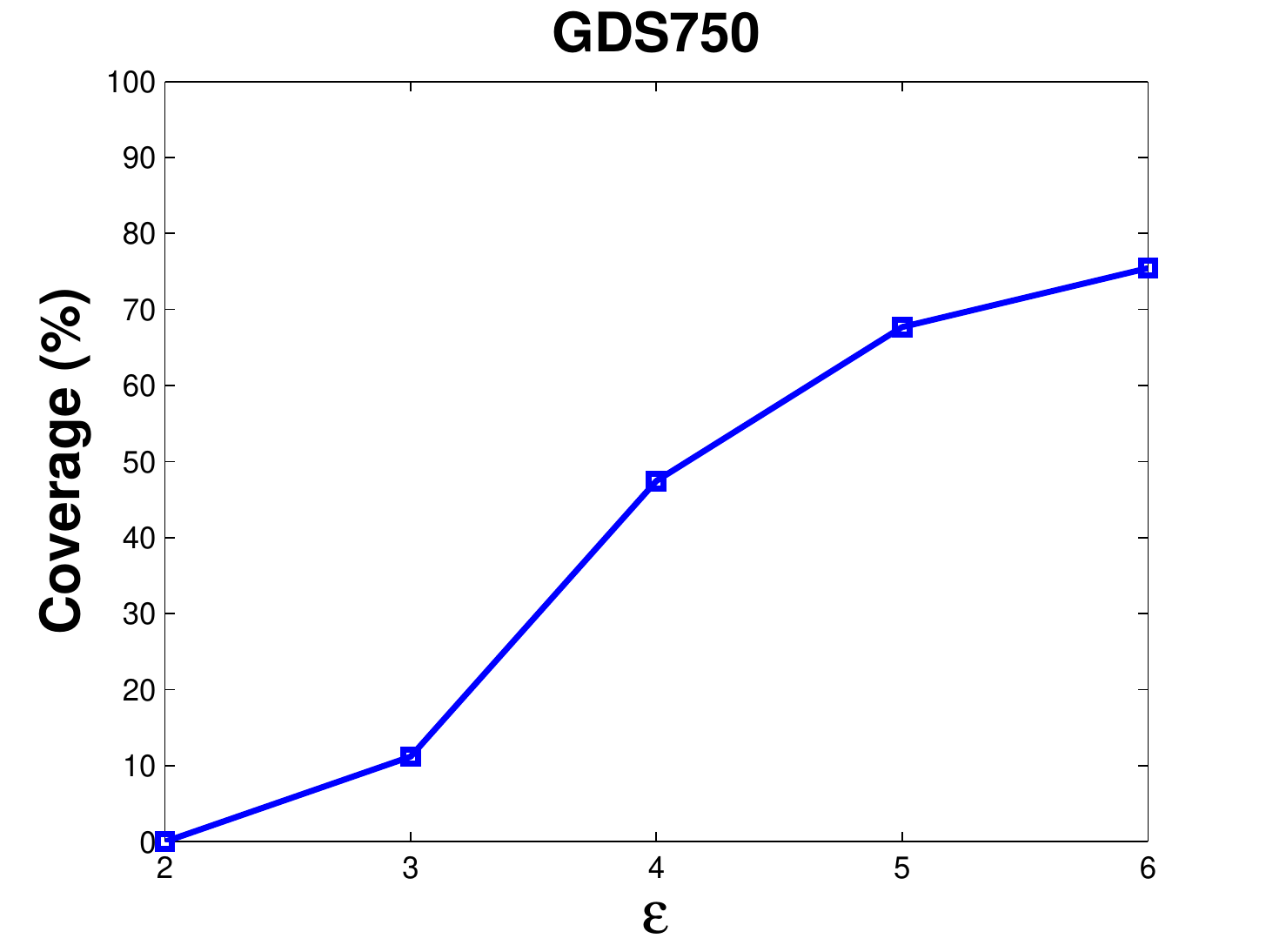}
	}
	\subfigure[]{
		\includegraphics[trim=0.3cm 0.1cm 0.9cm 0.1cm, clip, scale=0.26]{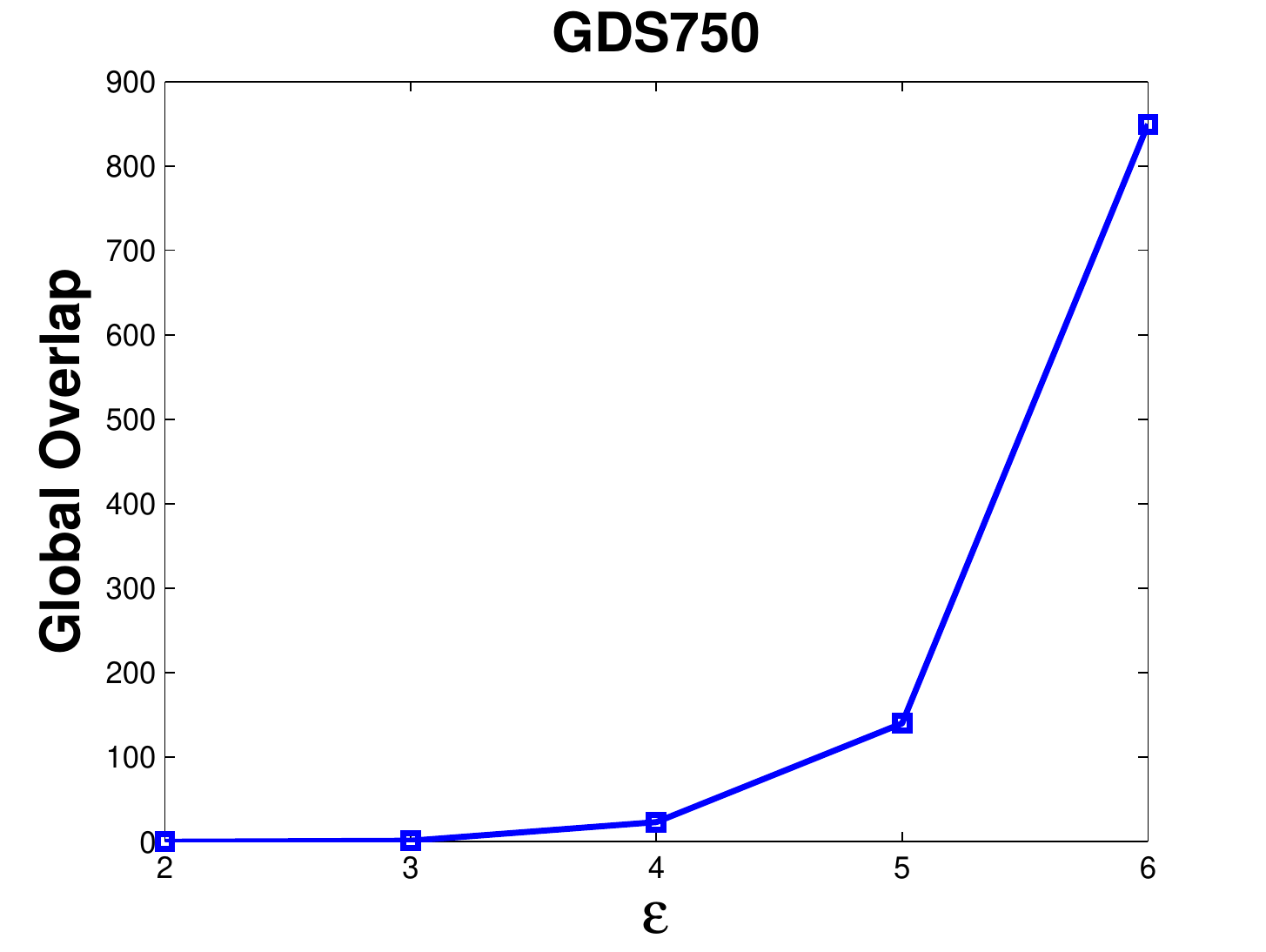}
	}

	\subfigure[]{
		\includegraphics[trim=0.3cm 0.1cm 0.9cm 0.1cm, clip, scale=0.26]{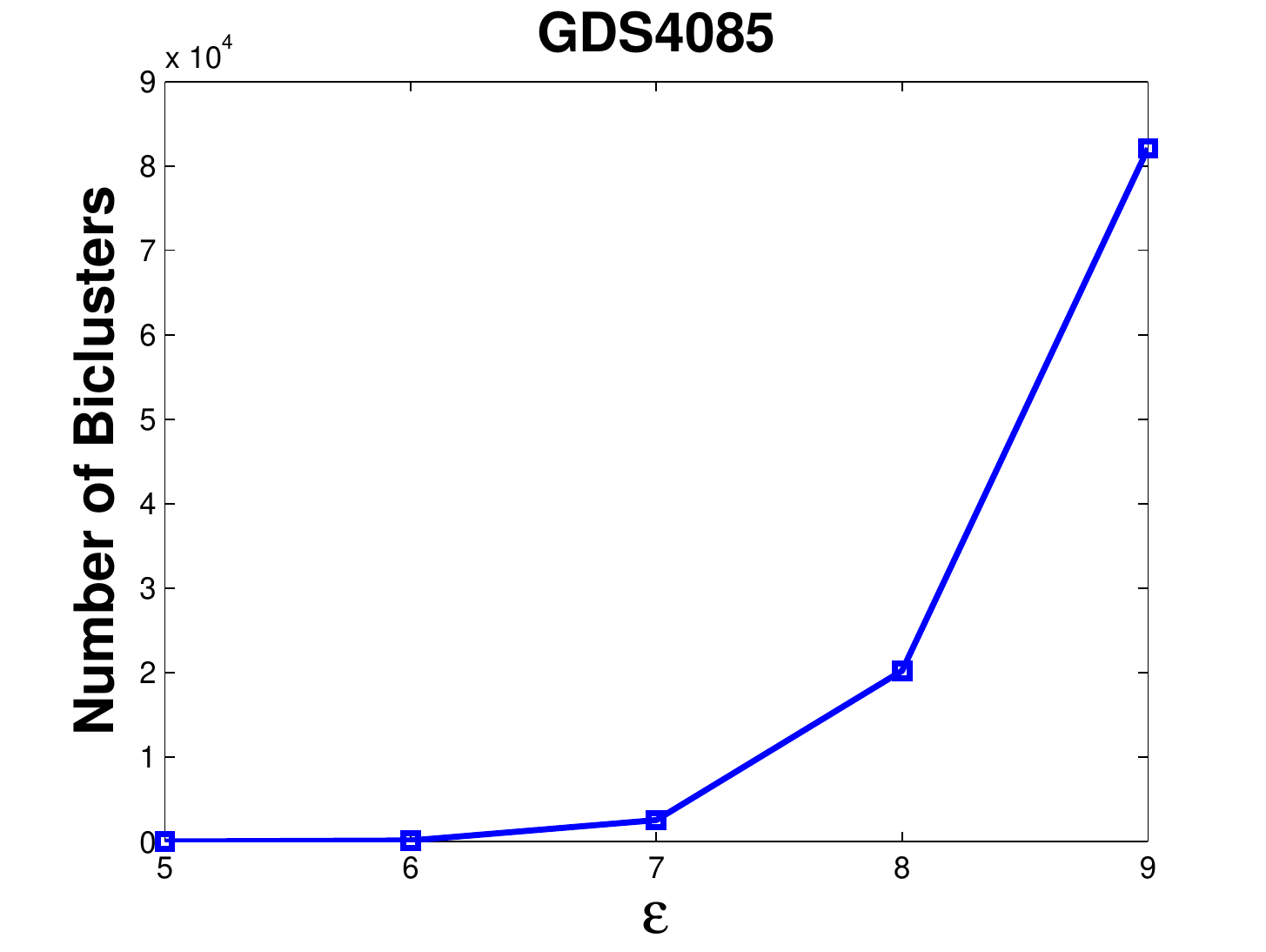}
	}
	\subfigure[]{
		\includegraphics[trim=0.3cm 0.1cm 0.9cm 0.1cm, clip, scale=0.26]{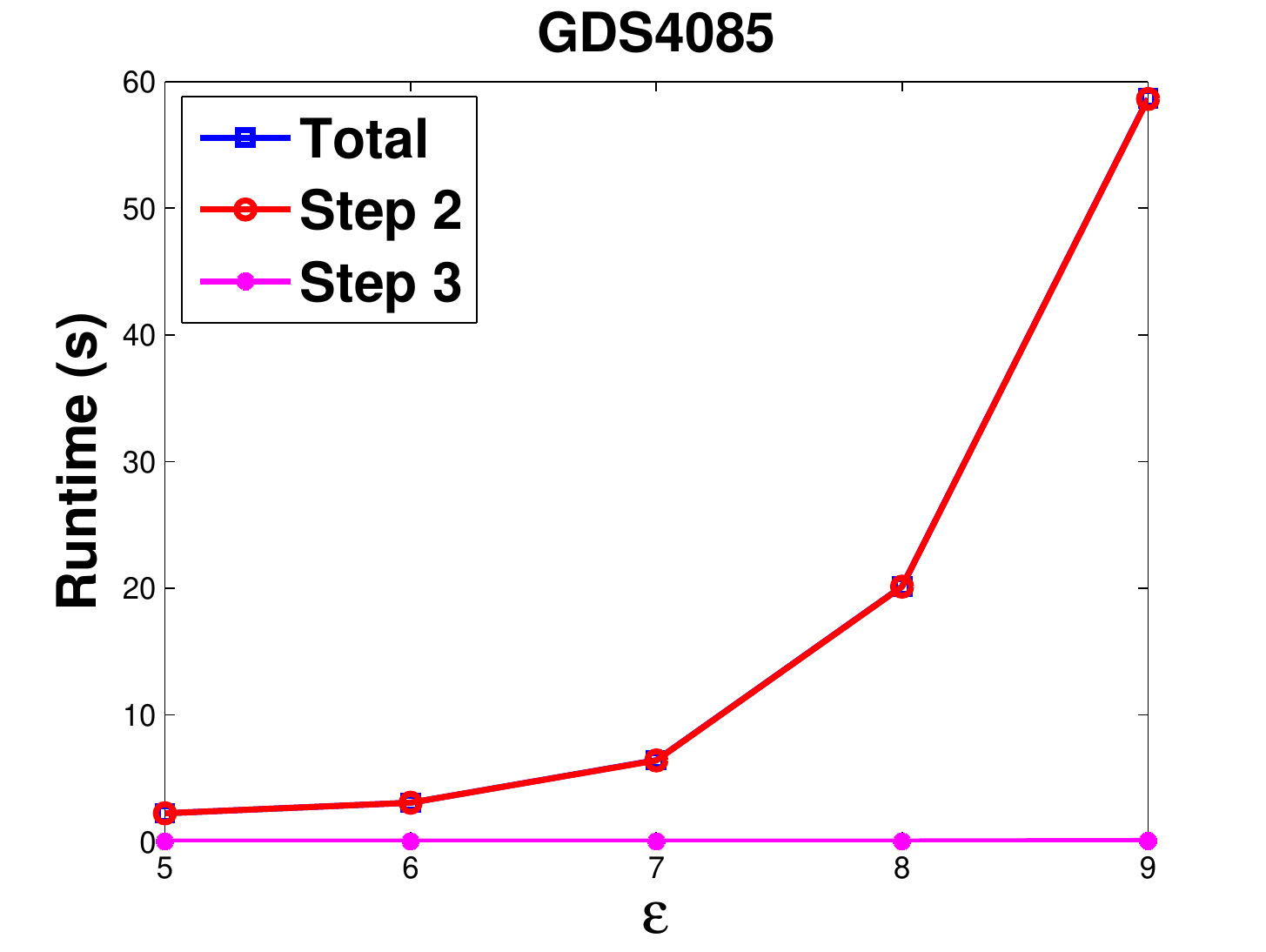}
	}
	\subfigure[]{
		\includegraphics[trim=0.3cm 0.1cm 0.9cm 0.1cm, clip, scale=0.26]{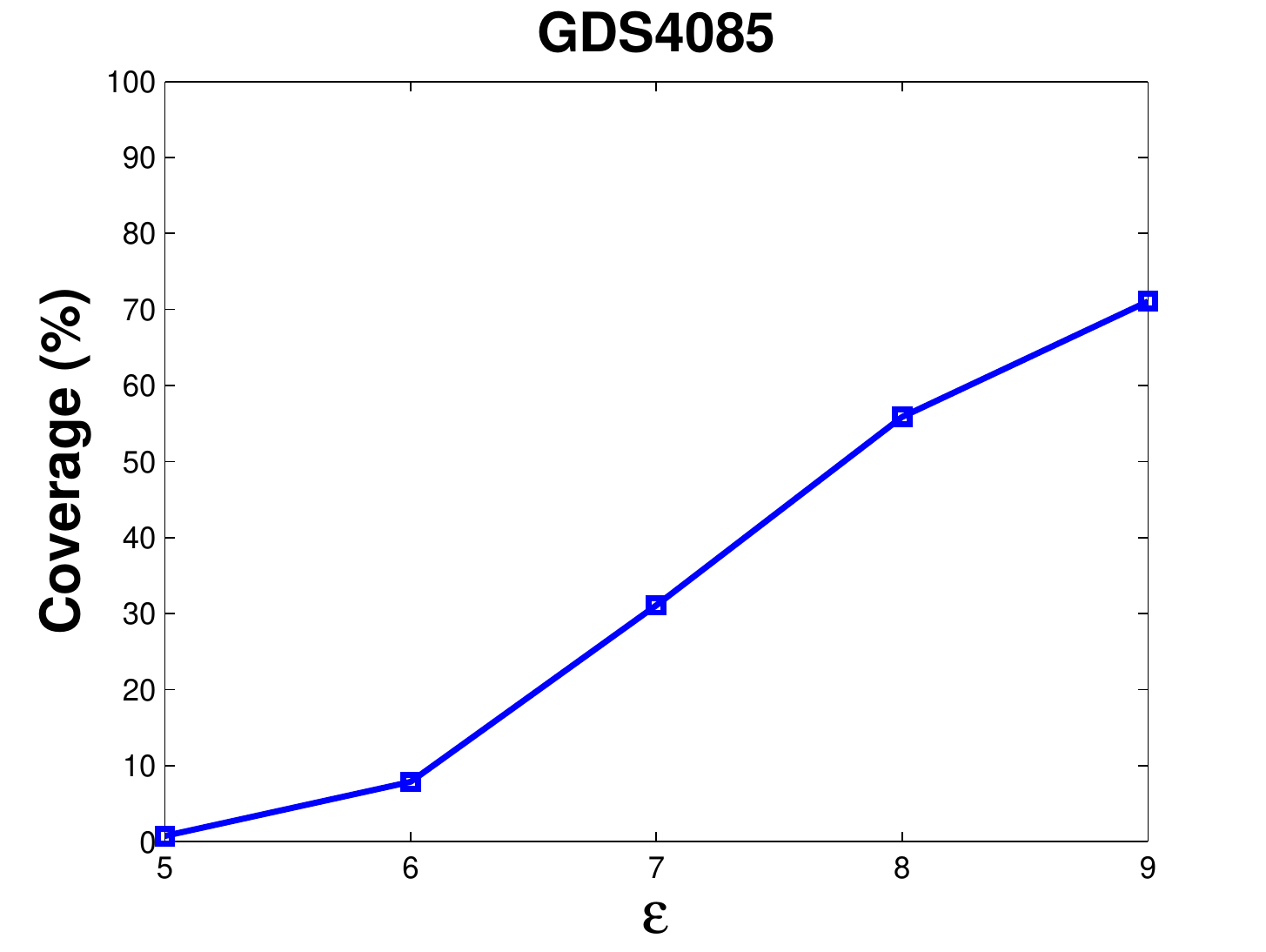}
	}
	\subfigure[]{
		\includegraphics[trim=0.3cm 0.1cm 0.9cm 0.1cm, clip, scale=0.26]{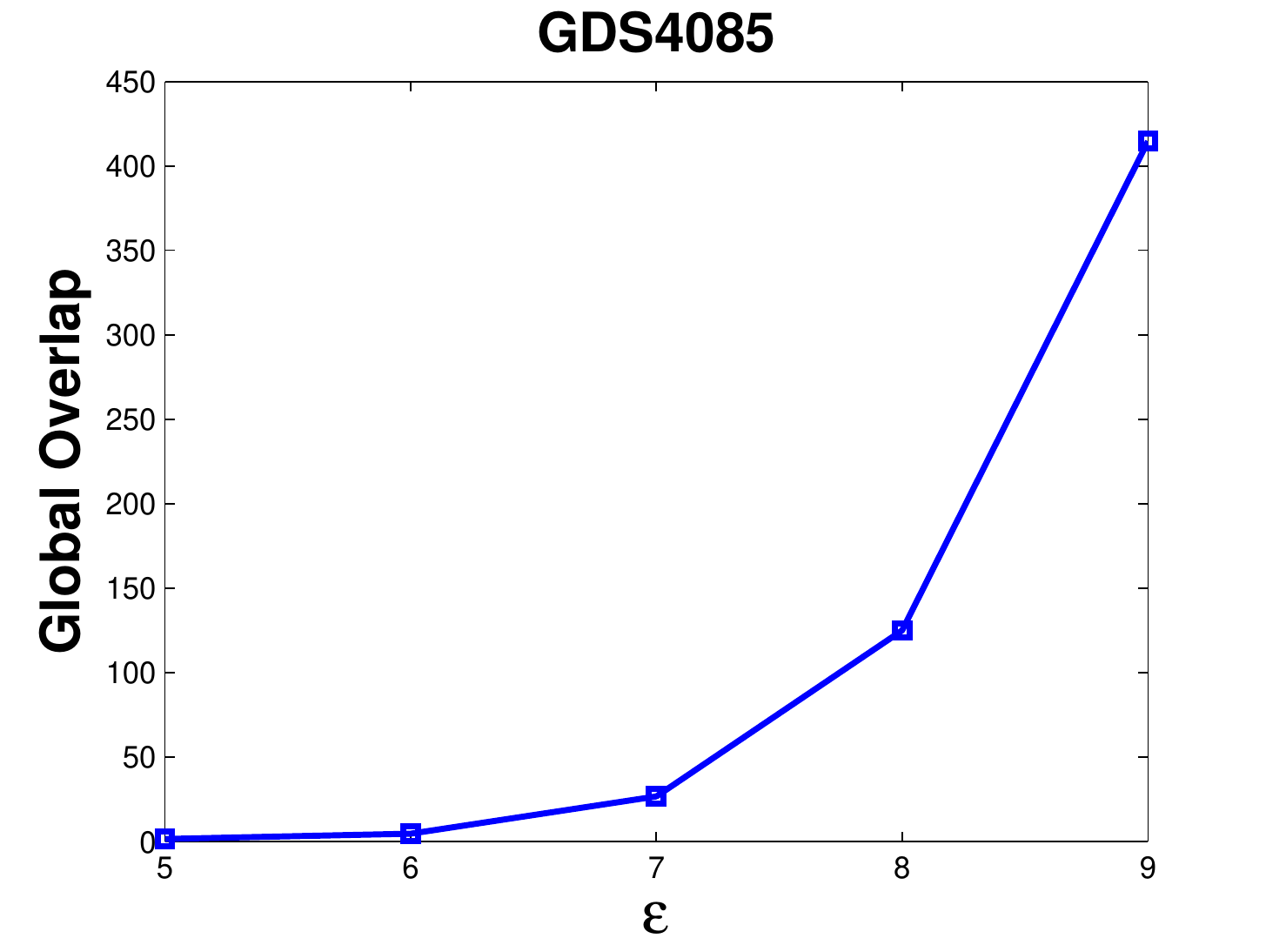}
	}
		
  \caption{Results of RIn-Close\_CHV's sensitivity to the parameter $\epsilon$. The parameter $minRow$ was set to: 144 for Yeast; 59 for GDS232; 795 for GDS750; and 23 for GDS4085.}
  \label{fig:expSensAnalCHV_e}
\end{figure*}

Figs.~\ref{fig:expSensAnalCVC_e} and \ref{fig:expSensAnalCHV_e} shows respectively the sensitivity of RIn-Close\_CVC and RIn-Close\_CHV to the parameter $\epsilon$. Usually, the number of biclusters, the runtime, the coverage, and the global overlap increases with $\epsilon$. The only exception is the global overlap of the dataset Yeast in Fig.~\ref{fig:expSensAnalCVC_e}(d). The coverage will always increase with $\epsilon$, because all portions of the dataset explored with $\epsilon = x$, will be explored with $\epsilon > x$, as stated in Property~\ref{prop:cov2} (see Subsection~\ref{sec:bicMaxProp}). However, as we stated in Subsection~\ref{sec:bicMaxProp}, the number of biclusters will not always increase with $\epsilon$. The global overlap depends on the coverage and the number of biclusters. If we increase $\epsilon$ and find more biclusters, but we explore pretty much the same portions of the dataset, the global overlap will increase. On the other hand, if these new biclusters bring a significant gain in coverage, the global overlap tends to decrease. Therefore, when the global overlap's growth rate is higher than the coverage's growth rate, it indicates that we are finding new biclusters in portions of the dataset explored with lower values of $\epsilon$. Oliveira \textit{et al}. \cite{OliveiraEtAl2015} illustrated how the noise is responsible for fragmenting each true bicluster into many with high overlapping. As it complicates the analysis of the results, the aggregation of these biclusters is recommended \cite{OliveiraEtAl2015, ZhaoZaki2005}. Given that the runtime is proportional to the number of biclusters, the choice of $\epsilon$ must consider the gain in the coverage and the gain in the global overlap. If we only have a considerable gain in the global overlap, then it might be more practicable to use a lower $\epsilon$.

\begin{figure*}
  \centering
	
	\subfigure[]{
		\includegraphics[trim=0.3cm 0.1cm 0.9cm 0.1cm, clip, scale=0.26]{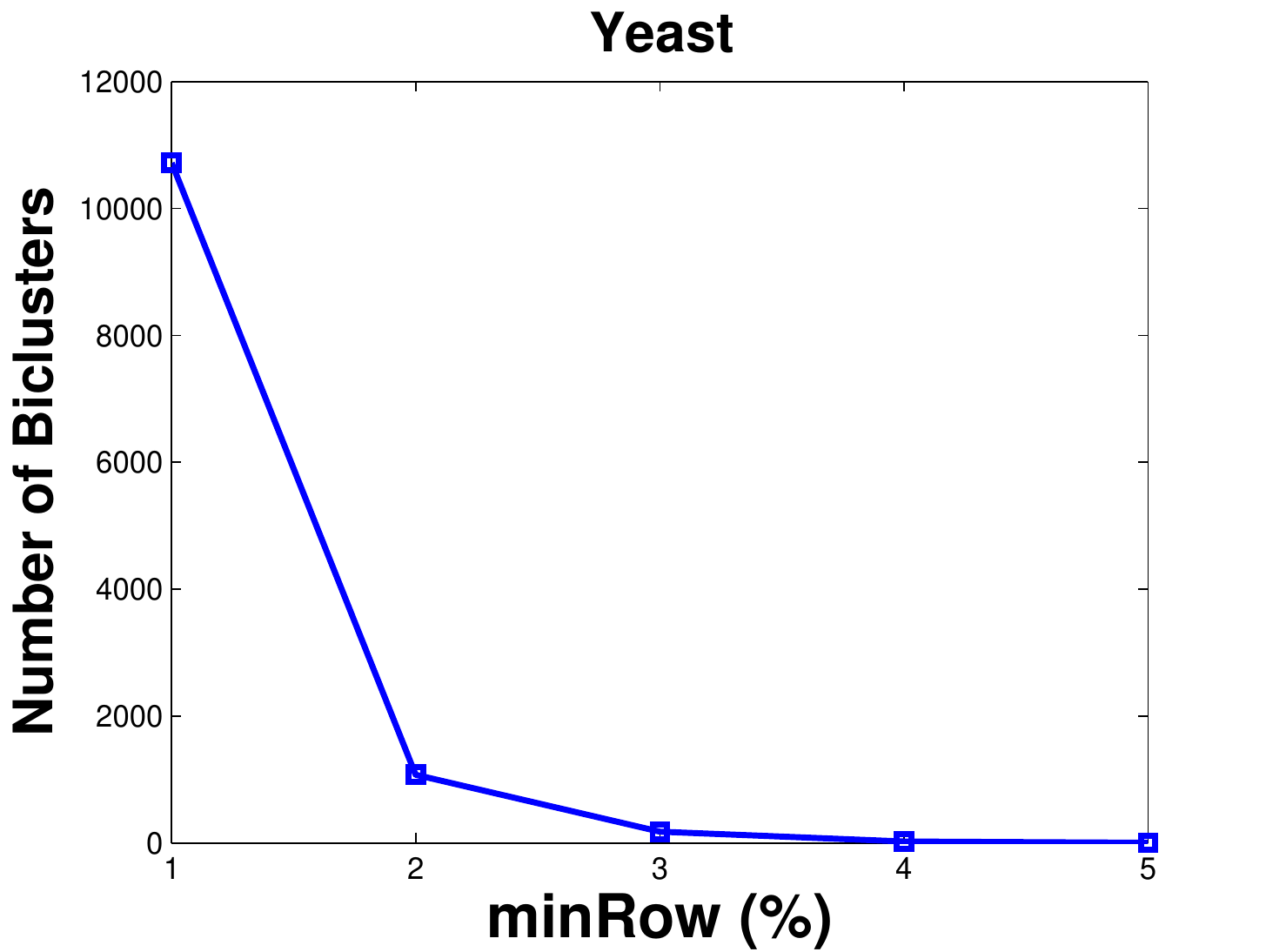}
	}
	\subfigure[]{
		\includegraphics[trim=0.3cm 0.1cm 0.9cm 0.1cm, clip, scale=0.26]{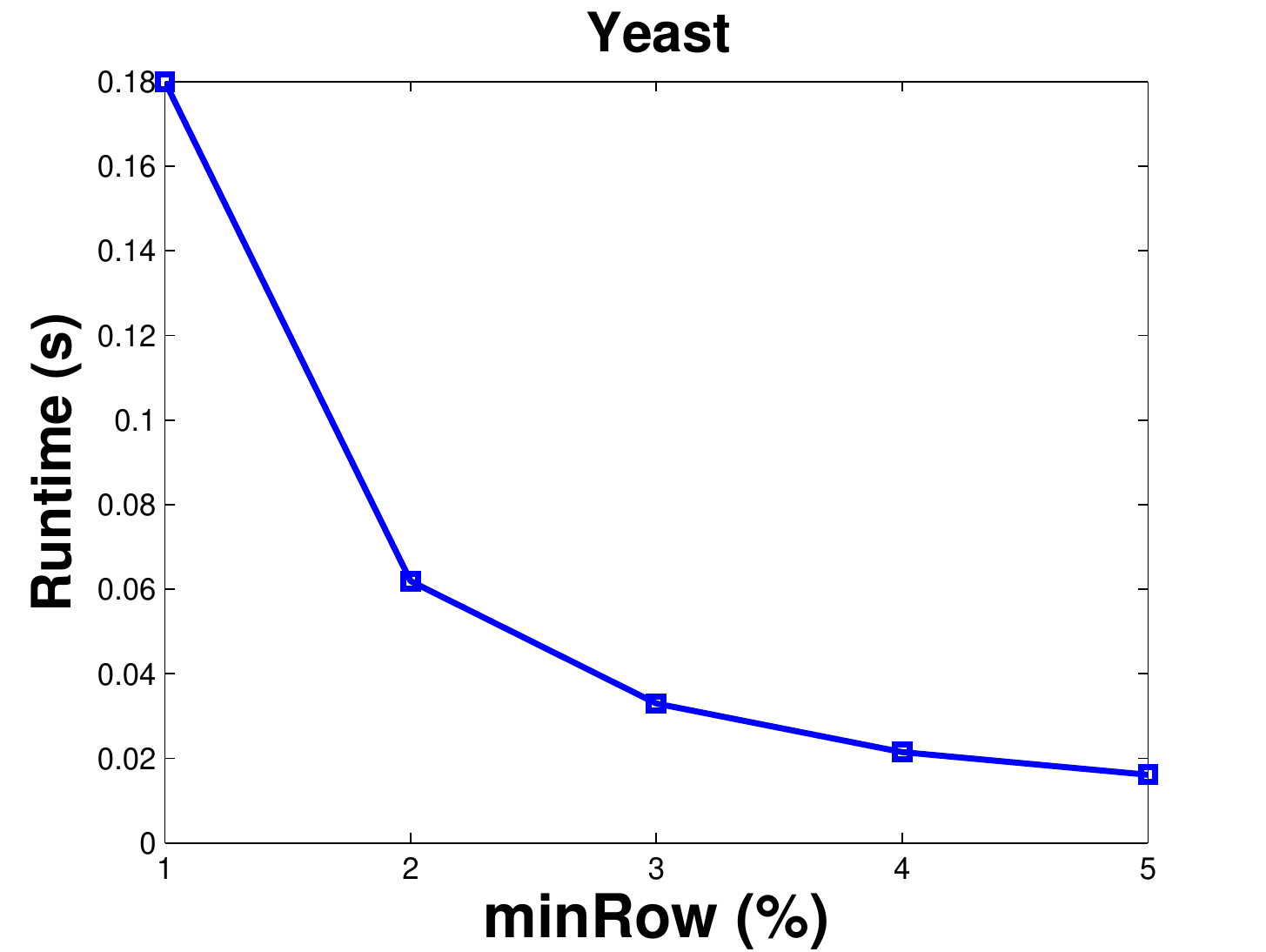}
	}
	\subfigure[]{
		\includegraphics[trim=0.3cm 0.1cm 0.9cm 0.1cm, clip, scale=0.26]{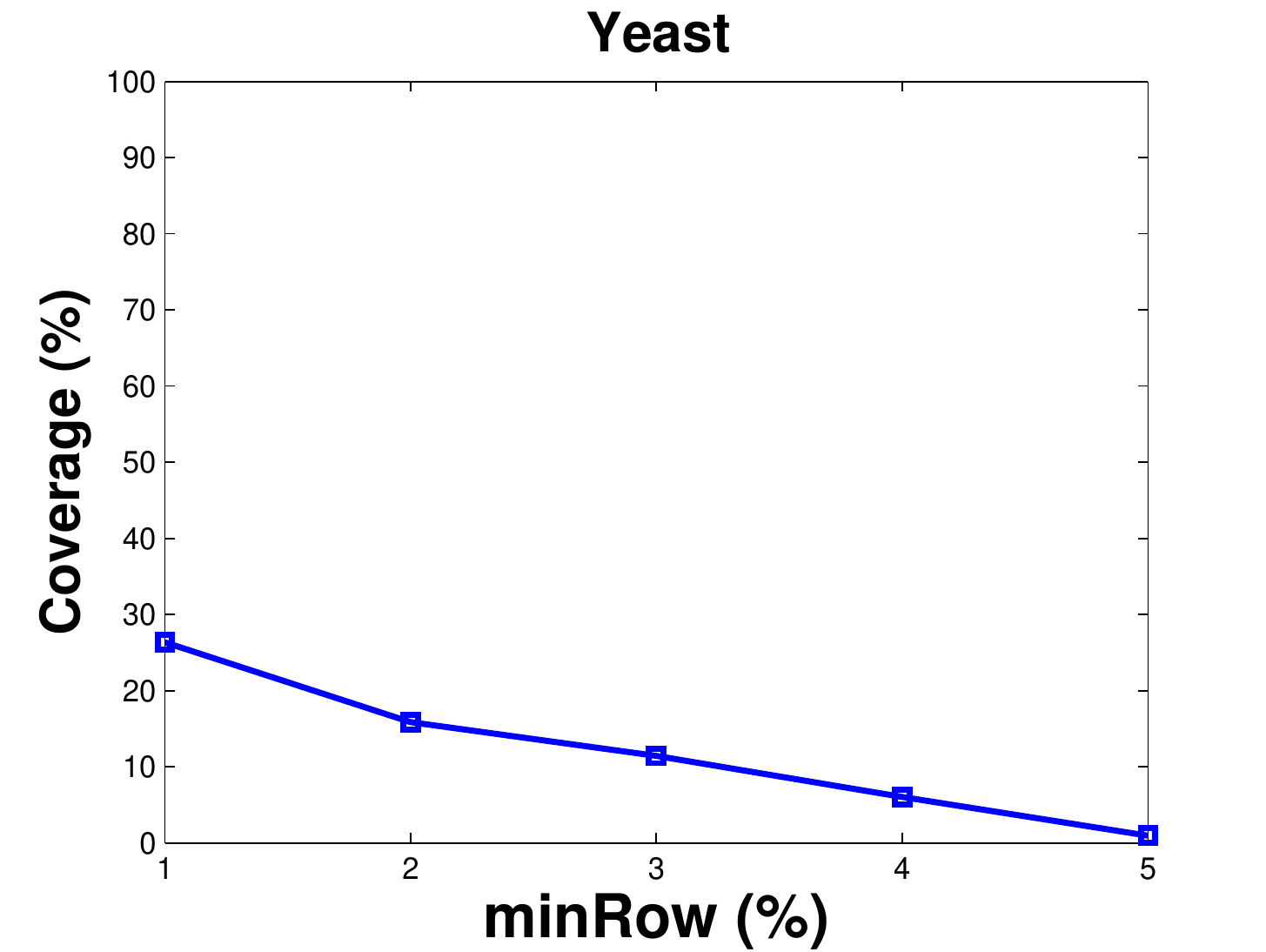}
	}
	\subfigure[]{
		\includegraphics[trim=0.3cm 0.1cm 0.9cm 0.1cm, clip, scale=0.26]{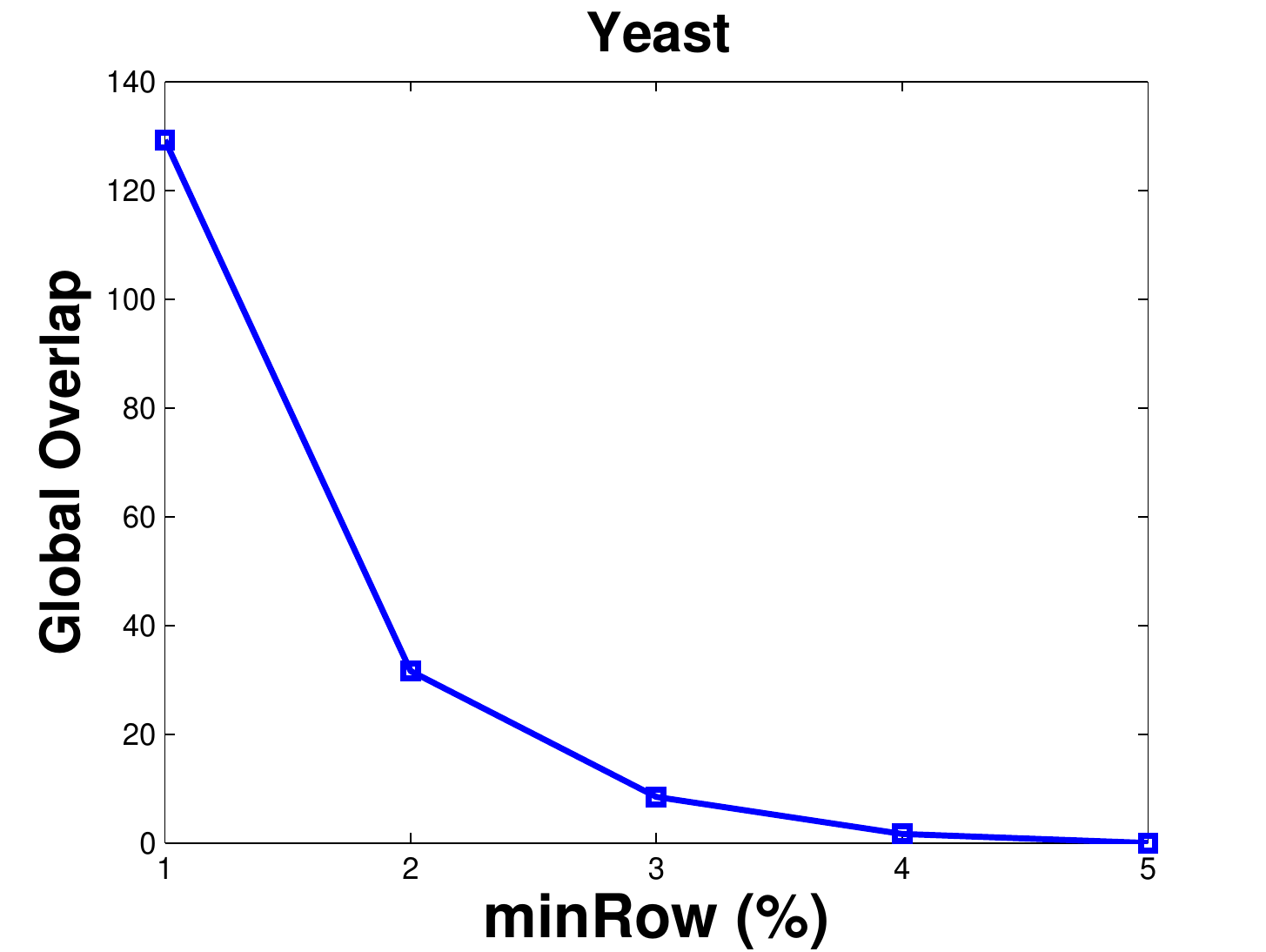}
	}
	
	\subfigure[]{
		\includegraphics[trim=0.3cm 0.1cm 0.9cm 0.1cm, clip, scale=0.26]{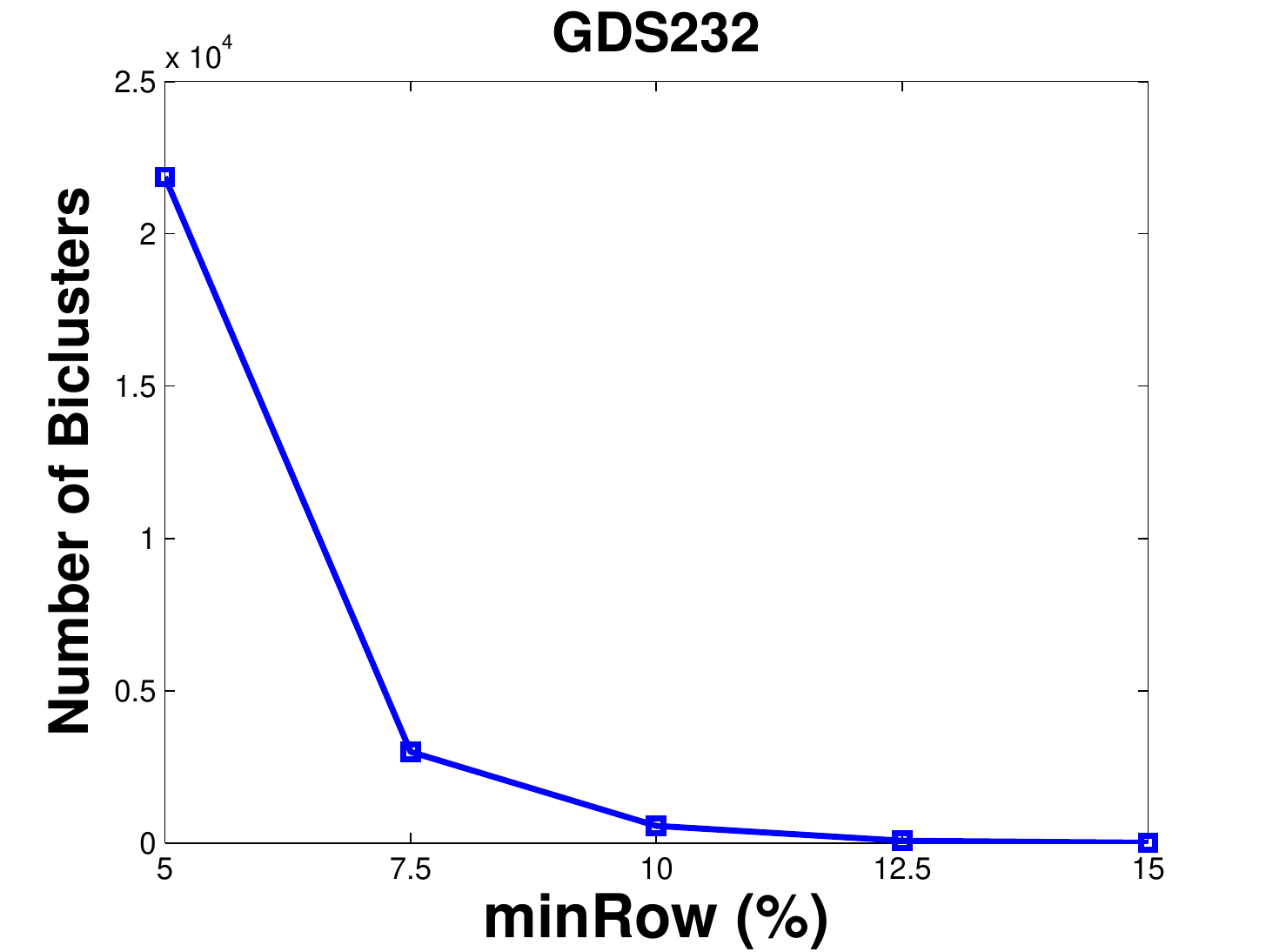}
	}
	\subfigure[]{
		\includegraphics[trim=0.3cm 0.1cm 0.9cm 0.1cm, clip, scale=0.26]{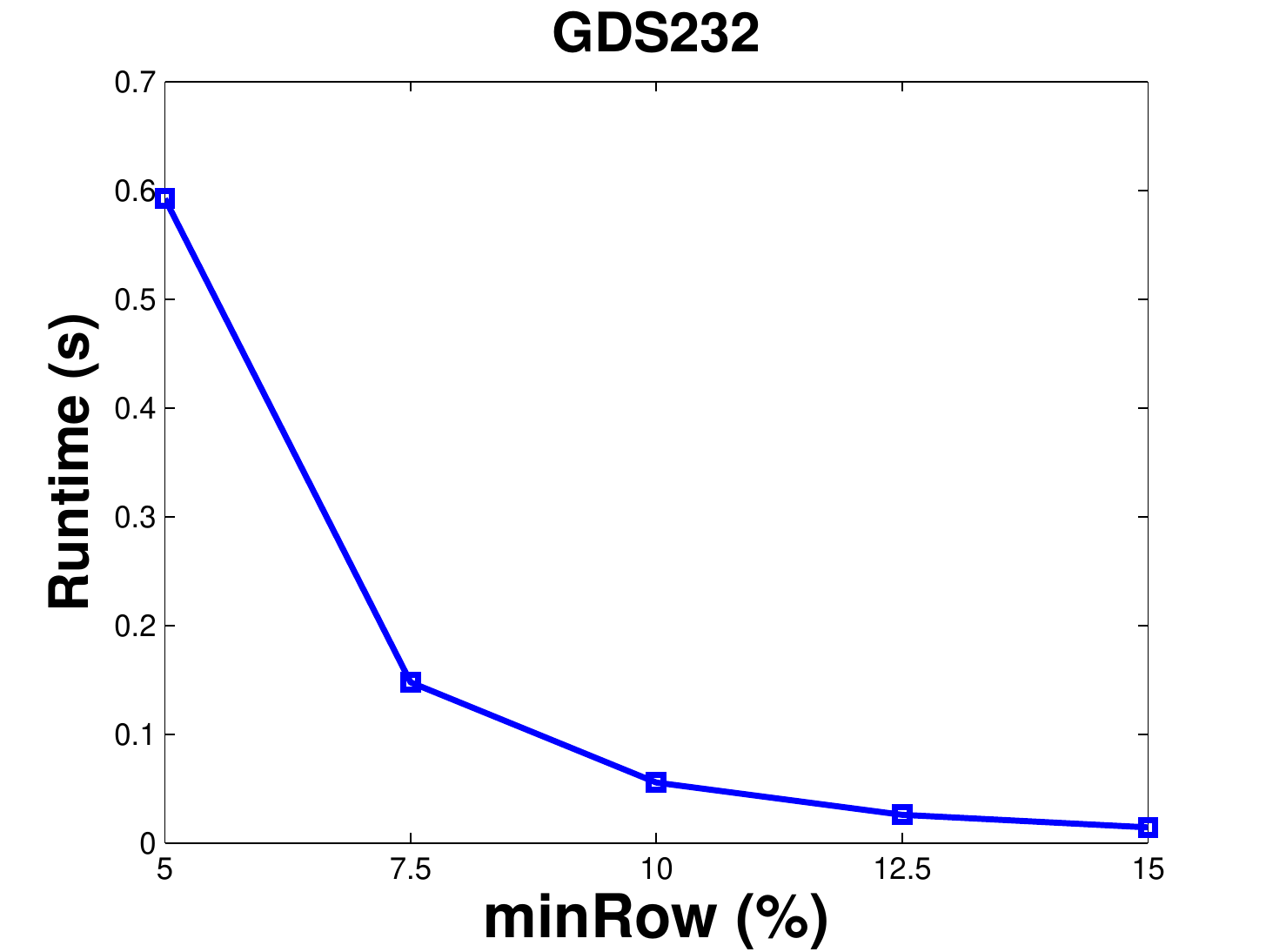}
	}
	\subfigure[]{
		\includegraphics[trim=0.3cm 0.1cm 0.9cm 0.1cm, clip, scale=0.26]{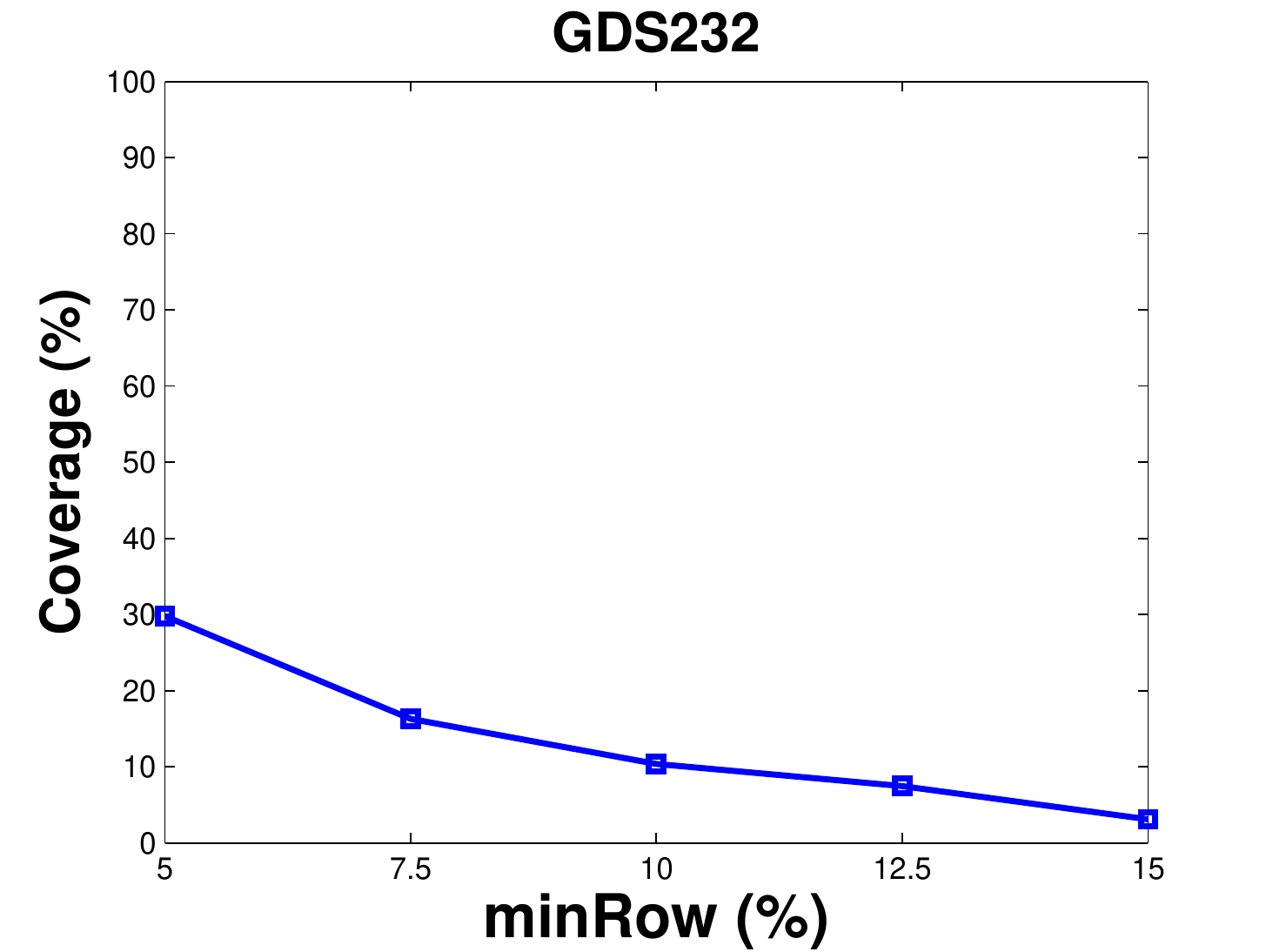}
	}
	\subfigure[]{
		\includegraphics[trim=0.3cm 0.1cm 0.9cm 0.1cm, clip, scale=0.26]{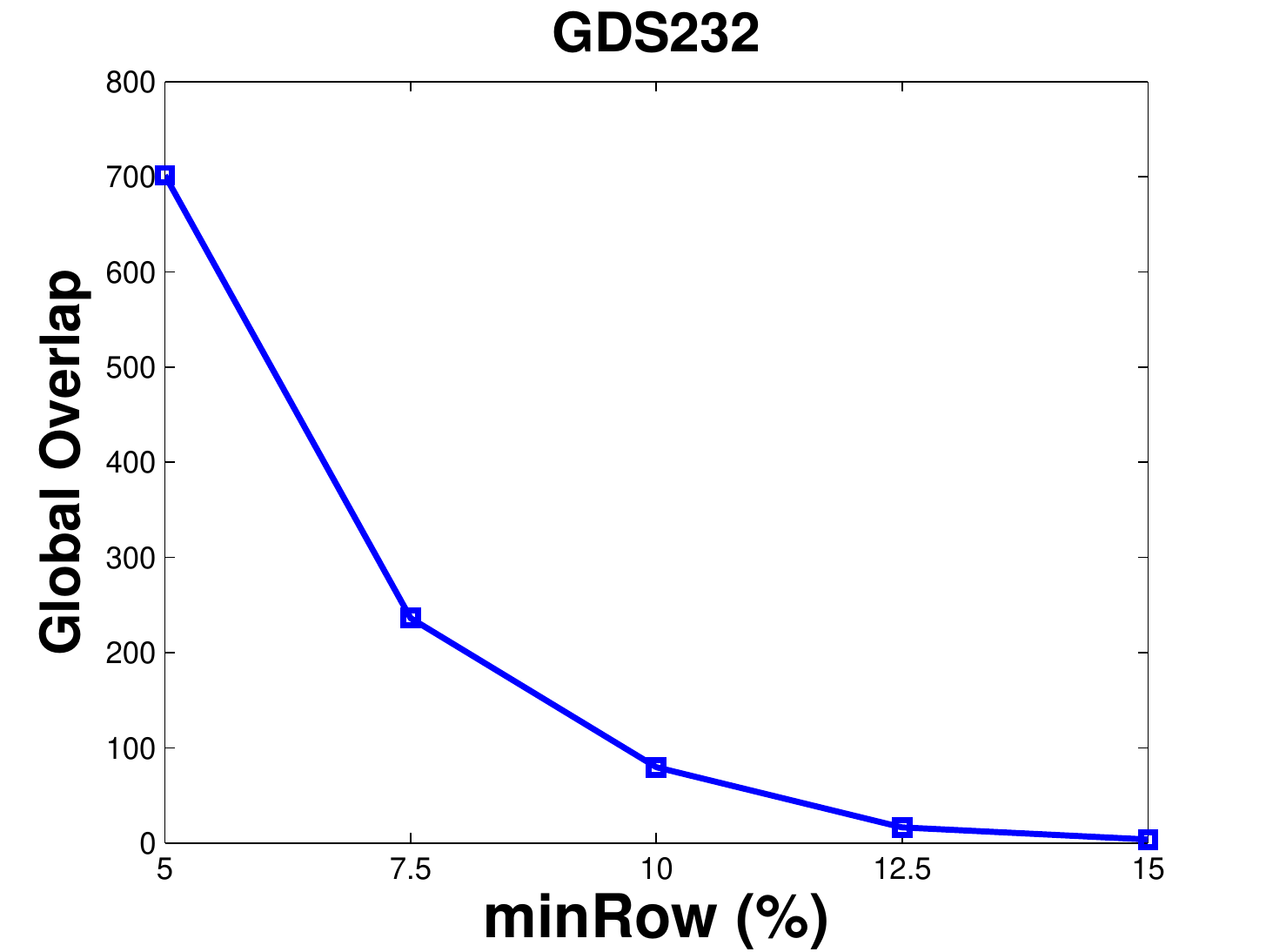}
	}
	
	\subfigure[]{
		\includegraphics[trim=0.3cm 0.1cm 0.9cm 0.1cm, clip, scale=0.26]{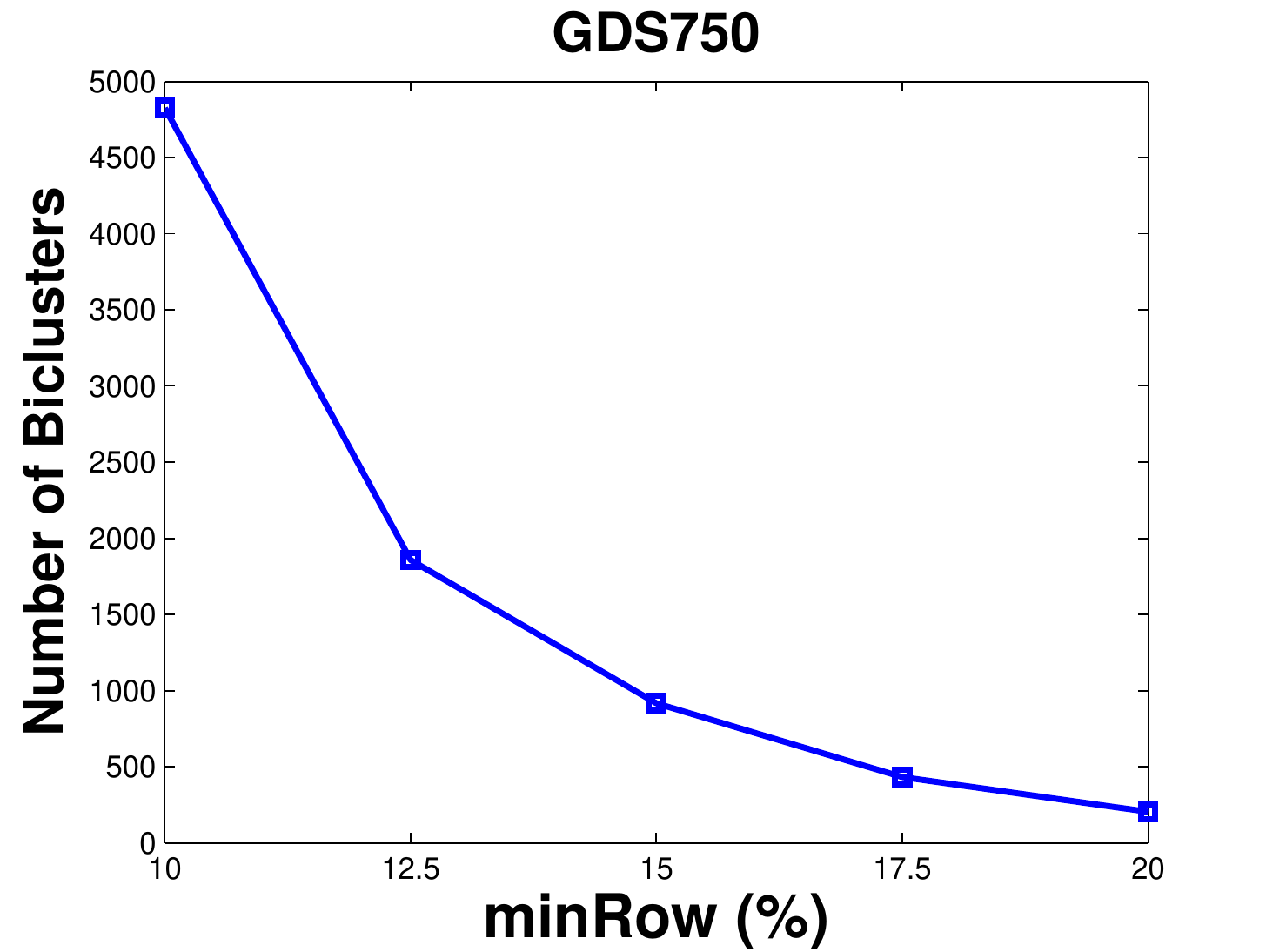}
	}
	\subfigure[]{
		\includegraphics[trim=0.3cm 0.1cm 0.9cm 0.1cm, clip, scale=0.26]{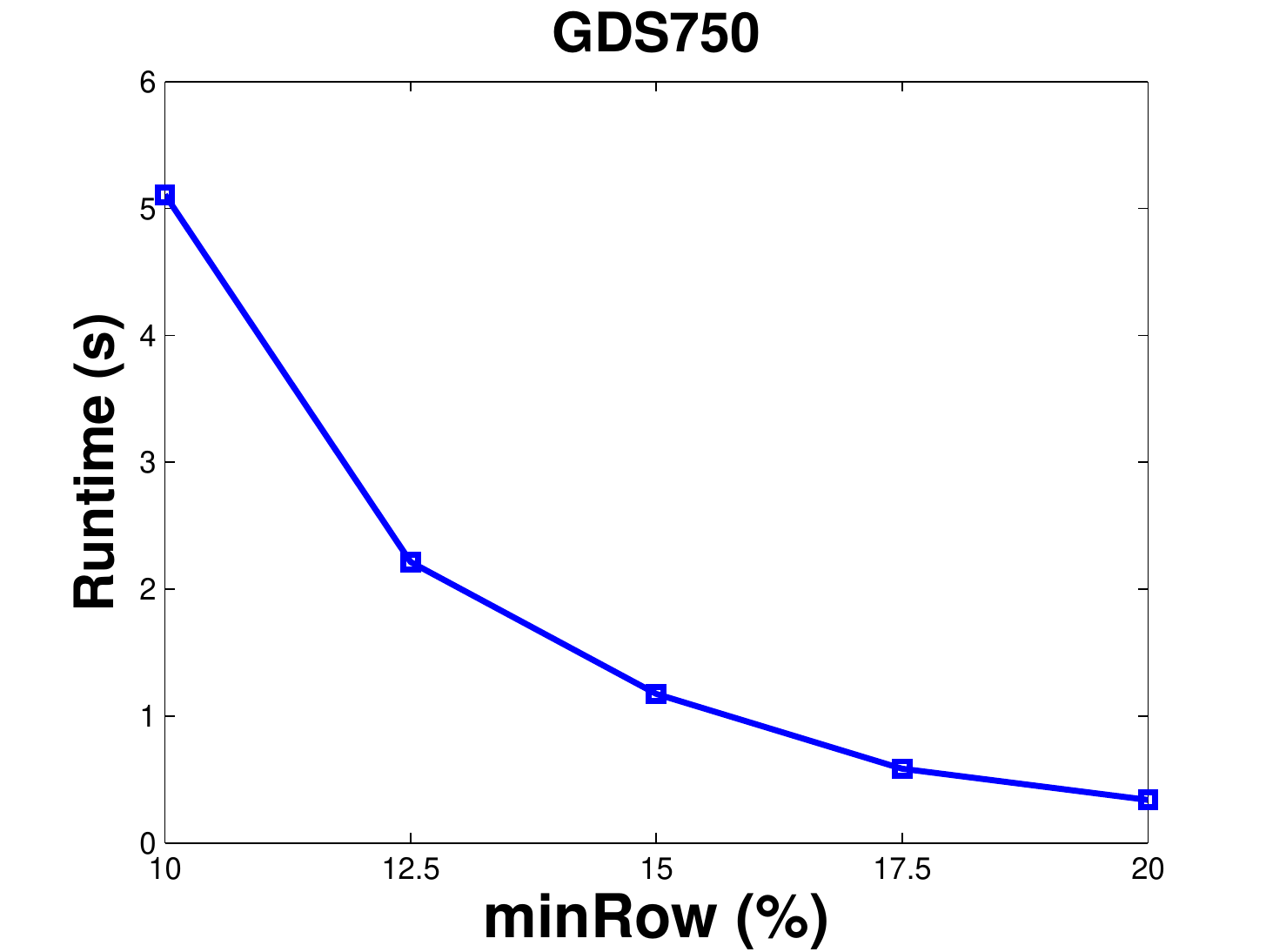}
	}
	\subfigure[]{
		\includegraphics[trim=0.3cm 0.1cm 0.9cm 0.1cm, clip, scale=0.26]{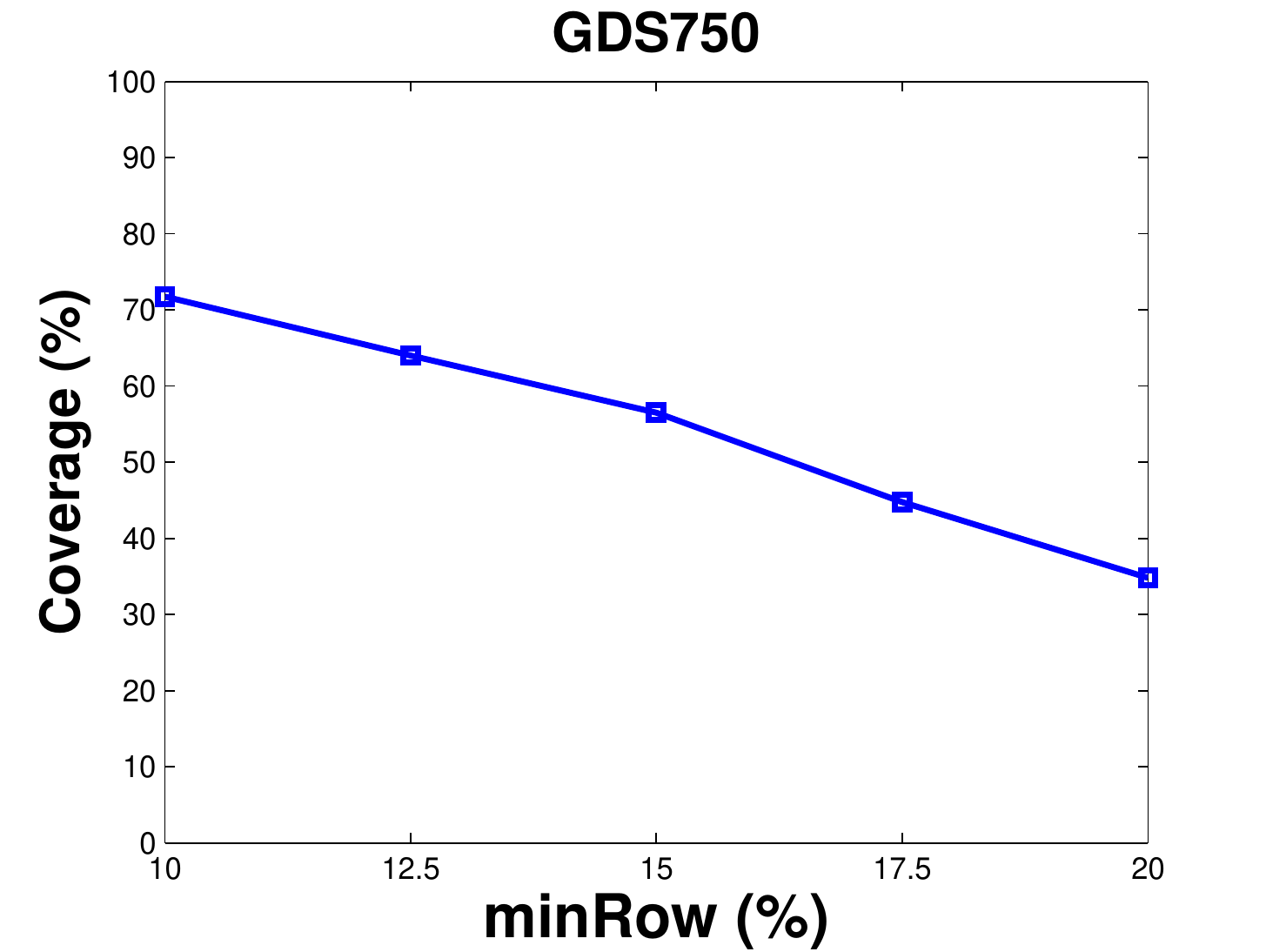}
	}
	\subfigure[]{
		\includegraphics[trim=0.3cm 0.1cm 0.9cm 0.1cm, clip, scale=0.26]{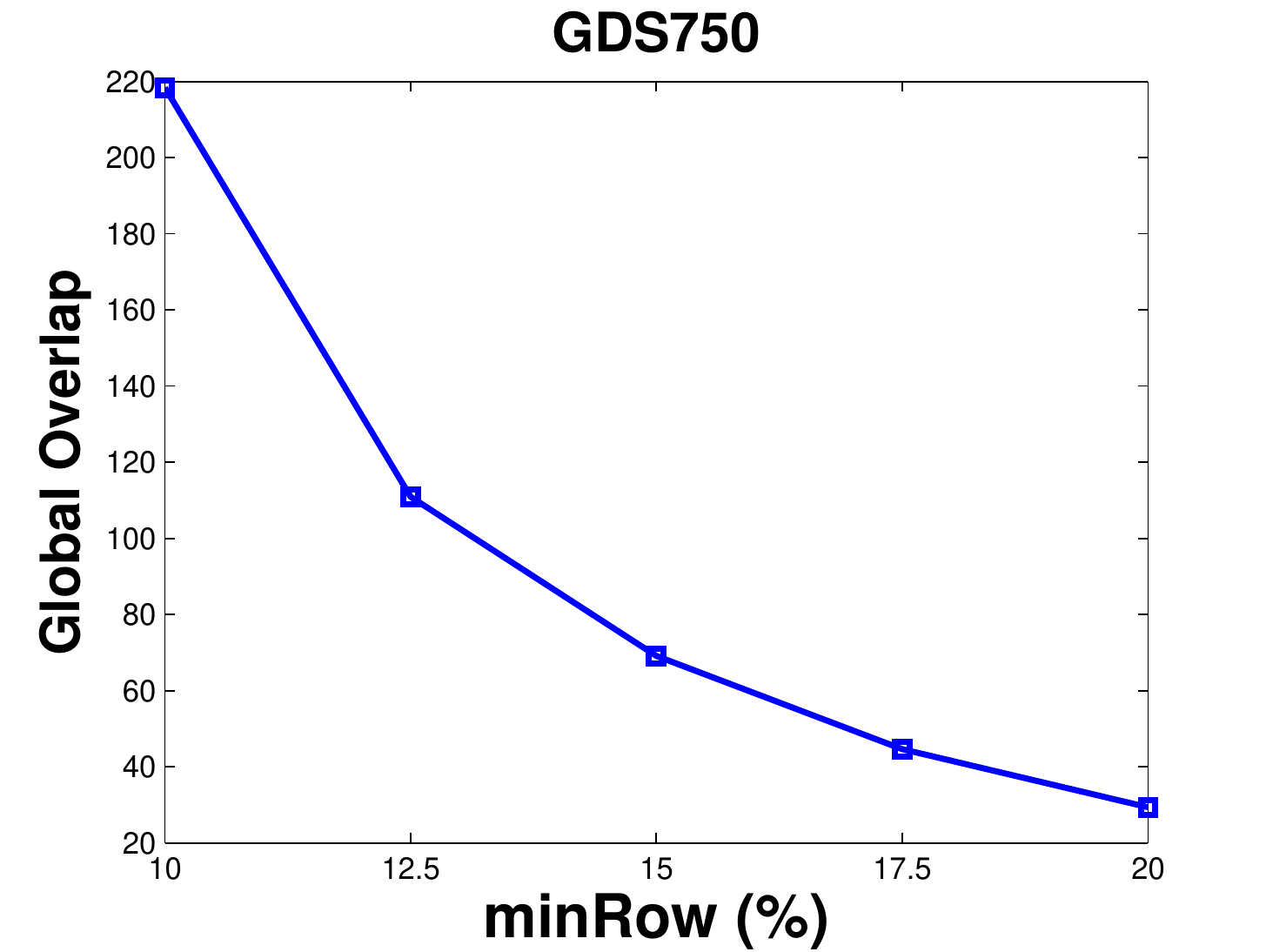}
	}

	\subfigure[]{
		\includegraphics[trim=0.3cm 0.1cm 0.9cm 0.1cm, clip, scale=0.26]{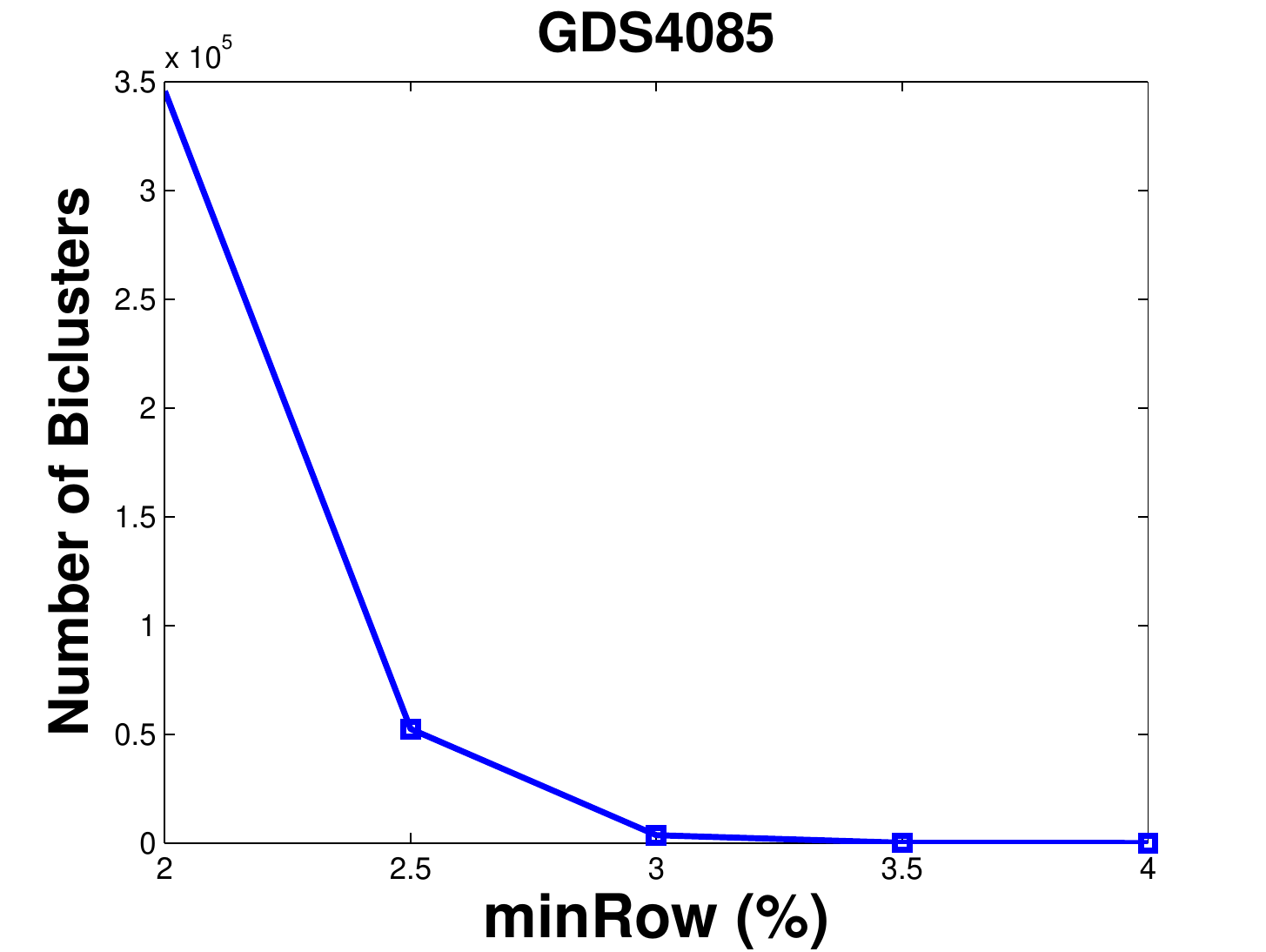}
	}
	\subfigure[]{
		\includegraphics[trim=0.3cm 0.1cm 0.9cm 0.1cm, clip, scale=0.26]{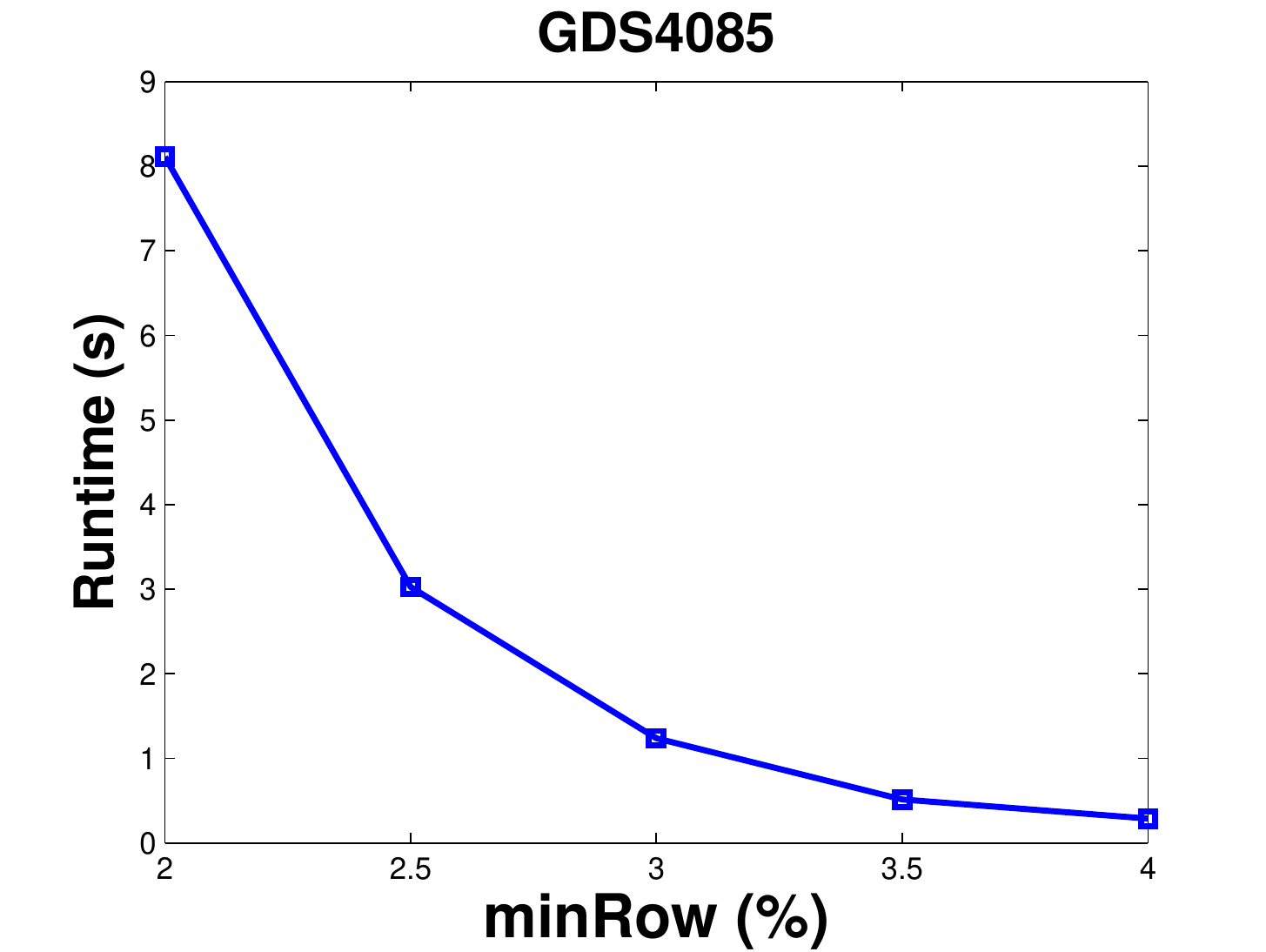}
	}
	\subfigure[]{
		\includegraphics[trim=0.3cm 0.1cm 0.9cm 0.1cm, clip, scale=0.26]{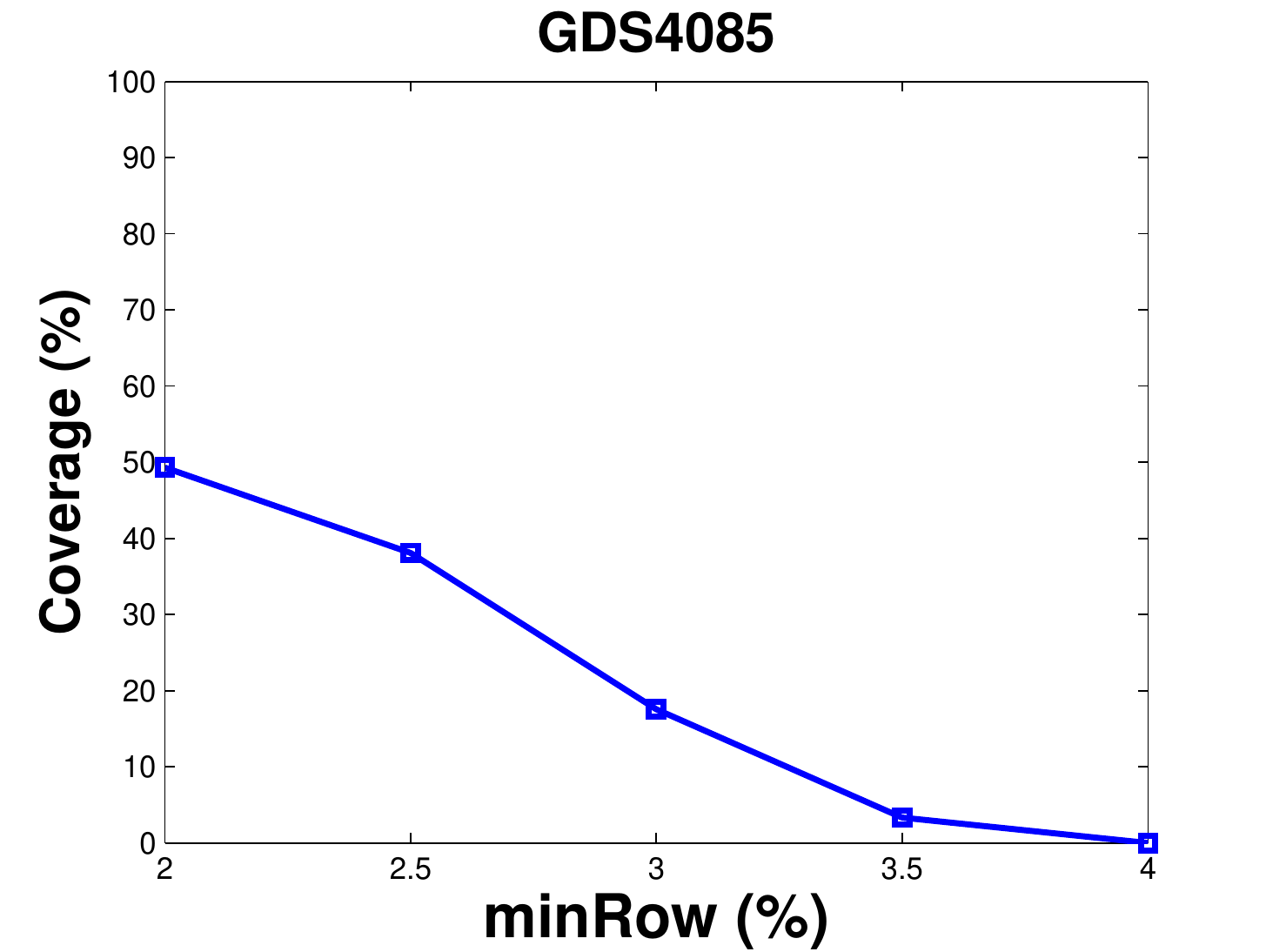}
	}
	\subfigure[]{
		\includegraphics[trim=0.3cm 0.1cm 0.9cm 0.1cm, clip, scale=0.26]{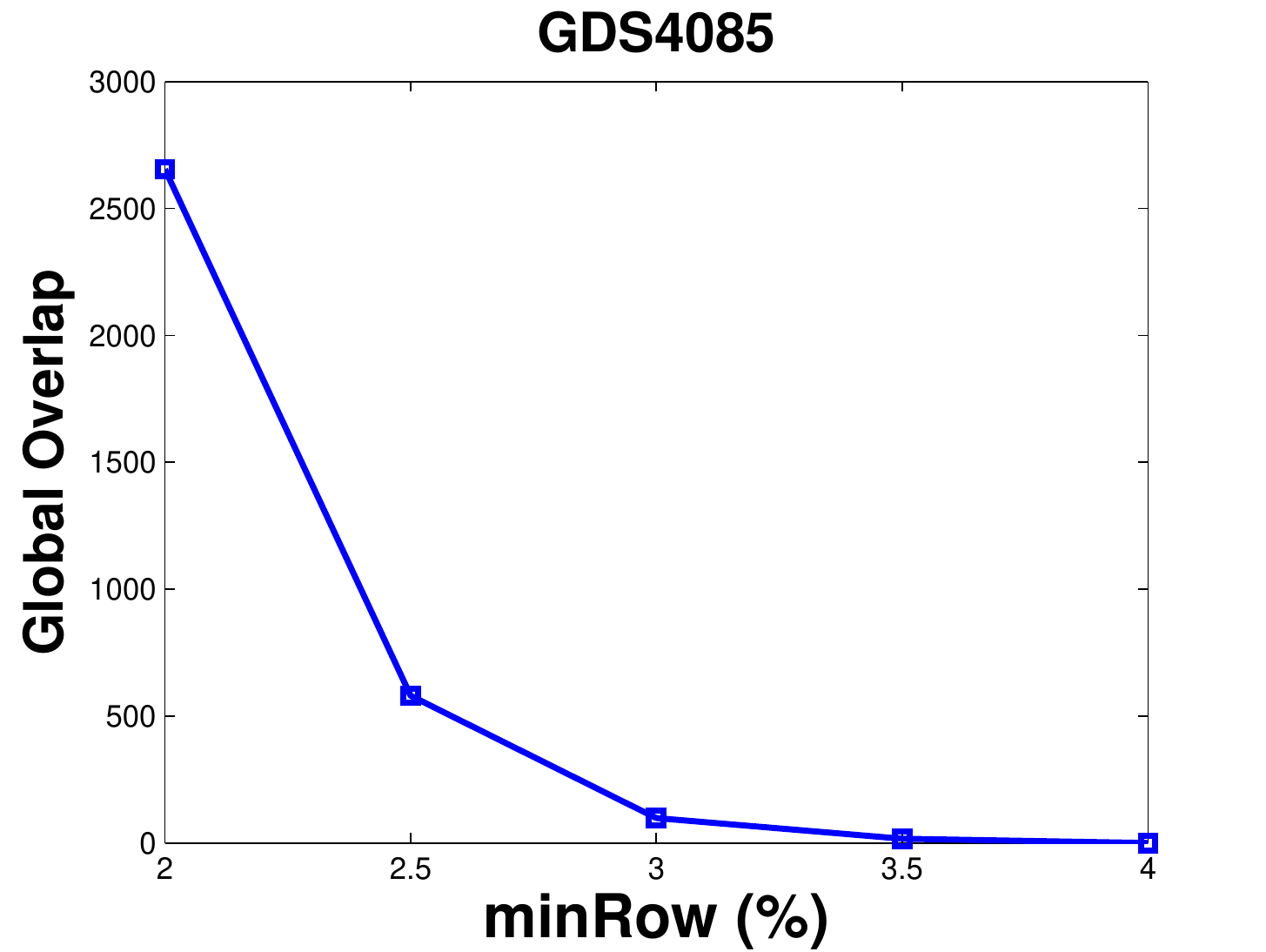}
	}
		
  \caption{Results of RIn-Close\_CVC's sensitivity to the parameter $minRow$. The parameter $\epsilon$ was set to: 5 for Yeast; 4 for GDS232; 4 for GDS750; and 37 for GDS4085.}
  \label{fig:expSensAnalCVC_mr}
\end{figure*}

\begin{figure*}
  \centering
	
	\subfigure[]{
		\includegraphics[trim=0.3cm 0.1cm 0.9cm 0.1cm, clip, scale=0.26]{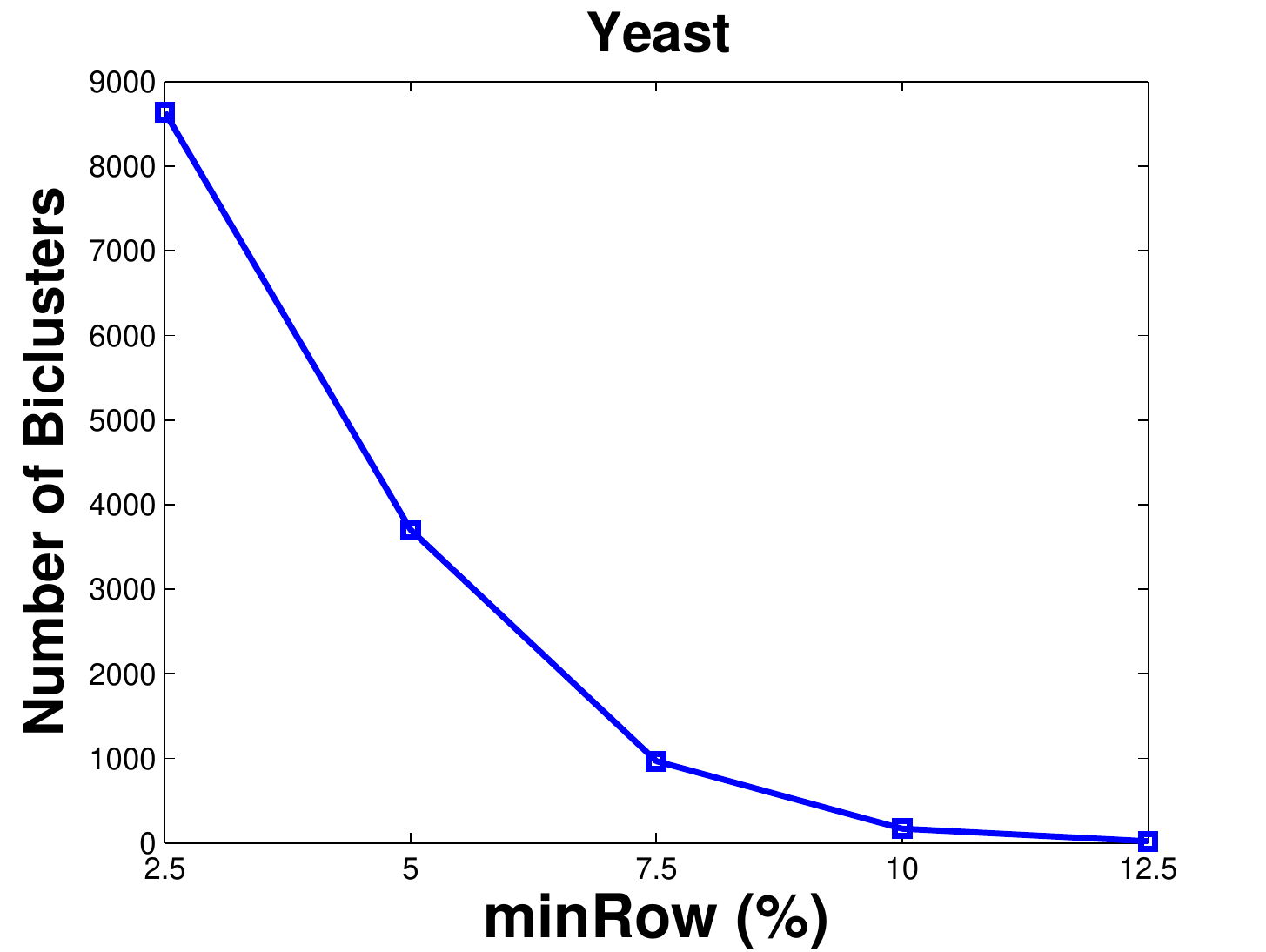}
	}
	\subfigure[]{
		\includegraphics[trim=0.3cm 0.1cm 0.9cm 0.1cm, clip, scale=0.26]{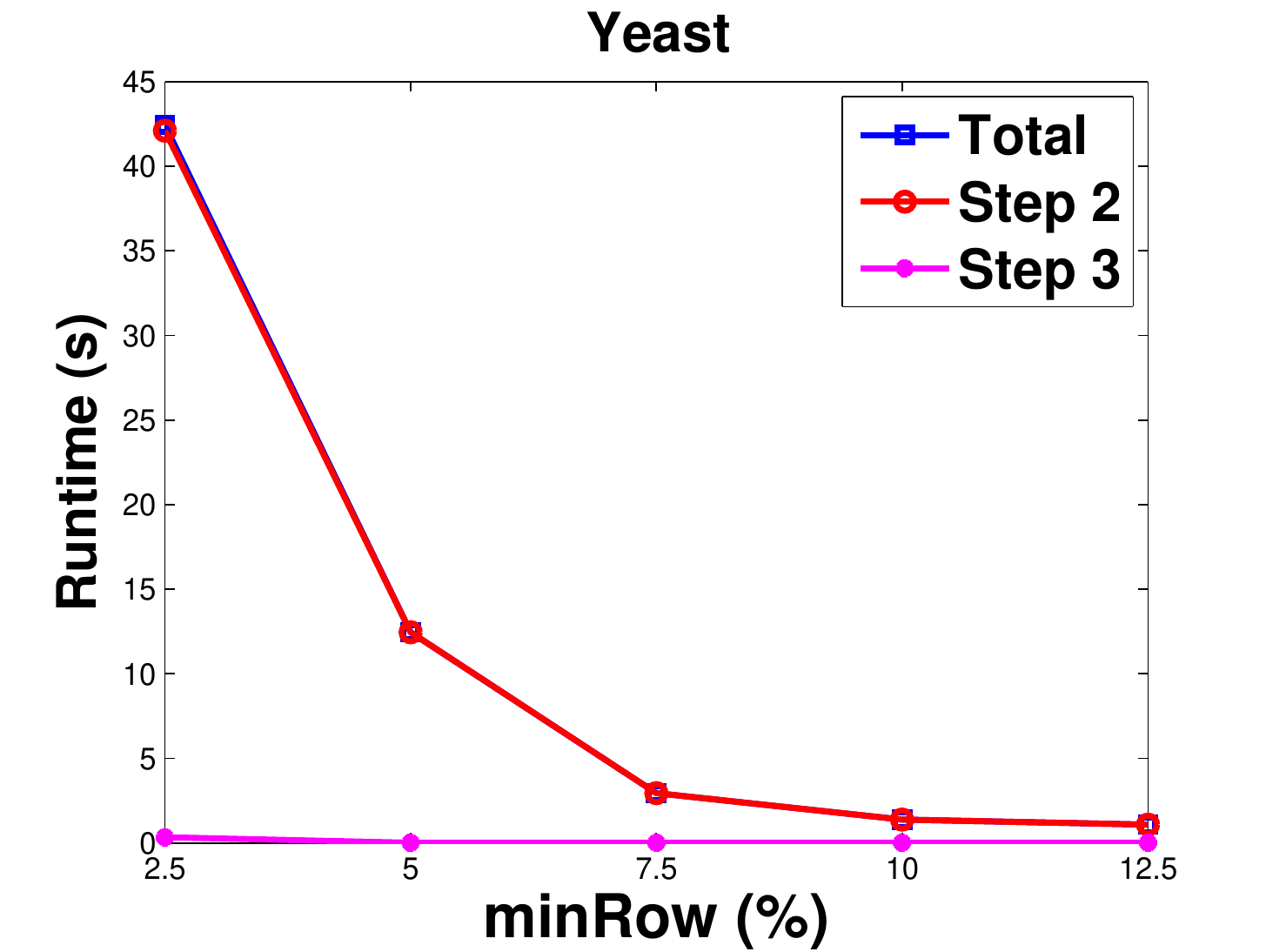}
	}
	\subfigure[]{
		\includegraphics[trim=0.3cm 0.1cm 0.9cm 0.1cm, clip, scale=0.26]{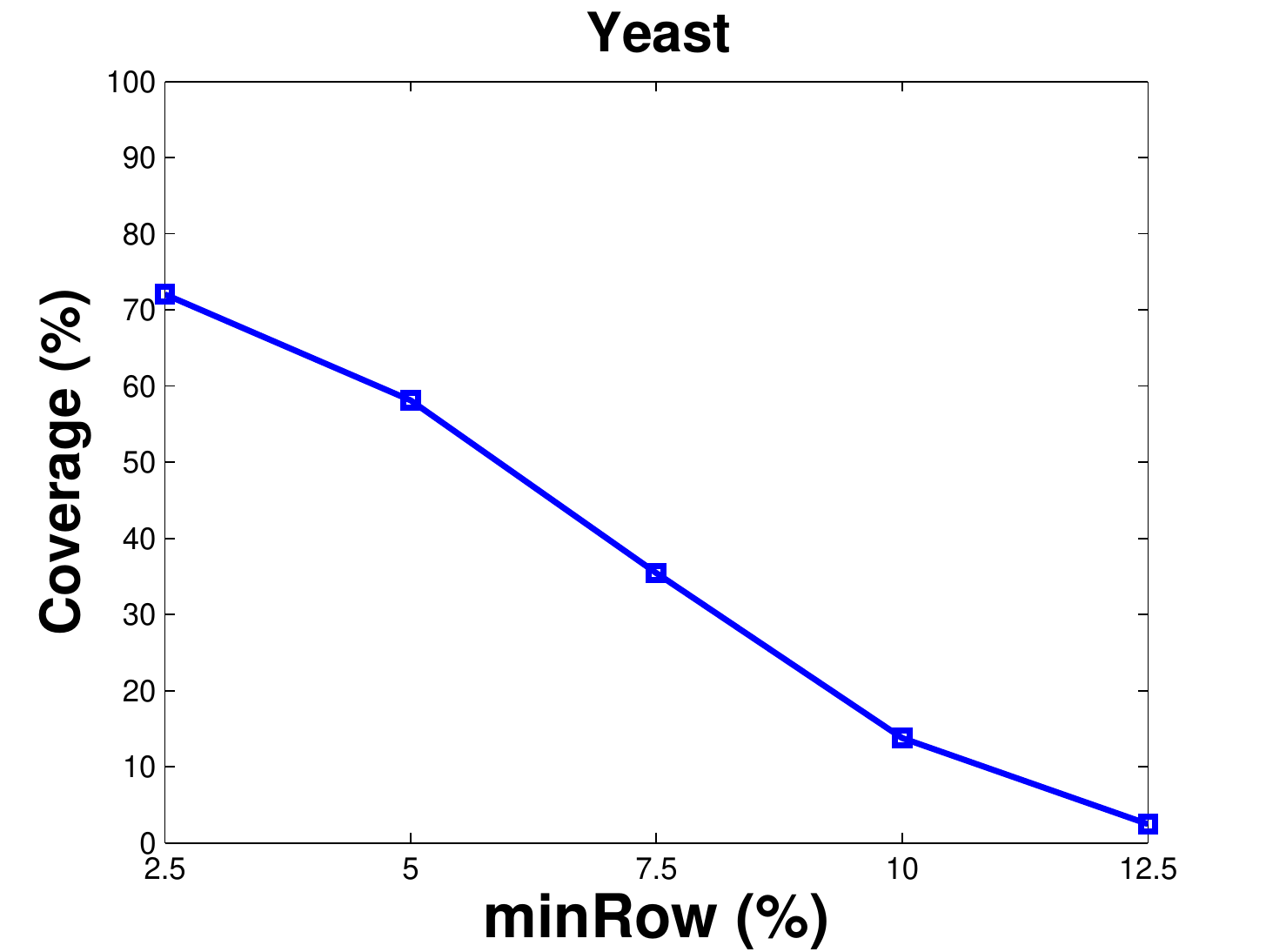}
	}
	\subfigure[]{
		\includegraphics[trim=0.3cm 0.1cm 0.9cm 0.1cm, clip, scale=0.26]{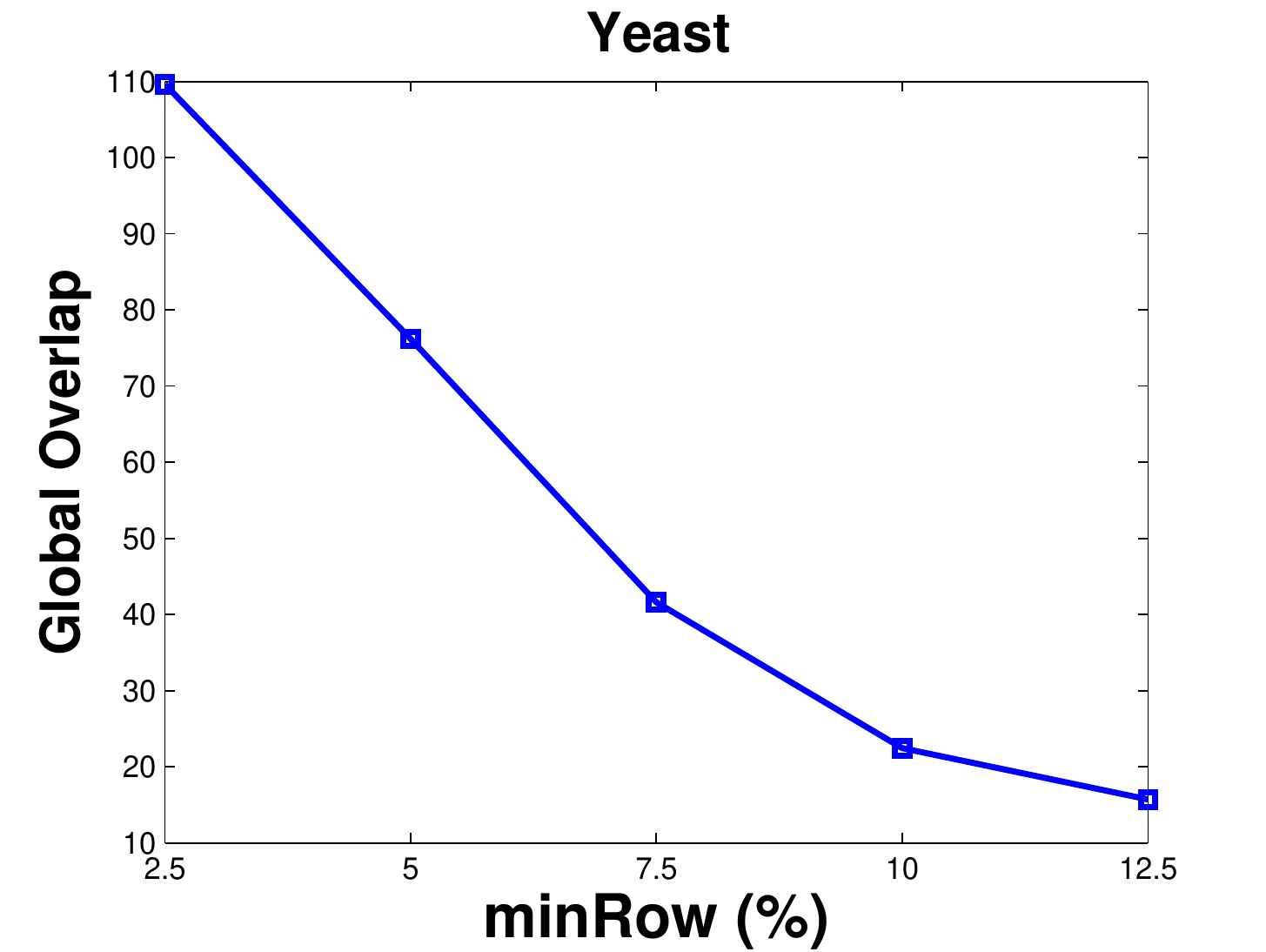}
	}
	
	\subfigure[]{
		\includegraphics[trim=0.3cm 0.1cm 0.9cm 0.1cm, clip, scale=0.26]{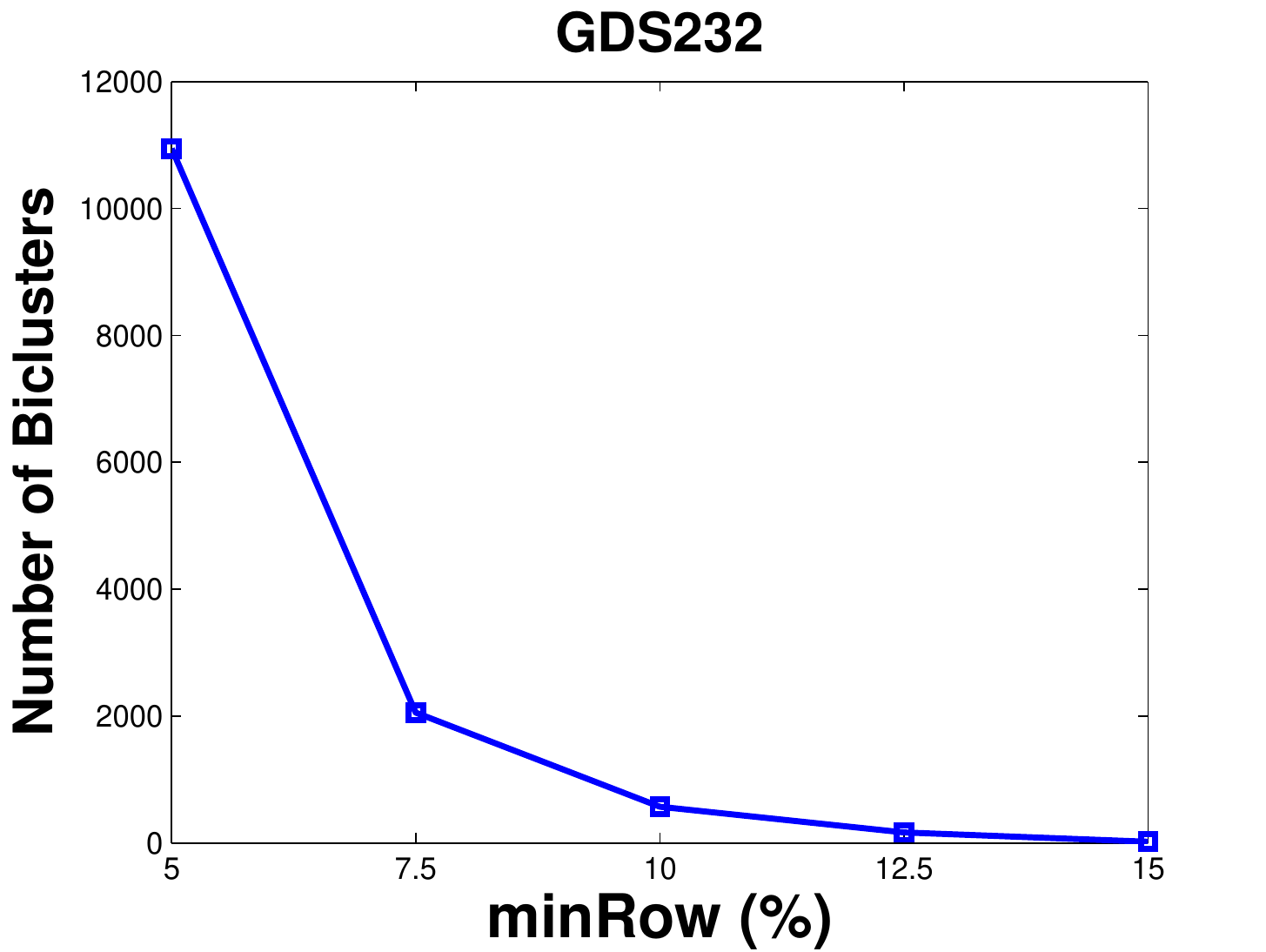}
	}
	\subfigure[]{
		\includegraphics[trim=0.3cm 0.1cm 0.9cm 0.1cm, clip, scale=0.26]{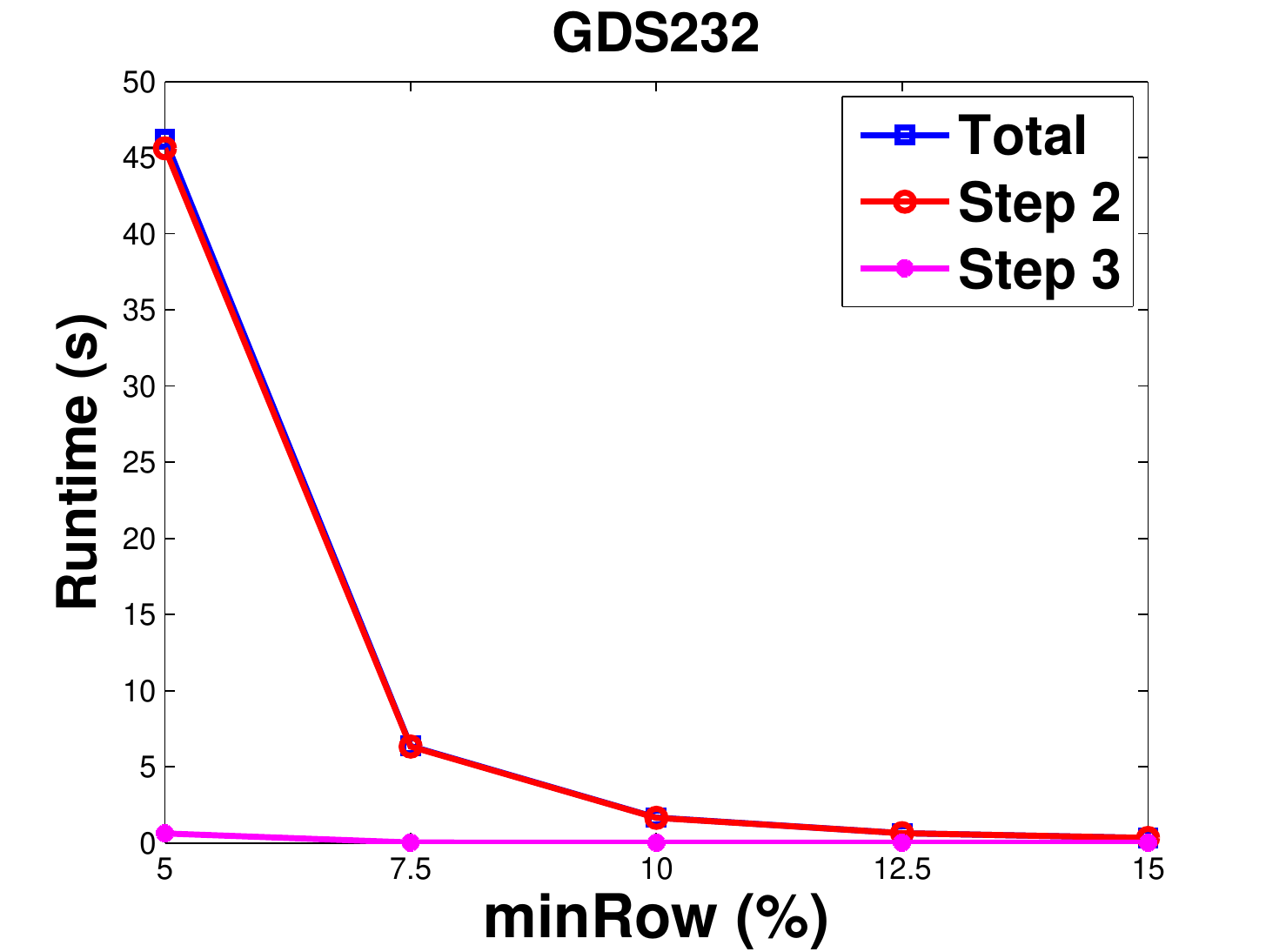}
	}
	\subfigure[]{
		\includegraphics[trim=0.3cm 0.1cm 0.9cm 0.1cm, clip, scale=0.26]{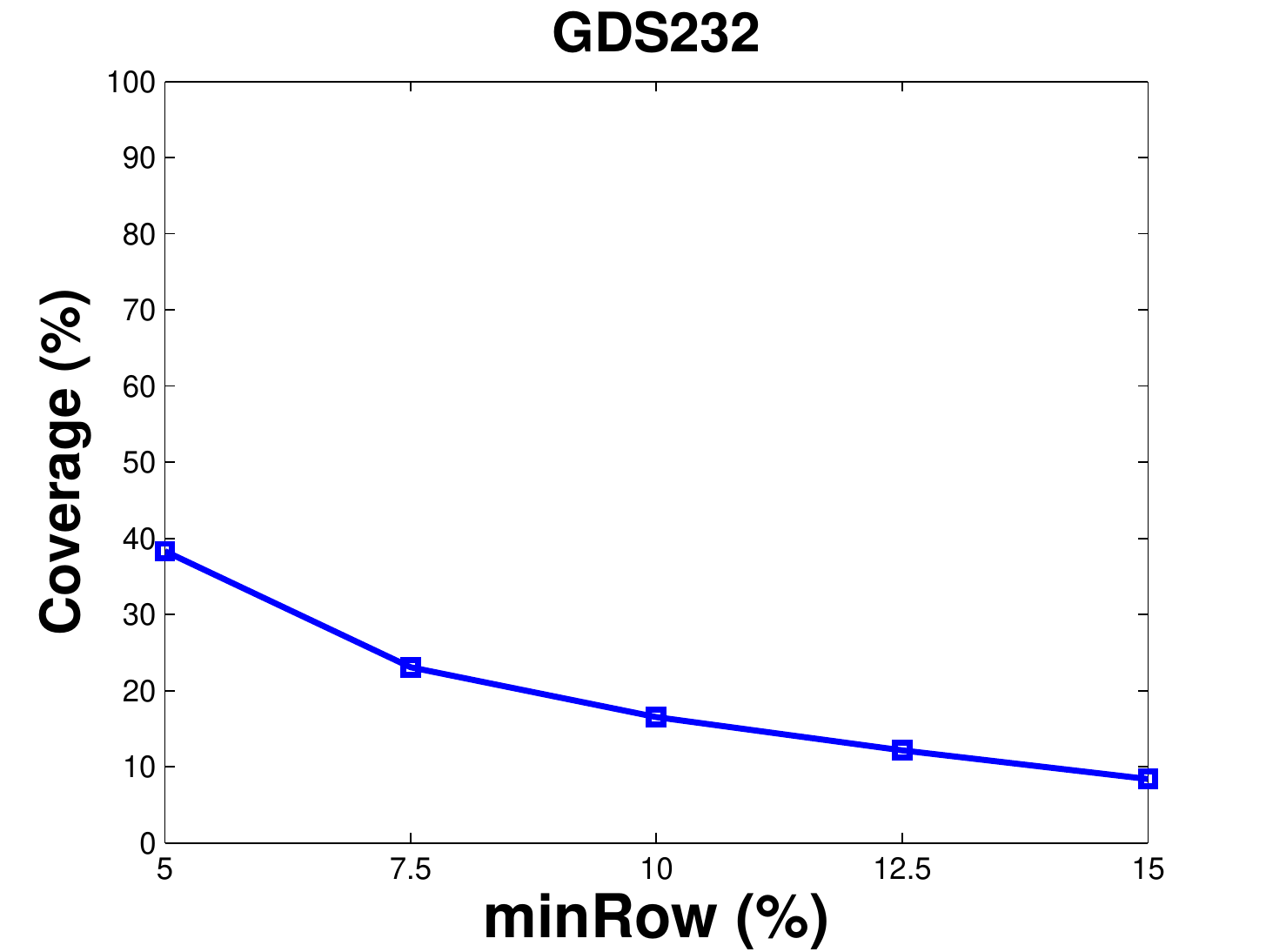}
	}
	\subfigure[]{
		\includegraphics[trim=0.3cm 0.1cm 0.9cm 0.1cm, clip, scale=0.26]{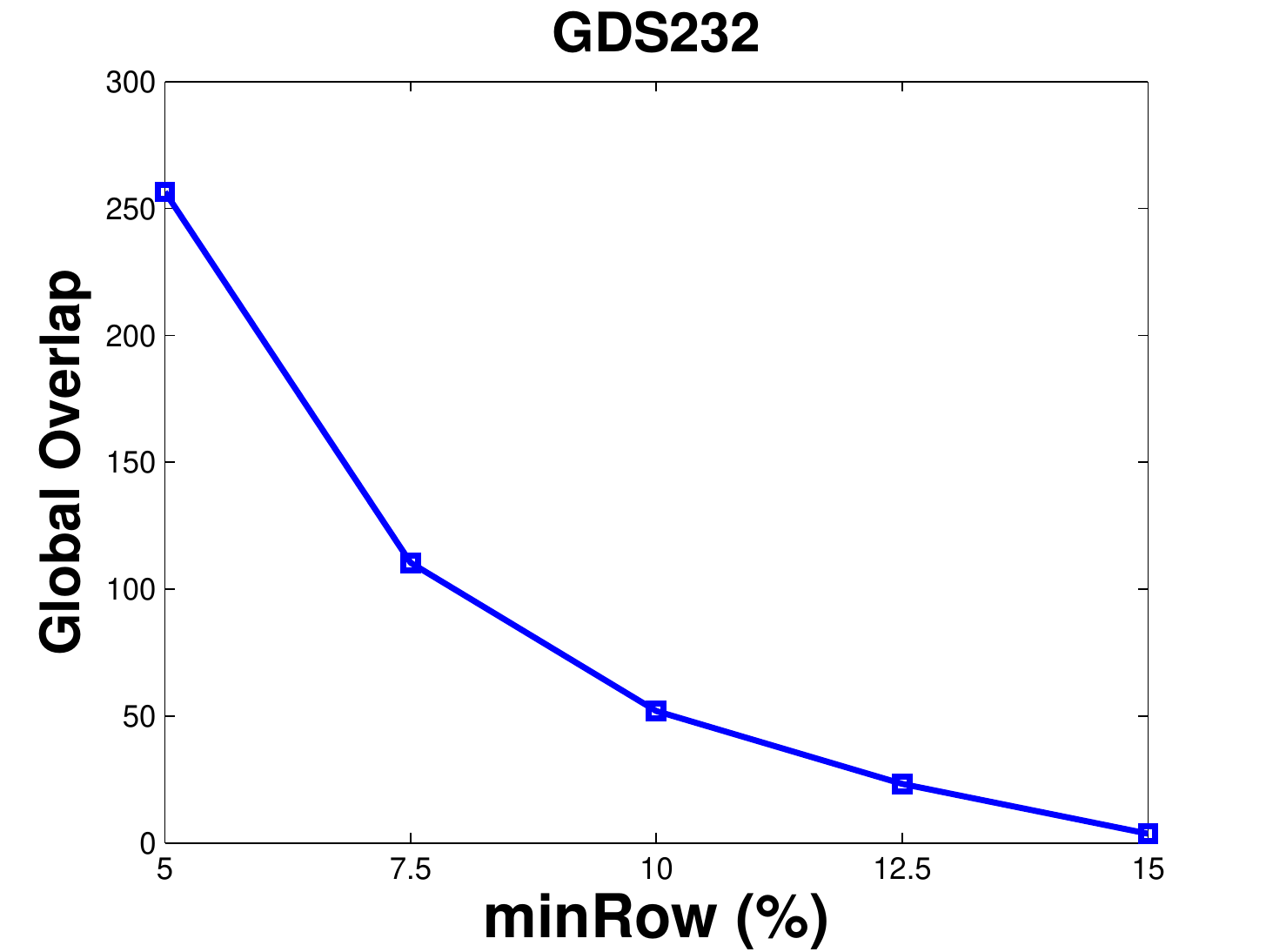}
	}
	
	\subfigure[]{
		\includegraphics[trim=0.3cm 0.1cm 0.9cm 0.1cm, clip, scale=0.26]{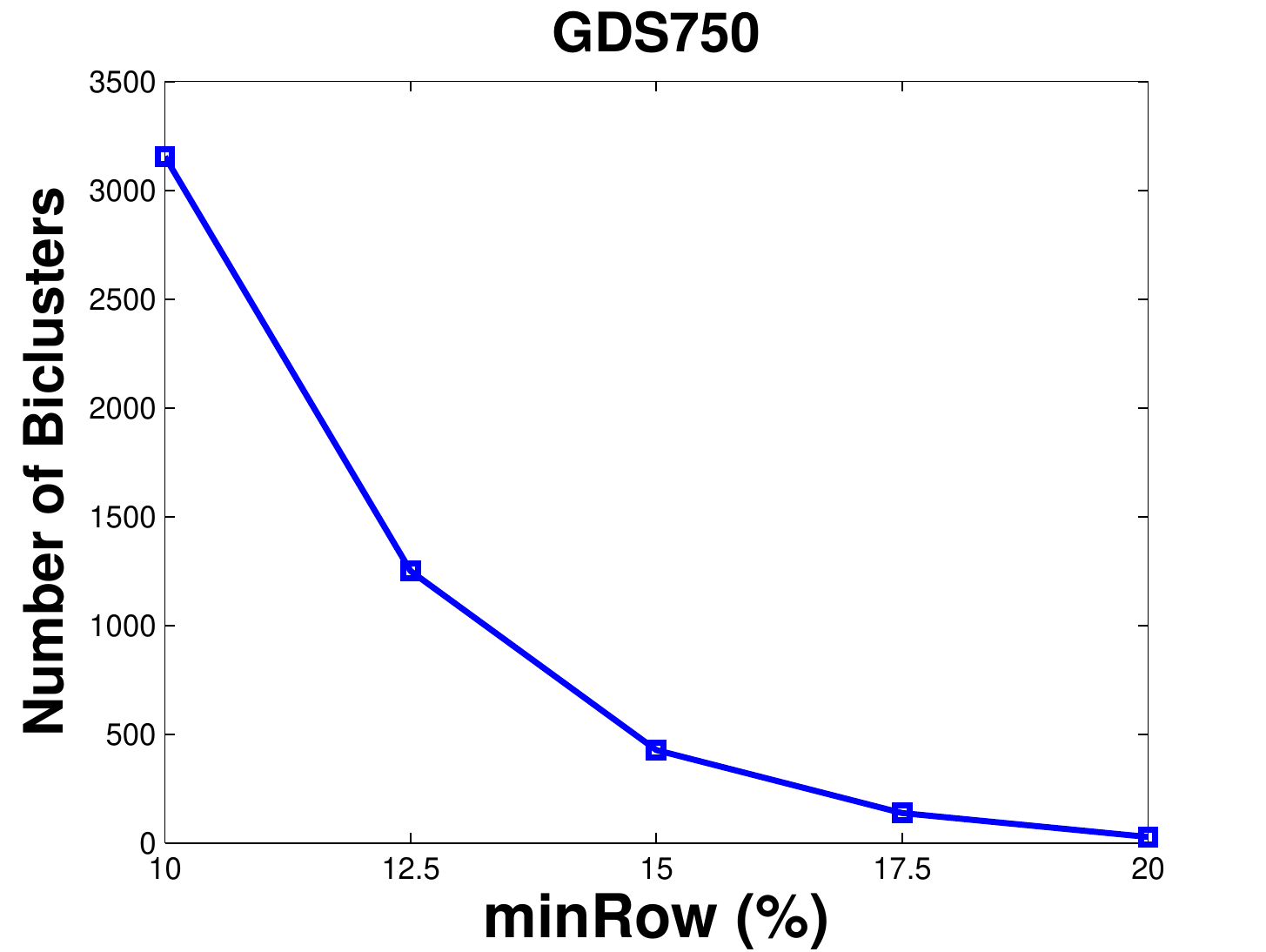}
	}
	\subfigure[]{
		\includegraphics[trim=0.3cm 0.1cm 0.9cm 0.1cm, clip, scale=0.26]{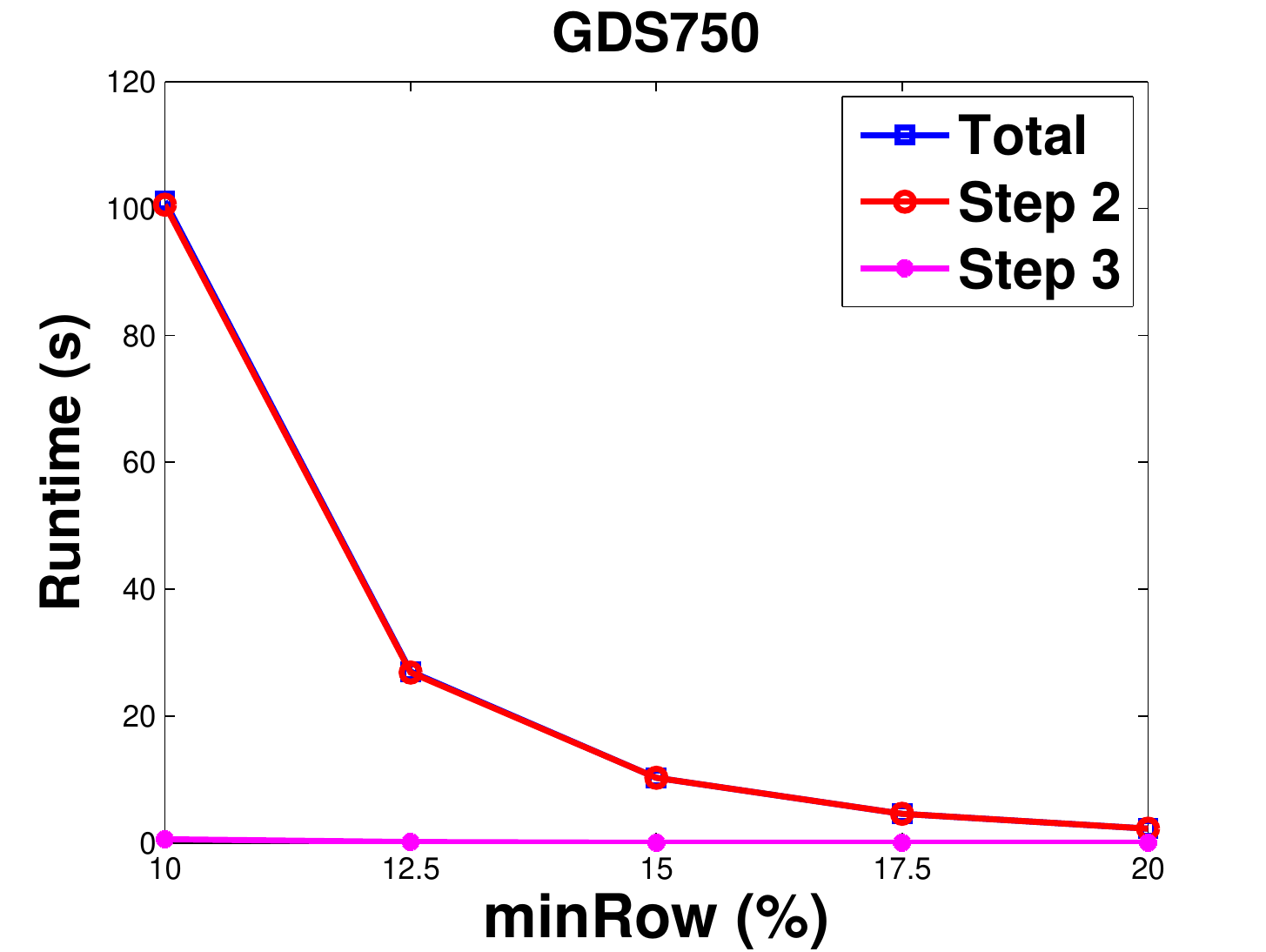}
	}
	\subfigure[]{
		\includegraphics[trim=0.3cm 0.1cm 0.9cm 0.1cm, clip, scale=0.26]{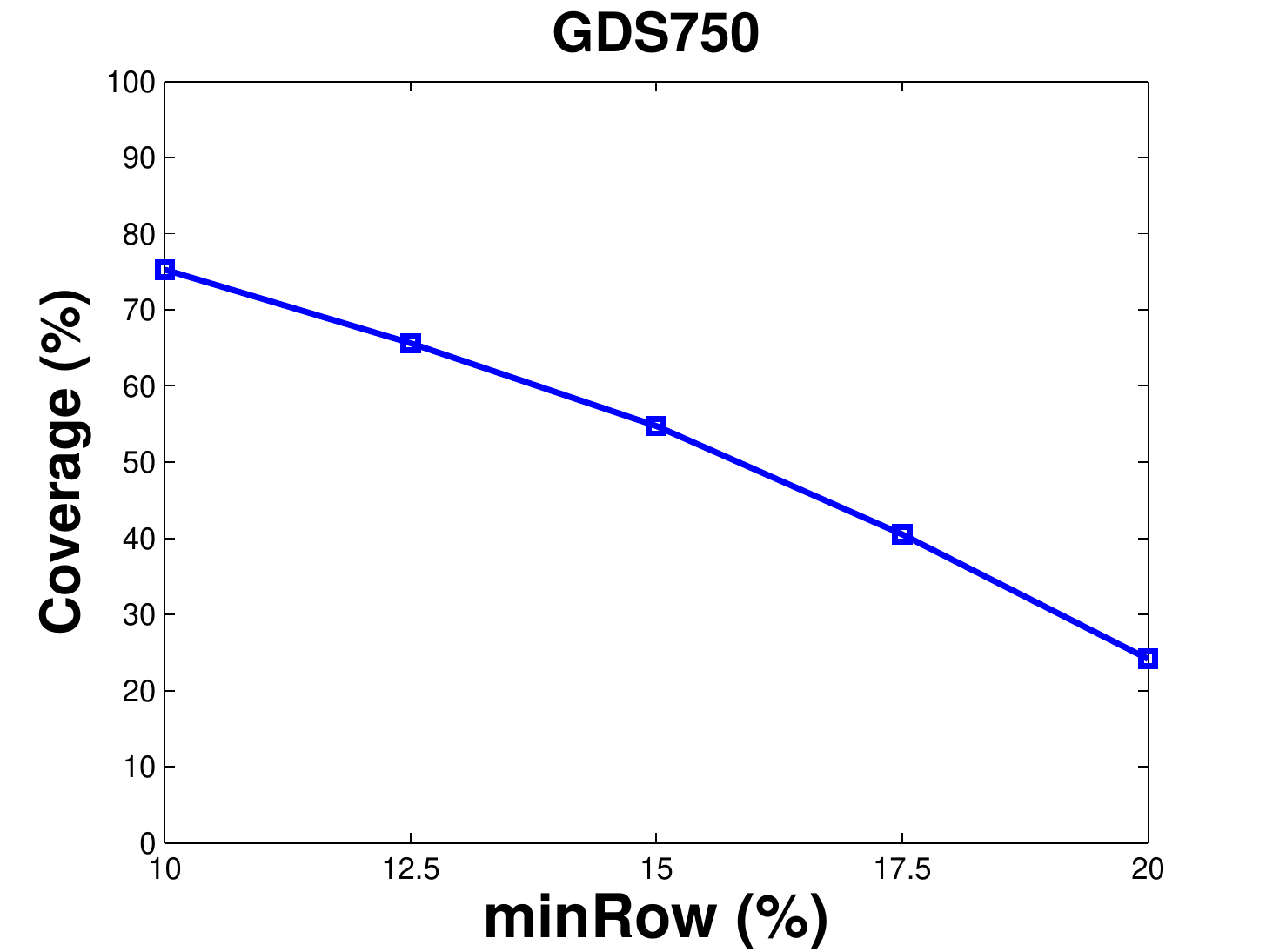}
	}
	\subfigure[]{
		\includegraphics[trim=0.3cm 0.1cm 0.9cm 0.1cm, clip, scale=0.26]{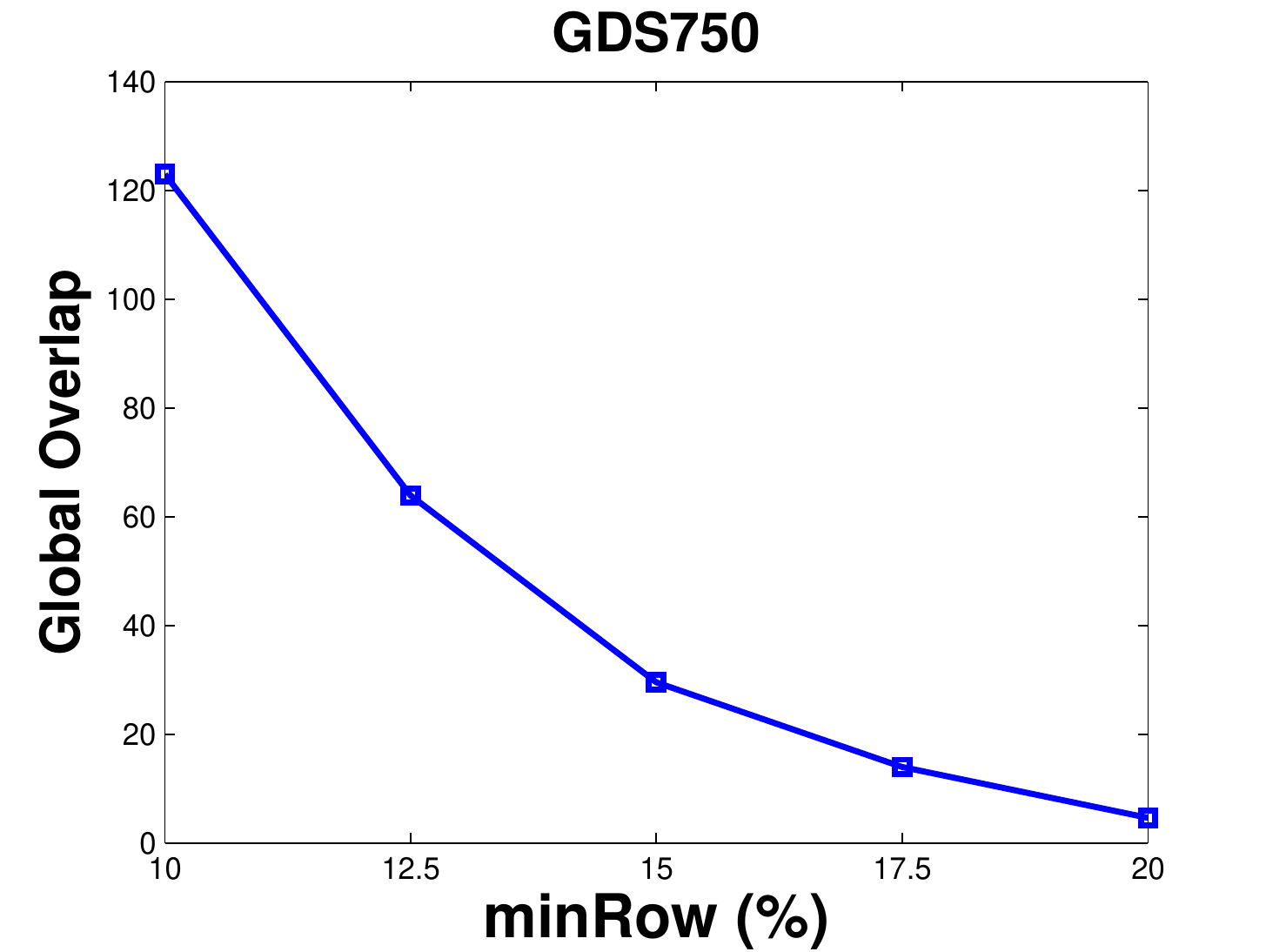}
	}

	\subfigure[]{
		\includegraphics[trim=0.3cm 0.1cm 0.9cm 0.1cm, clip, scale=0.26]{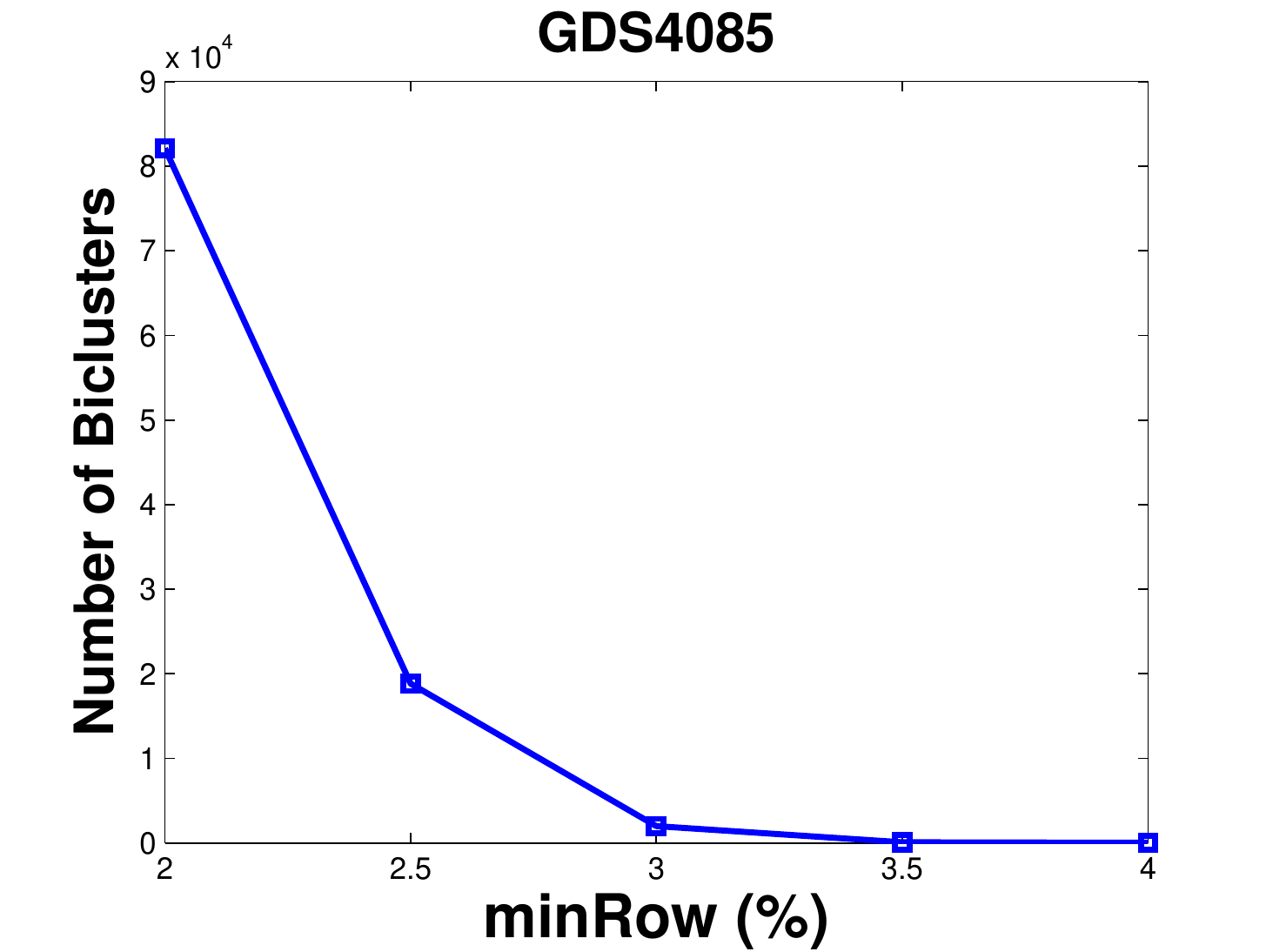}
	}
	\subfigure[]{
		\includegraphics[trim=0.3cm 0.1cm 0.9cm 0.1cm, clip, scale=0.26]{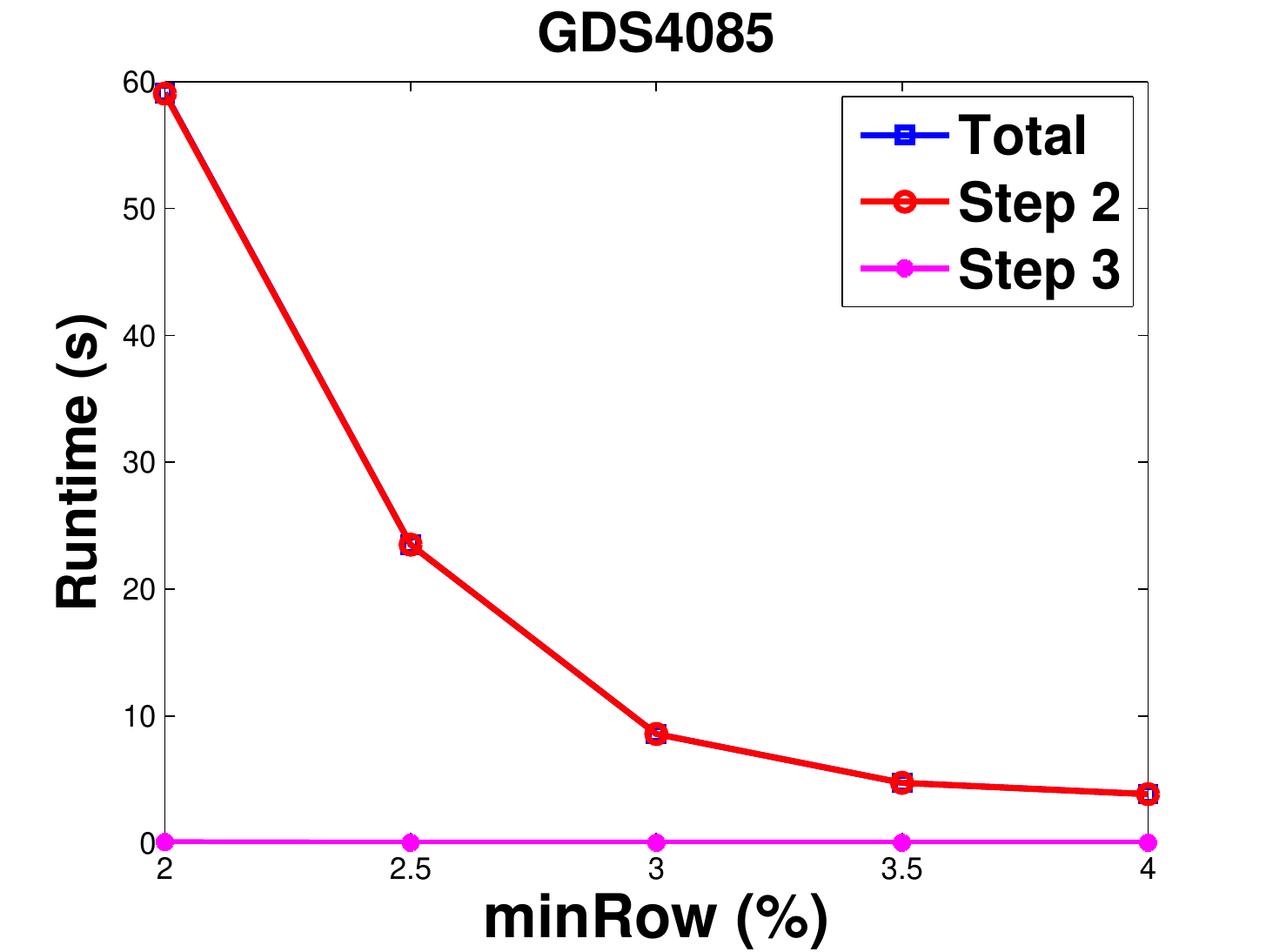}
	}
	\subfigure[]{
		\includegraphics[trim=0.3cm 0.1cm 0.9cm 0.1cm, clip, scale=0.26]{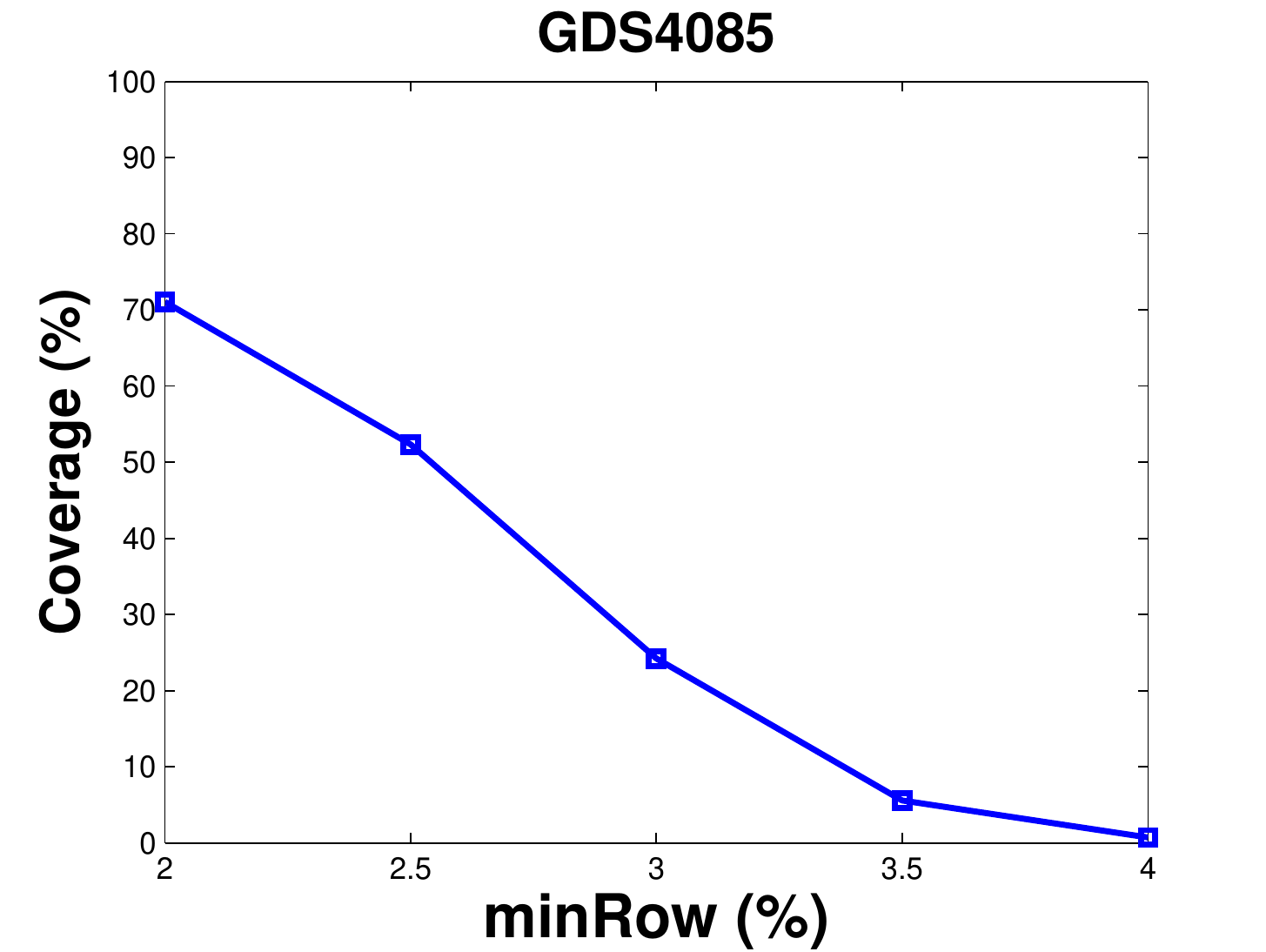}
	}
	\subfigure[]{
		\includegraphics[trim=0.3cm 0.1cm 0.9cm 0.1cm, clip, scale=0.26]{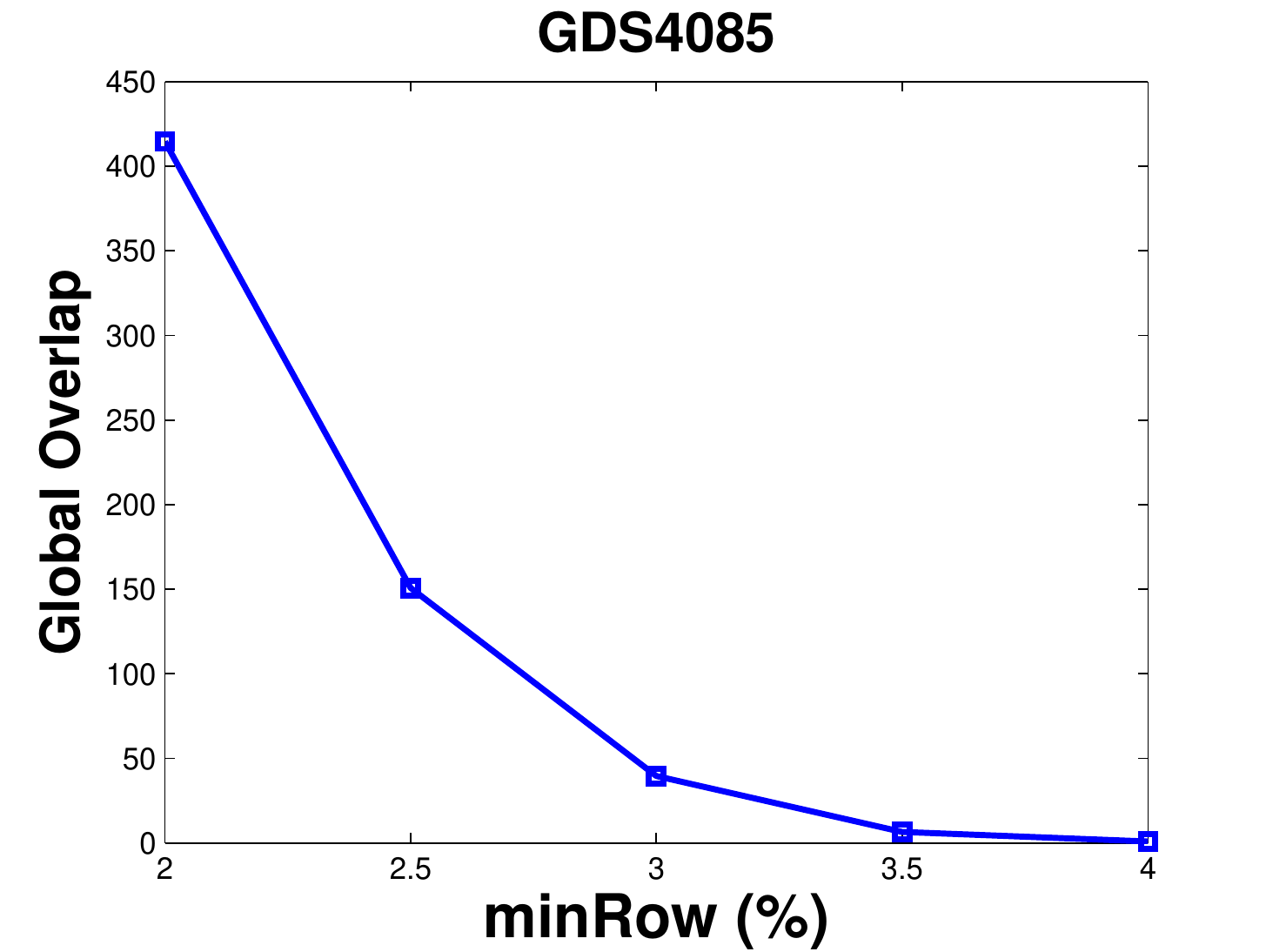}
	}
		
  \caption{Results of RIn-Close\_CHV's sensitivity to the parameter $minRow$. The parameter $\epsilon$ was set to: 5 for Yeast; 3 for GDS232; 3 for GDS750; and 9 for GDS4085.}
  \label{fig:expSensAnalCHV_mr}
\end{figure*}

Figs.~\ref{fig:expSensAnalCVC_mr} and \ref{fig:expSensAnalCHV_mr} shows respectively the sensitivity of RIn-Close\_CVC and RIn-Close\_CHV to the parameter $minRow$. The parameter $minRow$ also has a strong influence in the computational cost of RIn-Close. The higher its value, the smaller the search space for enumerating biclusters. Again, we see that the choice of $minRow$ must consider the relation between coverage and global overlap. If a larger value of $minRow$ has a small impact on the coverage and a significant impact on the global overlap, certainly it is a reasonable choice.

As we observed with this experiment, RIn-Close parameters can be set in a way that mitigates the explosion in the number of biclusters, with similar impact on the computational cost.

\subsection{Comparison with Baselines}

With this experiment, we are going to demonstrate that well-known heuristic-based approaches can fail spectacularly when trying to identify the existing biclusters in a dataset. We claim that the results to be presented turn to be a strong motivation for adopting enumerative algorithms, such as the ones proposed in this paper. The dataset is also carefully designed so that we can propose a suitable set of parameters for the heuristic-based approaches, given that we are aware of the main attributes of the existing biclusters. So, the disastrous behavior of the heuristic-based approaches can not be attributed to an unfortunate parameterization. We tested three heuristics that are specialized to mine CHV biclusters: CC, FLOC, and ROCC. These contenders were briefly described in Section~\ref{sec:relWork}. We chose to perform this experiment with the CHV type of biclusters because this is the most general type addressed in this work.

For this experiment, we used the 50 synthetic datasets described in Section~\ref{sec:exp_scalability} with the default parameters (i.e., $n = 5000$, $m = 60$, number of biclusters $= 10$, bicluster row size $= 200$, bicluster column size $= 8$, overlap = $0.2$, and Gaussian noise with $\mu = 0$ and $\sigma = 0.01$). These datasets represent a particular and controlled scenario, i.e., there is a very clear boundary between what should and what should not be part of a bicluster. Possibly, the boundaries are not so accurate in real-world applications. But in this way, these dataset allows us to clearly determine the parameters of the biclustering algorithms, and find out what they are able to mine when looking for the original biclusters. For CC and FLOC, we set the value of $\delta$ for each dataset considering its largest bicluster MSR. For both, the number of biclusters to be mined were set to 10. CC's threshold for multiple node deletion $\alpha$ was set to 1.2 (the value suggested by the authors). For FLOC, we set the probability to add a row/column to a seed (initial) bicluster based on the proportion of the minimum number of rows/columns of a bicluster and the total number of rows/columns in the dataset. So, we set these parameters to 0.04 and 0.13, respectively. For ROCC, we set $sr = 1586$ and $sc = 17$ because these are the number of distinct rows/columns covered by the biclusters. Based on the fraction of the number of rows/columns of a dataset over the number of rows/columns of a bicluster, we set $k = 25$ and $l = 8$.

Table~\ref{tab:comparisonBaselines} shows the results of this experiment. CC completely failed in finding the original biclusters. As the values of the parameter $\delta$ were very low, CC was unable to find so accurate submatrices. FLOC had a very poor result with low Precision and Recall. As the initial biclusters are generated at random, it is very unlike that FLOC can improve them to the original ones, that have a very clear boundary between what should and what should not be part of a bicluster, thus having a very low MSR. ROCC had better results than CC and FLOC. ROCC starts with a checkboard biclustering structure containing all rows and columns, so it is expected that ROCC would achieve a better Recall than the others. But its routine to refine this initial solution was not accurate, which led to a low Precision, even though its Precision was better than the others.

\linespread{1}

\begin{table}
  \centering
  \caption{Results of the comparison with baselines.}
    \begin{tabular}{lrr}
		\toprule
		  & Precision & Recall \\
    \midrule
		  RIn-Close & 1 & 1 \\
		  CC & - & 0 \\
			FLOC & 0.0768 (0.0241) & 0.0874 (0.0252) \\
			ROCC & 0.1831 (0.0355) & 0.3845 (0.0655) \\
    \bottomrule
    \end{tabular}
  \label{tab:comparisonBaselines}
\end{table}

\linespread{1.5}

As we have seen, although we knew how to choose good parameters for the heuristic-based algorithms, this case study was very challenging to them. On the other hand, RIn-Close easily accomplishes this task.

\section{Conclusion}
\label{sec:conclusion}

Biclustering is a very powerful data mining technique that overcomes several drawbacks of the well-known clustering technique. Due to its complexity, most of the proposed biclustering algorithms are heuristic-based. Nonetheless, there are several algorithms able to perform ($i$) efficient, ($ii$) complete, ($iii$) correct, and ($iv$) non-redundant enumeration of all maximal CTV biclusters of ones from a binary data matrix. These enumerative algorithms proved to be very useful and have been applied in various application domains. Nonetheless, the raw data matrix admits integer and/or real values in several other application domains, and to transform it into binary data leads to loss of information. Hence, there are some proposals capable of dealing directly with numerical data matrices, but none of these algorithms to enumerate CVC, CVR, or CHV biclusters keeps these four properties. 

In this paper, we proposed a family of algorithms, called RIn-Close, capable of preserving these four properties when enumerating perfect CVC (or CVR) biclusters, perturbed CVC (or CVR) biclusters, and perfect CHV biclusters, and capable of preserving the last three of these properties when enumerating perturbed CHV biclusters. As far as we know, our algorithms are the most complete from the literature.

Our experimental results provided a valuable insight into the scalability of RIn-Close and its sensitivity to the user-defined measure of similarity and minimum number of rows allowed in a bicluster. As the larger the dataset, the greater tends to be the number of biclusters, these parameters are critical to feasibility of the biclustering solution. We also showed that well-known heuristic-based algorithms can fail spectacularly when trying to identify the existing biclusters in a simple and controlled scenario, thus pointing to the necessity of having efficient enumerative biclustering algorithms.

In future works, we are planing to extend the proposal to handle data matrices with missing values, and to enumerate biclusters with coherent evolutions. We also intend to investigate the extension of RIn-Close to enumerate only the top $k$ biclusters in terms of volume, thus reducing RIn-Close computational cost. Still looking at the reduction of their computational cost, we will also implement parallelized versions of RIn-Close algorithms.

\section*{Acknowledgments}

R. Veroneze and F. J. Von Zuben would like to thank CAPES and CNPq for the financial support. A. Banerjee acknowledges support of NSF grants IIS-1447566, IIS-1422557, CCF-1451986, CNS-1314560, IIS-0953274, IIS-1029711, IIS-0916750,  and NASA grant NNX12AQ39A.

\section*{References}
\bibliography{tese}

\vfill

\textbf{Rosana Veroneze} is a PhD candidate in Electrical and Computer Engineering at the University of Campinas (Unicamp). Her research interests includes data mining and machine learning areas.\\

\textbf{Arindam Banerjee} is an Associate Professor at the Department of Computer and Engineering and a Resident Fellow at the Institute on the Environment at the University of Minnesota, Twin Cities. His research interests are in machine learning, data mining, convex analysis and optimization, and their applications in complex real-world problems.\\

\textbf{Fernando J. Von Zuben} is an Associate Professor at the Department of Computer Engineering and Industrial Automation, School of Electrical and Computer Engineering, University of Campinas (Unicamp). The main topics of his research are computational intelligence, bioinspired computing, multivariate data analysis, and machine learning.

\end{document}